\documentclass[]{aa}  

\usepackage{tabularx}
\usepackage[figuresright]{rotating}
\usepackage{txfonts}
\usepackage{graphicx}
\usepackage{nicefrac,xfrac}
\usepackage{upgreek}
\graphicspath{{images/}{figs/}}
\usepackage[breaklinks=true]{hyperref}
\usepackage{mathrsfs}
\usepackage{tikz}

\begin{document} 

   \title{Diffuse radio emission from non-Planck galaxy clusters in the LoTSS-DR2 fields}
	\author{D. N. Hoang\inst{\ref{Hamburg}}
	\and
	 M. Br\"uggen \inst{\ref{Hamburg}} 
    \and
    A. Botteon \inst{\ref{Leiden}}	 
    \and
    T. W. Shimwell \inst{\ref{ASTRON},\ref{Leiden}}          
    \and
    X. Zhang \inst{\ref{Leiden},\ref{SRON}}
    \and
    A. Bonafede \inst{\ref{Bologna},\ref{INAF}} 
    \and
     L. Bruno \inst{\ref{INAF},\ref{Bologna}}
    \and
    E. Bonnassieux \inst{\ref{Bologna}}
    \and
    R. Cassano \inst{\ref{INAF}} 
    \and
    V. Cuciti \inst{\ref{Hamburg},\ref{INAF}}
    \and
    A. Drabent \inst{\ref{TLS}}
    \and
    F. de Gasperin \inst{\ref{Hamburg},\ref{INAF}}
    \and
    F. Gastaldello \inst{\ref{IASF}}
    \and
    G. Di Gennaro \inst{\ref{Hamburg}}    
    \and
    M. Hoeft \inst{\ref{TLS}}
    \and    
    A. Jones \inst{\ref{Hamburg}}
    \and
    G. V. Pignataro \inst{\ref{Bologna},\ref{INAF}}
     \and
     H. J. A. R\"ottgering \inst{\ref{Leiden}}
    \and
    A. Simionescu \inst{\ref{SRON},\ref{Leiden},\ref{Kavli}}
    \and
    R. J. van Weeren \inst{\ref{Leiden}}
	}
	\authorrunning{D. N. Hoang et al.}
	\titlerunning{Diffuse radio emission from non-Planck galaxy clusters in LoTSS-DR2}
	%
	\institute{Hamburger Sternwarte, Universit\"at Hamburg, Gojenbergsweg 112, 21029 Hamburg, Germany\label{Hamburg}
	\and
	Leiden Observatory, Leiden University, PO Box 9513, NL-2300 RA Leiden, The Netherlands \label{Leiden}
    \and
    Netherlands Institute for Radio Astronomy (ASTRON), P.O. Box 2, 7990 AA Dwingeloo, The Netherlands \label{ASTRON}
	\and
    SRON Netherlands Institute for Space Research, Niels Bohrweg 4, 2333 CA Leiden, The Netherlands \label{SRON}
    \and
    Dipartimento di Fisica e Astronomia, Universit\"a di Bologna, via Gobetti 93/2, 40122 Bologna, Italy \label{Bologna}
    \and
    INAF - Istituto di Radioastronomia di Bologna, Via Gobetti 101, I-40129 Bologna, Italy \label{INAF}
    \and
    Th\"uringer Landessternwarte, Sternwarte 5, D-07778 Tautenburg, Germany \label{TLS}    
    \and
    INAF - IASF Milano, via A. Corti 12, I-20133 Milano, Italy \label{IASF}    
    \and
    Kavli Institute for the Physics and Mathematics of the Universe (WPI), The University of Tokyo, Kashiwa, Chiba 277-8583, Japan \label{Kavli}
	}

   \date{Received: January 13, 2022; accepted: May 31, 2022}

  \abstract
  {The presence of large-scale magnetic fields and ultra-relativistic electrons in the intra-cluster medium (ICM) is confirmed through the detection of diffuse radio synchrotron sources, so-called radio halos and relics. Due to their steep-spectrum nature, these sources are rarely detected at frequencies above a few GHz, especially in low-mass systems.}
   {The aim of this study is to discover and characterise diffuse radio sources in low-mass galaxy clusters in order to understand their origin and their scaling with host cluster properties.}
   {We searched for cluster-scale radio emission from low-mass galaxy clusters in the Low Frequency Array (LOFAR) Two-metre Sky Survey - Data Release 2 (LoTSS-DR2) fields. We made use of existing optical (Abell, DESI, WHL) and X-ray (comPRASS, MCXC) catalogues.
   The LoTSS-DR2 data were processed further to improve the quality of the images that are used to detect and characterize diffuse sources.}
   {We have detected diffuse radio emission in 28 galaxy clusters. The number of confirmed (candidates) halos and relics are six (seven) and 10 (three), respectively. Among these, 11 halos and 10 relics, including candidates, are newly discovered by LOFAR. Beside these, five diffuse sources are detected in tailed radio galaxies and are probably associated with mergers during the formation of the host clusters. We are unable to classify other 13 diffuse sources. We compare our newly detected, diffuse sources to known sources by placing them on the scaling relation between the radio power and the mass of the host clusters.} 
   {}


   \maketitle
%

\section{Introduction}
\label{sec:intro}

To date large-scale ($\sim$Mpc) radio emission has been primarily detected in dynamically disturbed, massive clusters of galaxies. These synchrotron  sources trace the distribution of relativistic electrons and magnetic fields on cluster scales. The short lifetime of the radio-emitting relativistic electrons ($10-100$ Myrs in a typical cluster environment) implies that they are locally \mbox{(re-)accelerated} throughout the intra-cluster medium (ICM) volume.  Depending on location, size, morphology and spectro-polarimetric properties, the diffuse radio sources in clusters are often classified as halos or relics. The different classes have been linked to different origins of the wide-scale and in-situ particle acceleration \cite[][for a review]{VanWeeren2019b}.

Radio halos are Mpc-scale, steep spectrum\footnote{The radio spectrum is defined as $S\propto \nu^\alpha$, where $S$ is the source flux density at the frequency $\nu$ and $\alpha$ is the spectral index.} ($\alpha\lesssim-1$), unpolarised sources with a morphology that approximately follows thermal X-ray emission from the ICM. Clusters hosting radio halos are predominantly dynamically-disturbed systems, indicating a connection between the formation of halos and the dynamical state  \citep[e.g.][]{Cassano2010}. The observed fraction of clusters with radio halos increases with cluster mass and is as high as 70 percent for the most massive systems ($M_{500}>8\times10^{14}M_{\odot}$; \citealt[][]{Cuciti2015,Cuciti2021}). Approximately 30 percent of radio halos are found in X-ray luminous ($L_{0.1-2.4\,{\rm keV}} > 5\times10^{44}\,{\rm erg\,s^{-1}}$), nearby ($0.2<z<0.4$) clusters \citep{Venturi2008a,Kale2013}. The synchrotron power of radio halos correlates with several other cluster properties, such as the cluster mass, the X-ray luminosity and the Sunyaev-Zeldovich (SZ) signal \cite[e.g.][]{Liang2000,Brunetti2009,Basu2012a,Cassano2013a,Sommer2014,Cuciti2021}.

Two main models have been proposed to explain the formation of radio halos: a turbulent \mbox{re-acceleration} scenario and models based on secondary electrons \citep[see ][for a review]{Brunetti2014}. In the turbulent \mbox{re-acceleration} scenario, radio halos trace turbulent regions in the ICM where radio-emitting particles are \mbox{re-accelerated} by multi-scale turbulence during cluster-merger events \citep{Brunetti2001,Petrosian2001a,Brunetti2007a}. In the \emph{secondary} model, relativistic electrons and gamma rays are generated as secondary products of inelastic collisions between relativistic protons and thermal ions that permeate the ICM. The \mbox{re-acceleration} model is favoured by observational evidence, including the detection of ultra-steep spectrum radio halos \citep[USSRH; ][]{Brunetti2008}, the presence of radio halos only in merging clusters \citep[e.g.][]{Cassano2013a}, and the absence of gamma rays from the Coma cluster \citep[][]{Brunetti2012,Ackermann2016b, Brunetti2017}. A combination of the two processes where turbulence \mbox{re-accelerates}, both, \emph{primary} and \emph{secondary} electrons can still be possible as this hybrid model can produce fainter gamma-rays emission that is below the detection limit of current observations \citep{Brunetti2004,Brunetti2011b,Zandanel2014,Pinzke2017a}.


Radio relics are Mpc-sized, elongated, steep spectrum ($\alpha\lesssim-1$) sources that are found in the peripheral regions of merging clusters. Some relics are highly polarised  \citep[up to 70  percent; e.g.][]{VanWeeren2010a,DiGennaro2021,Rajpurohit2021b,deGasperin2022a}. Observational evidence including X-ray surface brightness (SB) discontinuity and/or temperature jump at the location of radio relics supports the connection between the formation of relics and merger shocks \citep[e.g.][]{VanWeeren2016b,Pearce2017}. This connection is further supported by (\emph{i}) the presence of polarised radio emission at shock fronts implying the alignment of the shock-compressed magnetic fields and (\emph{ii}) the observed steepening of radio spectral index in the regions behind the shocks due to the synchrotron and inverse-Compton (IC) losses. However, the origin of the relativistic particles emitting the synchrotron emission in relics has not been fully understood. For instance, it is unclear whether the relativistic particles are accelerated via Fermi-I diffusive shock acceleration from the thermal pool \citep[DSA; ][]{EnBlin1998,Roettiger1999a} or are \mbox{re-accelerated} via a DSA-like process from a mildly relativistic population of pre-existing relativistic electrons \citep{Markevitch2005,Kang2011a,Kang2012}.

The diffuse radio sources in clusters (i.e. halos, relics) are characterized by steep ($\alpha\lesssim-1$) synchrotron spectra, which make them brighter at low frequencies. The LOFAR Two-metre Sky Survey \citep[LoTSS;][]{Shimwell2017} is a survey of the entire northern sky and aims at producing radio maps at $\sim$6$\arcsec$ resolutions between 120 and 168 MHz with $\sim$100~$\upmu{\rm Jy\,beam^{-1}}$ sensitivity noise level. This survey has enabled the discovery of new diffuse radio sources in clusters as well as the study of large samples of sources. Using the LoTSS first data release \citep[LoTSS-DR1;][]{Shimwell2019}, \cite{vanWeeren2021} studied the diffuse emission in 41 galaxy clusters located in the 424~deg$^2$ of the HETDEX Spring region. More recently, \cite{Botteon2022a} analysed all the galaxy clusters reported in the second catalogue of SZ detected sources \citep[PSZ2;][]{Planck2016} within the 5634 deg$^2$ area covered by the LoTSS - Second Data Release \citep[LoTSS-DR2;][]{Shimwell2022a}. \cite{Botteon2022a} found that among the 309 galaxy clusters selected from the PSZ2 catalogue with $0.016 < z < 0.9$ and $1.1 \times 10^{14}M_\odot < M_{500} < 11.7 \times 10^{14} M_\odot$, 83 clusters host a radio halo and 26 clusters hosting one or more radio relics (including candidates). These two studies have focused on the PSZ2 clusters, which effectively constitute  a mass-selected sample that opens the possibility to perform statistical studies (Cassano et al.; Cuciti et al.; Jones et al., in prep.).

In this paper, we present our search for radio diffuse emission in galaxy clusters that are missed by these studies \citep[i.e.][]{vanWeeren2021,Botteon2022a} in non-PSZ2 clusters in the LoTSS-DR2 area. As the SZ signal is proportional to the mass of the ICM gas \citep[e.g.][]{Bender2016}, the galaxy clusters that are not detected with the PSZ2 observations are likely to be low-mass systems. Cluster-scale radio sources in low-mass clusters are particularly interesting because theoretical models predict that the occurrence of the diffuse radio sources depends strongly on the mass of the host clusters \citep[e.g.][]{deGrasperin2014,Bruggen2020,Cassano2013a}. Thus, the presence of diffuse radio sources in low-mass clusters is crucial to constrain the models in this regime poorly explored by current observations.

Throughout this paper we adopt a flat $\Lambda$CMD (Lambda cold dark matter) cosmology with $H_0=70\,{\rm km\,s^{-1} \,Mpc^{-1}}$, $\Omega_{\rm m}=0.3$ and $\Omega_{\Lambda}=0.7$. The images are plotted in the J2000 coordinate system.

\section{The cluster sample}
\label{sec:sample}

We have visually inspected the images covered by the LoTSS-DR2 area spanning 5635 square degrees \citep{Shimwell2022a} to search for cluster-scale emission in the galaxy clusters that are detected with optical, X-ray, and SZ observations. To this end, we made use of optical (i.e. Abell: \citealt{Abell1989}; DESI: \citealt{Zou2021a}; WHL: \citealt{Wen2012,Wen2015}), X-ray (i.e. MCXC: \citealt{Piffaretti2011a}) and X-ray--SZ (i.e. comPRASS: \citealt{Tarrio2019}) cluster catalogues. In Table~\ref{tab:sample}, we present a list of 28 galaxy clusters hosting cluster-scale radio emission in the LoTSS-DR2 images. Throughout the paper, we assume that the centres of the galaxy clusters are the coordinates given in the respective catalogues, and they are reproduced in Table~\ref{tab:sample}.

We note that our sample by no means represents a complete sample of the non-PSZ2 clusters hosting cluster-scale radio emission in the LoTSS-DR2 fields. These clusters with diffuse radio emission span masses between $M_{500}=0.58\times 10^{14}M_\odot$ and $8.14\times 10^{14}M_\odot$, with a median of $2.83\times 10^{14}M_\odot$. Our study thus represents a first attempt to extend the sample of diffuse radio sources to clusters of lower masses (i.e. $\lesssim2\times 10^{14}M_\odot$). We note that the cluster masses $M_{500}$ presented in this paper are obtained from multi-wavelength (optical, X-ray, SZ) catalogues and are not derived from weak lensing analyses. Hence, $M_{500}$ here should be considered as mass proxies.

\begin{figure}
\centering
\includegraphics[width=1\columnwidth]{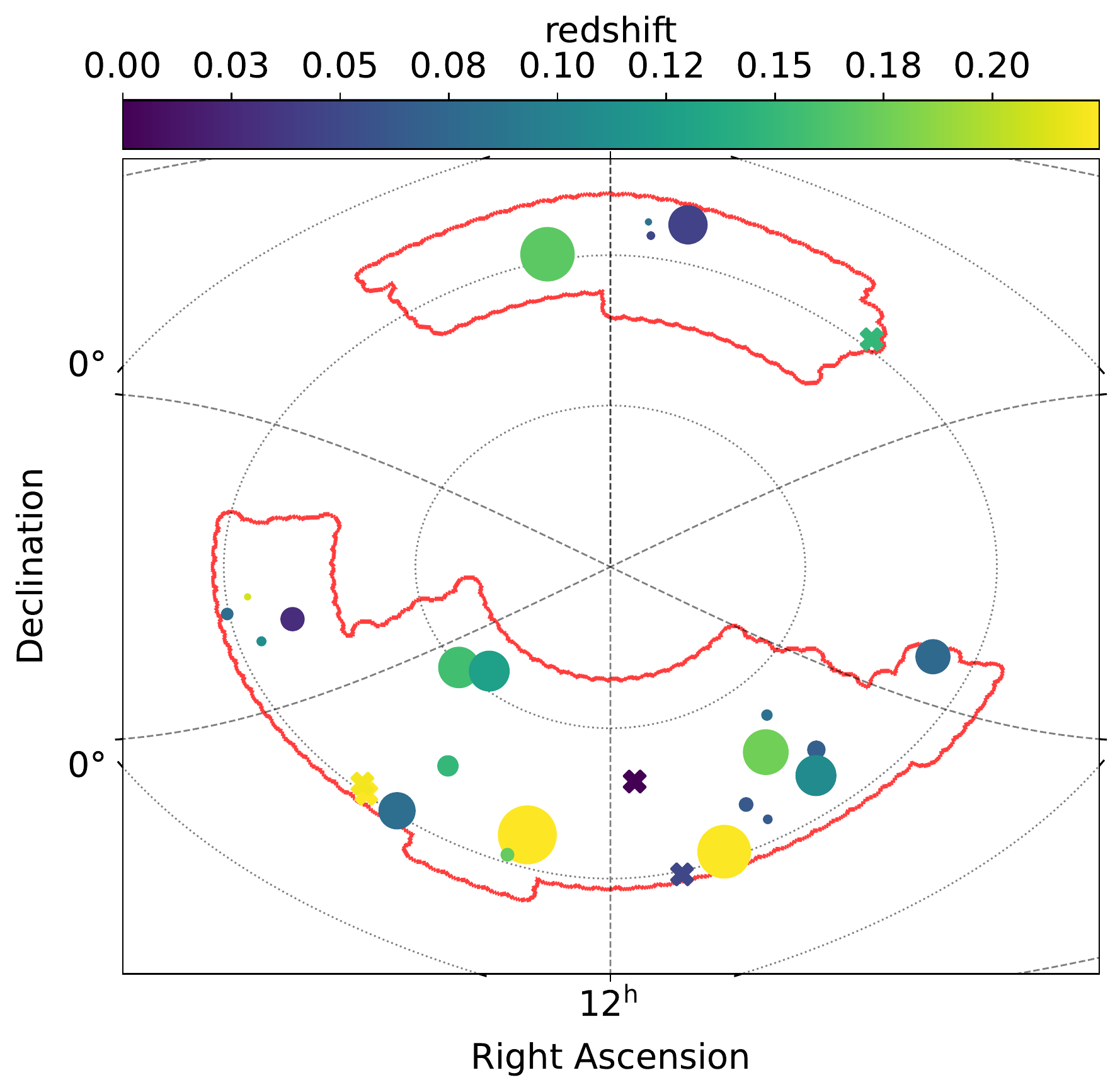}
  \caption{Location of the clusters in the LoTSS-DR2 fields shown with the red lines. The colour code presents the redshift of the clusters. The size of the points indicates the relative mass of the cluster, $M_{500}$. Clusters without $M_{500}$ are shown with crossed marks.
          }
     \label{fig:location}
\end{figure}

\begin{table*}[t]
	\centering
	\scriptsize
	\caption{A sample of  non-PSZ2 galaxy clusters that host diffuse sources in the LoTSS-DR2 fields.}
	\begin{tabular}{llcccccccccc}  
\hline\hline \\
No. & Name & RA    & DEC   & $z$ & $R_{500}$ & $M_{500}$ & $M_{500}^{\rm scaled}$ & Scale & X-ray & Ref. \\
  &    & [deg] & [deg] &   &  [Mpc] & $[\times10^{14}M_\odot]$   & $[\times10^{14}M_\odot]$ &  [kpc/\arcmin]  &  & \\ \hline
1 & Abell~84 & 10.4794 & 21.3771 & 0.102 & $0.96\pm0.02$ & $2.88\pm0.43$ & $2.86\pm0.41$ &  114 & R &  A89, W15 \\
2 & Abell~373 & 40.8613 & 27.9950 & 0.149 & -- & -- &-- & 157 & R & A89, G00 \\
3 & Abell~1213 & 169.1211 & 29.2603 & 0.048 & -- & -- & -- & 57 & X & A89, W98 \\
4 & Abell~1330 & 174.5984 & 49.5231 & 0.278 & 1.13 & 5.80$^b$ & 5.54 & 253 & -- & A89, W15 \\
5 & Abell~1889 & 214.1619 & 30.7358 & 0.185 & $0.91\pm0.05$ & $2.58\pm0.39$ & $2.58\pm0.37$ & 189 & C$^a$, R & A89, W15 \\
6 & Abell~1943 & 219.4128 & 30.2341 & 0.336 & 1.31 & 8.70$^b$ & 8.14 & 289 & -- & A89, W15 \\
7 & Abell~1963 & 221.2036 & 31.4731 & 0.210 & 1.10 & 4.40$^b$ & 4.27 & 206 & -- & S99, W15  \\
8 & DESI~201 & 213.3731 & 43.9794 & 0.344 & $0.57\pm0.04$ & $0.77\pm0.16$ & $0.82\pm0.16$ & 297 & -- & Z21 \\
9 & DESI~296 & 138.2320 & 41.9373 & 0.155 & $0.55\pm0.04$ & $0.56\pm0.12$ & $0.61\pm0.12$ & 163 & X & Z21 \\
10 & MCXC~J0928.6+3747 & 142.1570 & 37.7990 & 0.247 & $0.95$ & $3.15$ & $3.90$  & 236 & R &  P11 \\
11 & MCXC~J0943.1+4659 & 145.7790 & 46.9980 & 0.406 & $0.97$ & $3.97$ &  $4.75$ & 330 & X & P11 \\
12 & MCXC~J1020.5+3922 & 155.1270 & 39.3800 & 0.143 & $0.47$ & $0.33$ & $0.58$ & 152 & -- & P11 \\
13 & MCXC~J1711.0+3941 & 257.7620 & 39.6940 & 0.065 & $0.70$ & $1.02$ & $1.50$ & 76 & R & P11 \\
14 & PSZRX~G095.27+48.27 & 230.8676 & 59.8819 & 0.362 & -- & $3.23$ & $3.34$ & 307 & -- & T19 \\
15 & PSZRX~G100.21-30.38 & 350.5707 & 28.5264 & 0.388 & -- & $5.71$ & $5.77$ & 320 & X & T19 \\
16 & PSZRX~G102.17+48.88 & 223.7294 & 62.9749 & 0.292 & -- & $3.13$ & $3.25$ & 266 & -- & T19 \\
17 & PSZRX~G116.06+80.14 & 194.3289 & 36.9139 & 0.518 & -- & $6.74$ & $6.76$ & 378 & X$^a$, R & T19 \\
18 & PSZRX~G181.53+21.43 & 110.3160 & 36.7426 & 0.175 & -- & $2.28$ & $2.40$ & 180 & R & T19 \\
19 & PSZRX~G195.91+62.83 & 162.0091 & 31.6423 & 0.515 & -- & $5.55$ & $5.62$ & 378 & -- & T19 \\
20 & WHL~J002056.4+221752 & 5.2352 & 22.2977 & 0.199 & $0.60$ & $0.86^b$ & $0.96$ & 200 & R & W15 \\
21 & WHL~J002311.7+251510 & 5.7987 & 25.2527 & 0.109 & $0.75$ & $1.38^b$ & $1.46$ & 121 & -- & W15 \\
22 & WHL~J085608.5+541855 & 134.0355 & 54.3152 & 0.251 & -- & $2.93^b$ & $2.80$ & 238 & R & W15 \\
23 & WHL~J091721.4+524607 & 139.3391 & 52.7685 & 0.192 & -- & $4.10^b$ & $3.76$ & 194 & -- & W15 \\
24 & WHL~J101350.8+344251 & 153.4616 & 34.7142 & 0.146 & -- & $2.56^b$ & $2.49$ & 155 & -- & W15 \\
25 & WHL~J130503.5+314255 & 196.2644 & 31.7153 & 0.396 & $0.90$ & $3.52^b$ & $3.29$ & 325 & -- & W15 \\
26 & WHL~J165540.4+334422 & 253.9183 & 33.7394 & 0.255 & $0.72$ & $1.25^b$ & $1.33$ & 241 & -- & W15 \\
27 & WHL~J172125.4+294144 & 260.3557 & 29.6957 & 0.184 & -- & $2.32^b$ & $2.29$ & 187 & -- & W15 \\
28 & WHL~J173424.0+332526 & 263.5999 & 33.4239 & 0.485 & $0.60$ & $1.26^b$ & $1.34$ & 366 & -- & W15 \\

\hline \\
		\end{tabular}
	\\Notes: Col. 1: cluster number in the sample. Col. 2: cluster name. Col. 3: Right Ascension (J2000). Col. 4: Declination (J2000). Col. 5: cluster redshift. Col. W15: the radius enclosing the cluster volume at which the mean over-density is 500 times the critical density of the Universe at the cluster redshift. Col. 7: cluster mass within $R_{500}$. Col. 8: cluster mass that is scaled to that of the PSZ2 clusters (see Sec.~\ref{sec:mass}). Col. 9: angular scale at the cluster redshift. Col. 10: available X-ray data. Available X-ray data is denoted as C for Chandra, X for XMM-Newton, and R for ROSAT. Col. 11: reference. A89: \cite{Abell1989}, G00: \cite{Gal2000}, P11: \cite{Piffaretti2011a}, S99: \cite{Struble1999}, T19: \cite{Tarrio2019}, W98: \cite{Wu1998a}, W15: \cite{Wen2015}, Z21: \cite{Zou2021a},. $^a$: proprietary data. $^b$: $M_{500}$ is derived using the $M_{500}$-richness scaling relation \citep[Eq. 17 in][]{Wen2015}.
	\label{tab:sample}
\end{table*}
	

\section{Data reduction}
\label{sec:reduction}

\subsection{Calibration}
\label{sec:cal}

LoTSS datasets for each pointings are calibrated for direction-independent (DI) and direction-dependent (DD) effects using the automatic standard pipelines $\mathtt{PREFACTOR}$\footnote{\url{https://github.com/lofar- astron/prefactor}} \citep{VanWeeren2016a,Williams2016,DeGasperin2019} and $\mathtt{ddf-pipeline}$\footnote{\url{https://github.com/mhardcastle/ddf-pipeline}} \citep{Tasse2021}, respectively. The pipelines are developed for the calibration of the Surveys Key Science Projects (SKSP) data. For details of the data processing of the LoTSS-DR2 data, we refer to \cite{Shimwell2022a}.

To improve the image fidelity towards the target direction, we follow the ''extraction'' and self-calibration procedure that post processes the archived $\mathtt{ddf-pipeline}$ products and is described in \cite{vanWeeren2021,Botteon2022a}. To maximise the signal-to-noise of the detected sources, we include all data from LoTSS-DR2 pointings that are centred within $2.2^\circ$ from the target cluster. Within this radius, the sensitivity of the primary beam is at least 50 percent of that at the pointing centre. The extracted regions covering the targets are typically $0.3-0.7\,{\rm deg^2}$ in size and within this region the primary beam response is assumed to be constant. Finally, multiple self-calibration loops including phase and phase-amplitude are performed. For details, we refer to \cite{vanWeeren2021,Botteon2022a}.

\subsection{Imaging and source subtraction}
\label{sec:imaging}

In order to search for cluster-scale emission, we follow the imaging procedure for the PSZ2 clusters in \cite{Botteon2022a} to obtain images of our targets at several different angular resolutions using  \texttt{WSClean}\footnote{\url{https://gitlab.com/aroffringa/wsclean}} (version 2.10.0). To alter the resolution we image with a Briggs' weighting robust of $-0.5$ without and with a tapering of outer baselines (i.e. \texttt{taper-gaussian}) with two-dimensional Gaussian functions of $10\arcsec$, $15\arcsec$, $30\arcsec$, and $60\arcsec$ FWHM (full-width at half maximum). For our datasets, this typically corresponds to resolutions of $8\arcsec$, $19\arcsec$, $25\arcsec$, $38\arcsec$,  and $74\arcsec$, respectively. The \texttt{uv} data below 80~$k\lambda$ are not used in the imaging as they generally more contaminated with radio frequency interference (RFI) and we do not require the very large-scale emission that they give sensitivity to. For instance, these short baselines map diffuse emission with angular scales of larger than $53\arcmin$ that corresponds to physical scales that are larger than $11$~Mpc at $z=0.19$ (i.e. the redshift median for our sample) or 3~Mpc for the nearest cluster of galaxies (i.e. Abell~1213 at $z=0.048$) in the sample. To accurately deconvolve faint diffuse emission across the full bandwidth, the deconvolution is performed with the multiscale with multifrequency cleaning algorithm. We split the bandwidth of 48 MHz into 6 narrower sub-bands which are jointly deconvolved (i.e. \texttt{join-channels} and \texttt{channels-out=6}) and we deconvolve with scales (i.e. \texttt{multiscale-scales}) of $[0,4,8,16,32,64]$ in pixel unit, where the zero corresponds to a delta function for modelling of point sources.

Diffuse radio emission from the ICM is often contaminated by the emission from individual radio galaxies that are located in the clusters or along the line of sight. We remove the contamination from the discrete sources by subtracting their models from the \textit{uv} data. These models are created by imaging with a Briggs' weighting robust parameter of $-0.5$ and by cutting the shortest baselines in order to remove large-scale cluster emission that extends beyond physical scales of 250~kpc. The \textit{uv}-data are then used to make images at multiple resolutions with identical settings to those made prior to source subtraction.

\subsection{Flux scale correction}
\label{sec:fluxscale}

The flux scale of the LoTSS-DR2 \textit{uv}-data is calibrated according to \cite{Scaife2012} using the bookended observations of 3C radio sources (e.g. 3C~196, 3C~295). The calibration does not take into account errors of the LOFAR beam model. As a result, radio sources across the LOFAR wide-field images have non-uniform uncertainties in their flux densities, depending on their relative locations with respect to the centres of the grid pointings. Another source of uncertainty in the integrated flux densities of radio sources comes from phase calibration errors. The phase errors scatter flux and can systematically impact the measured values. To assess and partially correct these, \cite{Shimwell2022a} align the flux scale of the LoTSS-DR2 mosaics and catalogue with the \cite{Roger1973} flux scale through a comparison with 6C and the NRAO Very Large Array Sky Survey \cite[NVSS; see also][]{Hardcastle2021}.

Following \cite{Botteon2022a}, we align the flux scale of our LOFAR images with the LoTSS-DR2 catalogue which gives us comparable errors to that catalogue. The scaling factors for each cluster field are obtained by comparing the integrated flux densities of the compact sources in our images (i.e. those with robust of $-0.5$ without tapering of outer baselines) with those in the LoTSS-DR2 catalogue. A linear relation between the flux densities is fitted using three techniques including Theil-Sen \citep{Sen1968}, Huber \citep{Huber1981}, and traditional linear regressions that are built in the \texttt{scikit-learn} package\footnote{\url{https://scikit-learn.org}}. The different fitting procedures are used because they have different outlier rejection criteria and for each field we use the fitting method that has the lowest mean absolute error. To align the flux scale, our LOFAR images are multiplied by the scaling factors. For our sample, the scaling factors range between 0.7 and 1.2, with the mean value of 1.0 and a standard deviation of 0.1. In this paper, we adopt a flux scale uncertainty of 10 percent which is consistent with variations in the LoTSS-DR2 flux density scale \citep{Shimwell2022a}.

\subsection{Flux density and power measurements}
\label{sec:flux}


We measure the flux density of diffuse cluster sources by integrating over those pixels that are above $2\sigma$. The region where the pixel values above 2 times the root mean square (RMS) of the local background noise ($\sigma$) and within a manually defined region that is shown with the cyan dashed lines in the corresponding figures (e.g. see Fig.~\ref{fig:abell84} for Abell~84). In cases where diffuse cluster emission is contaminated by the emission from discrete sources, we identify the contaminated regions by eye and manually mask out the emission, shown with black dashed lines. We account for this missing flux density in the masked regions by extrapolating the measured diffuse cluster emission flux density by the area ratio between the masked and unmasked ($>2\sigma$) regions. The uncertainty in the flux density is estimated as follows,
\begin{equation}
    \Delta S = \sqrt {N\times\sigma^2 + (f_{\rm err}\times S)^2 +(e_{\rm mask}\times S_{\rm mask})^2},
\end{equation}
where $N$ is the area of diffuse sources in beam units; $\sigma$ (Jy~beam$^{-1}$) is the RMS of the background noise; $f_{\rm err}$ is the flux scale uncertainty (i.e. assuming 10 percent); $S$ (Jy) is the integrated flux density of the diffuse sources; $e_{\rm mask}$ and $S_{\rm mask}$ are the uncertainty, assuming 20 percent, and the extrapolated flux density for the masked region. Another source of uncertainty is due to the missing of short baselines (i.e. below 80~k$\lambda$) that is extensively studied through the simulations for radio halos in Bruno et al. (in prep.). They found that the flux density recovered by the LoTSS observations is more than 90 percent for halos with an angular size of smaller than $10.5\arcmin$ (i.e. 2~Mpc at $z=0.2$). As the real size of the halos in our sample is unknown, the estimates of the flux density we report in this paper do not include this type of uncertainty.

For detected and candidate radio halos, we additionally estimate the flux densities by fitting their SB profile with 2D models using Halo-Flux Density CAlculator (\texttt{Halo-FDCA}) code\footnote{\url{https://github.com/JortBox/Halo-FDCA}} \citep{Boxelaar2021}. The SB profile of radio halos is modelled with an exponential function, i.e.:
\begin{equation}
    I(\textbf{r}) = I_0 e^{-G(\textbf{r})},
\end{equation}
where $I_0$ is the SB at the halo centre and $G(\textbf{r})$ is the function that defines the morphology of the halos (e.g. circular, elliptical, or skewed). For the explicit mathematical form of the morphological function $G(\textbf{r})$, we refer to \cite{Boxelaar2021}. Depending on the models, the number of free parameters vary. For instance, the circular model has four free parameters describing the central brightness ($I_0$), the location (two for RA and Dec.) and the extension ($e$-folding radius). The elliptical model has two additional free parameters for the extension and the orientation of the fitted ellipsoid. The skewed model has two more free parameters compared to the elliptical model. The \texttt{Halo-FDCA} code employs Markov Chain Monte Carlo (MCMC) algorithm to search for the best-fit parameters and their uncertainties. Once the best-fit parameters are found, the flux densities of the halos can be integrated out to a radius defined by user. In this study, we fit the SB of the detected/candidate halos with the circular, elliptical, and skewed models. The integrated flux density is calculated to a radius of three times the $e$-folding radius which has been commonly used in literature \citep[e.g.][]{Murgia2009,vanWeeren2021}. We found that the models result in consistent flux density estimates within $1\sigma$ for all detected or candidate halos in our sample that is also consistent with the analysis in \cite{Botteon2022a}. Hence, we only report the flux densities obtained from the elliptical model in Table \ref{tab:sources}. The best-fit parameters are summarized in Appendix~\ref{sec:bestfit}. 

The radio power of all diffuse cluster sources is estimated using the corresponding flux density estimates,
\begin{equation}
    P = 4\pi D_L^2\,K(z)\,S ,
\end{equation}
where $D_L$ is the luminosity distance to the target; $K(z)=(1+z)^{-(1+\alpha)}$ is the standard radio \emph{k}-correction term for a source at redshift $z$; here we assume a spectral index of $\alpha=-1.2$ that is commonly found for diffuse radio sources in galaxy clusters \citep[e.g.][]{Feretti2012a,VanWeeren2019b}.

\subsection{Ancillary data}
\label{sec:anc_data}

We make use of optical and X-ray archival data to assist with the identification of the diffuse radio sources and to characterise the dynamical status of the clusters. In the optical band, we make use of the \textit{g}, \textit{r}, and \textit{i} filter images from the Sloan Digital Sky Survey (SDSS) survey \citep{Alam2015}. In the X-ray band, we found that only 13 (46 percent) of the clusters where we detect radio emission have been observed and are usable in the X-rays with Chandra, XMM-Newton, or ROSAT All Sky Survey (RASS; see Table \ref{tab:sample}). The Chandra and XMM-Newton X-ray data are calibrated in a similar manner as done in \cite{Botteon2022a}.

\section{Results}
\label{sec:res}

In this section we present LOFAR 144~MHz images and diffuse emission classifications making use of the multi-wavelength data of the galaxy clusters listed in Table \ref{tab:sample}. The multiple-resolution images show the presence of diffuse emission at different scales. In clusters where diffuse emission is co-located in projection with discrete sources, we remove these contaminating sources (see Sec. \ref{sec:imaging}) and present the source-subtracted images that is indicated with "Subtracted" on the top right corner of the images. The classification and flux measurements for the diffuse sources in the cluster sample is presented in Table \ref{tab:sources}.

\subsection{Abell~84}
\label{sec:abell84}

\begin{figure*}[!ht]
	\centering
	\begin{tikzpicture}
		\draw (0, 0) node[inner sep=0] {\includegraphics[width=0.33\textwidth]{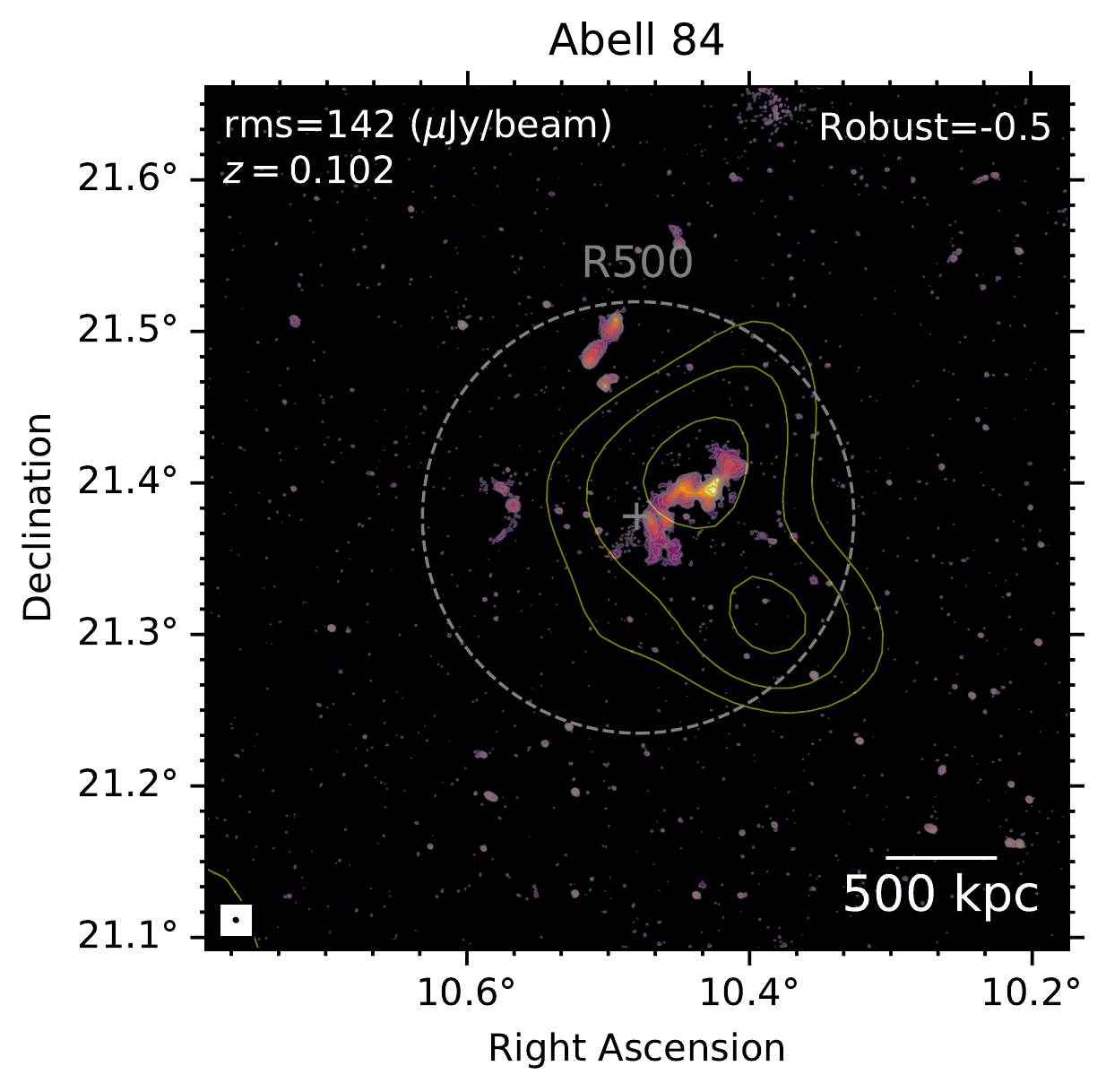}  \hfil
			\includegraphics[width=0.33\textwidth]{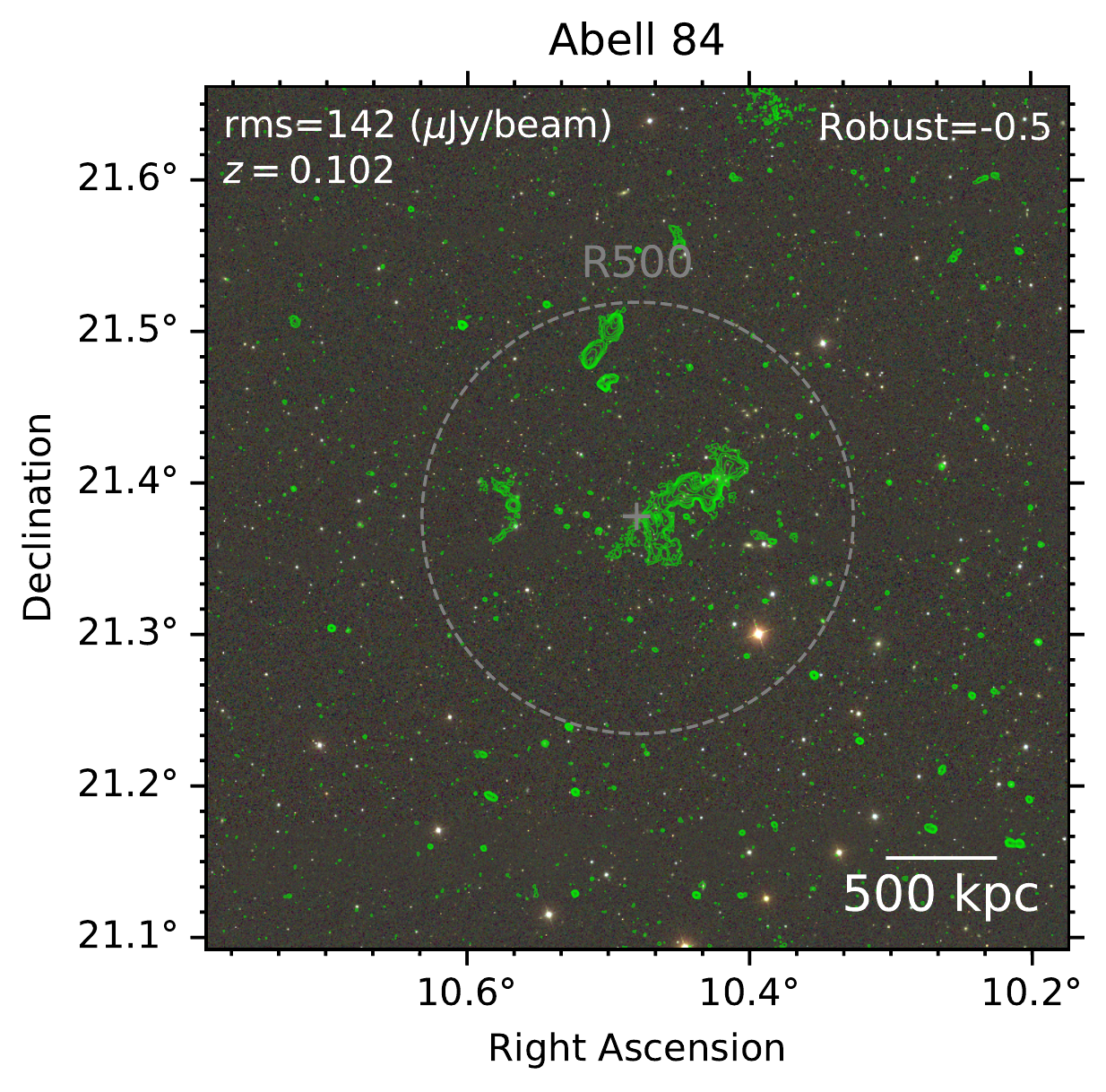}  \hfil
    			\includegraphics[width=0.33\textwidth]{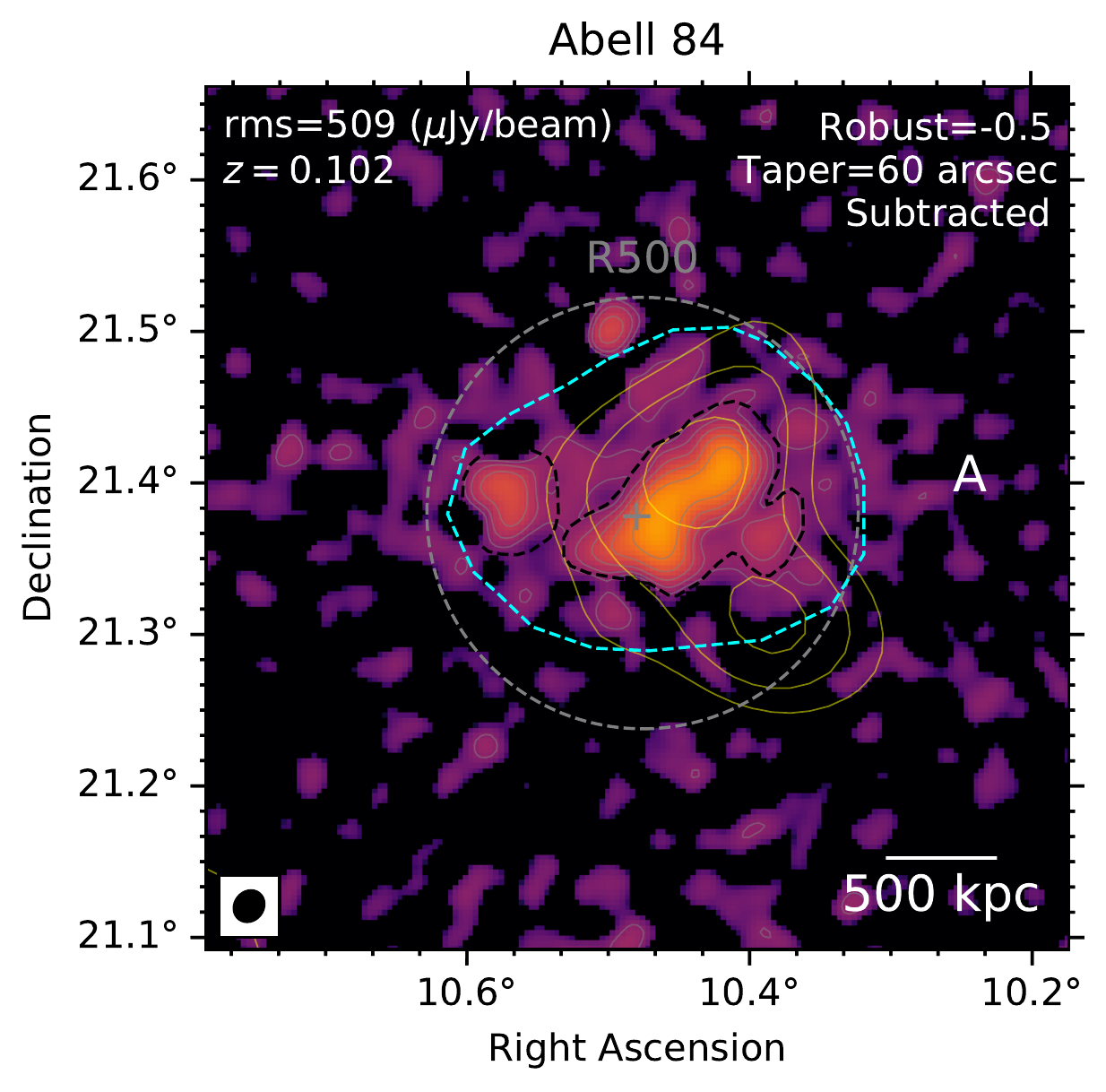}   };
		\draw (-7.7, -1.7) node {\color{white} (a)};
		\draw (-1.5, -1.7) node {\color{white} (b)};
		\draw (4.6, -1.55) node {\color{white} (c)};
	\end{tikzpicture}
	\caption{Abell~84. LOFAR~144~MHz images (a, c) at different resolutions shown in the bottom-left corners. The optical SDSS image (b) is overlaid with the LOFAR contours. In all images, the LOFAR first contour is drawn at $3\sigma$, where $\sigma$ is shown on the top-left corners. The next contours are multiplied by a factor of 2. In the panels (a, c), the X-ray yellow contour is drawn at $3\sigma$, and subsequently spaced by a factor of $\sqrt{2}$. In the right panel (c), the cyan dashed line shows the region where source flux density is measured. The black dashed line region is masked out. All pixels below $2\sigma$ are blanked. The cluster centre is marked with the grey cross ($+$) sign. The grey dashed circle has a radius of $R_{500}$.
	}
	\label{fig:abell84}
\end{figure*}

As seen in the LOFAR images in Fig.~\ref{fig:abell84}, multiple discrete radio sources are detected in the centre and outskirts (N and E) of the galaxy cluster Abell~84 ($z=0.102$). In Fig.~\ref{fig:abell84} (b), the central radio sources roughly follow the distribution of optical cluster galaxies that is in line with the cluster galaxy number map presented in \citep{Strazzullo2005}. In the low-resolution, source-subtracted image in Fig.~\ref{fig:abell84} (c), faint diffuse emission is newly detected at $2\sigma$ around the central discrete sources and roughly follows the main part of the ROSAT X-ray emission. The projected size of the diffuse source is 1500~kpc$\times$740~kpc with the major axis in the E-W direction. Based on the location and the extension of the diffuse source, we classify it as a candidate radio halo. We estimate the flux density of the candidate halo by (i) integrating the $>2\sigma$ pixels within the cyan region (Fig.~\ref{fig:abell84}, c) and (ii) fitting the SB with the elliptical model (see Sec.~\ref{sec:flux}). The resulting estimates are given in Table \ref{tab:sources}.

\subsection{Abell~373}
\label{sec:abell373}

\begin{figure*}[!ht]
		\centering
			\begin{tikzpicture}
				\draw (0, 0) node[inner sep=0] {\includegraphics[width=0.33\textwidth]{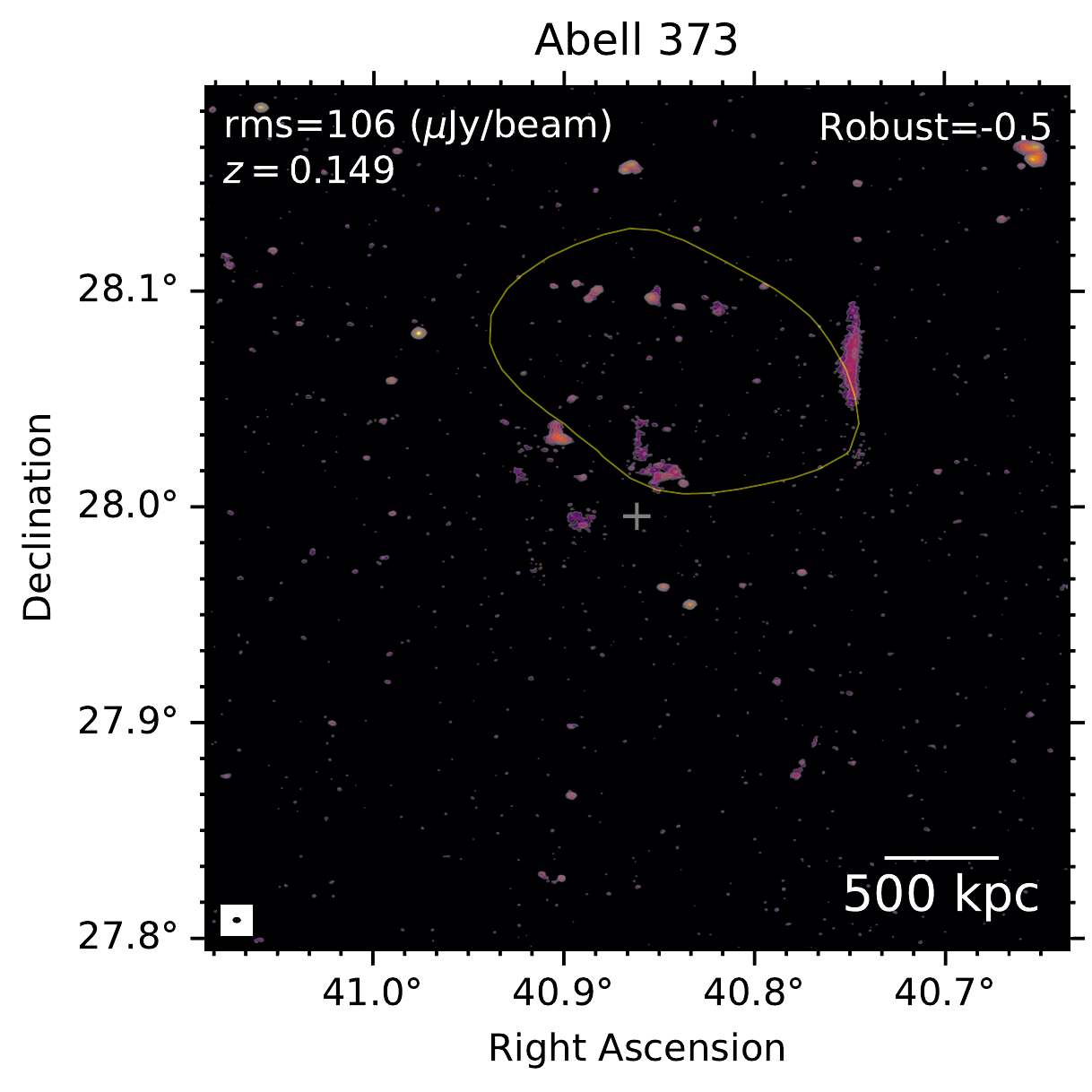}  \hfil
				\includegraphics[width=0.33\textwidth]{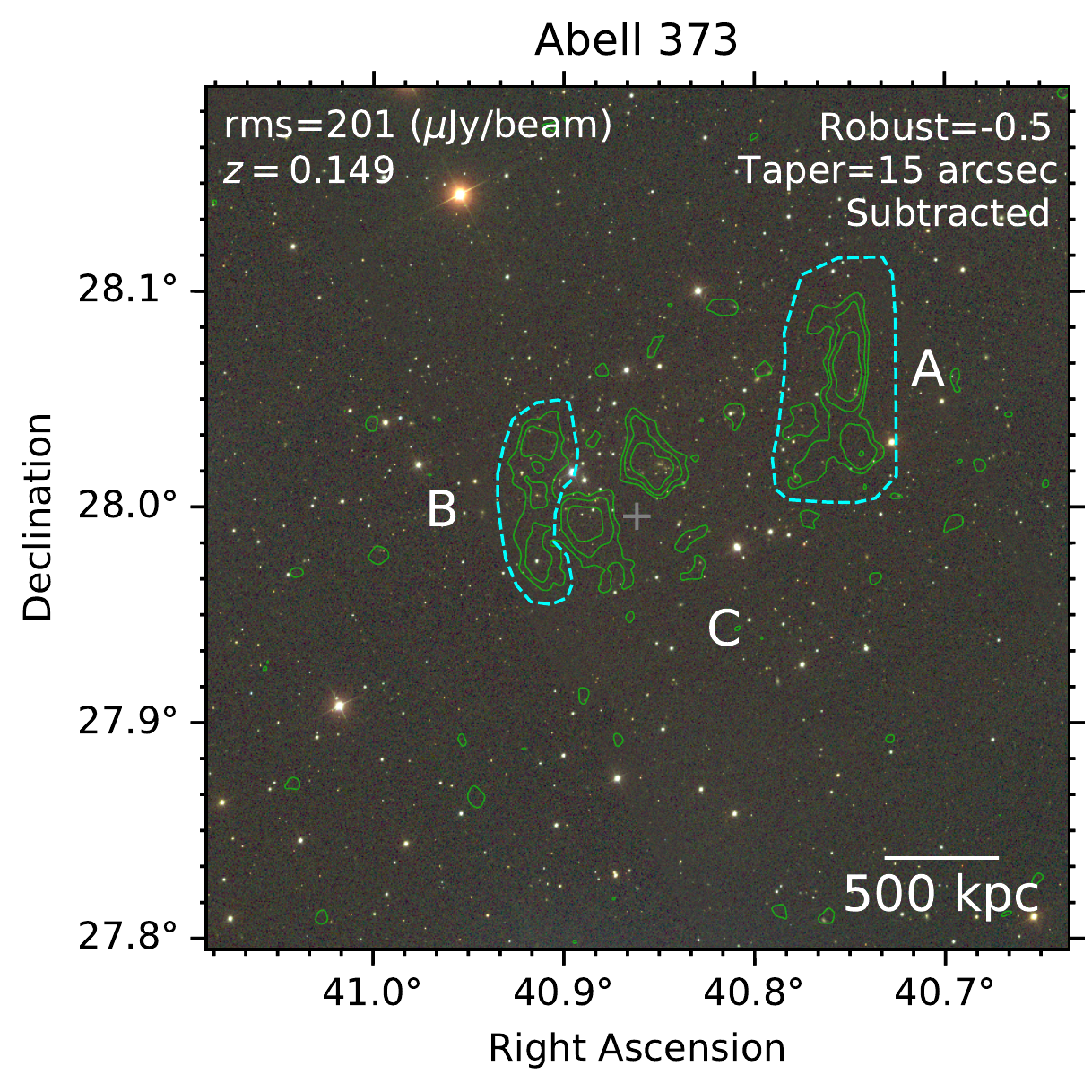}  \hfil
			    \includegraphics[width=0.33\textwidth]{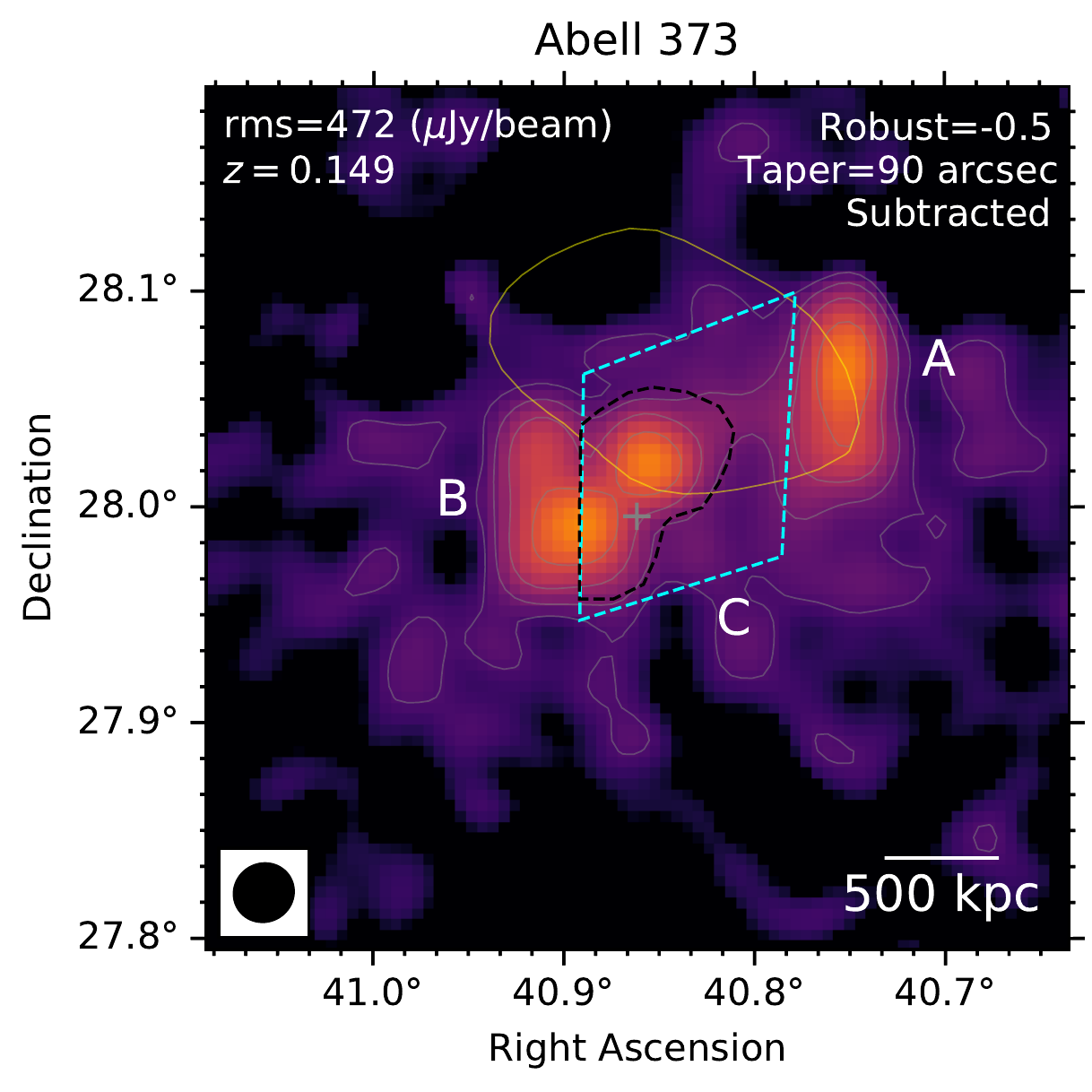}    };
        		\draw (-7.7, -1.7) node {\color{white} (a)};
        		\draw (-1.5, -1.7) node {\color{white} (b)};
        		\draw (4.6, -1.55) node {\color{white} (c)};
			\end{tikzpicture}
			\caption{Abell~373. Image description is the same as that in Fig.~\ref{fig:abell84}.
			}
			\label{fig:abell373}
\end{figure*}

In Fig.~\ref{fig:abell373}, LOFAR images show the new detection of diffuse radio sources, named A--C, in  Abell~373 ($z=0.149$). About 1.1~Mpc to the NW of the cluster centre (as reported in the Abell catalogue) a diffuse source A with a projected size of 770~kpc$\times$200~kpc elongated in the N-S direction is detected. The southern part of source A is only seen in the low-resolution images in Fig.~\ref{fig:abell373} (b, c). Source A has no optical counterpart in the SDSS image (see Fig.~\ref{fig:abell373}, b). On the eastern side of the cluster, at 1.5~Mpc from source A, an elongated source B is detected in the low-resolution image in Fig.~\ref{fig:abell373} (b). Source B has a similar projected size and orientation as those of A. The morphology and location of A and B suggest that they are radio relics. Fig.~\ref{fig:abell373} (c) shows the presence of a 800~kpc$\times$1000~kpc diffuse source C connecting sources A and B. Source C partly covers the ROSAT X-ray low-SNR (signal to noise ratio) emission. We classify source C as a radio halo. The 144 MHz flux densities of the sources A, B and C are given in Table~\ref{tab:sources}.

\subsection{Abell~1213}
\label{sec:abell1213}

\begin{figure*}[!ht]
	\centering
	\begin{tikzpicture}
		\draw (0, 0) node[inner sep=0] {\includegraphics[width=0.33\textwidth]{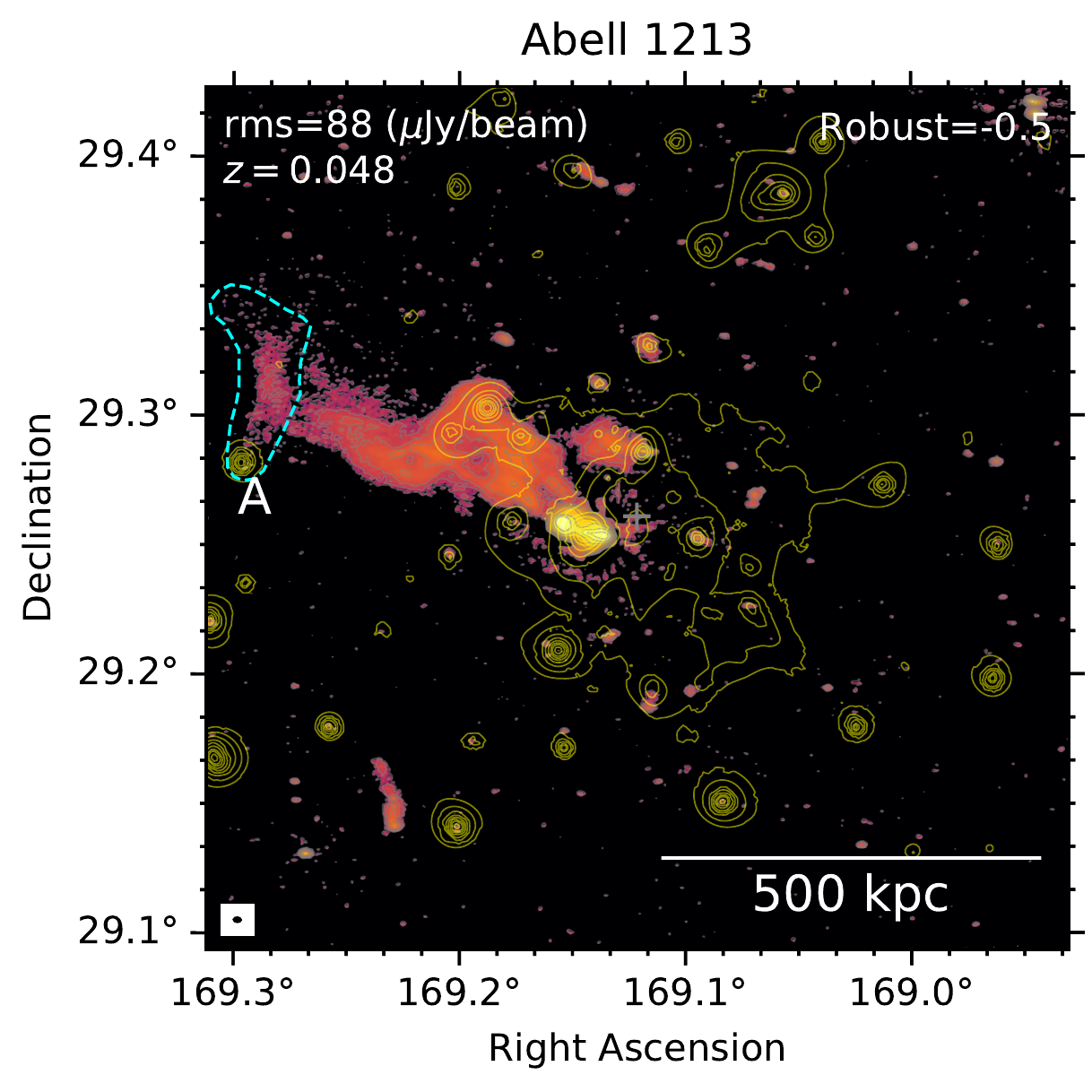}  \hfil
			\includegraphics[width=0.33\textwidth]{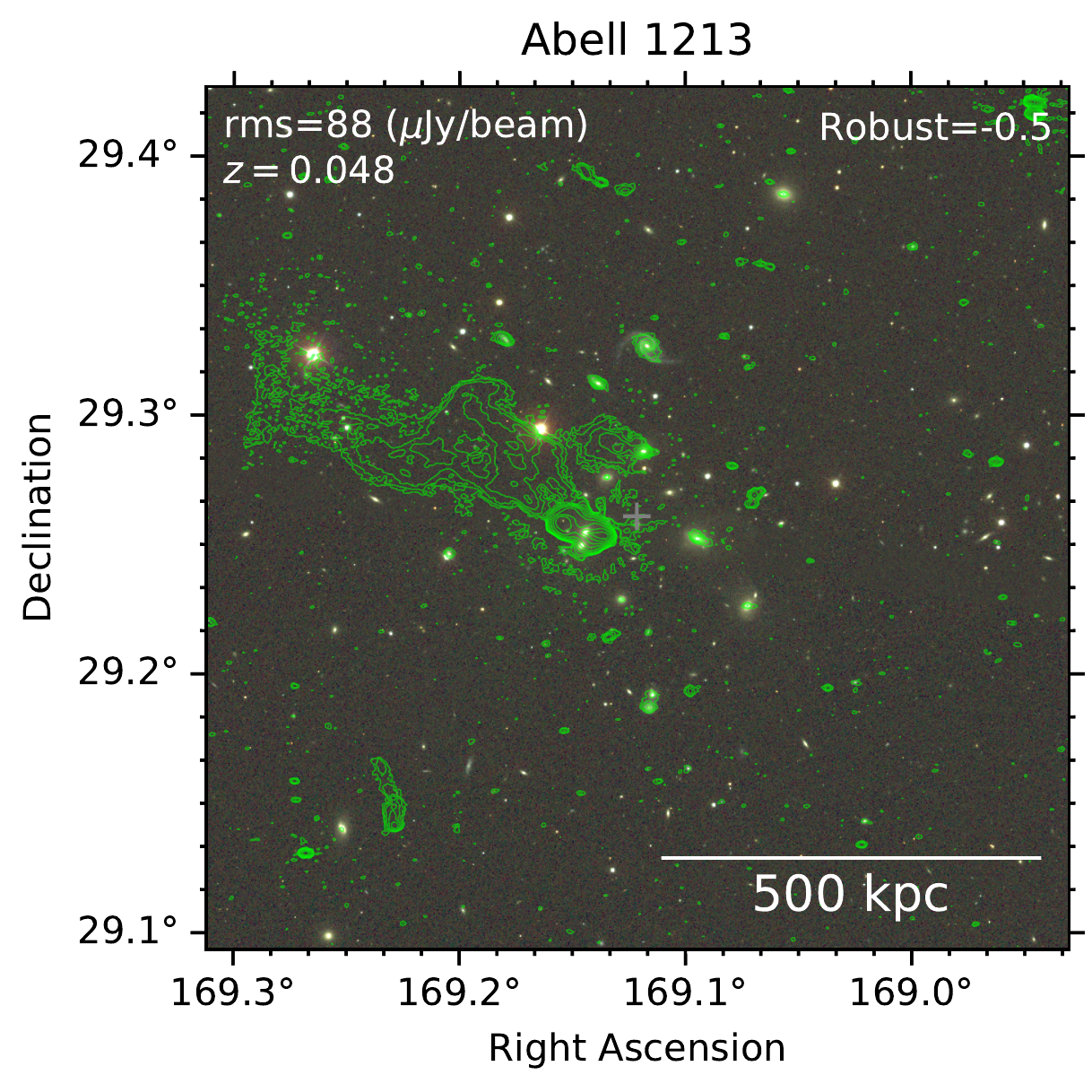}  \hfil
			\includegraphics[width=0.33\textwidth]{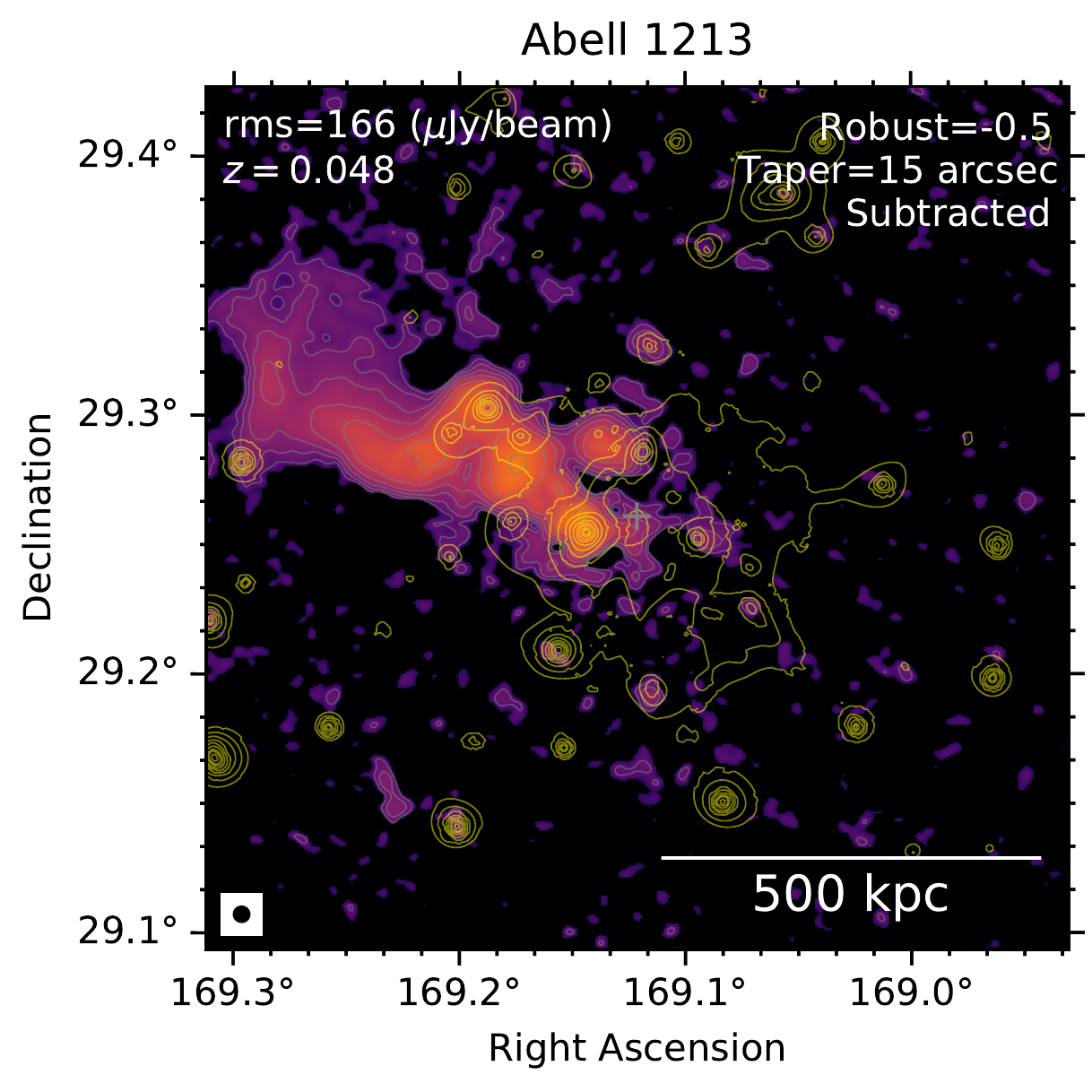}   };
		\draw (-7.7, -1.7) node {\color{white} (a)};
		\draw (-1.5, -1.7) node {\color{white} (b)};
		\draw (4.6, -1.7) node {\color{white} (c)};
	\end{tikzpicture}
	\caption{Abell~1213. Image description is the same as that in Fig.~\ref{fig:abell84}.
	}
	\label{fig:abell1213}
\end{figure*}

Abell~1213 ($z=0.048$, the lowest redshift in the sample) is a non-relax galaxy cluster hosting the brightest radio galaxy (BCG; 4C29.41) in the cluster centre \citep{Jones1999,Fanti1982,Ledlow2003,Giovannini2009}. VLA 1.4 GHz observations by \cite{Giovannini2009} show the detection of a diffuse source to the E direction of the BCG and they classified the source as a small-size radio halo. The power of the radio halo of Abell~1213 does not follow the correlation between radio halo power and X-ray luminosity \citep{Giovannini2011}.

In Fig.~\ref{fig:abell1213}, LOFAR images confirm the presence of the central radio galaxy and the diffuse emission to the eastern region. The diffuse emission is connected with the central radio galaxy; its major axis follows the distribution of optical cluster galaxies; and its SB does not follow the distribution of X-ray emission (see Fig. \ref{fig:abell1213}). These suggest that the diffuse emission is not a radio halo, but it is the tail of the central radio galaxy. We measure the projected size of the tailed galaxy to be 510~kpc. Interestingly, in the easternmost region of the tail galaxy excess diffuse emission, named A, is detected. There is no obvious counterparts of source A seen in the SDSS image in Fig. \ref{fig:abell1213} (b). The projected size of source A is 260~kpc$\times$80~kpc. Across source A, from the eastern direction, the SB increases rapidly towards the central region of the source, but the SB gradually decreases on the other side towards the cluster centre. The origin of the excess emission (source A) is still unknown. It could be associated with a merger that is occurring in the NE-SW direction as suggested by the disturbed morphology of the XMM-Newton X-ray emission and the distribution of the optical cluster galaxies in Fig. \ref{fig:abell1213}.

\subsection{Abell~1330}
\label{sec:abell1330}

\begin{figure*}[!ht]
	\centering
	\begin{tikzpicture}
		\draw (0, 0) node[inner sep=0] {\includegraphics[width=0.33\textwidth]{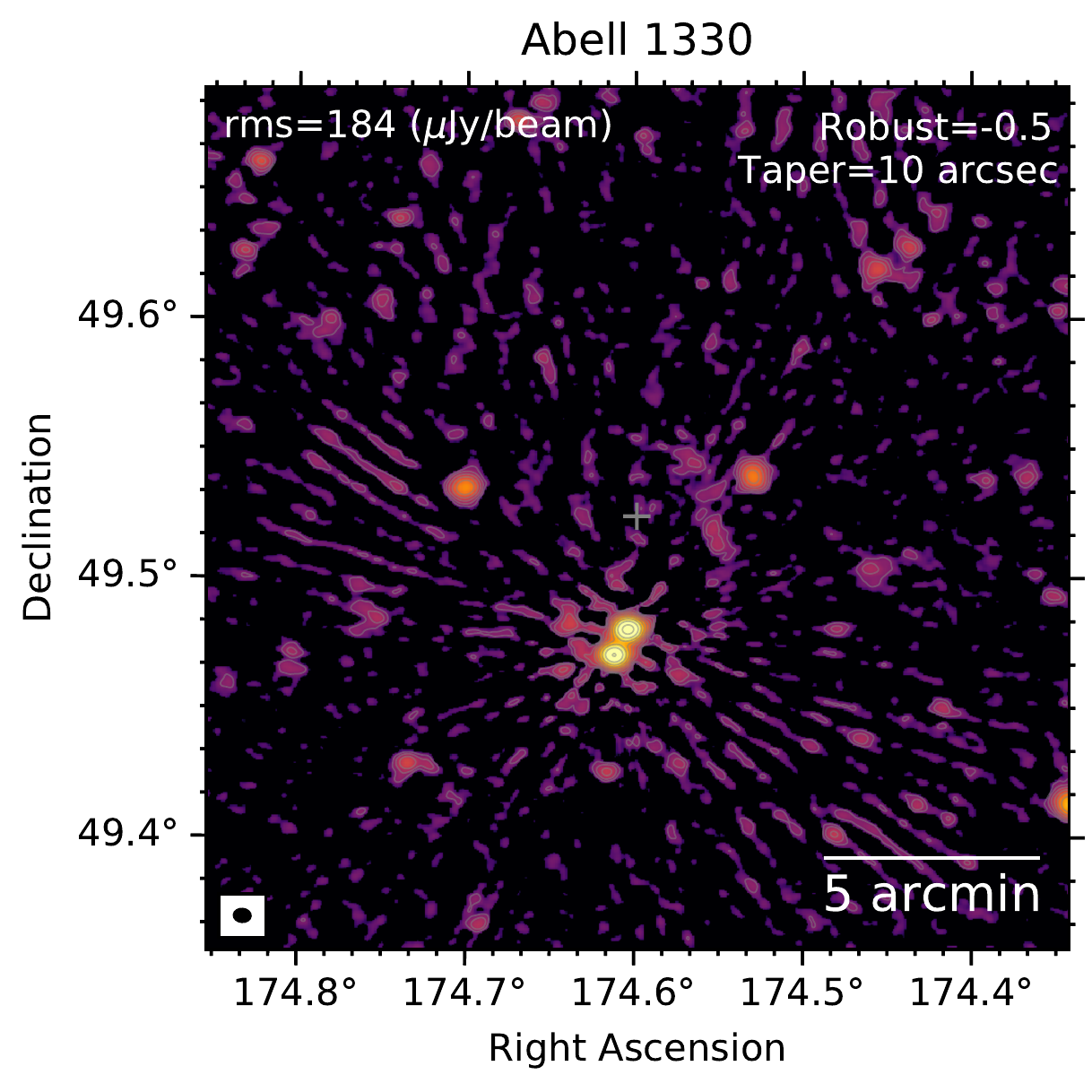}  \hfil
			\includegraphics[width=0.33\textwidth]{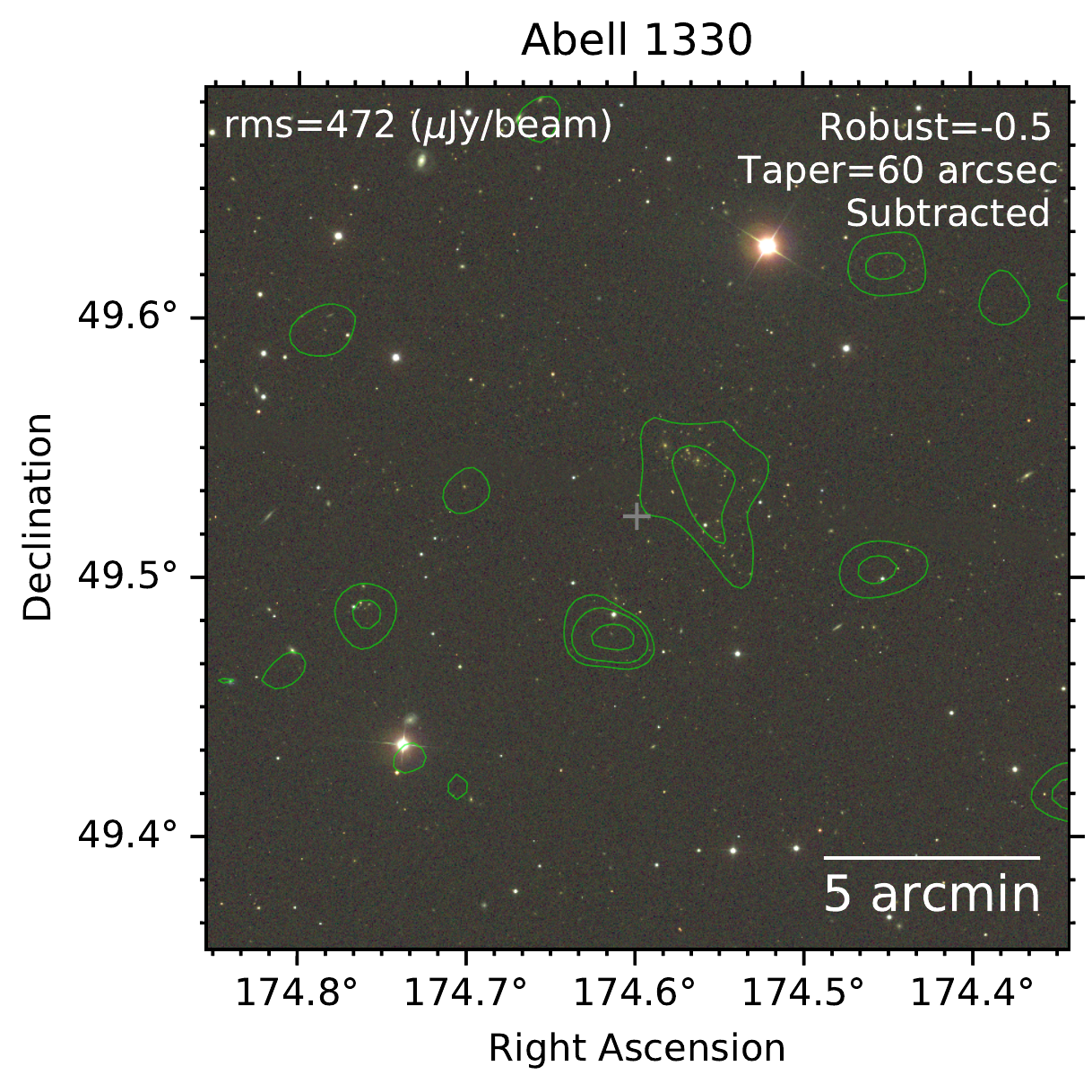}  \hfil
			\includegraphics[width=0.33\textwidth]{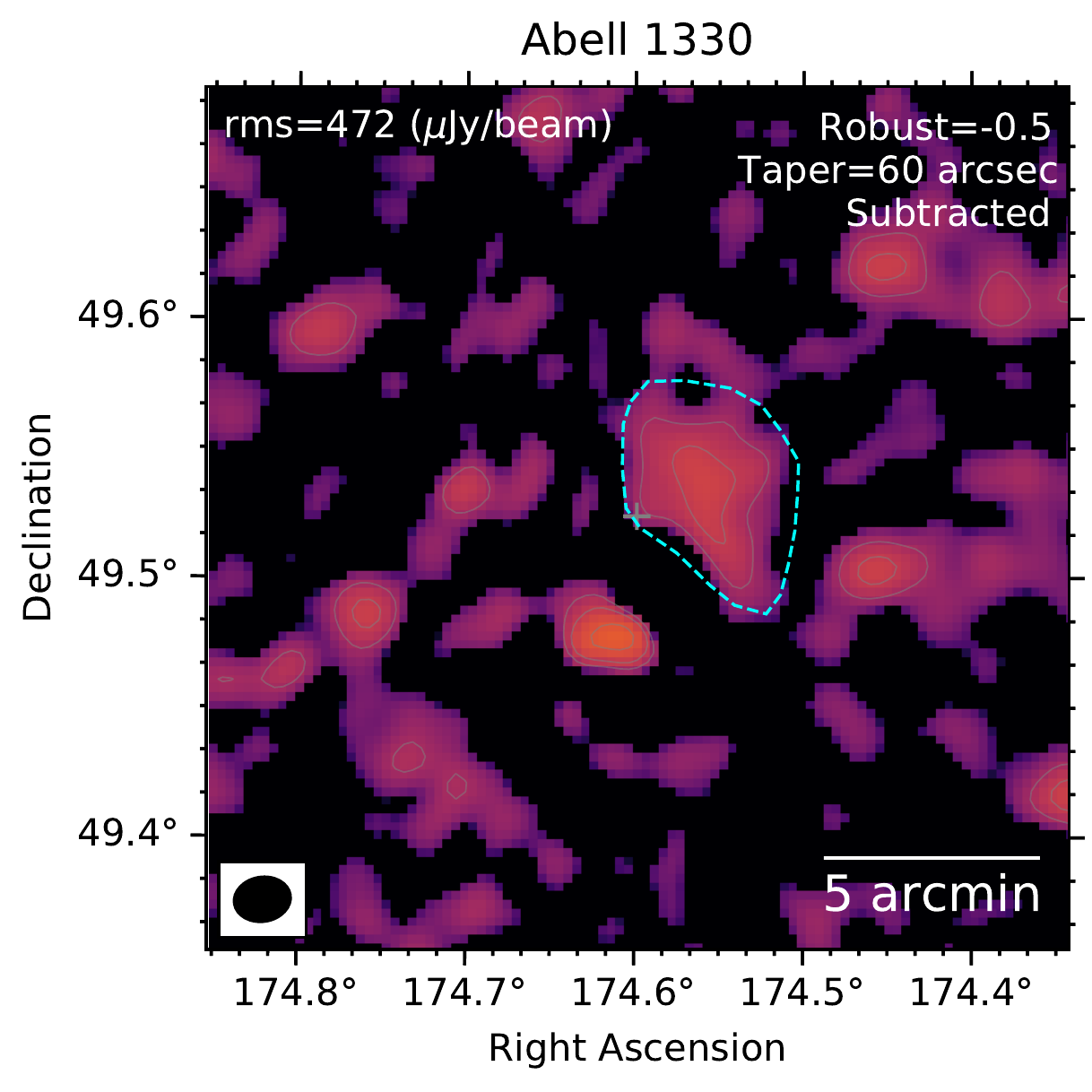}   };
		\draw (-7.7, -1.7) node {\color{white} (a)};
		\draw (-1.5, -1.7) node {\color{white} (b)};
		\draw (4.6, -1.7) node {\color{white} (c)};
	\end{tikzpicture}
	\caption{Abell~1330. Image description is the same as that in Fig.~\ref{fig:abell84}.
	}
	\label{fig:abell1330}
\end{figure*}

In Fig.~\ref{fig:abell1330} we present LOFAR images of Abell~1330. The high-resolution image in the left panel shows high-noise level around compact sources to the south of the cluster centre. The cluster has no redshift information reported in literature. In the panel (c), the source-subtracted low-resolution image shows the new detection of a diffuse radio source that is not visible in the high-resolution image in the left panel (a). The diffuse emission has a projected size of $3.5\arcmin$ (i.e. 700~kpc if $z=0.2$) and is located $1.5\arcmin$ in the northwestern direction from the centre (as reported in \cite{Abell1989}). Its SB follows the distribution of the galaxies in the SDSS image in the panel (b), and hence we classify the diffuse source as a candidate radio halo. The 144~MHz flux density of the diffuse source is given in Table~\ref{tab:sources}. Without redshift, we could not estimate the radio power of the source.

\subsection{Abell~1889}
\label{sec:abell1889}

\begin{figure*}[!ht]
	\centering
	\begin{tikzpicture}
		\draw (0, 0) node[inner sep=0] {\includegraphics[width=0.33\textwidth]{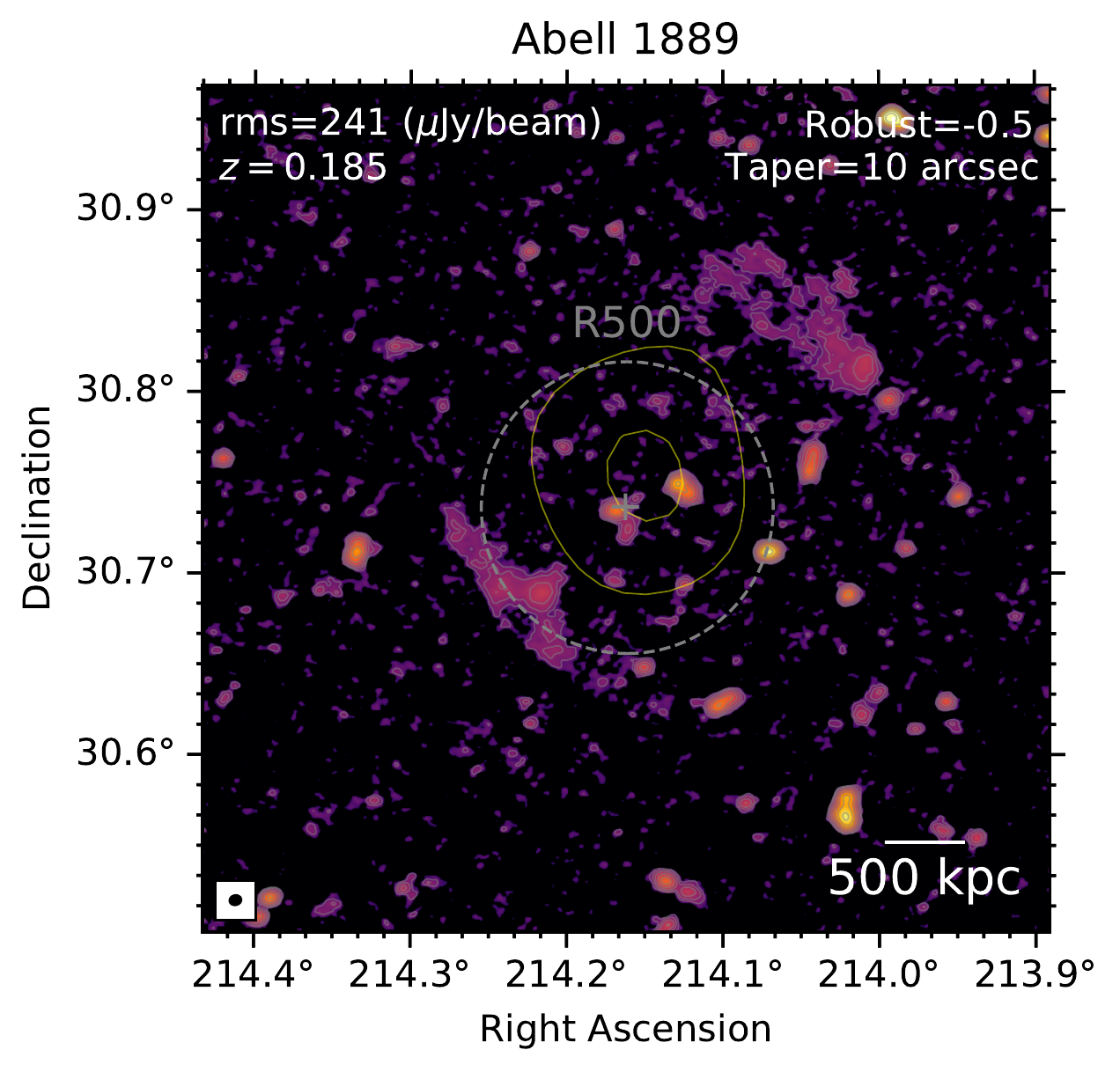}  \hfil
			\includegraphics[width=0.33\textwidth]{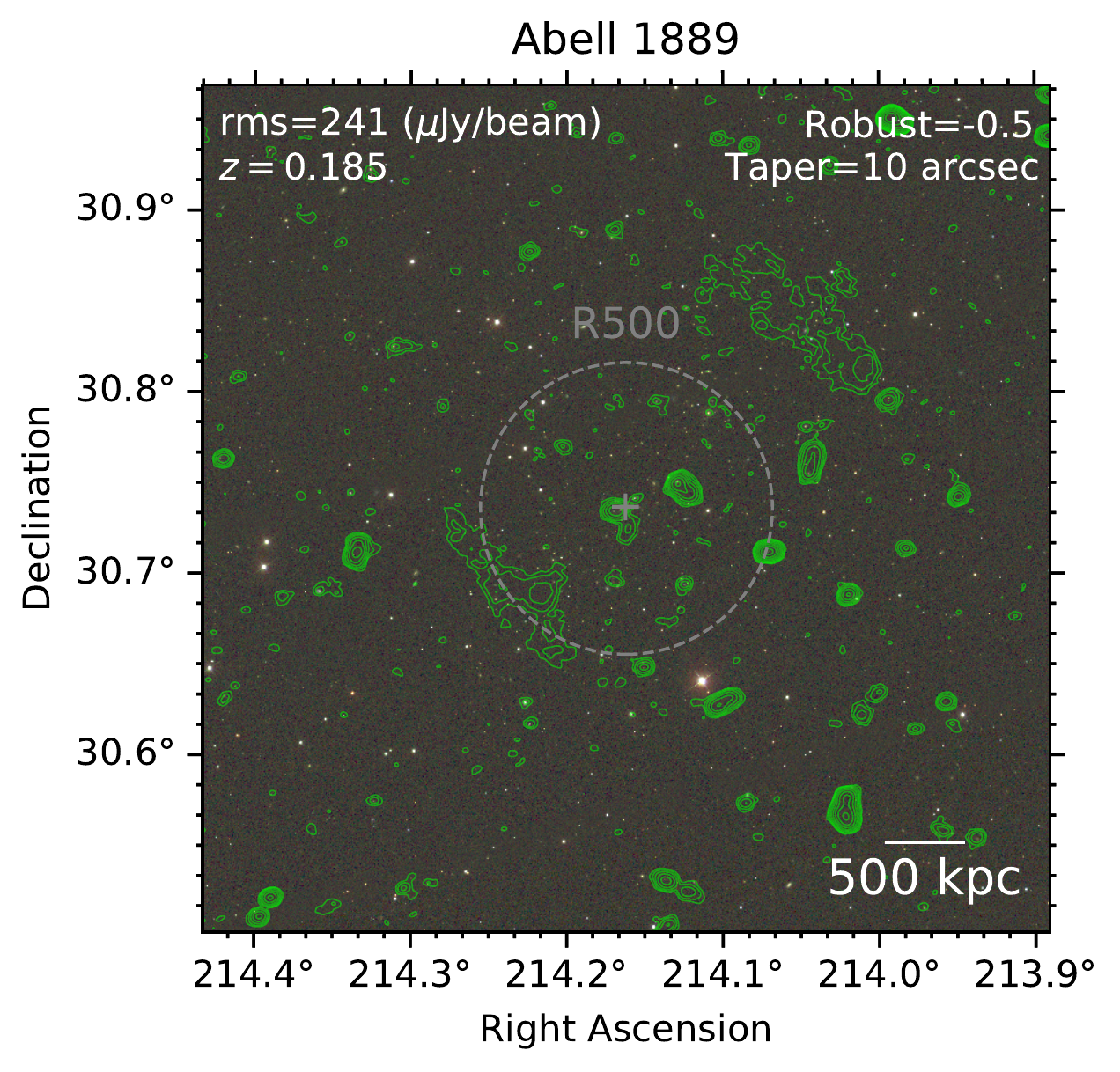}  \hfil
			\includegraphics[width=0.33\textwidth]{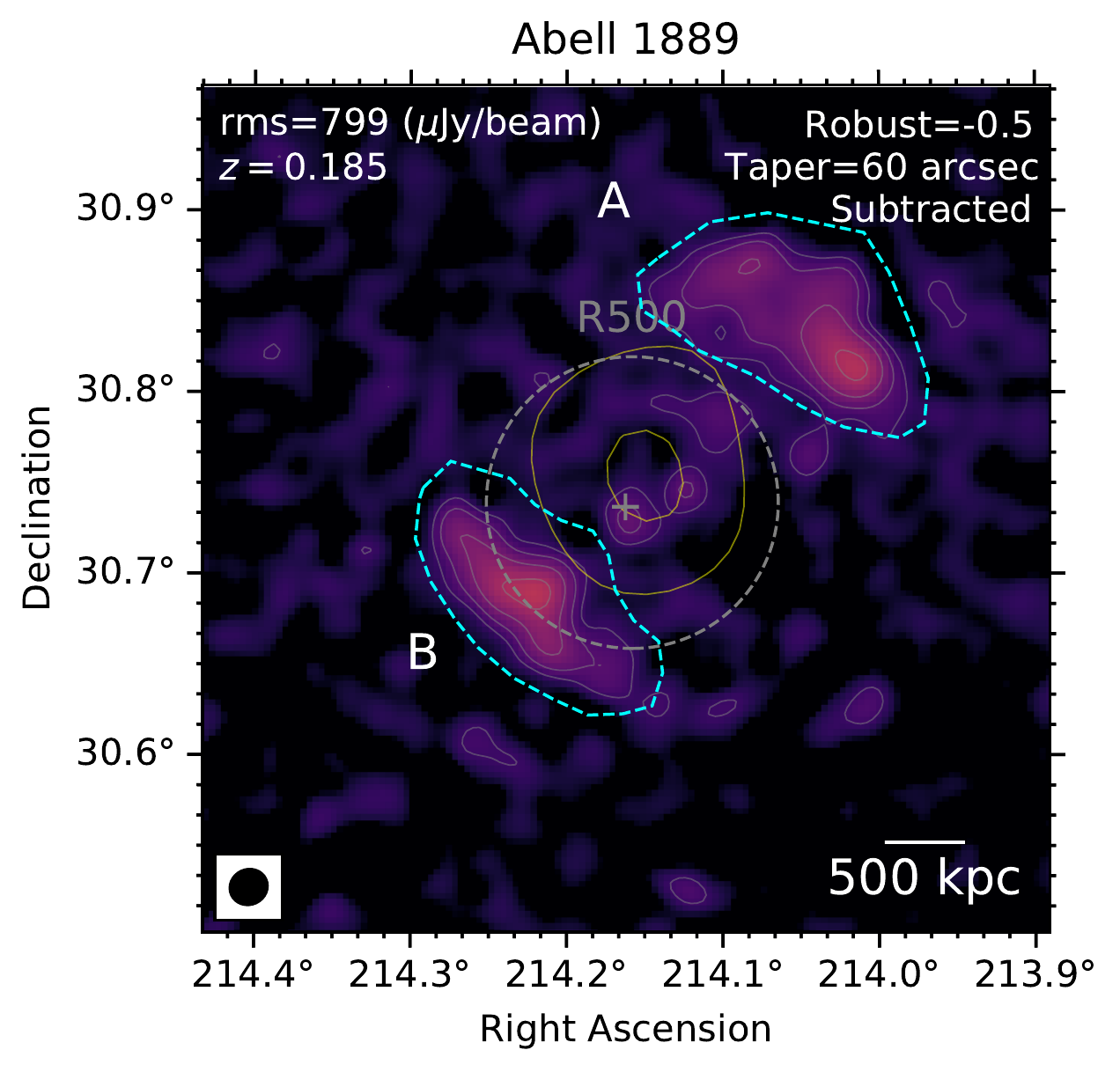}   };
		\draw (-7.7, -1.7) node {\color{white} (a)};
		\draw (-1.5, -1.7) node {\color{white} (b)};
		\draw (4.6, -1.55) node {\color{white} (c)};
	\end{tikzpicture}
	\caption{Abell~1889.  Image description is the same as that in Fig.~\ref{fig:abell84}.
	}
	\label{fig:abell1889}
\end{figure*}

LOFAR images of Abell~1889 ($z=0.185$) in Fig.~\ref{fig:abell1889} show the new detection of two diffuse sources, named A and B for the NW and SE sources, respectively. These sources are located on the boundaries of the ROSAT X-ray emission and are on the opposite sides (2.5~Mpc apart in projection) of the cluster centre. The projected sizes of these diffuse sources are roughly equal (i.e. 1650~kpc$\times$830~kpc and 1650~kpc$\times$570~kpc, respectively). The major axes of sources A and B are perpendicular to the line connecting the sources through the cluster centre. The brightness of source A increases along its length from NE to SW. The SB of source B is higher in the middle region. There is no clear optical counterparts for A and B seen in the SDSS optical image in the panel (b) of Fig.~\ref{fig:abell1889}. The morphology and location of sources A and B suggest that they are radio relics. We estimate the 144~MHz flux density and radio power of the relics and present them in Table~\ref{tab:sources}. In the region between the relics, we find small-scale diffuse emission behind the relic A (see the panel c), but we do not see large-scale emission connecting the relics where the ROSAT X-ray emission is detected. 

\subsection{Abell~1943}
\label{sec:abell1943}

\begin{figure*}[!ht]
	\centering
	\begin{tikzpicture}
		\draw (0, 0) node[inner sep=0] {\includegraphics[width=0.33\textwidth]{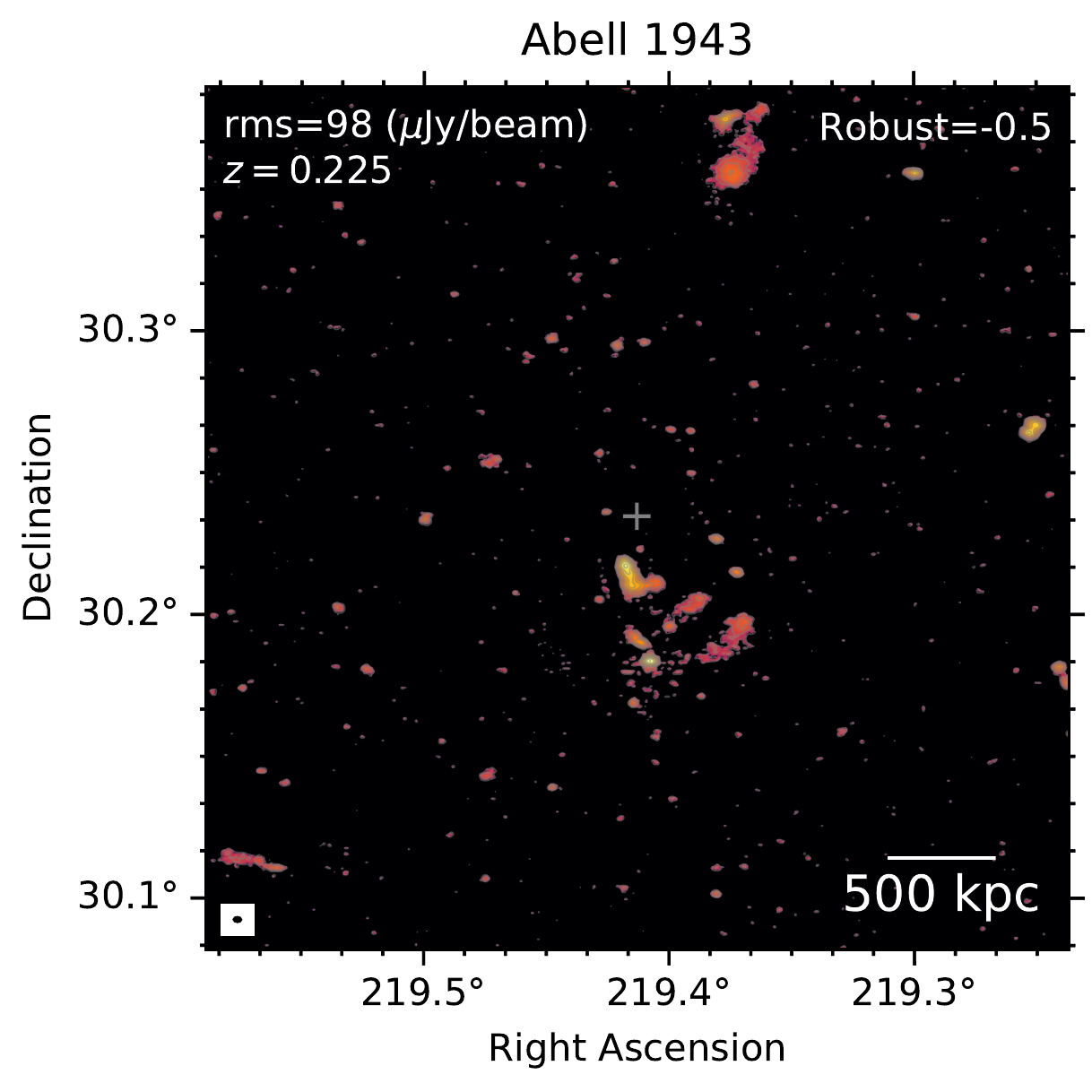}  \hfil
			\includegraphics[width=0.33\textwidth]{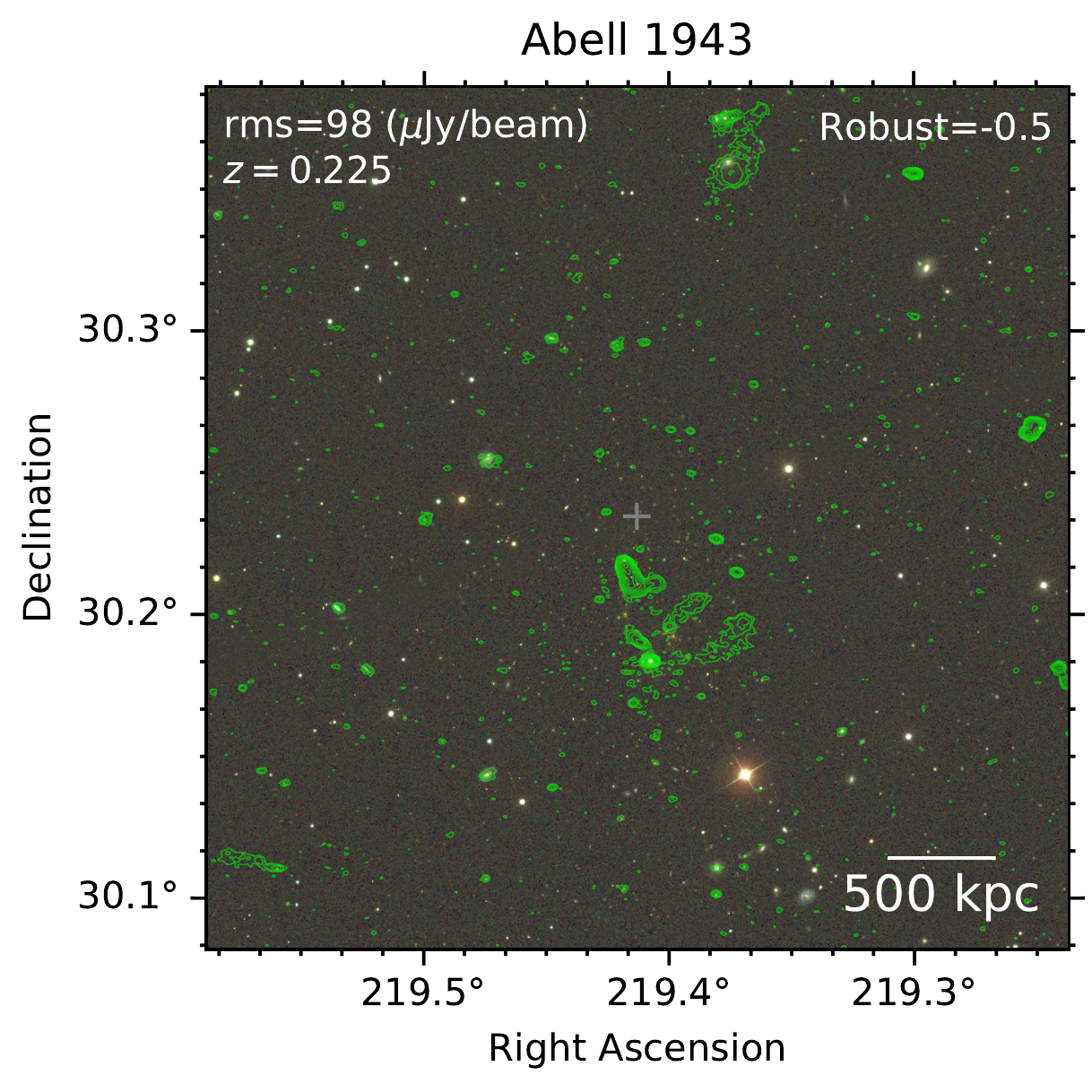}  \hfil
			\includegraphics[width=0.33\textwidth]{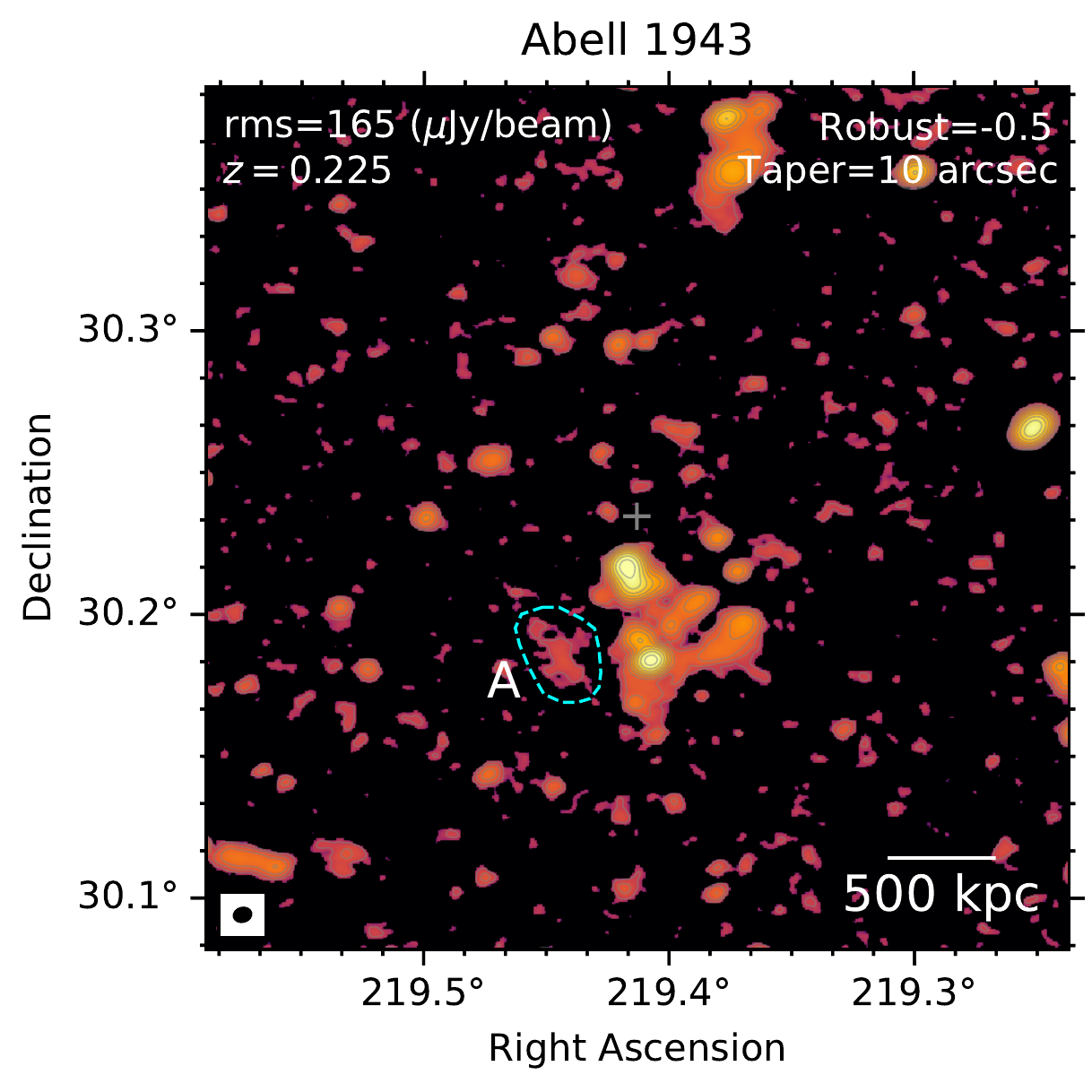}   };
		\draw (-7.7, -1.7) node {\color{white} (a)};
		\draw (-1.5, -1.7) node {\color{white} (b)};
		\draw (4.6, -1.55) node {\color{white} (c)};
	\end{tikzpicture}
	\caption{Abell~1943.  Image description is the same as that in Fig.~\ref{fig:abell84}.
	}
	\label{fig:abell1943}
\end{figure*}

In Fig.~\ref{fig:abell1943}, LOFAR detected a number of tailed radio galaxies in the central region of Abell~1943 ($z=0.225$). The orientation of these radio galaxies shows that they are moving in different directions with respect to the ICM, indicating that the cluster is in a dynamically-disturbed state. Towards the SE of the cluster centre, an elongated diffuse source with a projected size of 370~kpc$\times$170~kpc is newly detected in the low-resolution image (see the panel c). No optical counterpart associated with the diffuse source is seen in the SDSS image in the panel (b). The flux density and power of the source at 144~MHz are given Table~\ref{tab:sources}. The diffuse source could be a radio relic generated during the formation of the cluster or an active galactic nucleus (AGN) remnant. 

\subsection{Abell~1963}
\label{sec:abell1963}

\begin{figure*}[!ht]
	\centering
	\begin{tikzpicture}
		\draw (0, 0) node[inner sep=0] {\includegraphics[width=0.33\textwidth]{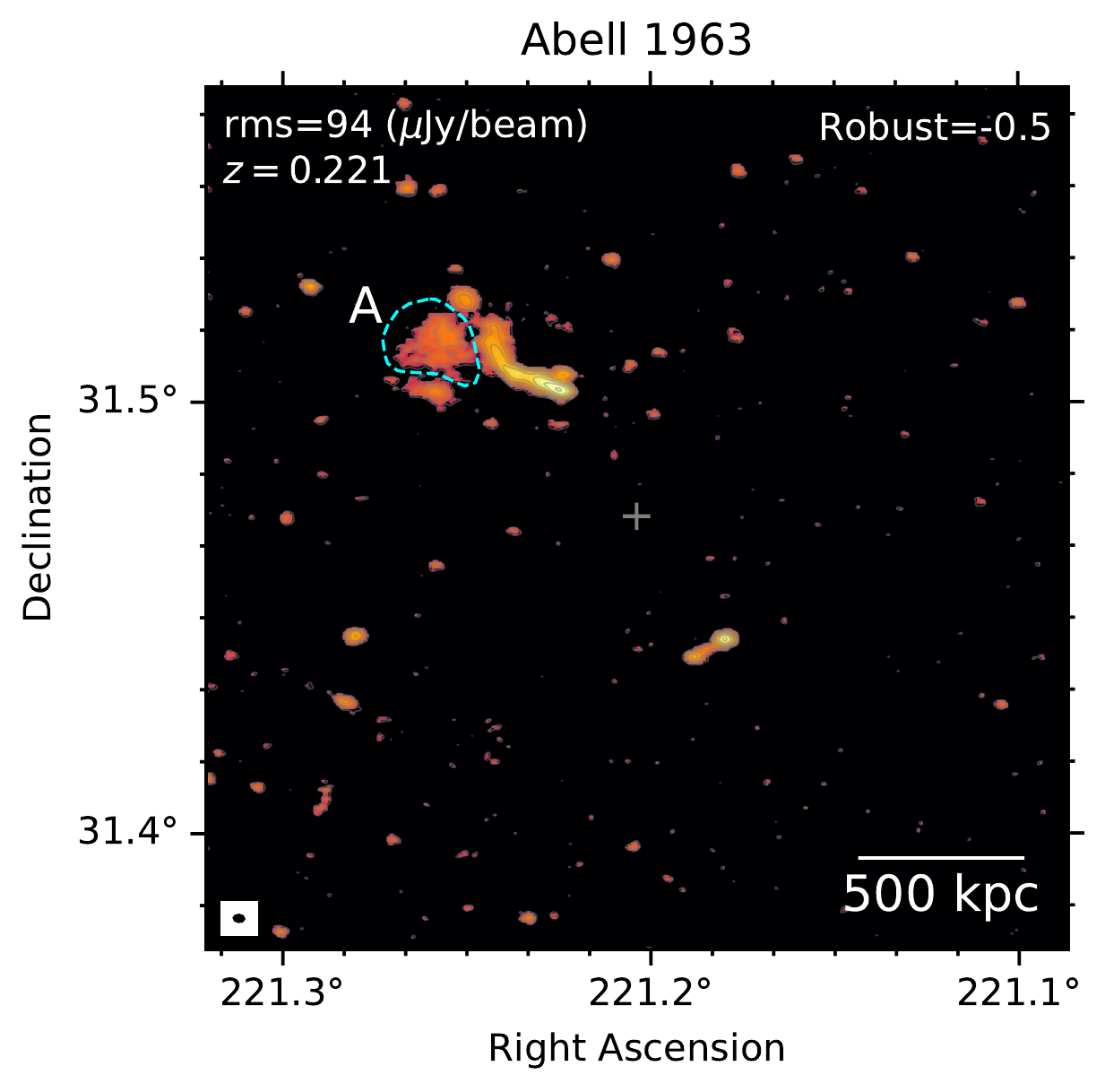}  \hfil
			\includegraphics[width=0.33\textwidth]{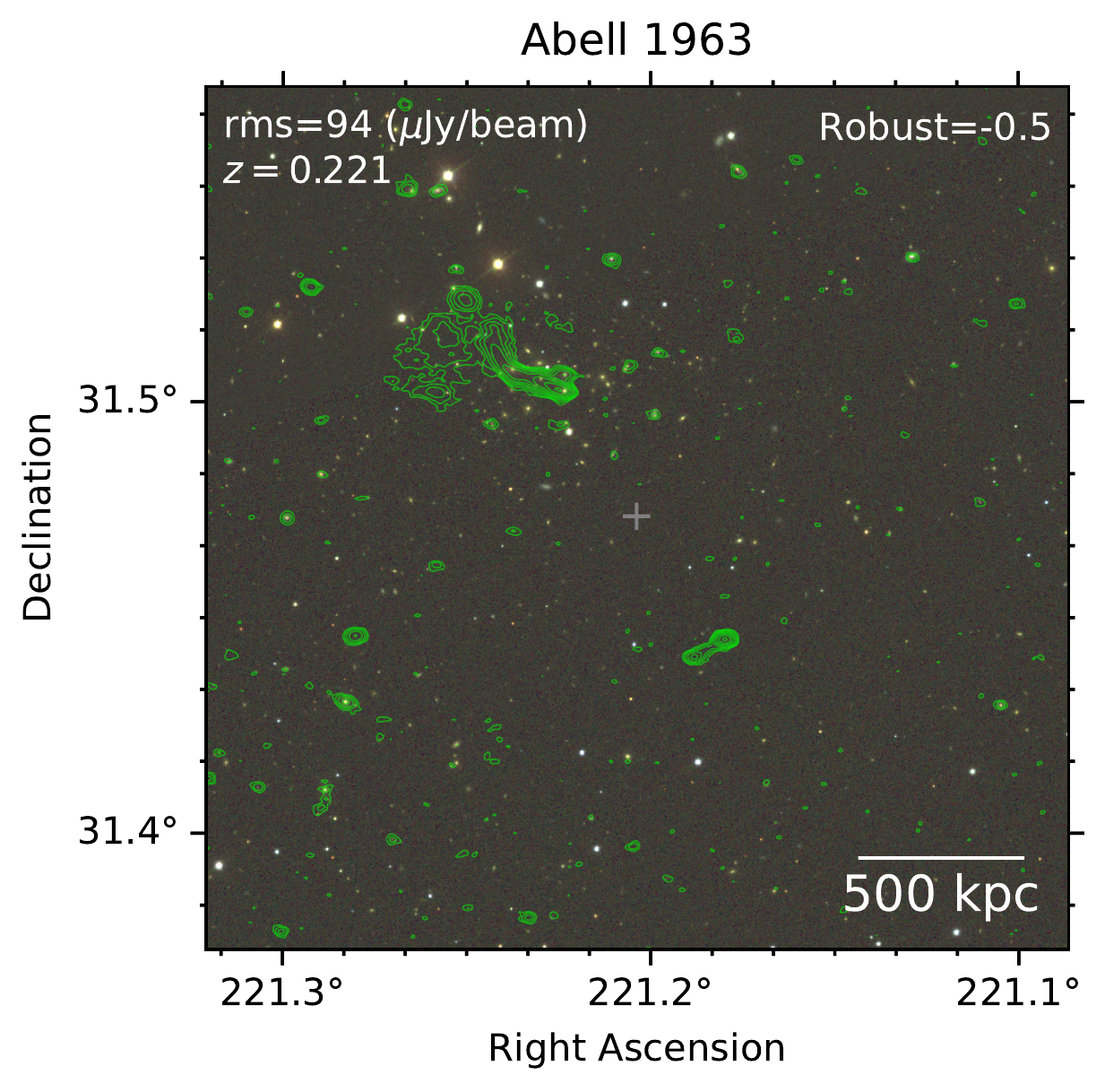}  \hfil
			\includegraphics[width=0.33\textwidth]{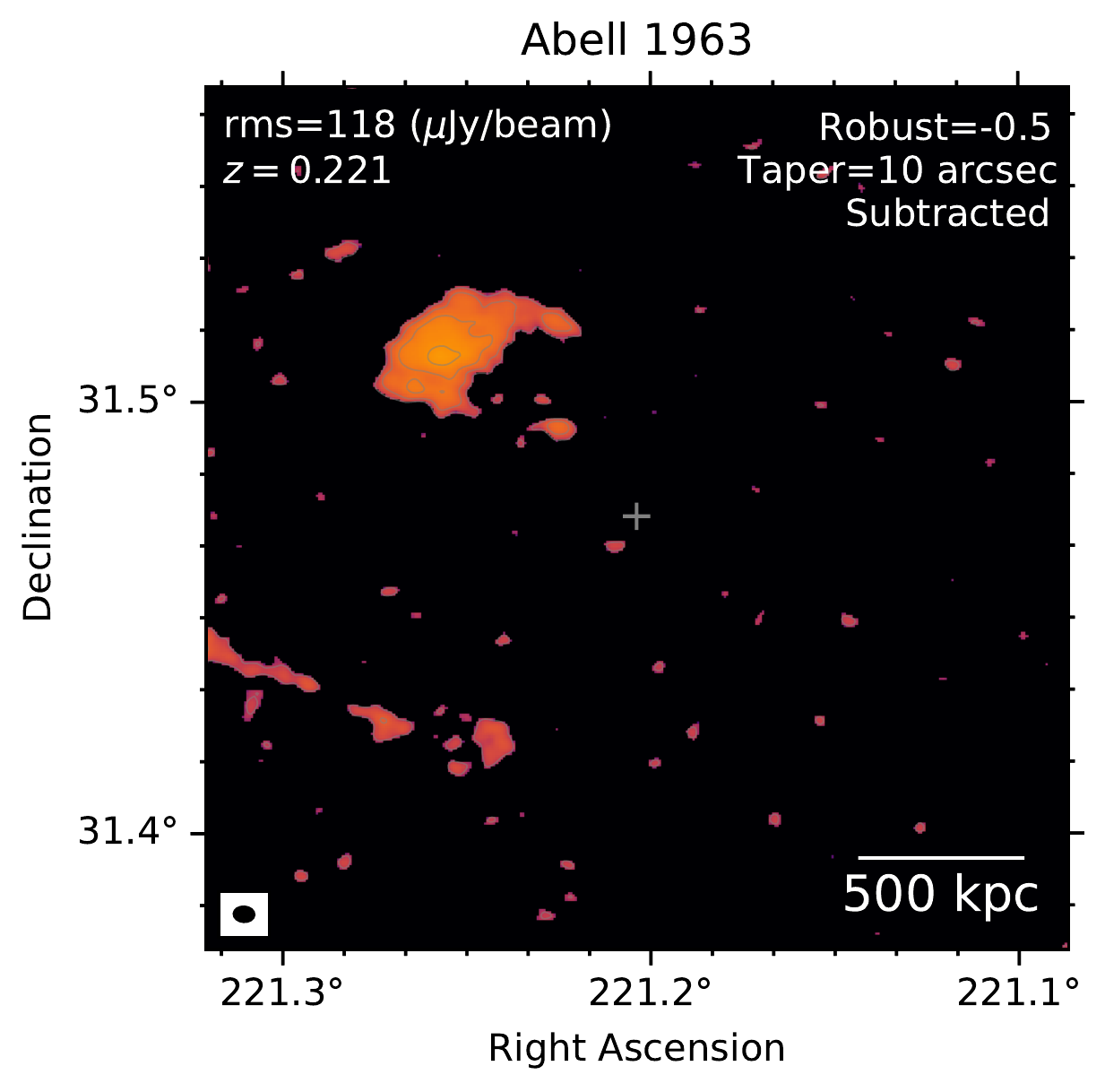}   };
		\draw (-7.7, -1.7) node {\color{white} (a)};
		\draw (-1.5, -1.7) node {\color{white} (b)};
		\draw (4.6, -1.55) node {\color{white} (c)};
	\end{tikzpicture}
	\caption{Abell~1963.  Image description is the same as that in Fig.~\ref{fig:abell84}.
	}
	\label{fig:abell1963}
\end{figure*}

In Fig.~\ref{fig:abell1963}, we present LOFAR and SDSS images of Abell~1963 ($z=0.221$). The cluster galaxies in the SDSS image are distributed in the NE-SW direction, indicating that Abell~1963 is a dynamically-disturbed system. The cluster centre is likely close to the location of the radio galaxies (i.e. $RA=221.2226^\circ$, $Dec.=31.5031^\circ$), where most of the galaxies seen in the optical image reside, rather than that reported in the Abell catalogue \citep{Abell1989}. The LOFAR images show the detection of two tailed radio galaxies in the centre. At the end of the galaxy tails, a diffuse source, named A, is newly detected. It has a projected size of 200~kpc and is connected with the radio galaxies through low-SB emission in its western direction. The diffuse source could be fossil plasma from radio galaxies or a radio relic that is generated during the formation of the cluster.

\subsection{DESI~201}
\label{sec:desi201}

\begin{figure*}[!ht]
	\centering
	\begin{tikzpicture}
		\draw (0, 0) node[inner sep=0] {\includegraphics[width=0.33\textwidth]{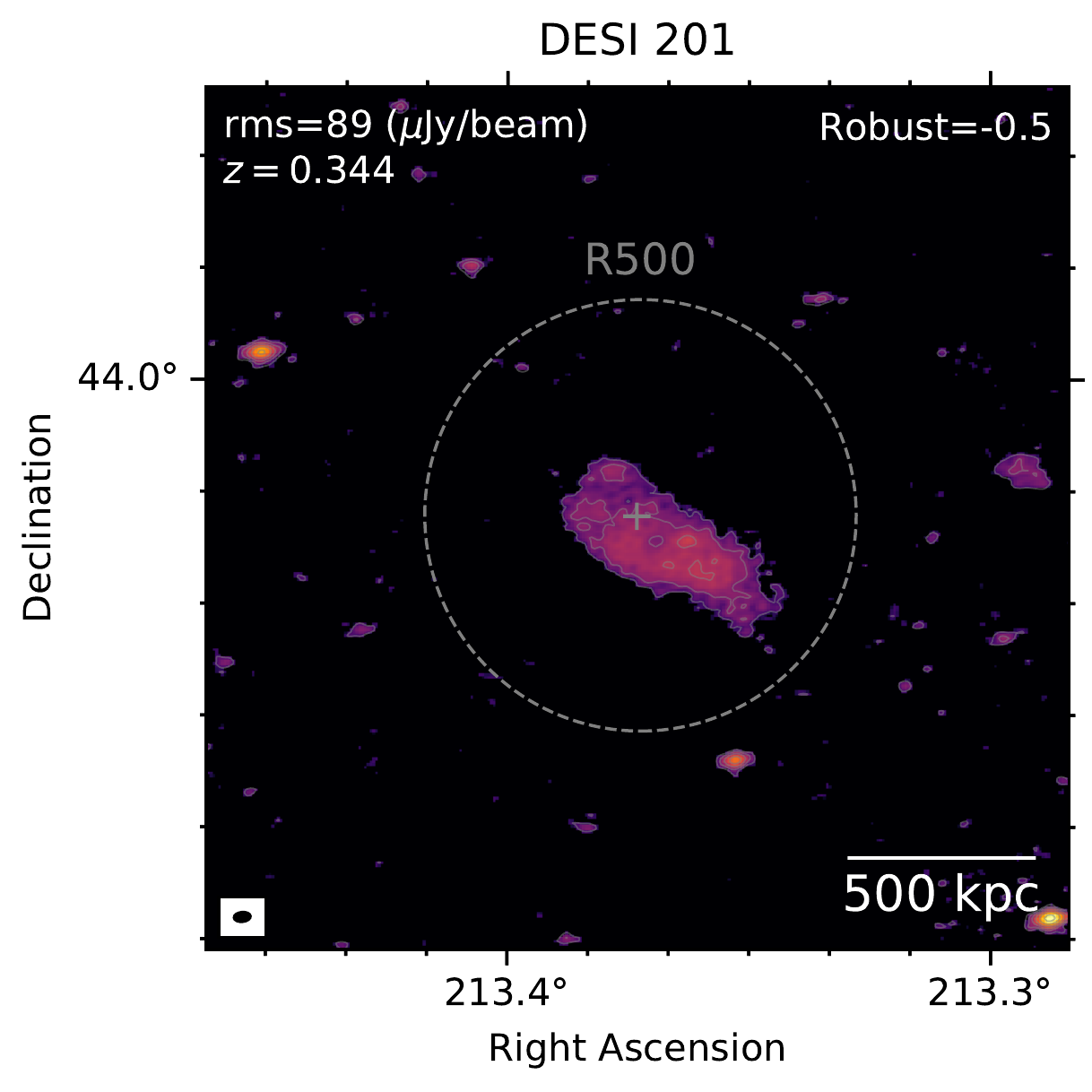}  \hfil
			\includegraphics[width=0.33\textwidth]{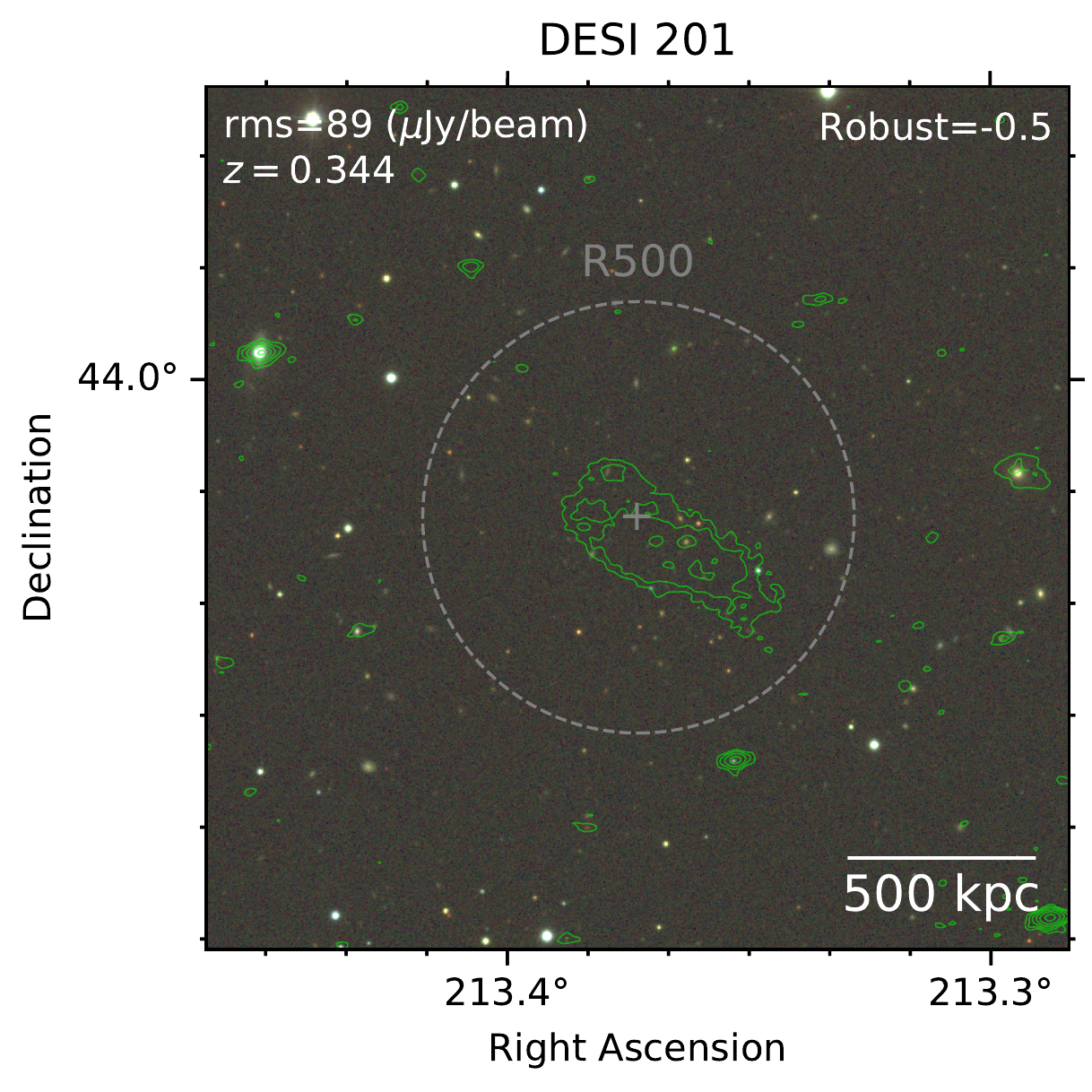}  \hfil
			\includegraphics[width=0.33\textwidth]{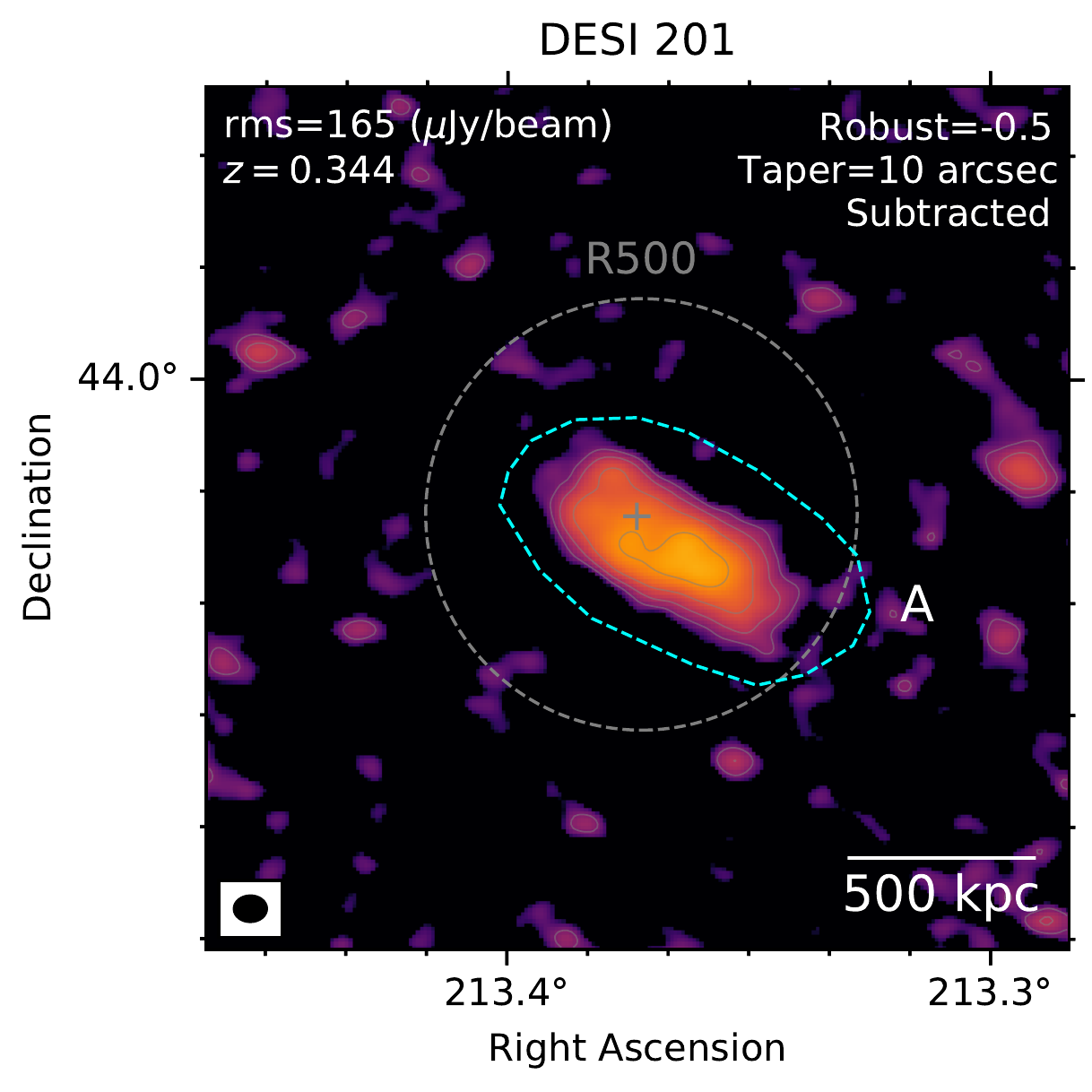}   };
		\draw (-7.7, -1.7) node {\color{white} (a)};
		\draw (-1.5, -1.7) node {\color{white} (b)};
		\draw (4.6, -1.55) node {\color{white} (c)};
	\end{tikzpicture}
	\caption{DESI~201.  Image description is the same as that in Fig.~\ref{fig:abell84}.
	}
	\label{fig:desi201}
\end{figure*}

LOFAR images in Fig.~\ref{fig:desi201} show the new detection of a diffuse emission in DESI~201 ($z=0.344$). The diffuse source has a projected size of 620~kpc$\times$270~kpc and has no SDSS optical counterpart. It could be associated with radio galaxies, but its true nature is still unknown.

\subsection{DESI~296}
\label{sec:desi296}

\begin{figure*}[!ht]
	\centering
	\begin{tikzpicture}
		\draw (0, 0) node[inner sep=0] {\includegraphics[width=0.33\textwidth]{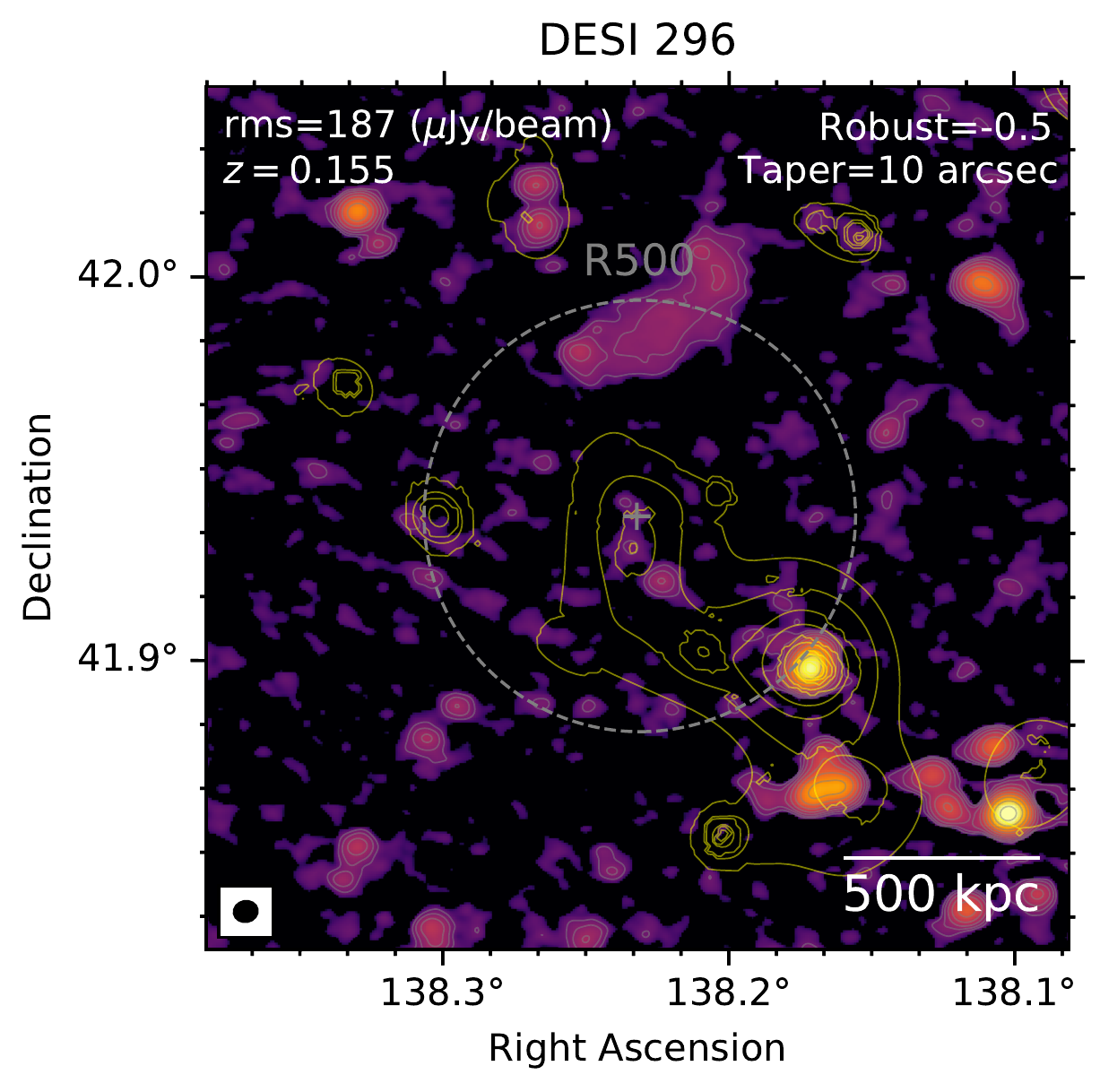}  \hfil
			\includegraphics[width=0.33\textwidth]{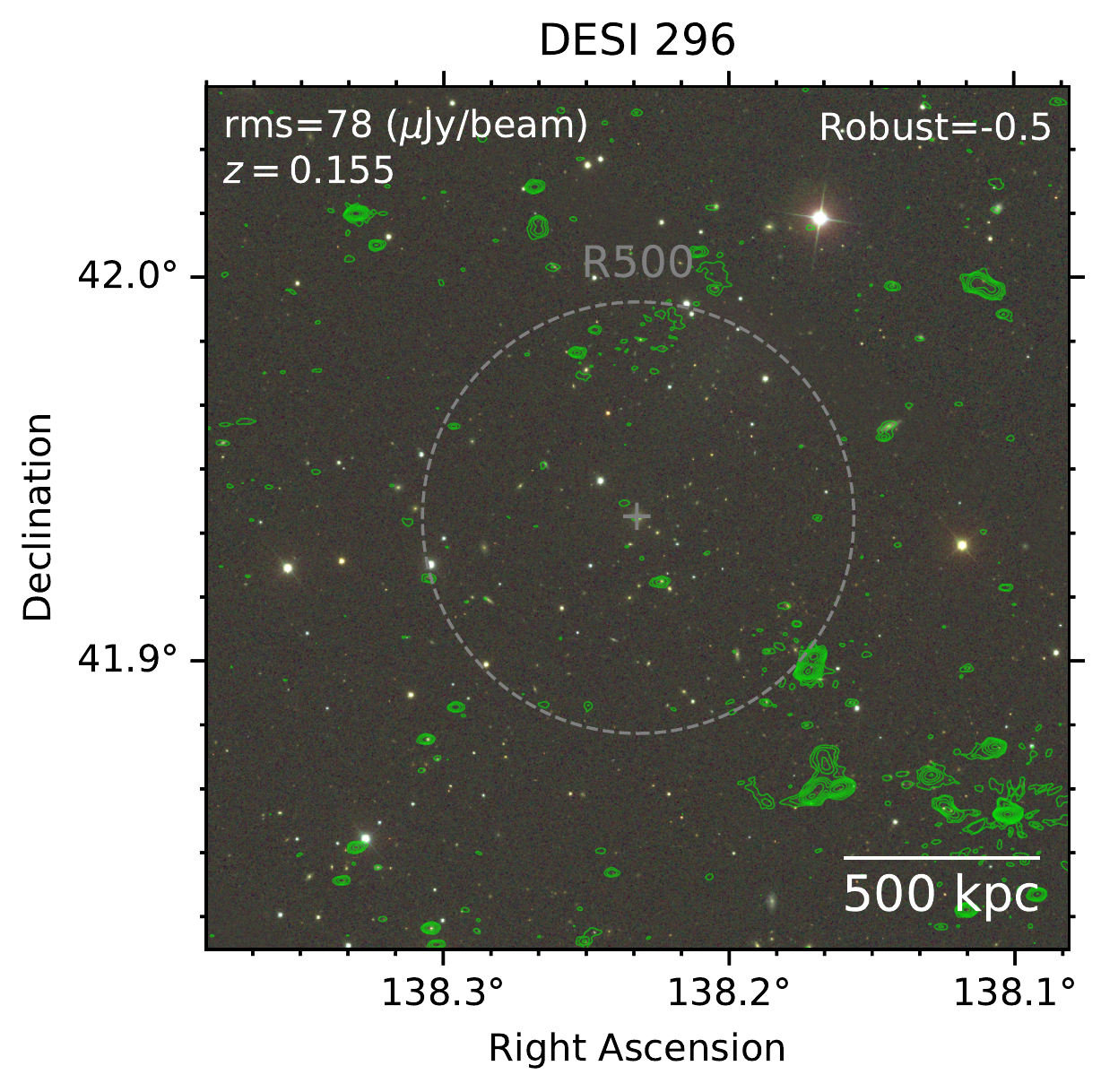}  \hfil
			\includegraphics[width=0.33\textwidth]{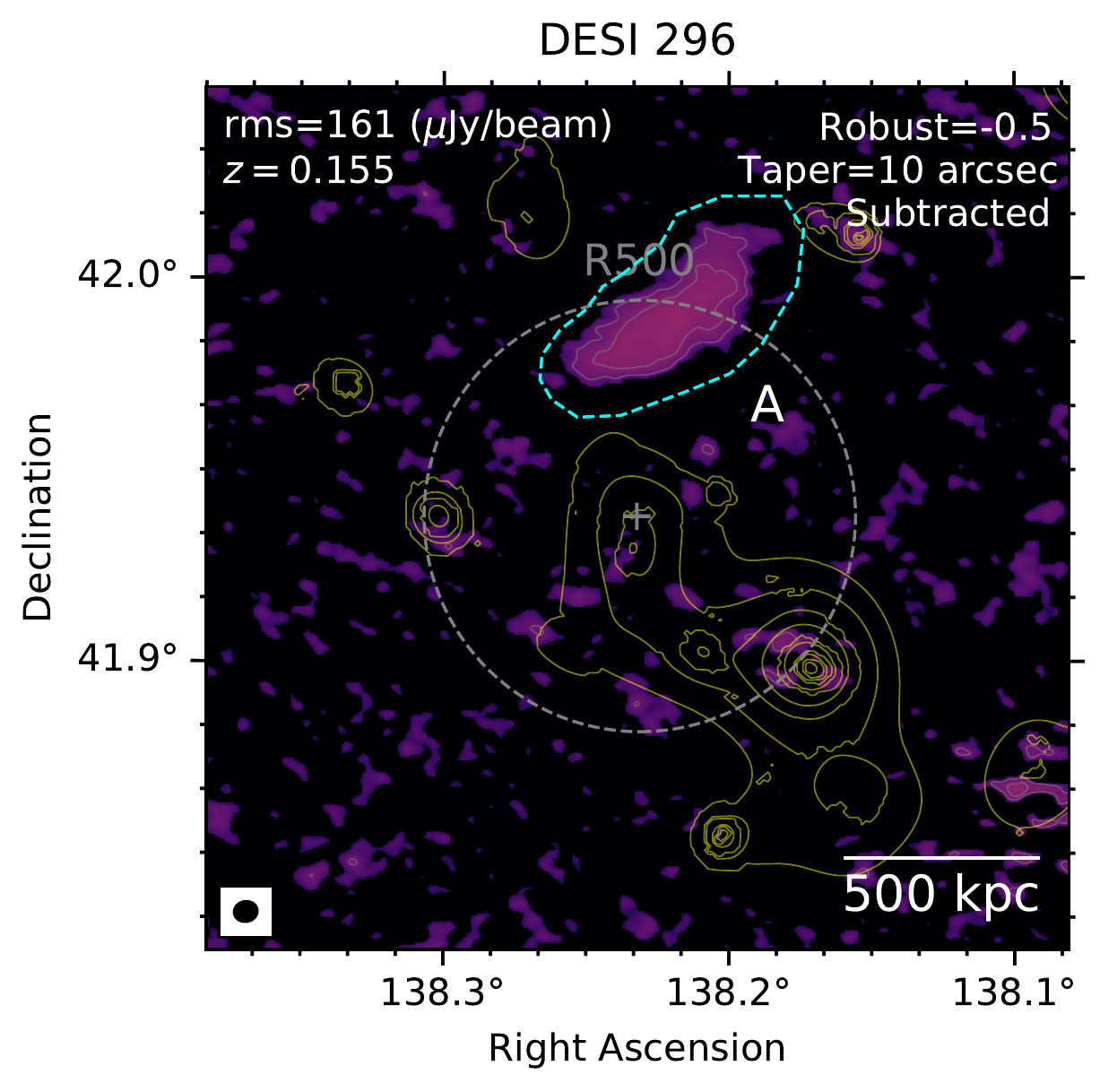}   };
		\draw (-7.7, -1.7) node {\color{white} (a)};
		\draw (-1.5, -1.7) node {\color{white} (b)};
		\draw (4.6, -1.55) node {\color{white} (c)};
	\end{tikzpicture}
	\caption{DESI~296.  Image description is the same as that in Fig.~\ref{fig:abell84}.
	}
	\label{fig:desi296}
\end{figure*}

LOFAR images displayed in Fig.~\ref{fig:desi296} show the new detection of a diffuse source, named A, in DESI~296 ($z=0.155$). The diffuse source with a projected size of 550~kpc$\times$200~kpc is located 500~kpc to the north of the cluster centre. The SDSS image in the panel (b) does not show a clear optical counterpart associated with source A. However, part of the radio emission to the northern and southern sides of source A is from individual galaxies as shown in the LOFAR high-resolution contours in the panel (b). The 144~MHz flux density and power of source A are given in Table~\ref{tab:sources}. Source A is about 600~kpc from the central X-ray emission, seen in the panel (a, c). The location and extension of source A are consistent with those for relics, but its morphology bending towards the outer direction of the cluster is unusual and the source does not show typical SB profile (i.e. sharp edge in the outside) as seen in other relics. Additional high-frequency radio (polarimetric) observations will be needed to understand the nature of the source.

\subsection{MCXC~J0928.6+3747}
\label{sec:MCXCJ0928.6+3747}

\begin{figure*}[!ht]
	\centering
	\begin{tikzpicture}
		\draw (0, 0) node[inner sep=0] {\includegraphics[width=0.33\textwidth]{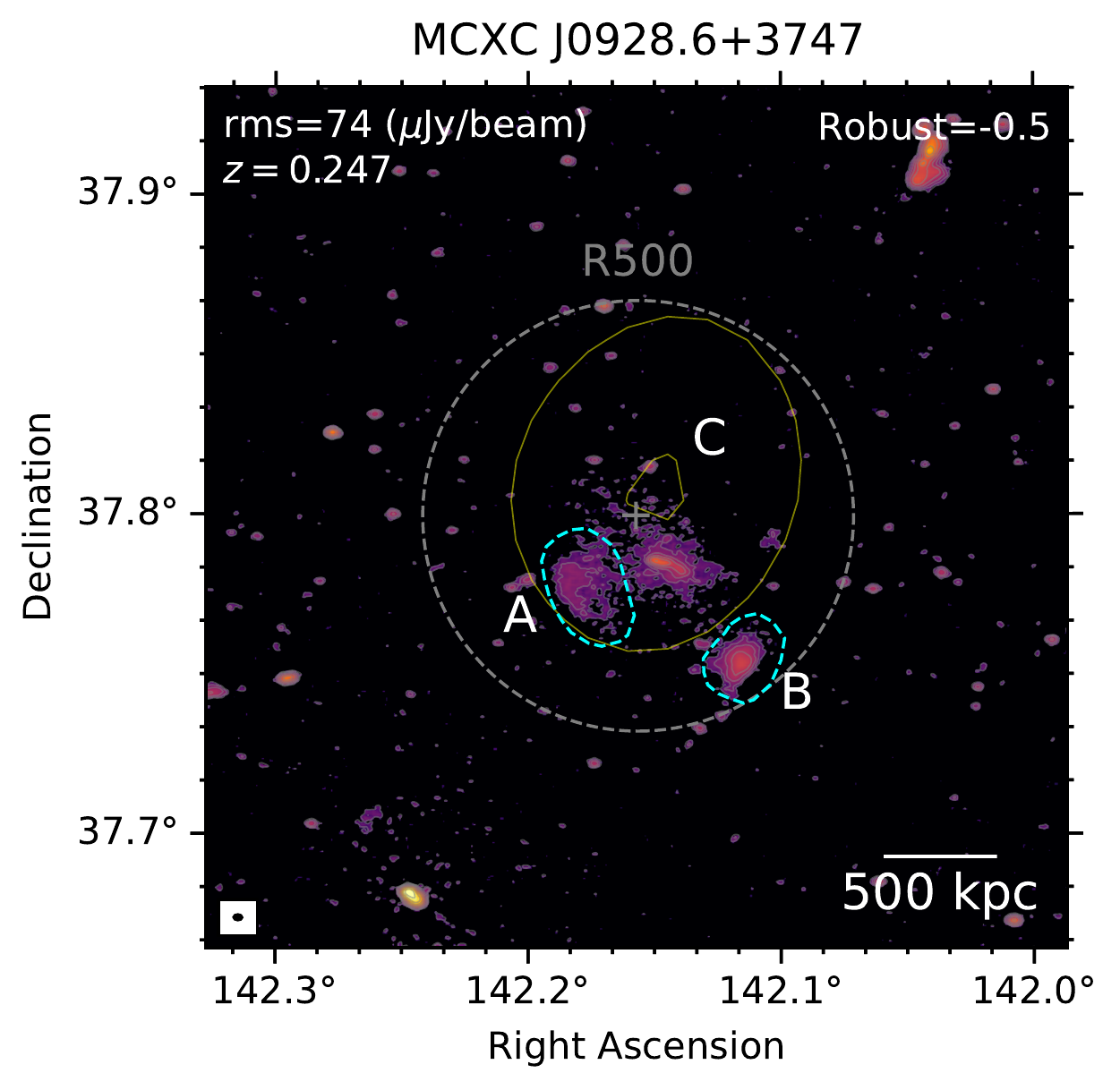}  \hfil
			\includegraphics[width=0.33\textwidth]{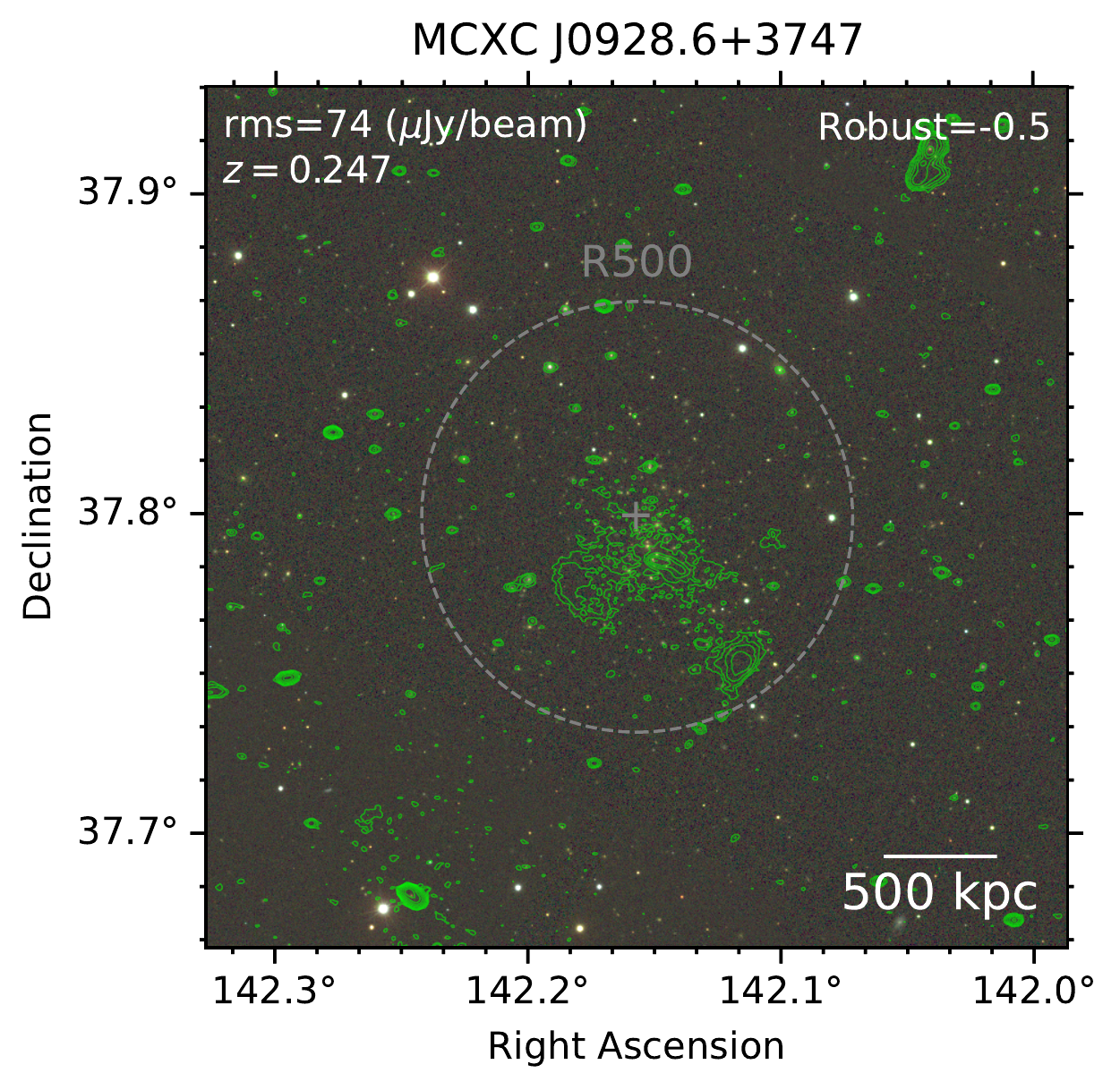}  \hfil
			\includegraphics[width=0.33\textwidth]{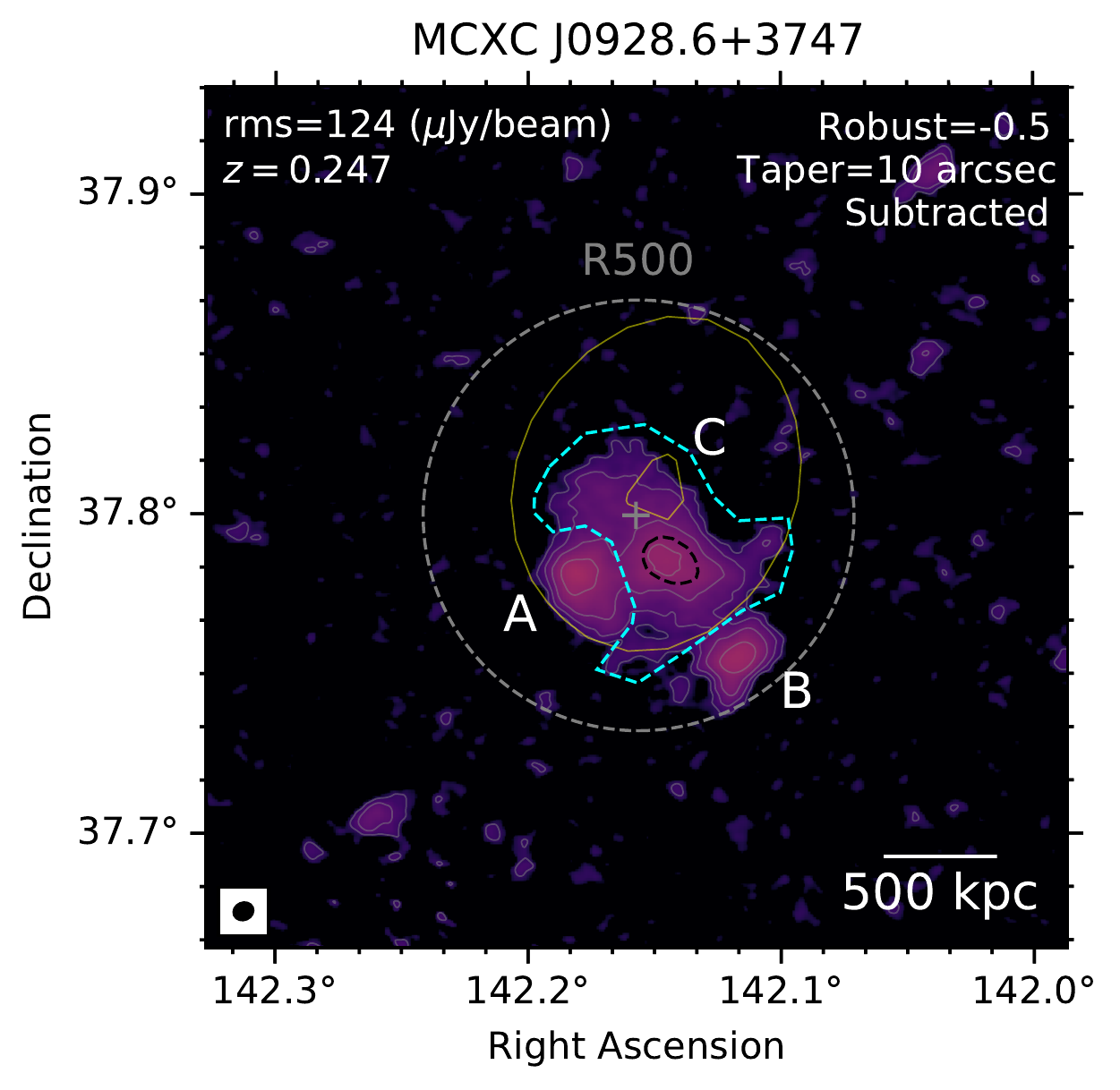}   };
		\draw (-7.7, -1.7) node {\color{white} (a)};
		\draw (-1.5, -1.7) node {\color{white} (b)};
		\draw (4.6, -1.55) node {\color{white} (c)};
	\end{tikzpicture}
	\caption{MCXC~J0928.6+3747. Image description is the same as that in Fig.~\ref{fig:abell84}.
	}
	\label{fig:MCXCJ0928.6+3747}
\end{figure*}

MCXC~J0928.6+3747 ($z=0.247$; also known as Abell~800) was reported by \cite{Govoni2012} to host diffuse radio source that they classified as a radio halo. Despite of the classification, the halo emission is patchy in their VLA low-resolution ($63\arcsec$) image and does not follow the ROSAT X-ray emission (see Fig.~1 in \citealt{Govoni2012}).

LOFAR high-resolution images in Fig.~\ref{fig:MCXCJ0928.6+3747} confirm the presence of diffuse emission in MCXC~J0928.6+3747. Owning to the high-resolution images of LOFAR, the diffuse source reported by \cite{Govoni2012} is resolved into three separated sources, labelled as A-C. In the SE direction, 350~kpc from the cluster centre, source A is detected with a projected size of 380~kpc$\times$250~kpc. The SB of A decreases gradually towards the NW, but its SB rapidly decreases towards the SE. Roughly 680~kpc towards the SW direction from the cluster centre, source B with a projected size of 260~kpc$\times$190~kpc is detected. The morphology of B is similar to that of A in which the radial SB gradually increases then sharply decreases in the outer region. Both A and B are not associated with SDSS optical sources (see Fig.~\ref{fig:MCXCJ0928.6+3747}, b). We classify A and B as radio relics. In the discrete-source-subtracted image (the panel c), a low SB emission (C) with a projected size of 850~kpc$\times$400~kpc is detected and partly covers the ROSAT X-ray emission in the cluster centre. The extension of source C implies that it is not likely associated with the tailed radio galaxy embedded in the cluster centre. We classify source C as a radio halo, confirming the classification by \cite{Govoni2012}.

\subsection{MCXC~J0943.1+4659}
\label{sec:MCXCJ0943.1+4659}

\begin{figure*}[!ht]
	\centering
	\begin{tikzpicture}
		\draw (0, 0) node[inner sep=0] {\includegraphics[width=0.33\textwidth]{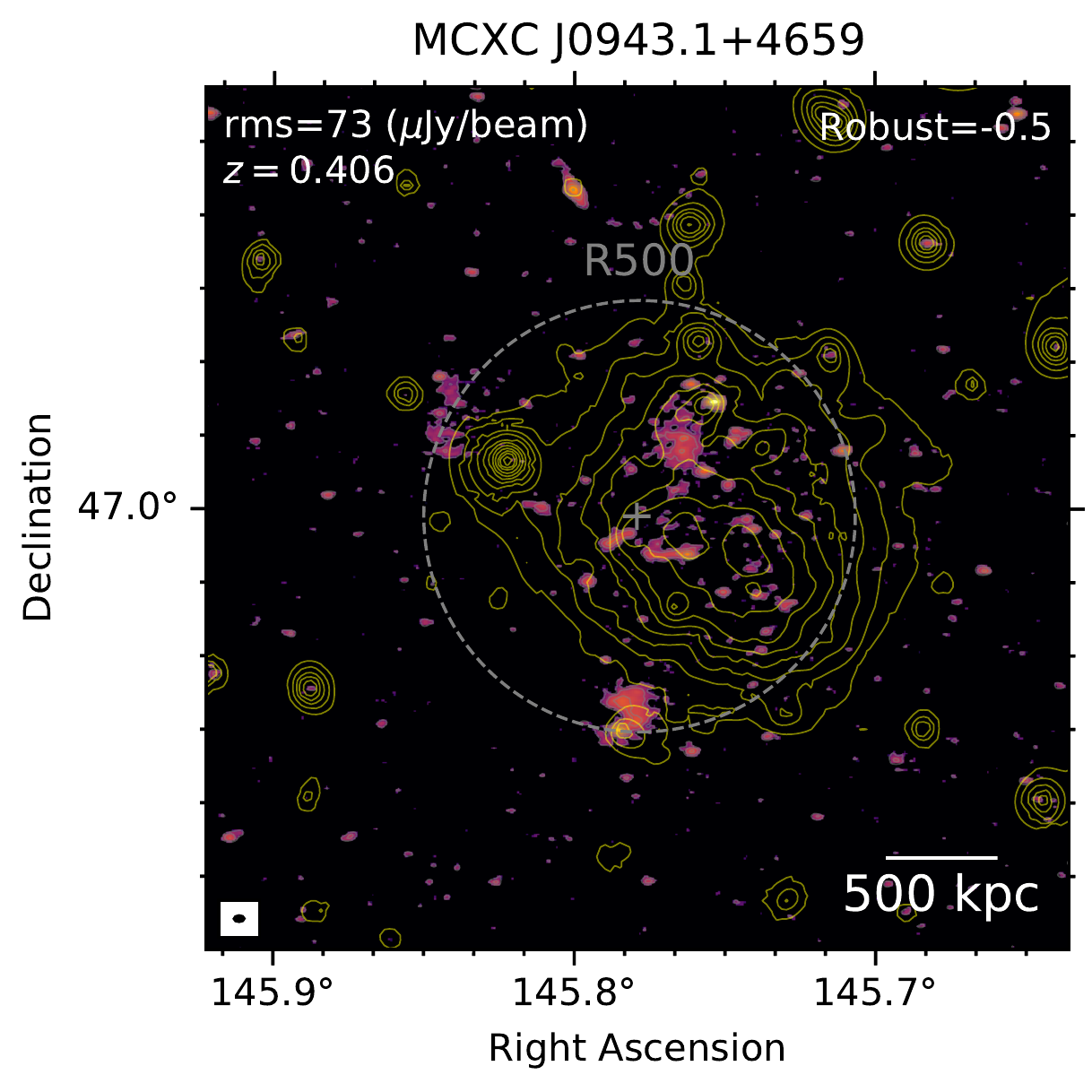}  \hfil
			\includegraphics[width=0.33\textwidth]{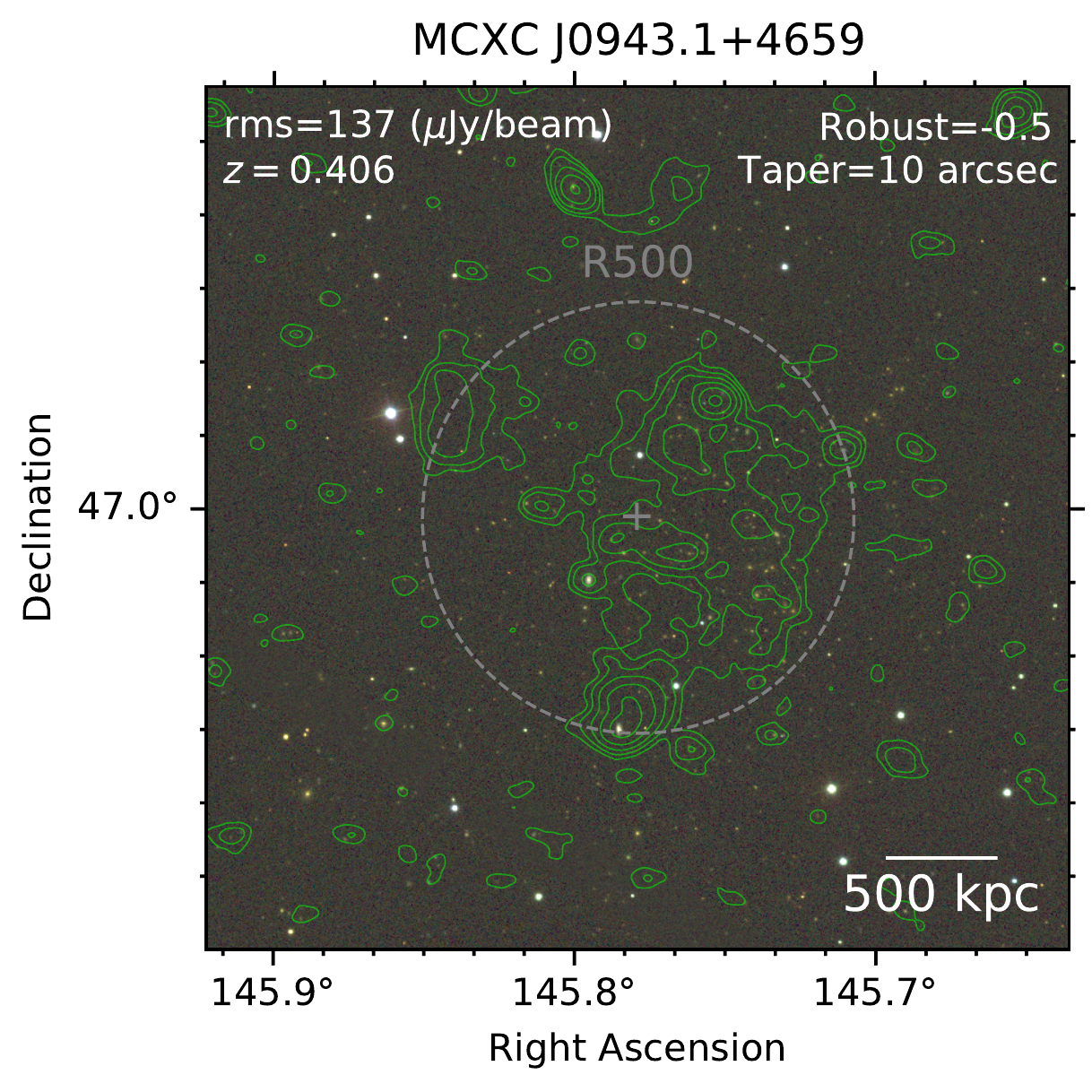}  \hfil
			\includegraphics[width=0.33\textwidth]{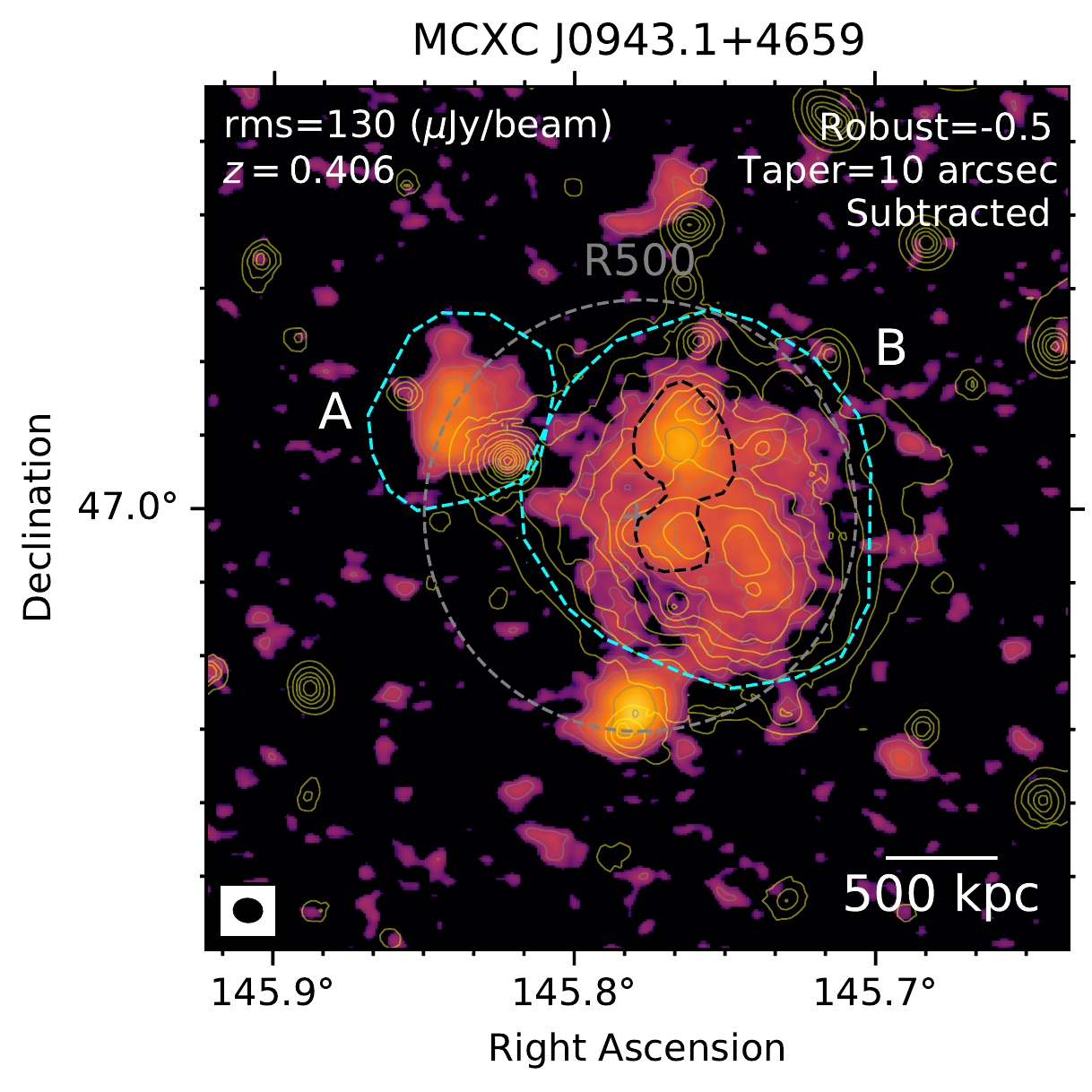}   };
		\draw (-7.7, -1.7) node {\color{white} (a)};
		\draw (-1.5, -1.7) node {\color{white} (b)};
		\draw (4.6, -1.55) node {\color{white} (c)};
	\end{tikzpicture}
	\caption{MCXC~J0943.1+4659. Image description is the same as that in Fig.~\ref{fig:abell84}.
	}
	\label{fig:MCXCJ0943.1+4659}
\end{figure*}

\cite{Giovannini2009} reported the presence of diffuse emission in the central region of MCXC~J0943.1+4659 ($z=0.406$). The diffuse emission has a size of 400~kpc and is offset from the ROSAT X-ray emission. They pointed out that the X-ray emission from the cluster has double structure separated within $2\arcmin$.

As seen in Fig.~\ref{fig:MCXCJ0943.1+4659}, the LOFAR images confirm the presence of the diffuse sources in the central region of MCXC~J0943.1+4659 and shows the new detection of a diffuse source in the NE region of the cluster. The diffuse source, named A, in the NE direction has a projected size of 530~kpc$\times$460~kpc and is elongated in the N-S direction. The SB of source A increases rapidly from E to W, before gradually decreases towards the W side of the source. In the cluster centre a diffuse source, named B, is seen in the LOFAR low-resolution images. Source B has a size of 1.2~Mpc in projection. A number of radio galaxies are also seen embedded in the centre of B. The radio emission from source B roughly follows the X-ray emission detected with the XMM-Newton observations, unlike what is seen with the VLA 1.4~GHz and ROSAT shallow observations in \cite{Giovannini2009}. We classify sources A and B as a radio relic and radio halo, respectively. This is supported by the fact that the X-ray morphology indicates that the cluster is disturbed, as seen in Fig. \ref{fig:MCXCJ0943.1+4659} and Fig.~4 in \cite{Giovannini2009}. 

\subsection{MCXC~J1020.5+3922}
\label{sec:MCXCJ1020.5+3922}

\begin{figure*}[!ht]
	\centering
	\begin{tikzpicture}
		\draw (0, 0) node[inner sep=0] {\includegraphics[width=0.33\textwidth]{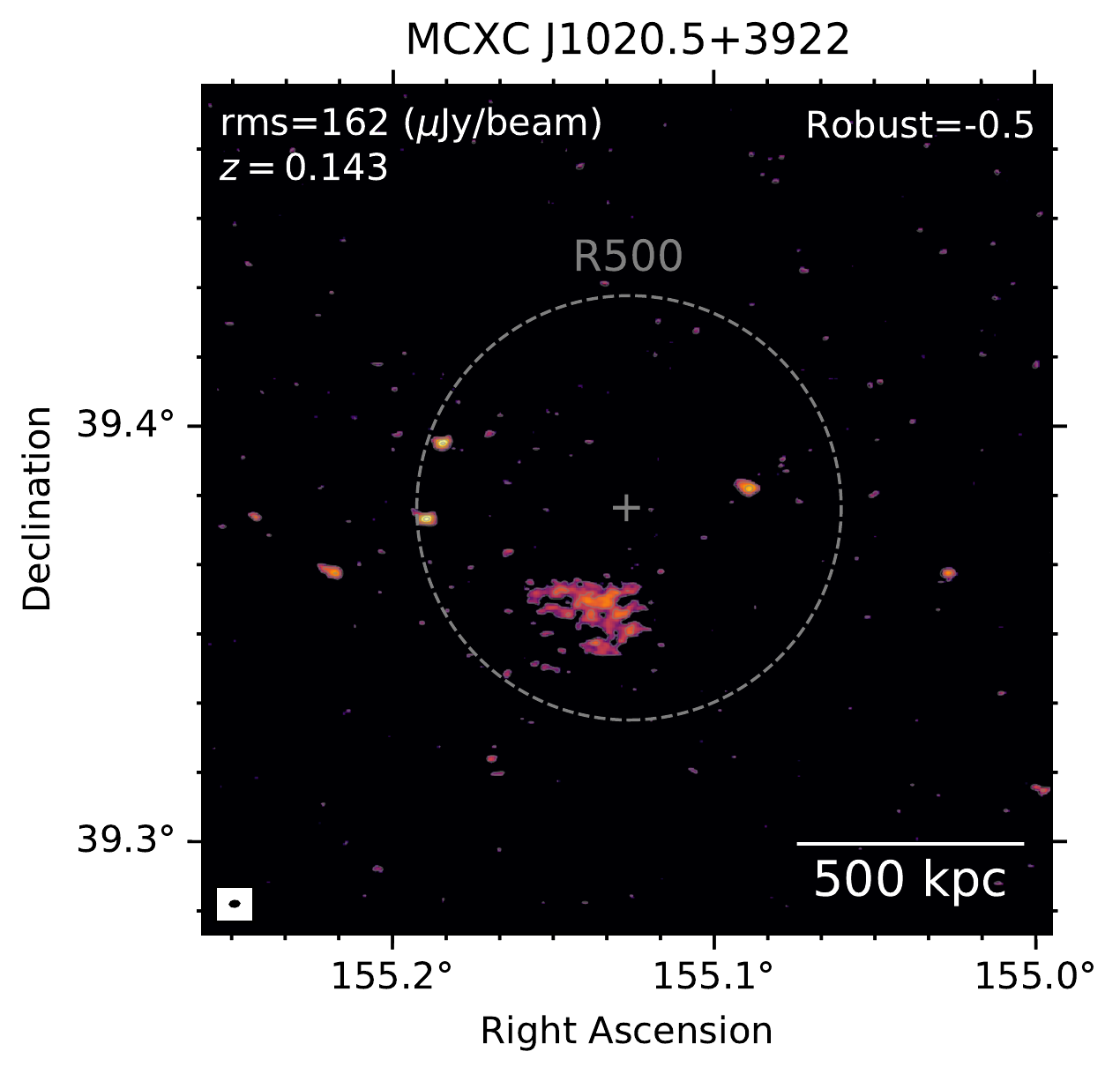}  \hfil
			\includegraphics[width=0.33\textwidth]{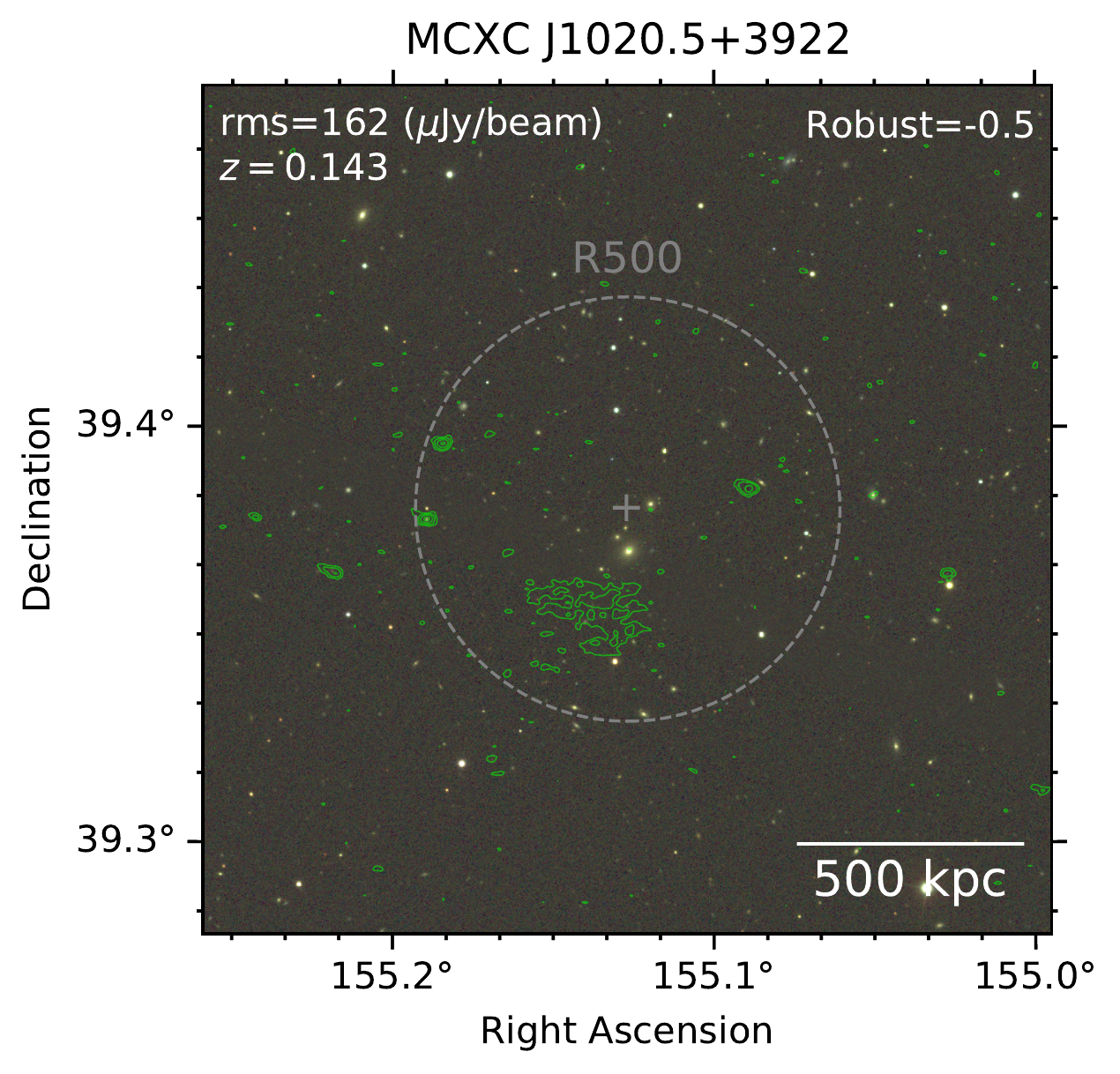}  \hfil
			\includegraphics[width=0.33\textwidth]{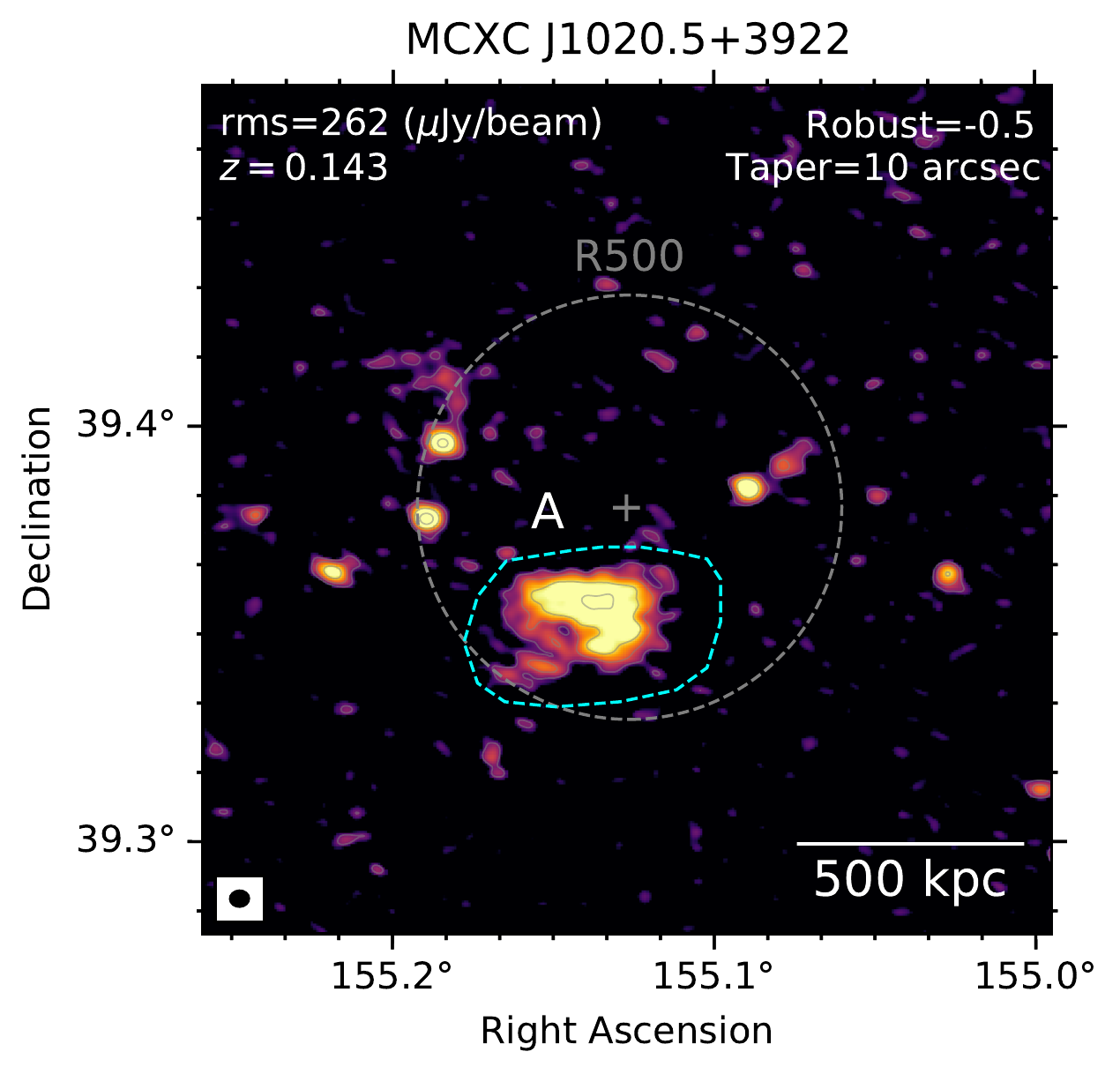}   };
		\draw (-7.7, -1.7) node {\color{white} (a)};
		\draw (-1.5, -1.7) node {\color{white} (b)};
		\draw (4.6, -1.55) node {\color{white} (c)};
	\end{tikzpicture}
	\caption{MCXC~J1020.5+3922. Image description is the same as that in Fig.~\ref{fig:abell84}.
	}
	\label{fig:MCXCJ1020.5+3922}
\end{figure*}

A diffuse source, labelled as A, is discovered towards the SE region of MCXC~J1020.5+3922 ($z=0.143$), seen in the LOFAR images in Fig.~\ref{fig:MCXCJ1020.5+3922}. The high-resolution radio emission from source A is patchy. Its projected size is 360~kpc$\times$230~kpc. The SDSS image in Fig.~\ref{fig:MCXCJ1020.5+3922} (b) shows no obvious connection between source A and optical sources in the field. Source A could be an AGN remnant (900~kpc) although further studies are required to confirm its nature. 

\subsection{MCXC~J1711.0+3941}
\label{sec:MCXCJ1711.0+3941}

\begin{figure*}[!ht]
	\centering
	\begin{tikzpicture}
		\draw (0, 0) node[inner sep=0] {\includegraphics[width=0.33\textwidth]{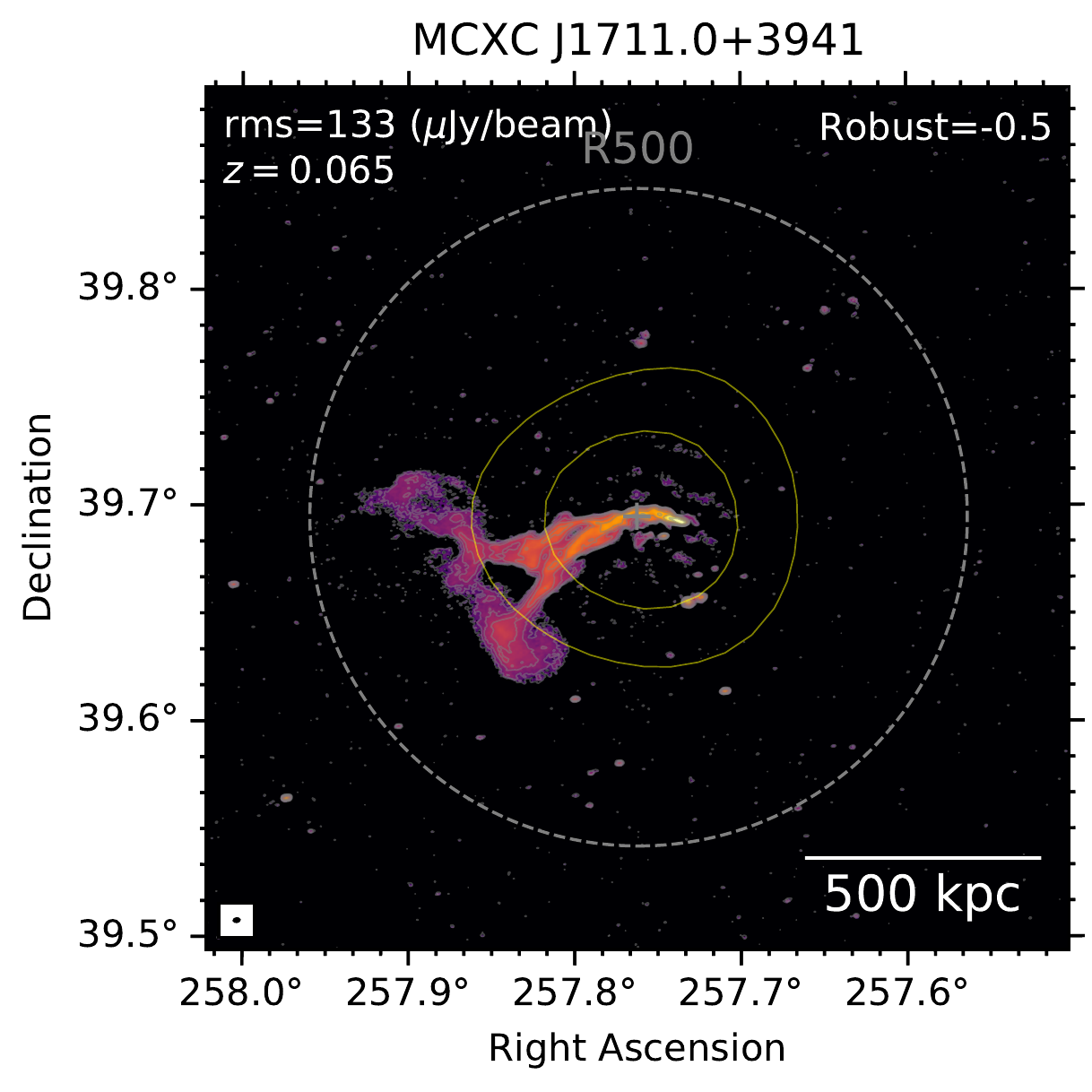}  \hfil
			\includegraphics[width=0.33\textwidth]{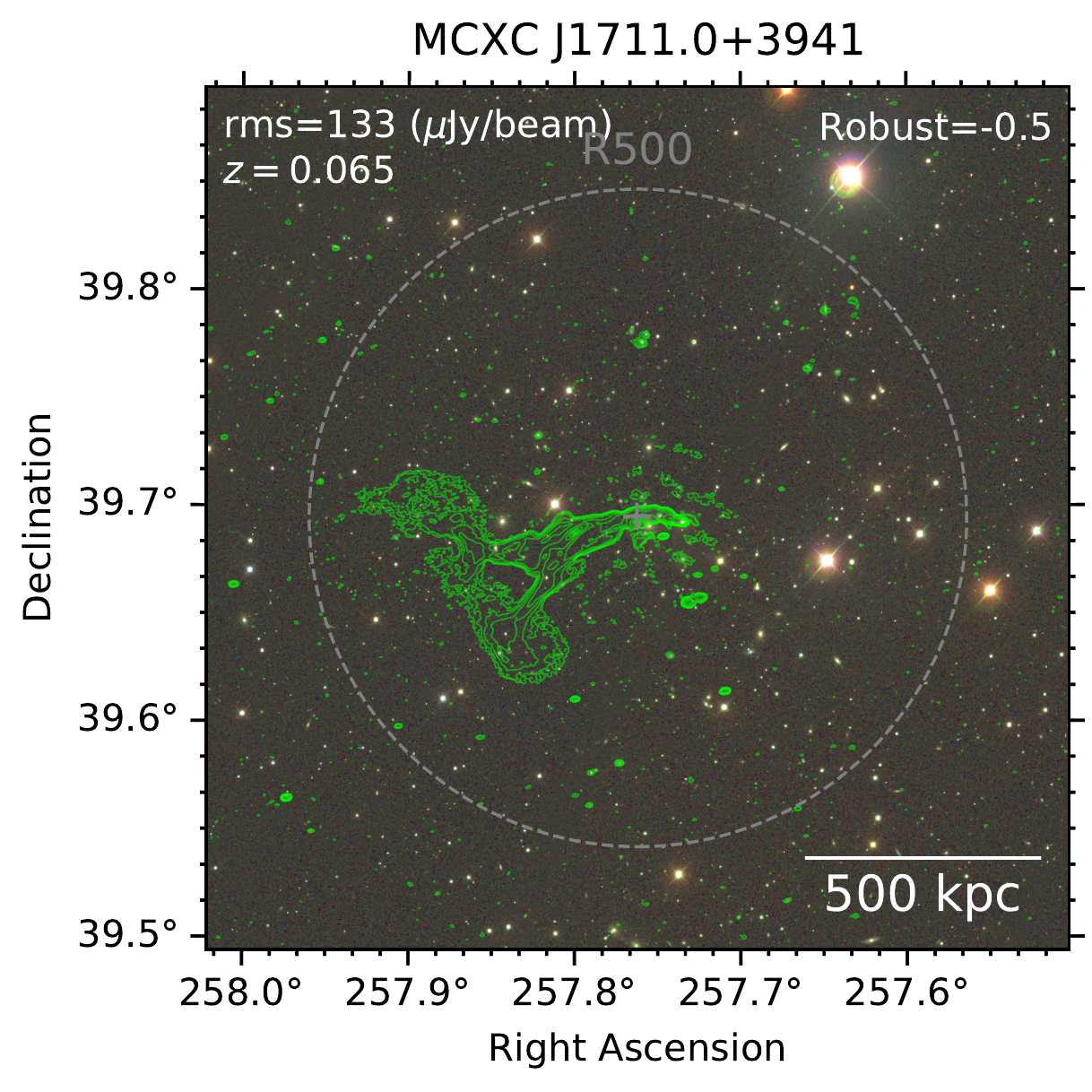}  \hfil
			\includegraphics[width=0.33\textwidth]{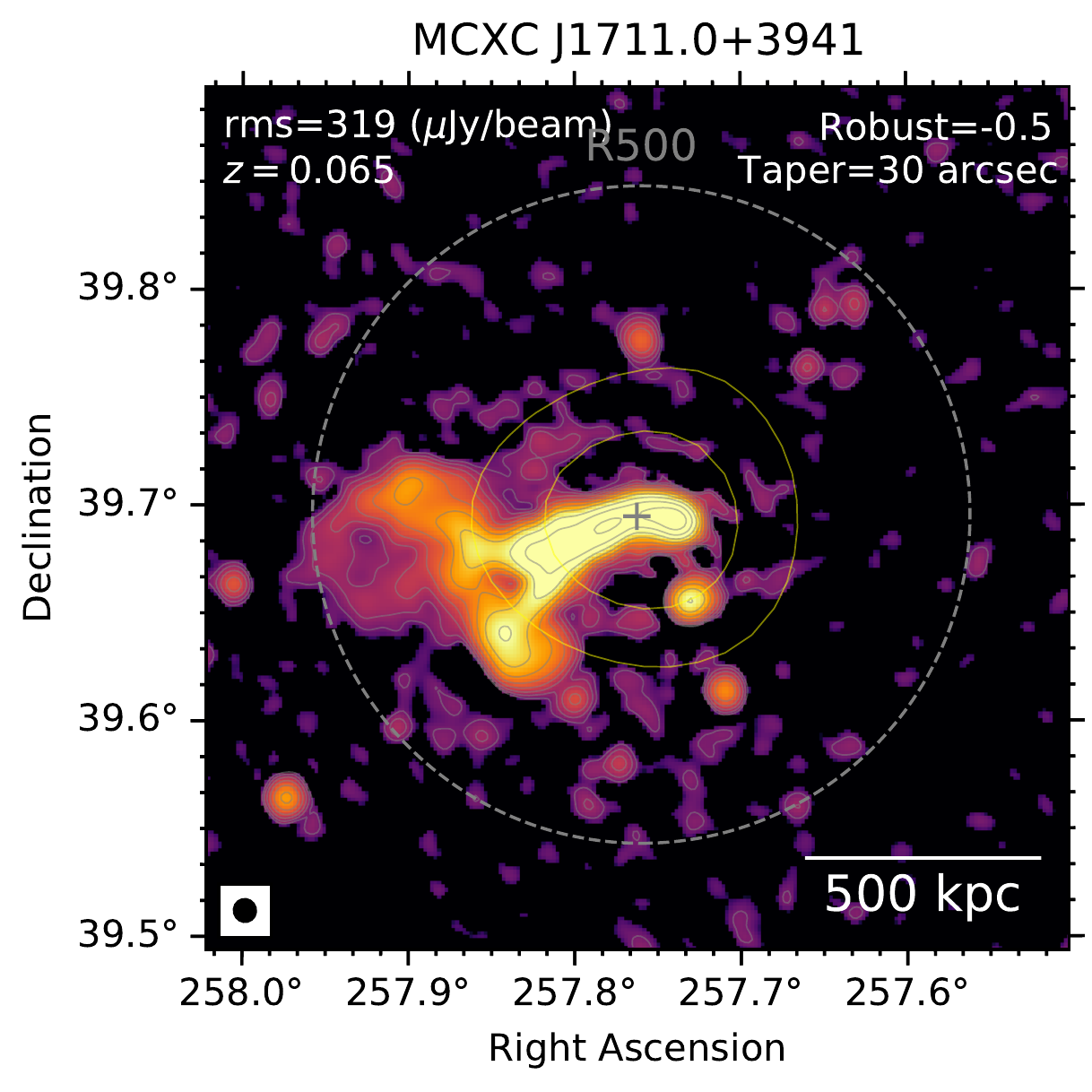}   };
		\draw (-7.7, -1.7) node {\color{white} (a)};
		\draw (-1.5, -1.7) node {\color{white} (b)};
		\draw (4.6, -1.55) node {\color{white} (c)};
	\end{tikzpicture}
	\caption{MCXC~J1711.0+3941. Image description is the same as that in Fig.~\ref{fig:abell84}.
	}
	\label{fig:MCXCJ1711.0+3941}
\end{figure*}

Fig.~\ref{fig:MCXCJ1711.0+3941} shows that LOFAR observations detect a 600~kpc long narrow-angle tailed (NAT) radio galaxy in MCXC~J1711.0+3941 ($z=0.065$). The source has an optical counterpart as seen in the SDSS image in the panel (b) of Fig.~\ref{fig:MCXCJ1711.0+3941}. The tail end of the source is highly extended, 630~kpc$\times$380~kpc, in the NE-SW direction and is located at the edge of the ROSAT X-ray emission. The formation of the disturbed tail of the radio galaxy is still unknown. It might originate from the interaction between the radio galaxy and the surrounding medium. We do not measure the flux density of the disturbed tail as the boundary between the diffuse emission and the NAT galaxy is unknown.

\subsection{PSZRX~G095.27+48.27}
\label{sec:PSZRXG095.27+48.27}

\begin{figure*}[!ht]
	\centering
	\begin{tikzpicture}
		\draw (0, 0) node[inner sep=0] {\includegraphics[width=0.33\textwidth]{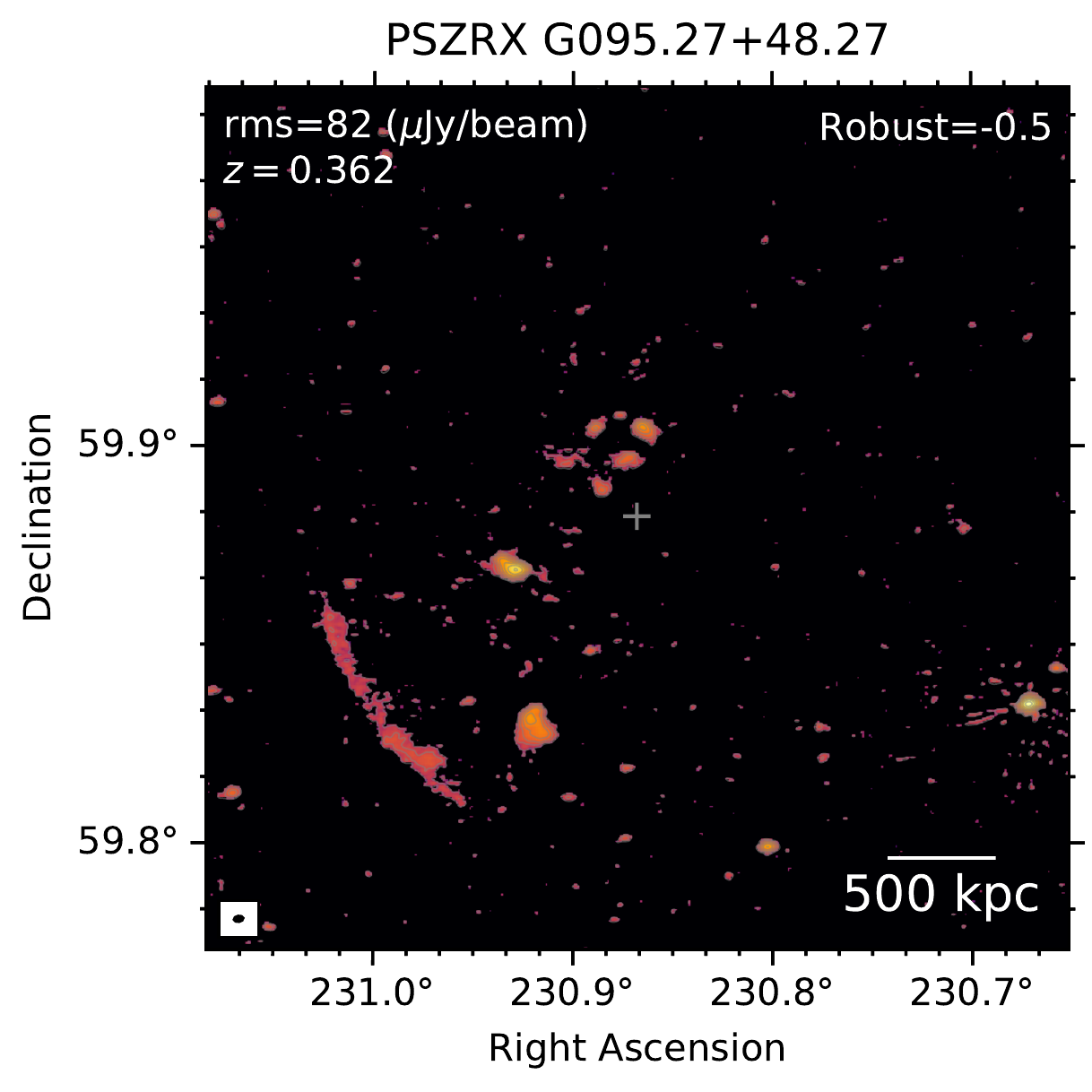}  \hfil
			\includegraphics[width=0.33\textwidth]{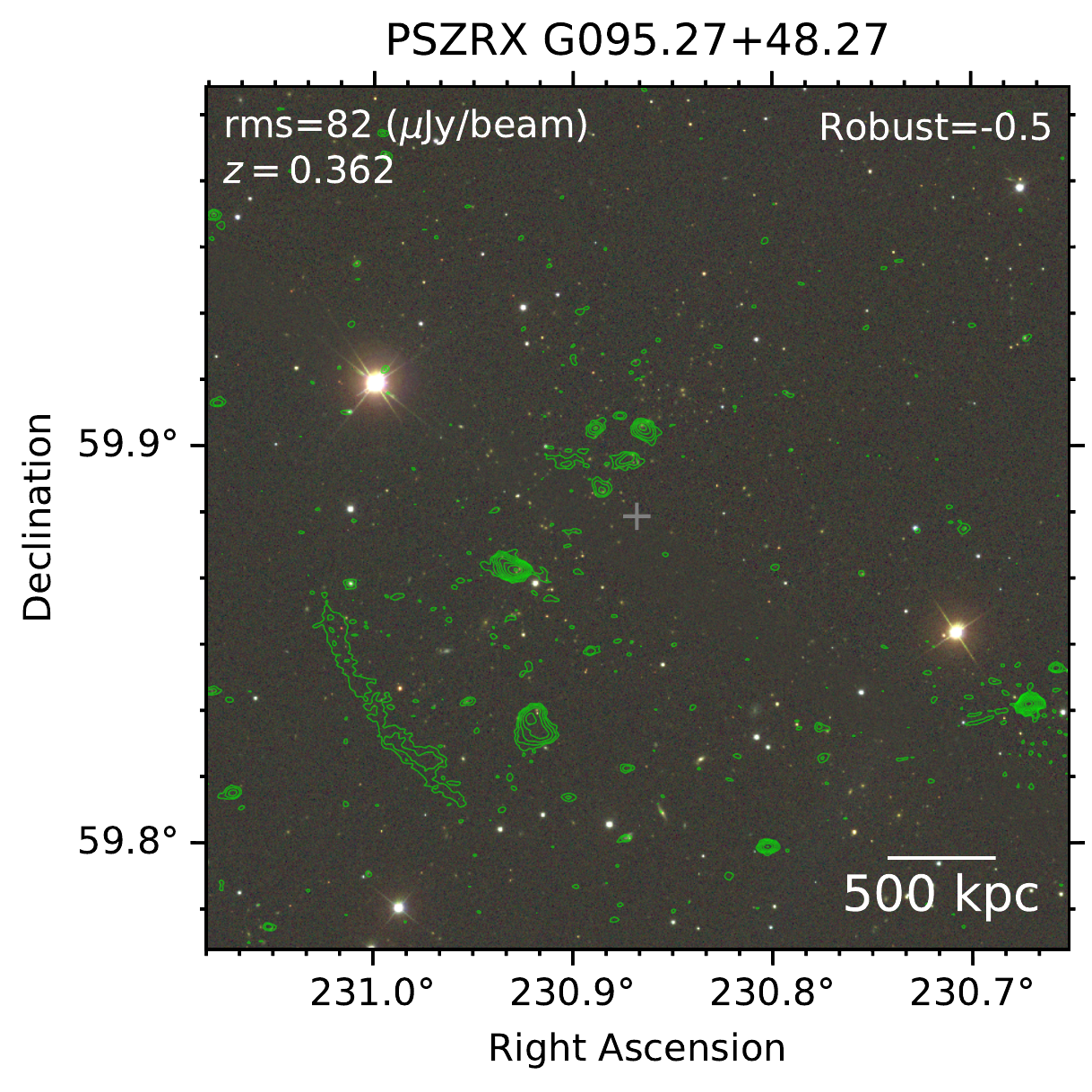}  \hfil
			\includegraphics[width=0.33\textwidth]{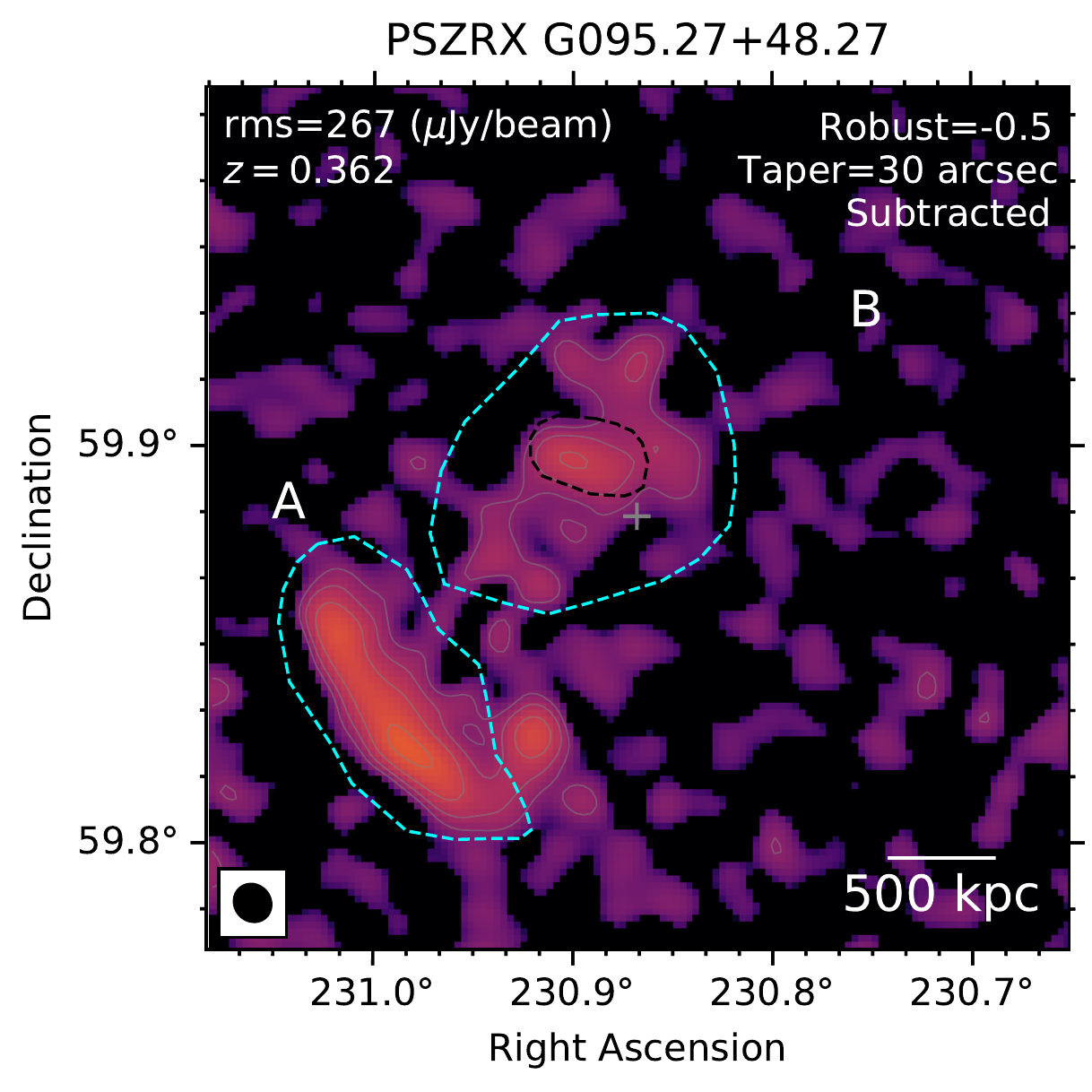}   };
		\draw (-7.7, -1.7) node {\color{white} (a)};
		\draw (-1.5, -1.7) node {\color{white} (b)};
		\draw (4.6, -1.55) node {\color{white} (c)};
	\end{tikzpicture}
	\caption{PSZRX~G095.27+48.27.  Image description is the same as that in Fig.~\ref{fig:abell84}.
	}
	\label{fig:PSZRXG095.27+48.27}
\end{figure*}

LOFAR images in Fig.~\ref{fig:PSZRXG095.27+48.27} show the new detection of diffuse sources in the SE and central regions of PSZRX~G095.27+48.27 ($z=0.362$), labelled as A and B. Source A in the SE region has an arc-like shape with a projected size of 1300~kpc$\times$450~kpc. The width of source A is 90~kpc as measured in the high-resolution image in the panel (a). The major axis of source A is along the NE-SW direction which is perpendicular to the line connecting the source and the cluster centre. The optical image in the panel (b) shows that source A is not related to any SDSS sources. The location and morphology of source A suggest that it is a radio relic that is generated in a dynamically-disturbed cluster with a SE-NW merger axis as confirmed by the SDSS data. In the central region of the cluster we detect a diffuse source (1200~kpc$\times$610~kpc) at $2\sigma$ that contains a number of radio galaxies and we classify it as a candidate radio halo.

\subsection{PSZRX~G100.21-30.38}
\label{sec:PSZRXG100.21-30.38}

\begin{figure*}[!ht]
	\centering
	\begin{tikzpicture}
		\draw (0, 0) node[inner sep=0] {\includegraphics[width=0.33\textwidth]{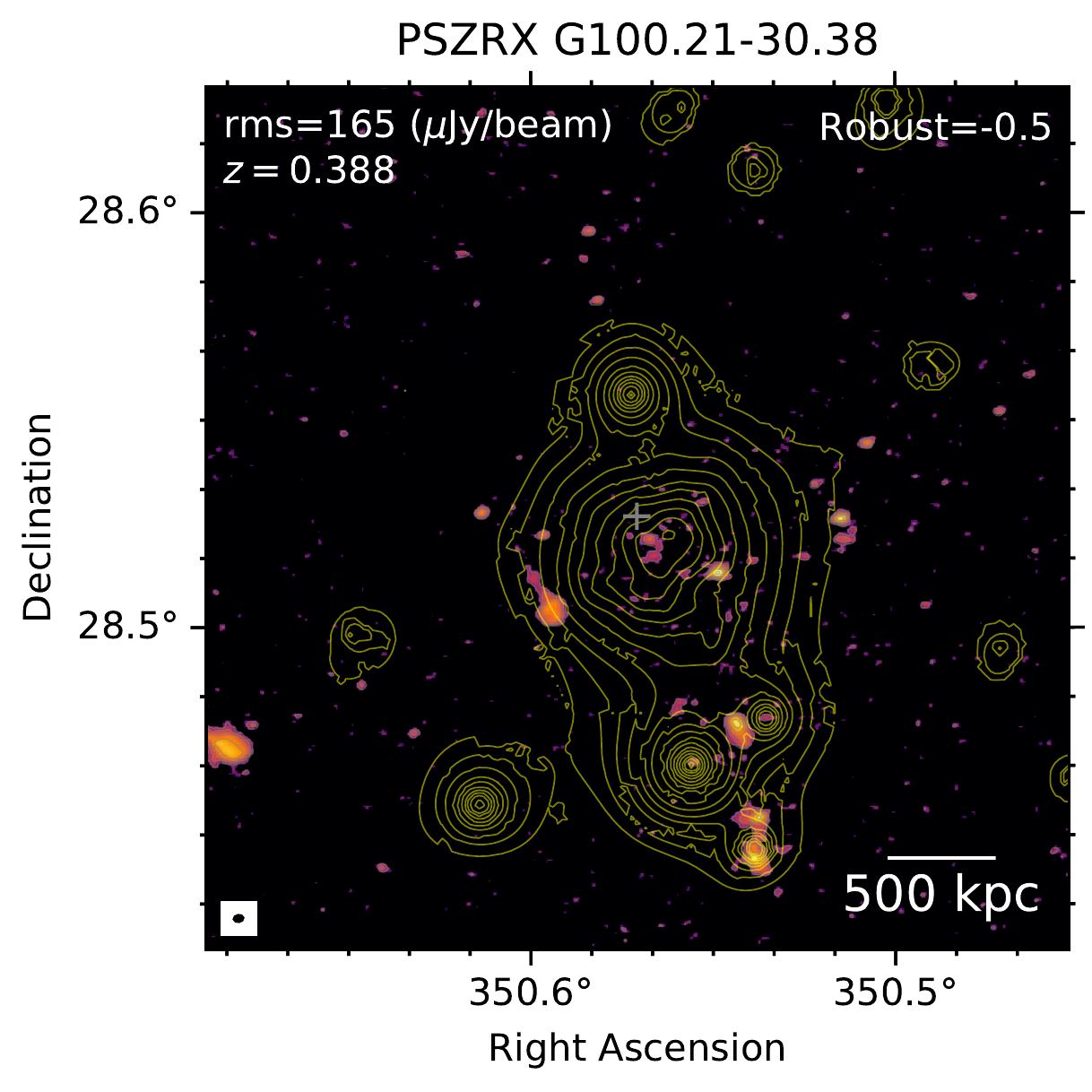}  \hfil
			\includegraphics[width=0.33\textwidth]{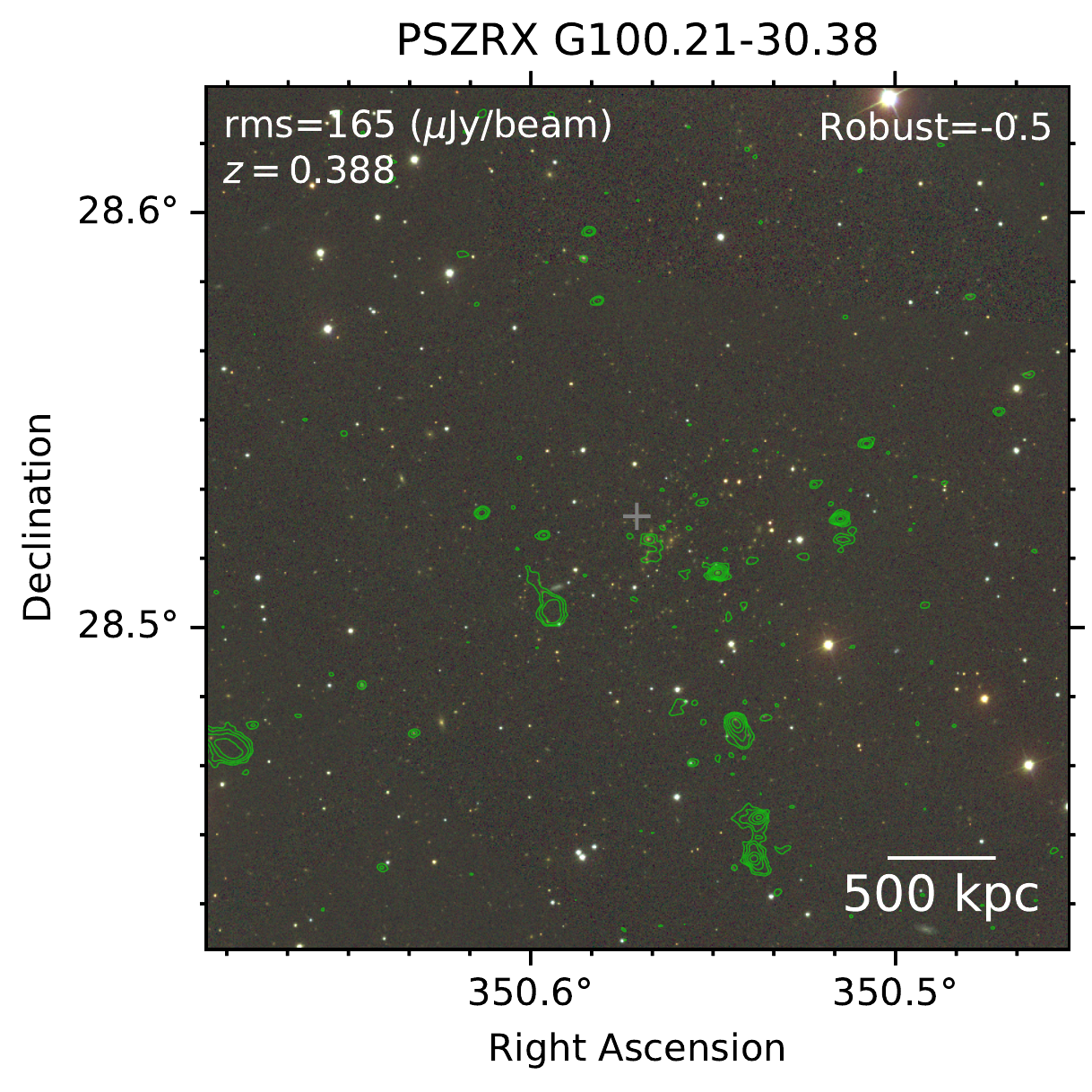}  \hfil
			\includegraphics[width=0.33\textwidth]{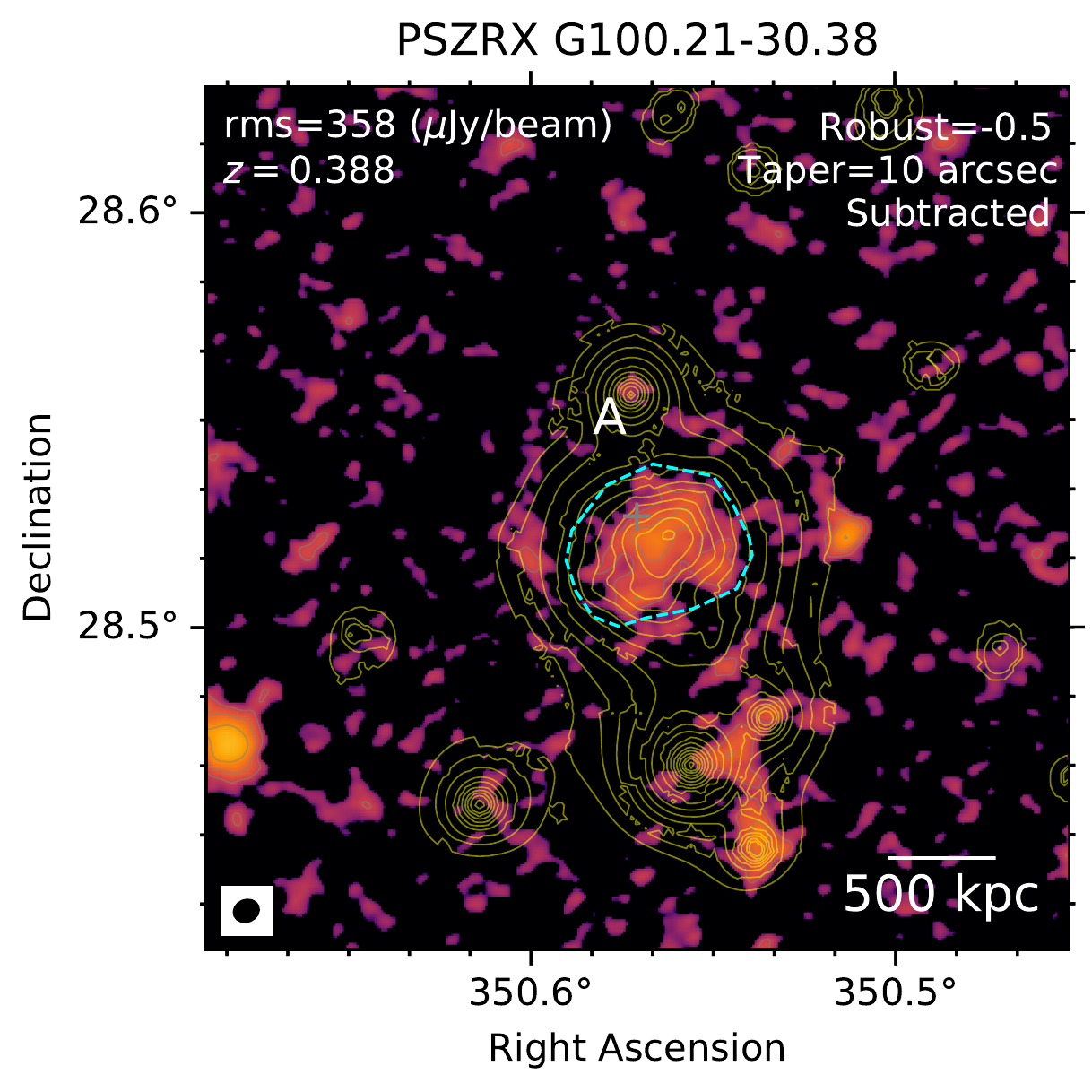}   };
		\draw (-7.7, -1.7) node {\color{white} (a)};
		\draw (-1.5, -1.7) node {\color{white} (b)};
		\draw (4.6, -1.55) node {\color{white} (c)};
	\end{tikzpicture}
	\caption{PSZRX~G100.21-30.38.  Image description is the same as that in Fig.~\ref{fig:abell84}.
	}
	\label{fig:PSZRXG100.21-30.38}
\end{figure*}

LOFAR observations shown in Fig.~\ref{fig:PSZRXG100.21-30.38} detect a new diffuse radio source in the centre of PSZRX~G100.21-30.38 ($z=0.388$). The diffuse source, named A, has a size of 600~kpc$\times$380~kpc and is elongated in the SE-NW direction that is aligned with the major axis of the galaxy distribution in the cluster, as seen in the SDSS image in panel (b). The radio emission from A roughly follows the XMM-Newton X-ray emission in the central region which is slightly elongated in the NW-SE direction. The morphology of the X-ray emission indicates that the cluster is in a dynamically-disturbed state. We classify source A as a radio halo due to its location, morphology, and its correlation with the X-rays.

\subsection{PSZRX~G102.17+48.88}
\label{sec:PSZRXG102.17+48.88}

\begin{figure*}[!ht]
	\centering
	\begin{tikzpicture}
		\draw (0, 0) node[inner sep=0] {\includegraphics[width=0.33\textwidth]{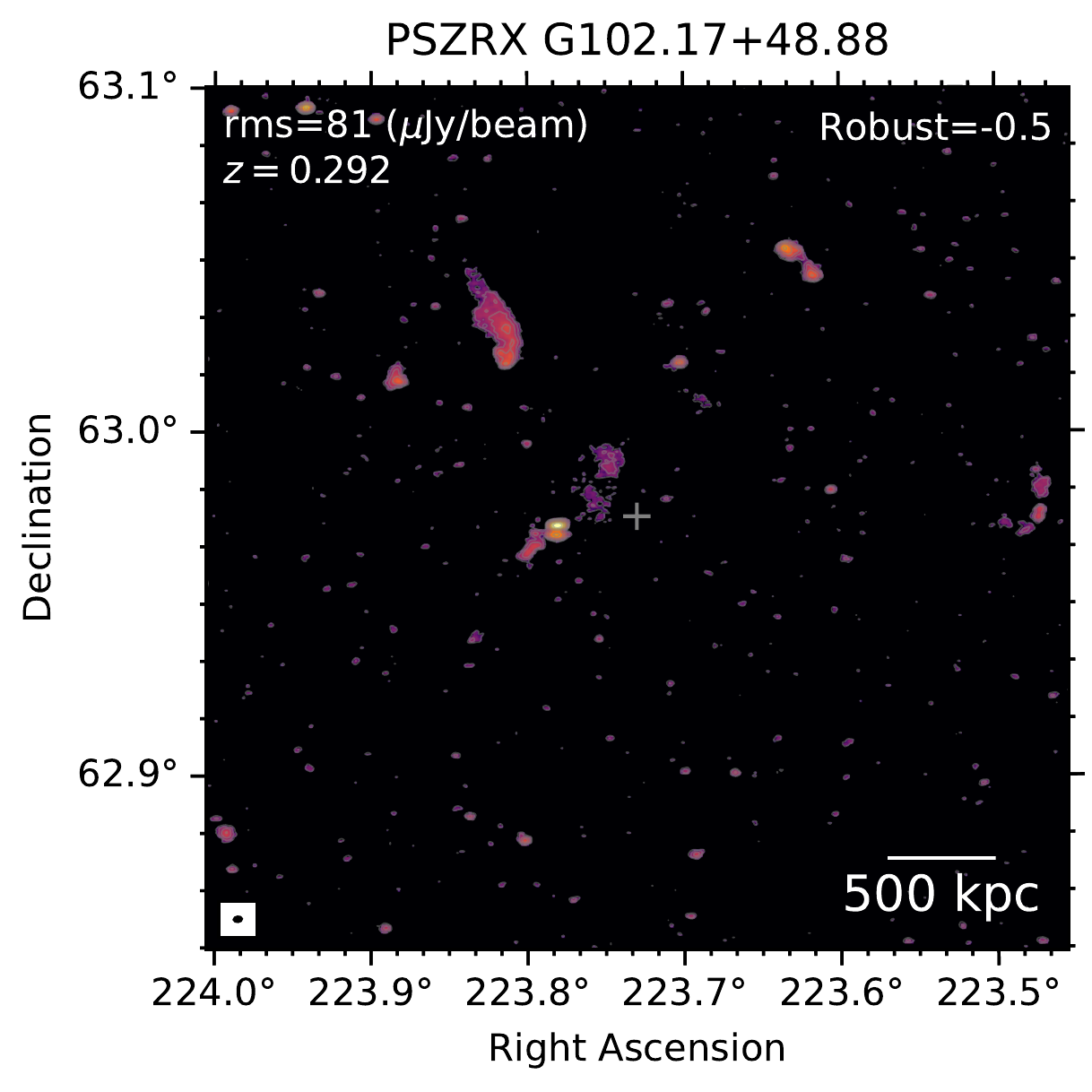}  \hfil
			\includegraphics[width=0.33\textwidth]{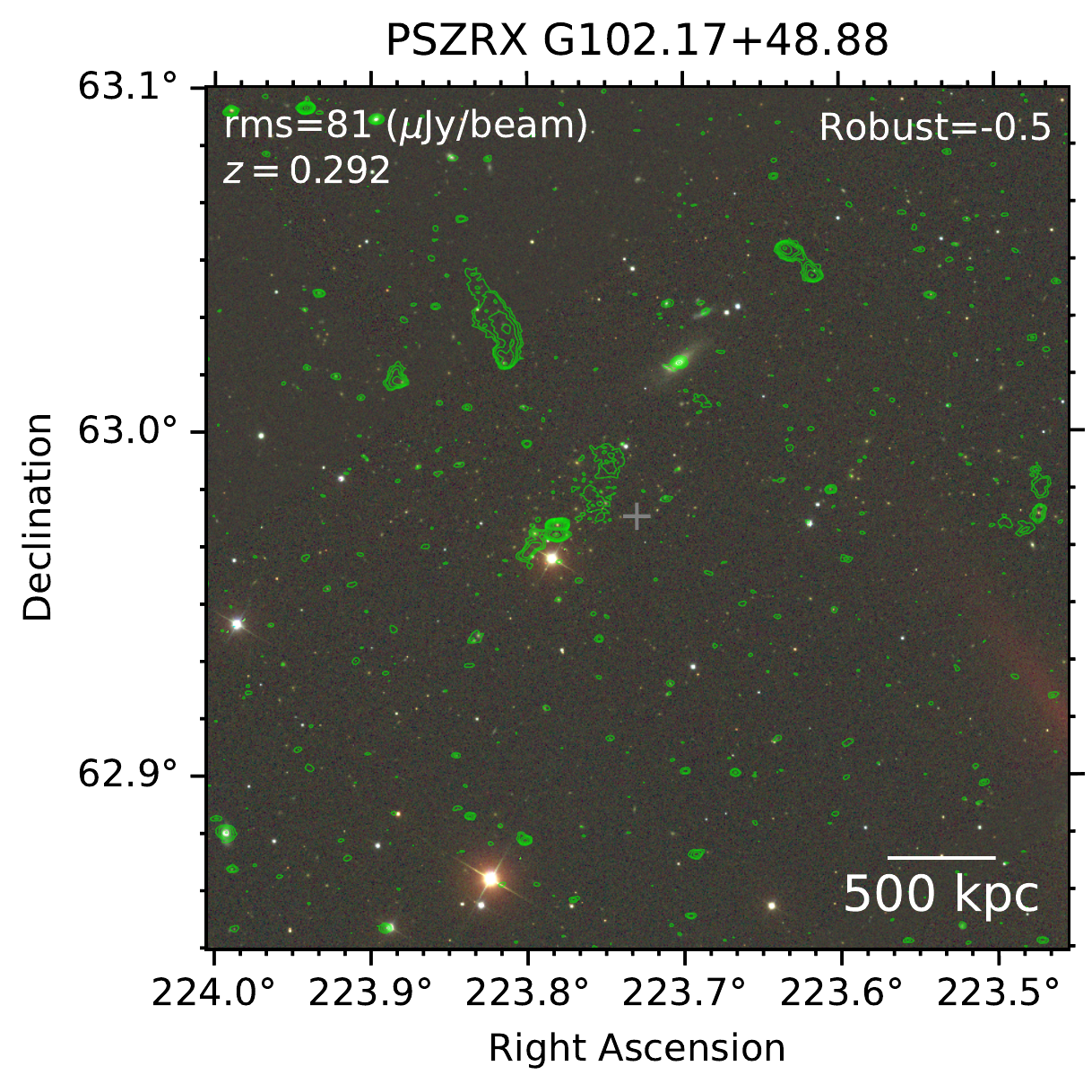}  \hfil
			\includegraphics[width=0.33\textwidth]{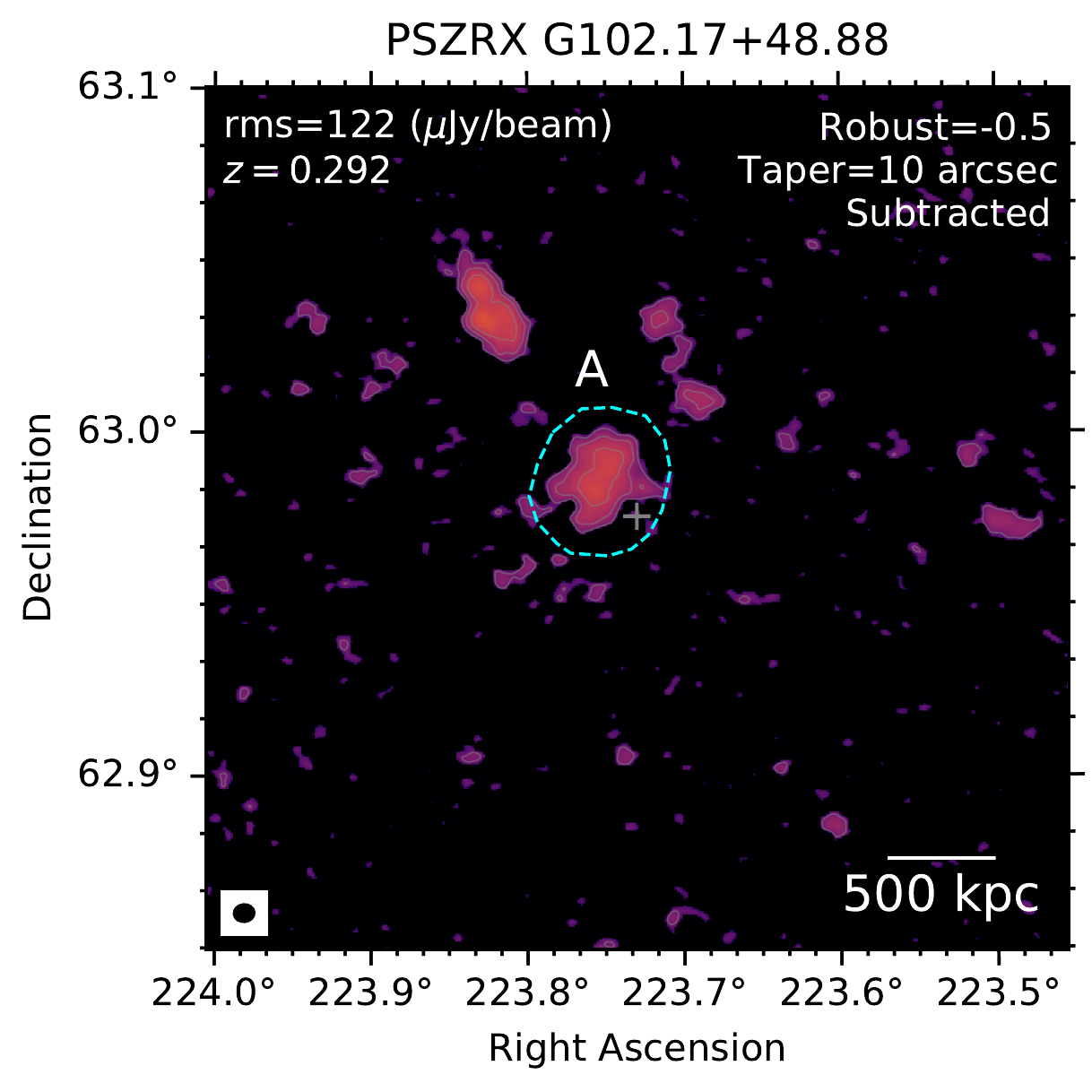}   };
		\draw (-7.7, -1.7) node {\color{white} (a)};
		\draw (-1.5, -1.7) node {\color{white} (b)};
		\draw (4.6, -1.55) node {\color{white} (c)};
	\end{tikzpicture}
	\caption{PSZRX~G102.17+48.88. Image description is the same as that in Fig.~\ref{fig:abell84}.}
	\label{fig:PSZRXG102.17+48.88}
\end{figure*}

In Fig.~\ref{fig:PSZRXG102.17+48.88} we show LOFAR images of PSZRX~G102.17+48.88 ($z=0.292$). In the low-resolution source-subtracted image (panel a), diffuse emission with a projected size of 460~kpc$\times$340~kpc, elongated in the north-south direction is newly detected. To the SE direction, compact radio sources are seen to be associated with SDSS optical counterparts. The galaxy distribution in the optical image is elongated in the NW-SE direction, similar to the orientation of the diffuse radio source. This implies a connection of the diffuse radio source to the morphology of the ICM, suggesting that it could be a radio halo. Although its size is smaller than typically found for radio halos in other systems, this could be due to the lower sensitivity of the observations that prevents the detection of the faint emission in the outer region. Alternatively, the diffuse emission could be related to AGN activities, which could explain the patchy emission in the high-resolution image.

\subsection{PSZRX~G116.06+80.14}
\label{sec:PSZRXG116.06+80.14}

\begin{figure*}[!ht]
	\centering
	\begin{tikzpicture}
    		\draw (0, 0) node[inner sep=0] {\includegraphics[width=0.33\textwidth]{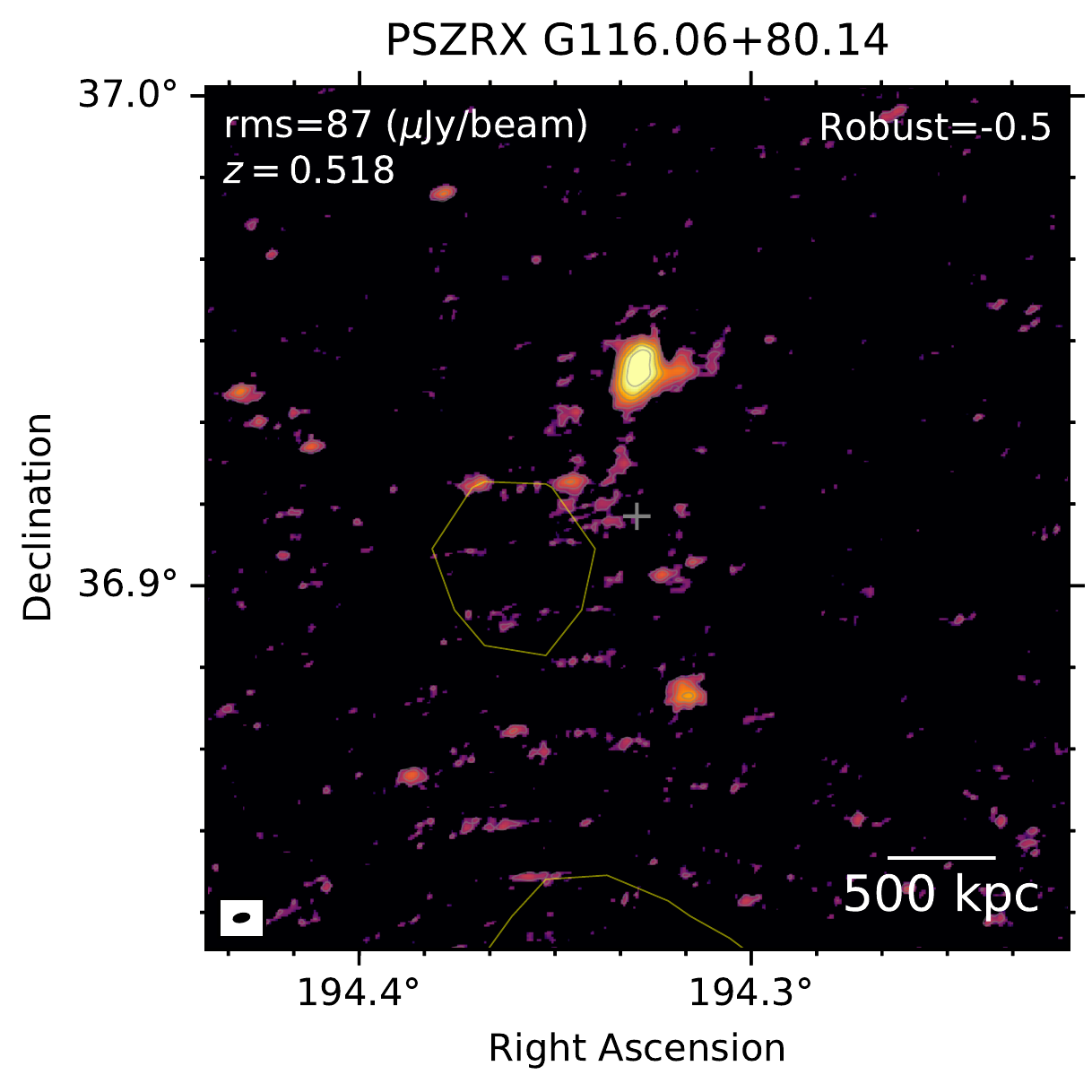}  \hfil
    			\includegraphics[width=0.33\textwidth]{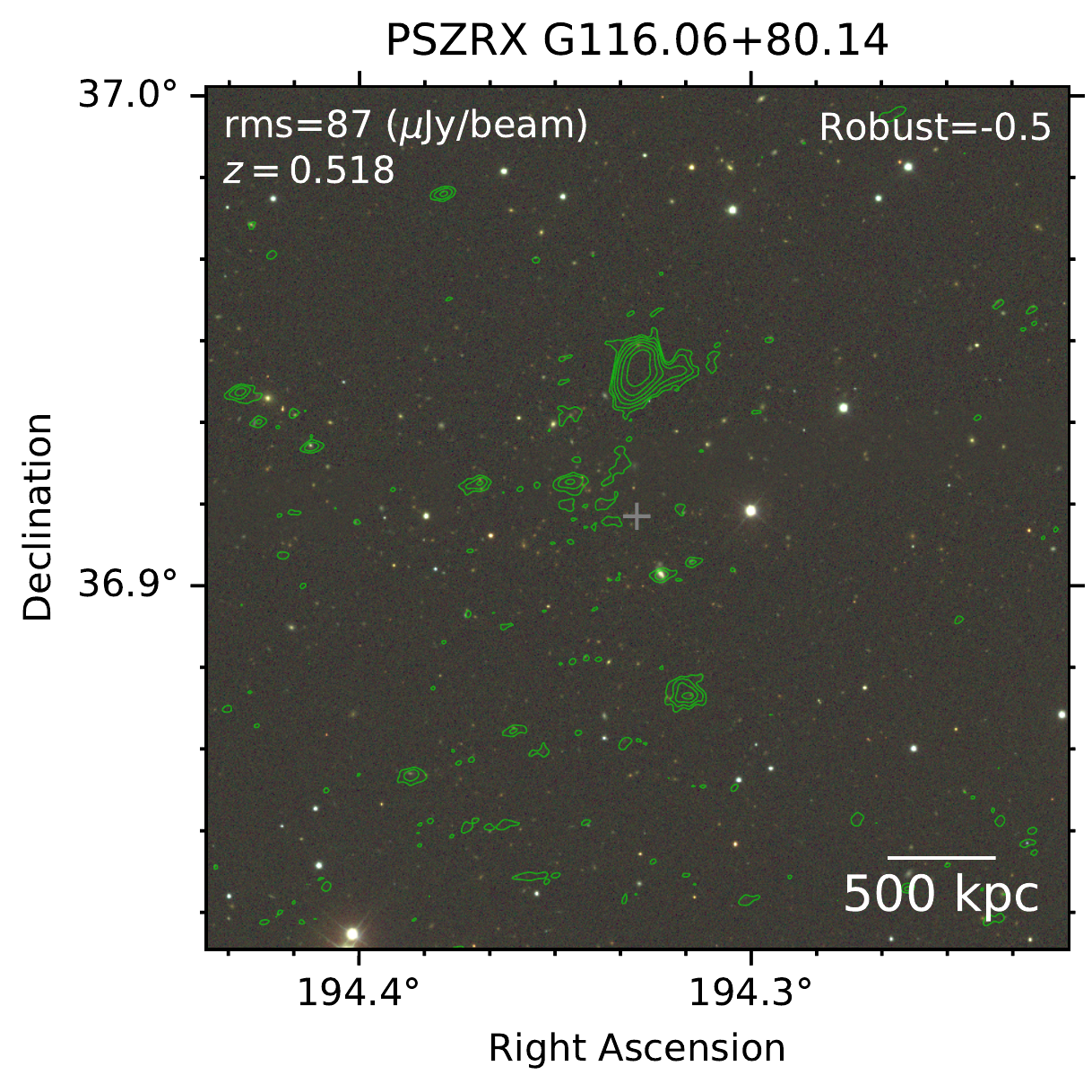}  \hfil
			\includegraphics[width=0.33\textwidth]{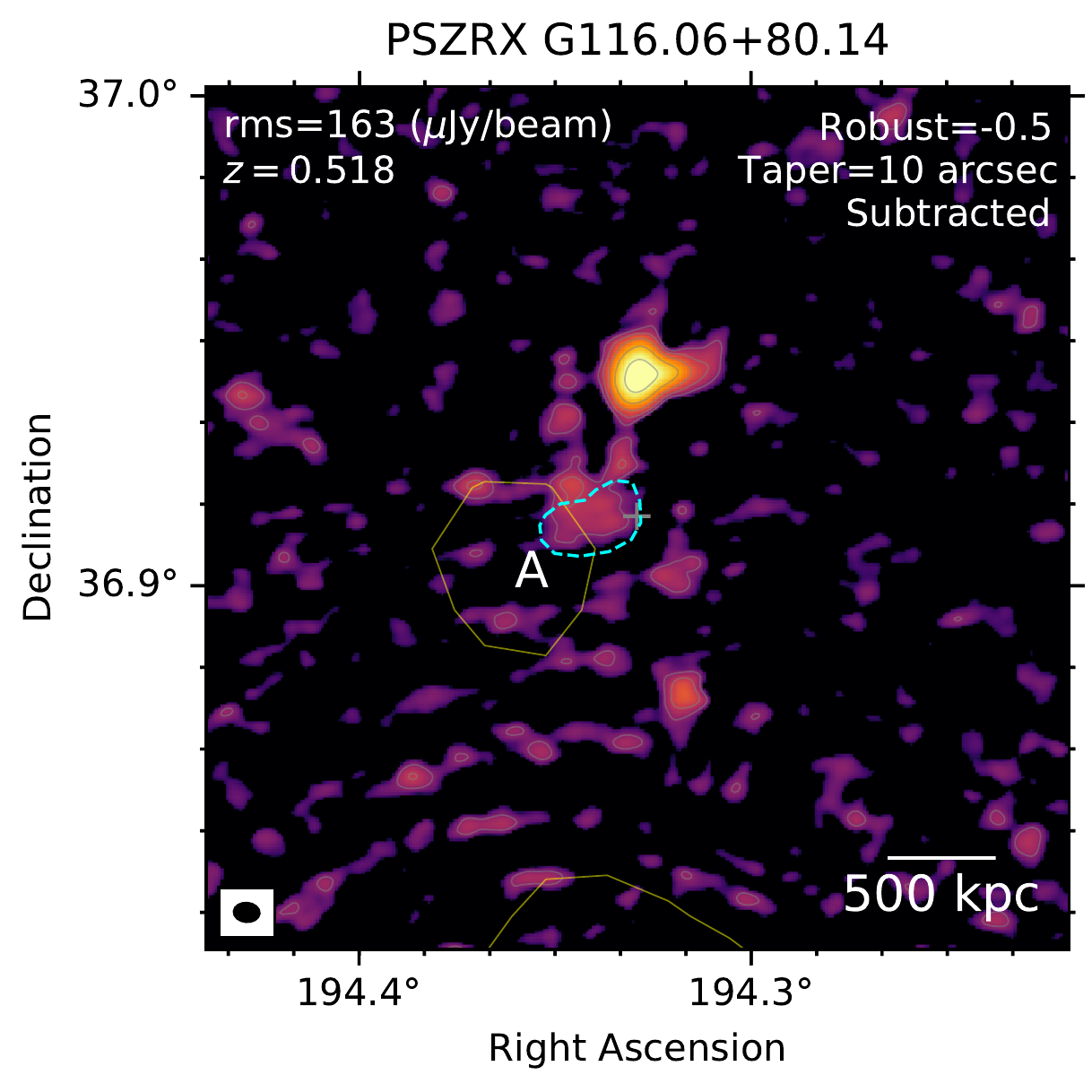}   };
		\draw (-7.7, -1.7) node {\color{white} (a)};
		\draw (-1.5, -1.7) node {\color{white} (b)};
		\draw (4.6, -1.55) node {\color{white} (c)};
	\end{tikzpicture}
	\caption{PSZRX~G116.06+80.14.  Image description is the same as that in Fig.~\ref{fig:abell84}.
	}
	\label{fig:PSZRXG116.06+80.14}
\end{figure*}

LOFAR observations detect faint diffuse emission, named A, in the central region of PSZRX~G116.06+80.14 ($z=0.518$, the highest-$z$ system in the sample), as seen in Fig.~\ref{fig:PSZRXG116.06+80.14}. The diffuse source has a projected size of 340~kpc. The high-resolution image shows patchy emission from this source. The radio emission is not spatially correlated with the X-ray low-SNR emission from ROSAT. It is unclear if source A is part of a larger radio halo or is associated with the compact radio source towards its NE direction.

\subsection{PSZRX~G181.53+21.43}
\label{sec:PSZRXG181.53+21.43}

\begin{figure*}[!ht]
	\centering
	\begin{tikzpicture}
		\draw (0, 0) node[inner sep=0] {\includegraphics[width=0.33\textwidth]{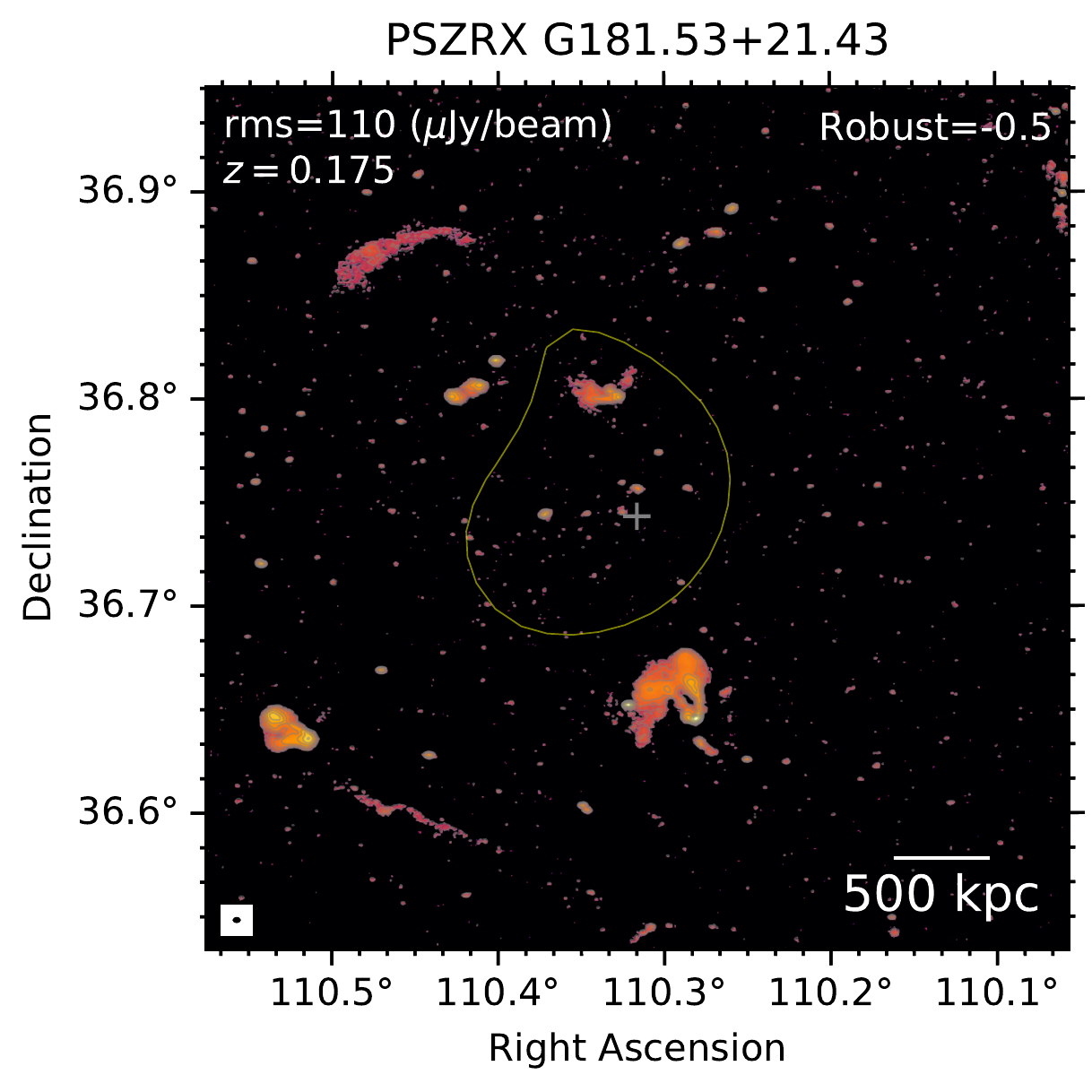}  \hfil
			\includegraphics[width=0.33\textwidth]{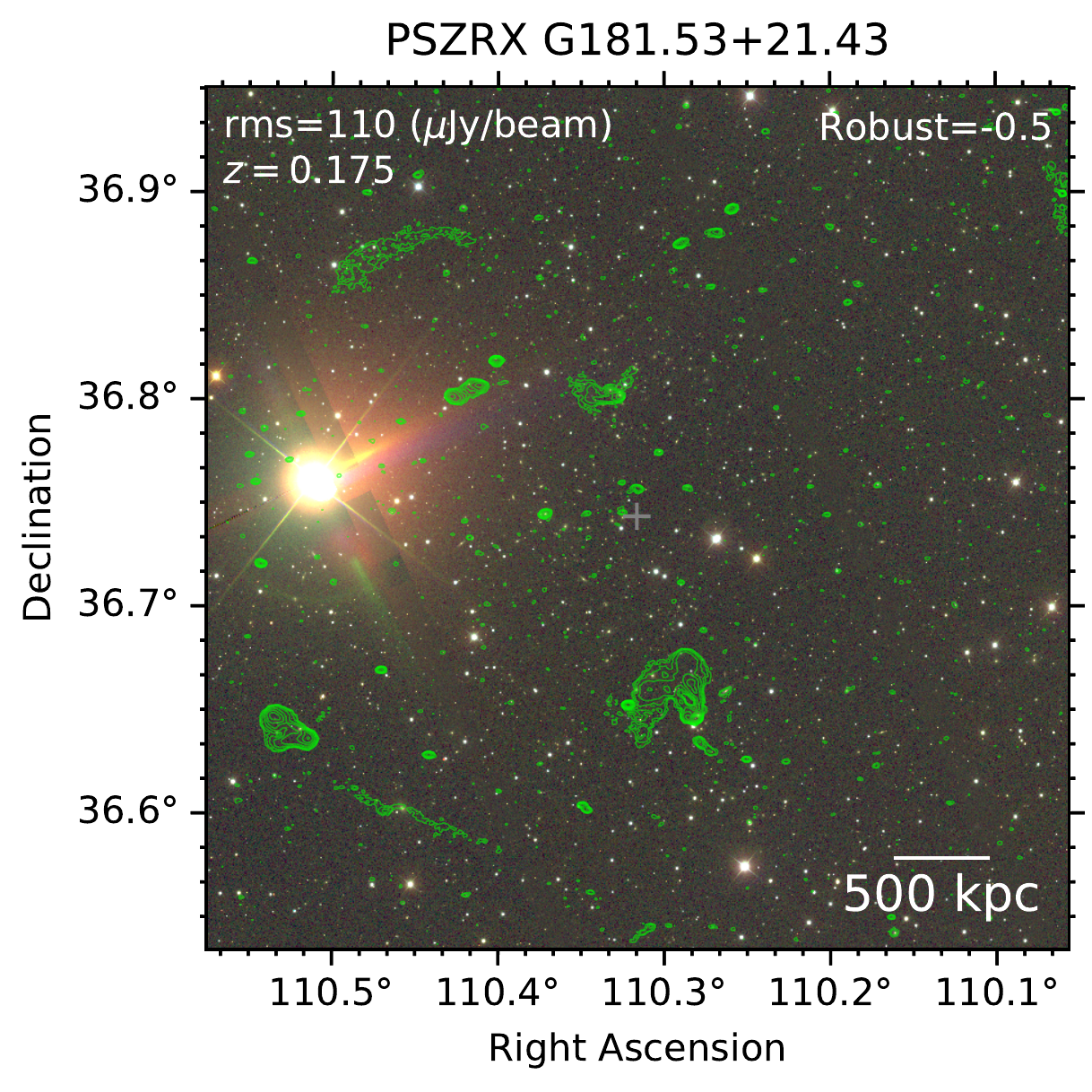}  \hfil
			\includegraphics[width=0.33\textwidth]{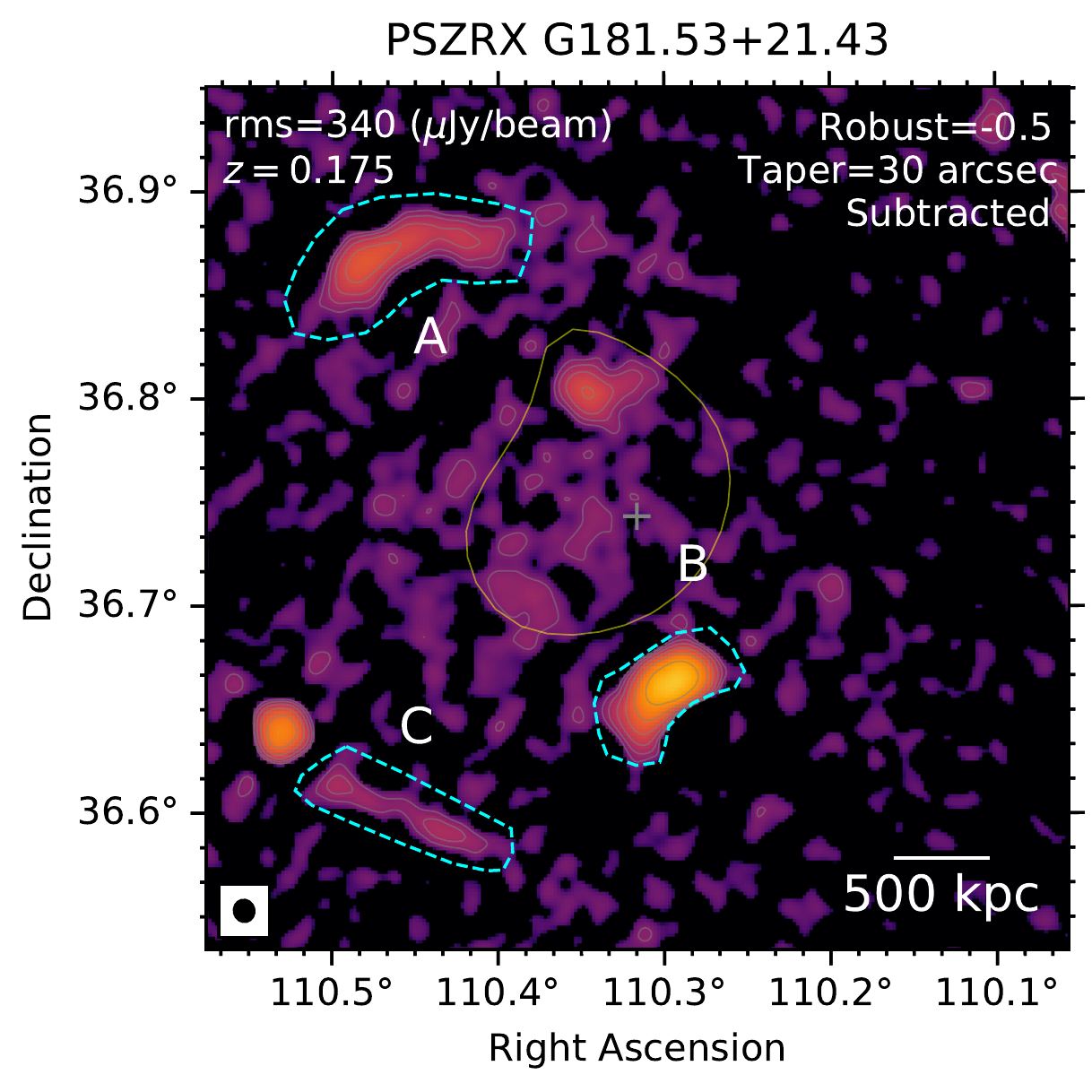}   };
		\draw (-7.7, -1.7) node {\color{white} (a)};
		\draw (-1.5, -1.7) node {\color{white} (b)};
		\draw (4.6, -1.55) node {\color{white} (c)};
	\end{tikzpicture}
	\caption{PSZRX~G181.53+21.43. Image description is the same as that in Fig.~\ref{fig:abell84}.
	}
	\label{fig:PSZRXG181.53+21.43}
\end{figure*}

In Fig.~\ref{fig:PSZRXG181.53+21.43}, diffuse sources are detected in the outskirts of PSZRX~G181.53+21.43 ($z=0.175$). The diffuse sources are labelled A--C. Towards the NE (1.7~Mpc) of the cluster centre, a new diffuse source (A) with a projected size of 1000~kpc$\times$250~kpc elongated in the NW-SE direction is detected. Source A is not obviously connected to any SDSS optical sources, as seen in the panel (b) of Fig.~\ref{fig:PSZRXG181.53+21.43}. The observed properties of source A suggest that it is a radio relic. In the SW direction (1~Mpc) from the cluster centre we find a wide-angle tailed (WAT) radio galaxy with the core in the outer region. The end tail of WAT radio galaxy (source B) is highly disturbed and is oriented in a direction similar to the major axis of source A. The optical SDSS image shows that most of the galaxies are distributed in the NE-SW direction, implying that A and B have been shaped by the dynamics of the cluster. In the SE region, a thread-like source (C; 900~kpc$\times$90~kpc in projection) is seen in the LOFAR images. An SDSS optical counterpart in the middle of C suggests that it is a Fanaroff-Riley class I (FR-I) radio galaxy. In the central region of the cluster, there is a hint of diffuse emission which might be a tentative faint radio halo. We do not measure the size and the flux density of the candidate halo as its boundary is unclear in the current images.

\subsection{PSZRX~G195.91+62.83}
\label{sec:PSZRXG195.91+62.83}

\begin{figure*}[!ht]
	\centering
	\begin{tikzpicture}
		\draw (0, 0) node[inner sep=0] {\includegraphics[width=0.33\textwidth]{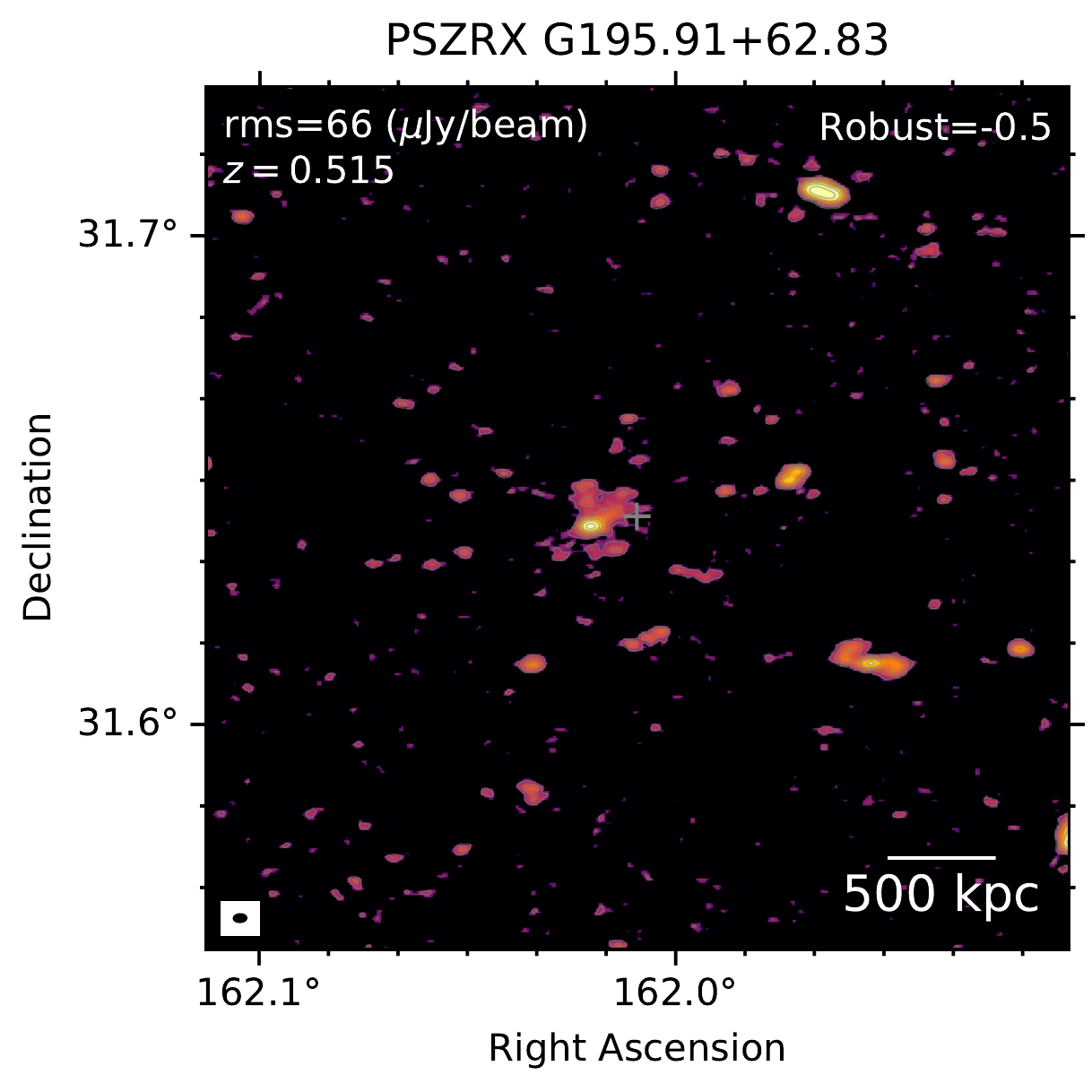}  \hfil
			\includegraphics[width=0.33\textwidth]{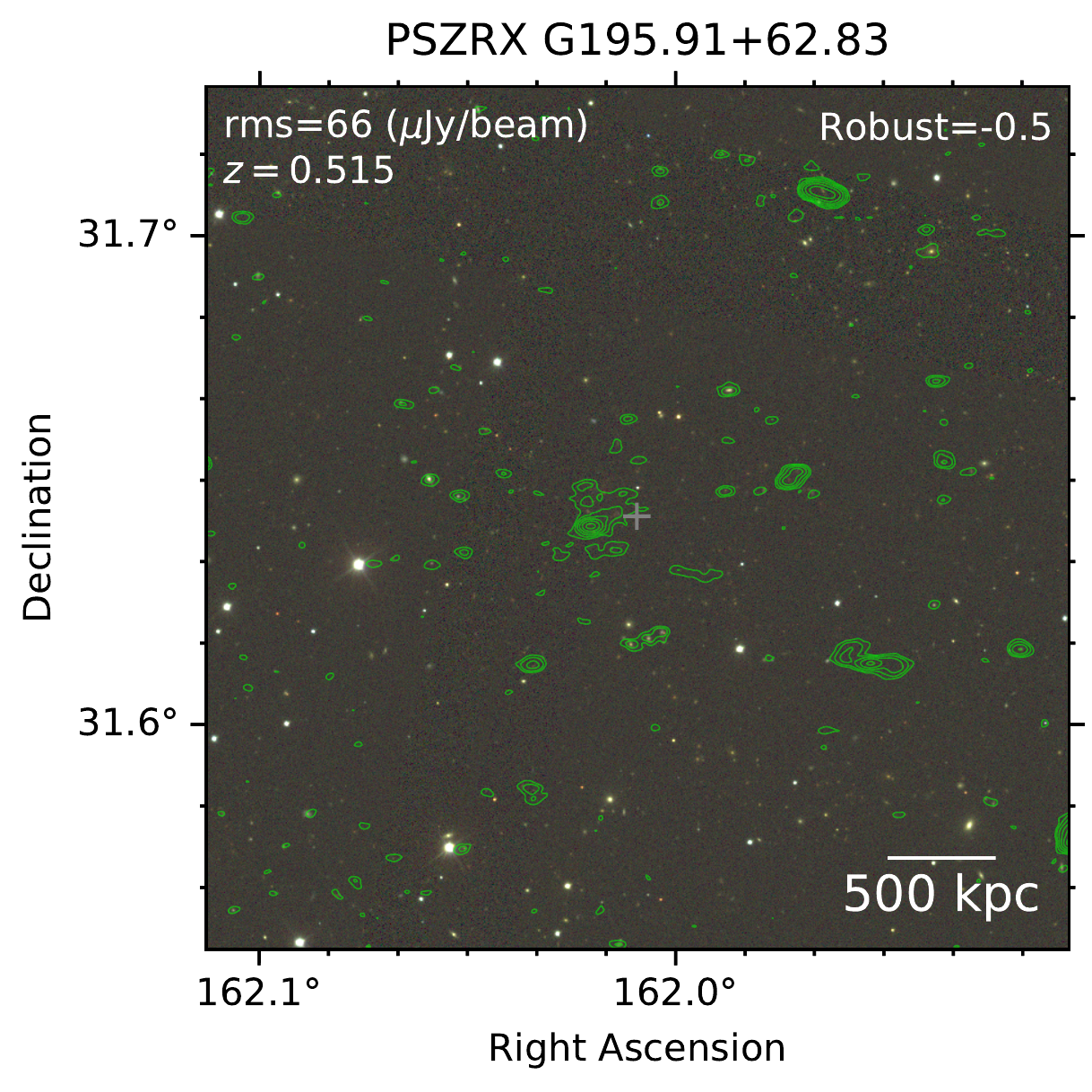}  \hfil
			\includegraphics[width=0.33\textwidth]{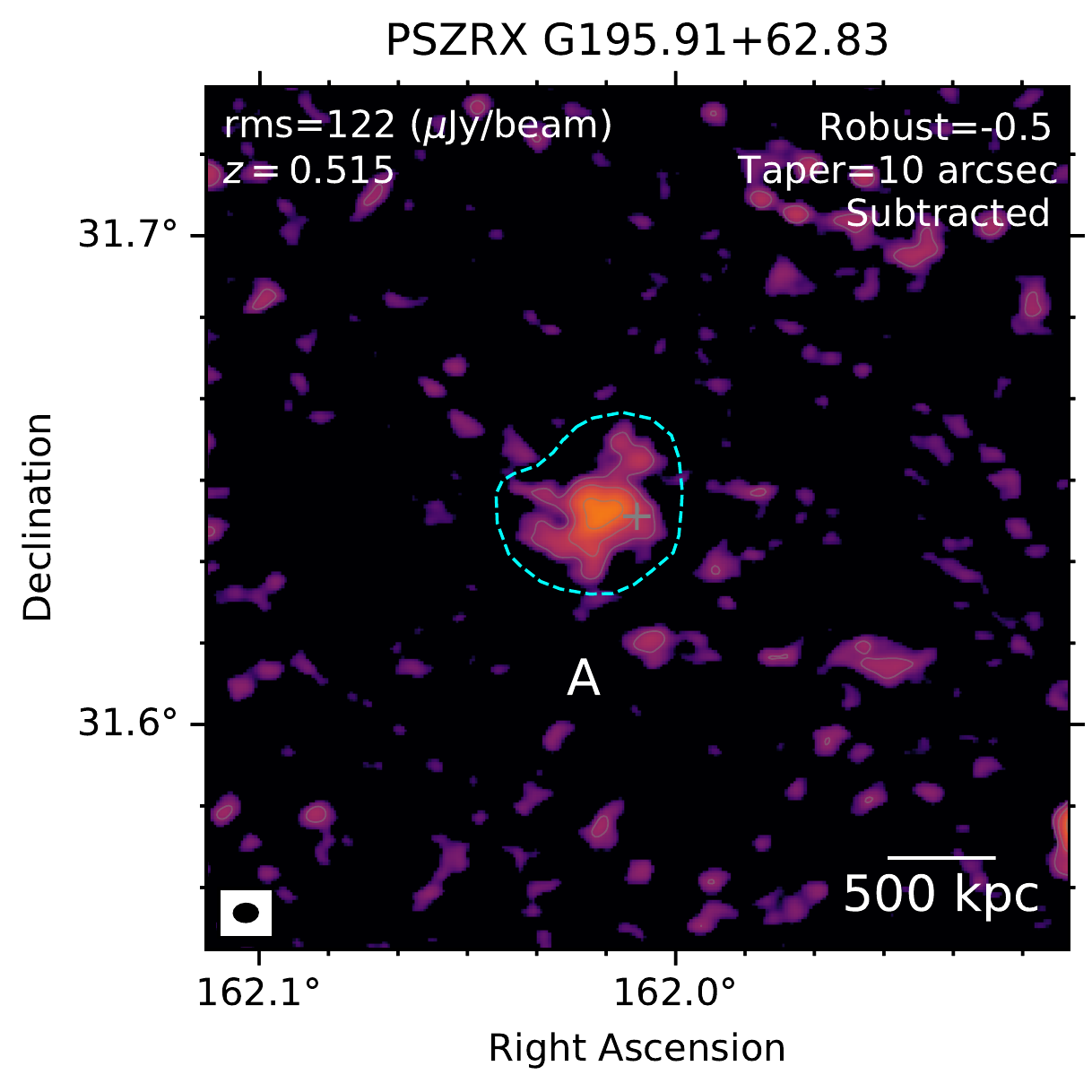}   };
		\draw (-7.7, -1.7) node {\color{white} (a)};
		\draw (-1.5, -1.7) node {\color{white} (b)};
		\draw (4.6, -1.55) node {\color{white} (c)};
	\end{tikzpicture}
	\caption{PSZRX~G195.91+62.83. Image description is the same as that in Fig.~\ref{fig:abell84}.
	}
	\label{fig:PSZRXG195.91+62.83}
\end{figure*}

In Fig.~\ref{fig:PSZRXG195.91+62.83}, LOFAR images show the presence of a new diffuse radio source around a radio galaxy in the centre of PSZRX~G195.91+62.83 ($z=0.515$). When removing the emission from the radio galaxy, a diffuse source is seen in the low-resolution image (i.e. the panel c). The diffuse source has a projected size of 600~kpc and we classify it as a radio halo.

\subsection{WHL~J002056.4+221752}
\label{sec:WHLJ002056.4+221752}

\begin{figure*}[!ht]
	\centering
	\begin{tikzpicture}
		\draw (0, 0) node[inner sep=0] {\includegraphics[width=0.33\textwidth]{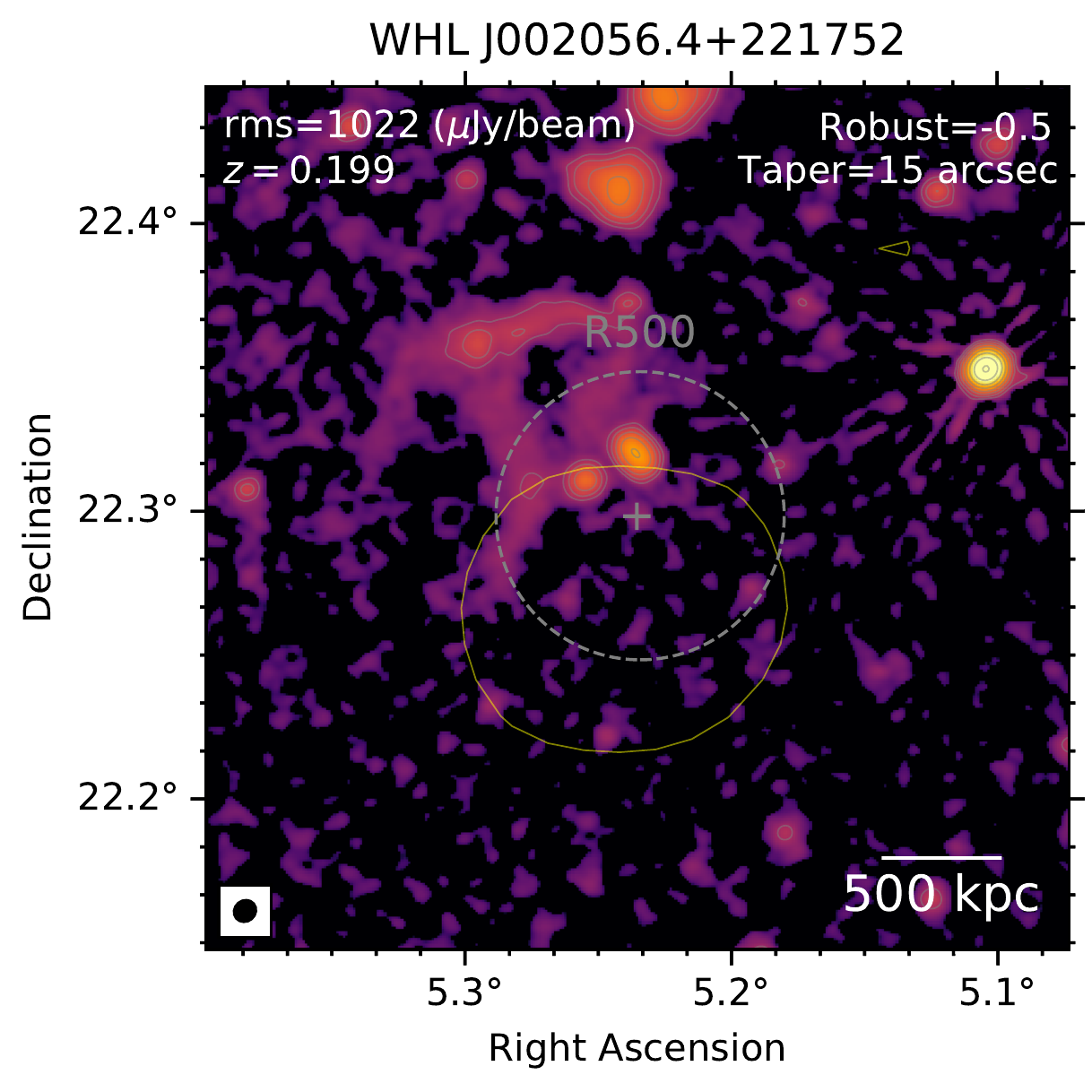}  \hfil
			\includegraphics[width=0.33\textwidth]{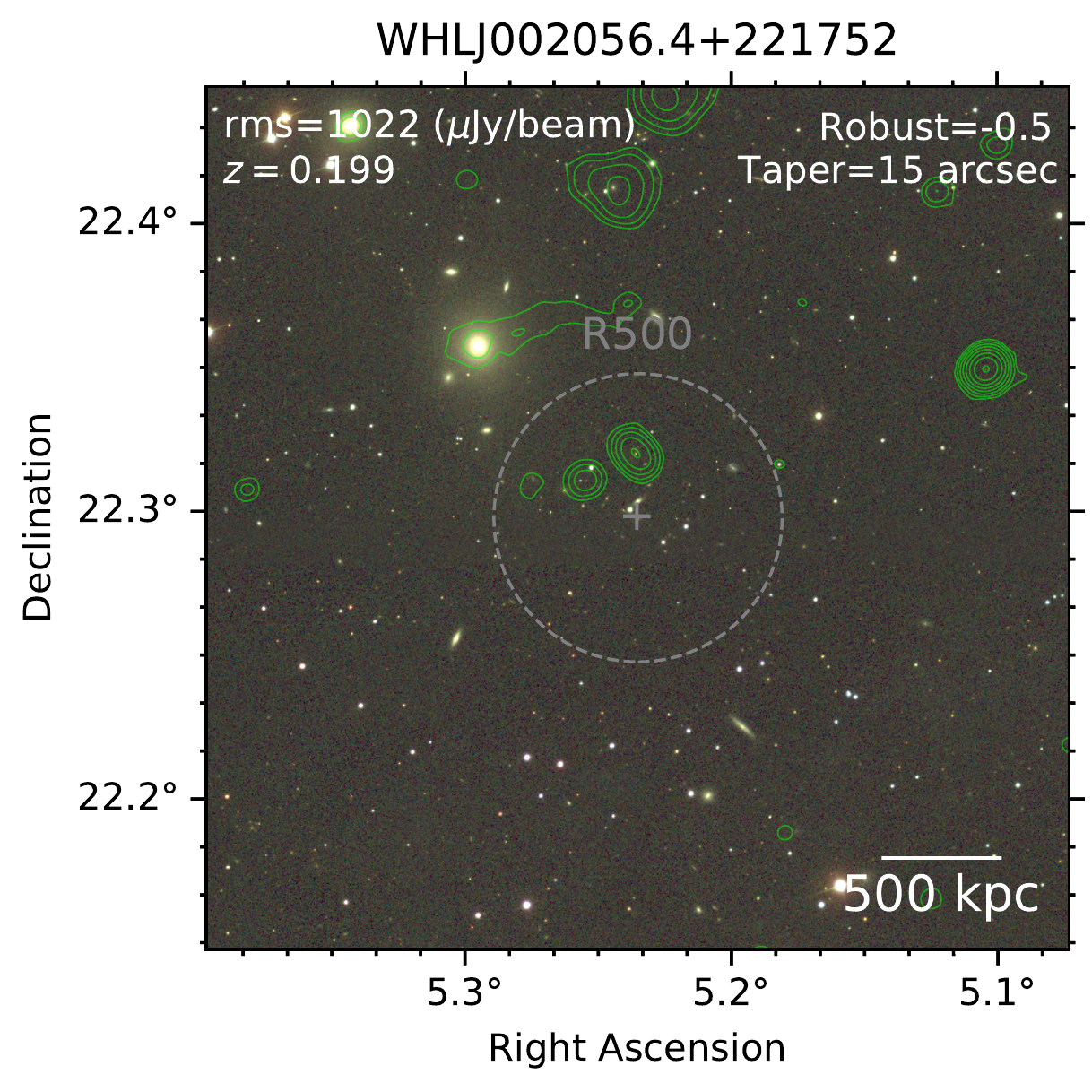}  \hfil
			\includegraphics[width=0.33\textwidth]{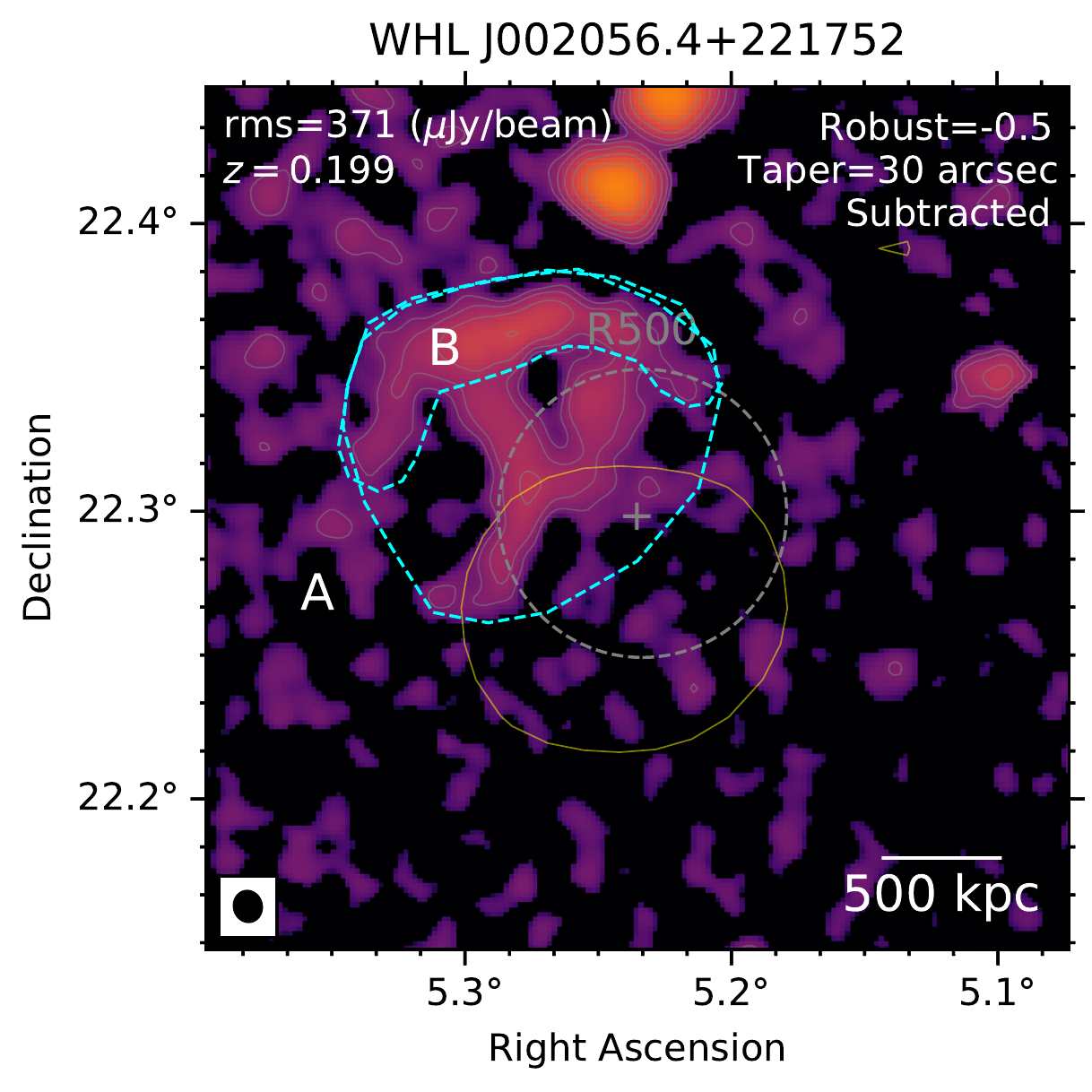}   };
		\draw (-7.7, -1.7) node {\color{white} (a)};
		\draw (-1.5, -1.7) node {\color{white} (b)};
		\draw (4.6, -1.55) node {\color{white} (c)};
	\end{tikzpicture}
	\caption{WHL~J002056.4+221752. Image description is the same as that in Fig.~\ref{fig:abell84}.
	}
	\label{fig:WHLJ002056.4+221752}
\end{figure*}

As seen in the LOFAR images in Fig.~\ref{fig:WHLJ002056.4+221752}, a new diffuse structure is seen in the NE region of the galaxy cluster WHL~J002056.4+221752 ($z=0.199$). The entire diffuse structure, labelled A, has a rectangular shape, spanning from the cluster centre to 750~kpc in the outskirts. The outer size of the source (1~Mpc) is smaller than that of the inner region (640~kpc). In the northern region, the source is brightest, named B. In the panels b and c, two compact radio galaxies are found in the southern side of the structure. It is unclear if the radio galaxies physically connect to the northern structure. The nature of the radio diffuse structure is unknown.

\subsection{WHL~J002311.7+251510}
\label{sec:WHLJ002311.7+251510}

\begin{figure*}[!ht]
	\centering
	\begin{tikzpicture}
		\draw (0, 0) node[inner sep=0] {\includegraphics[width=0.33\textwidth]{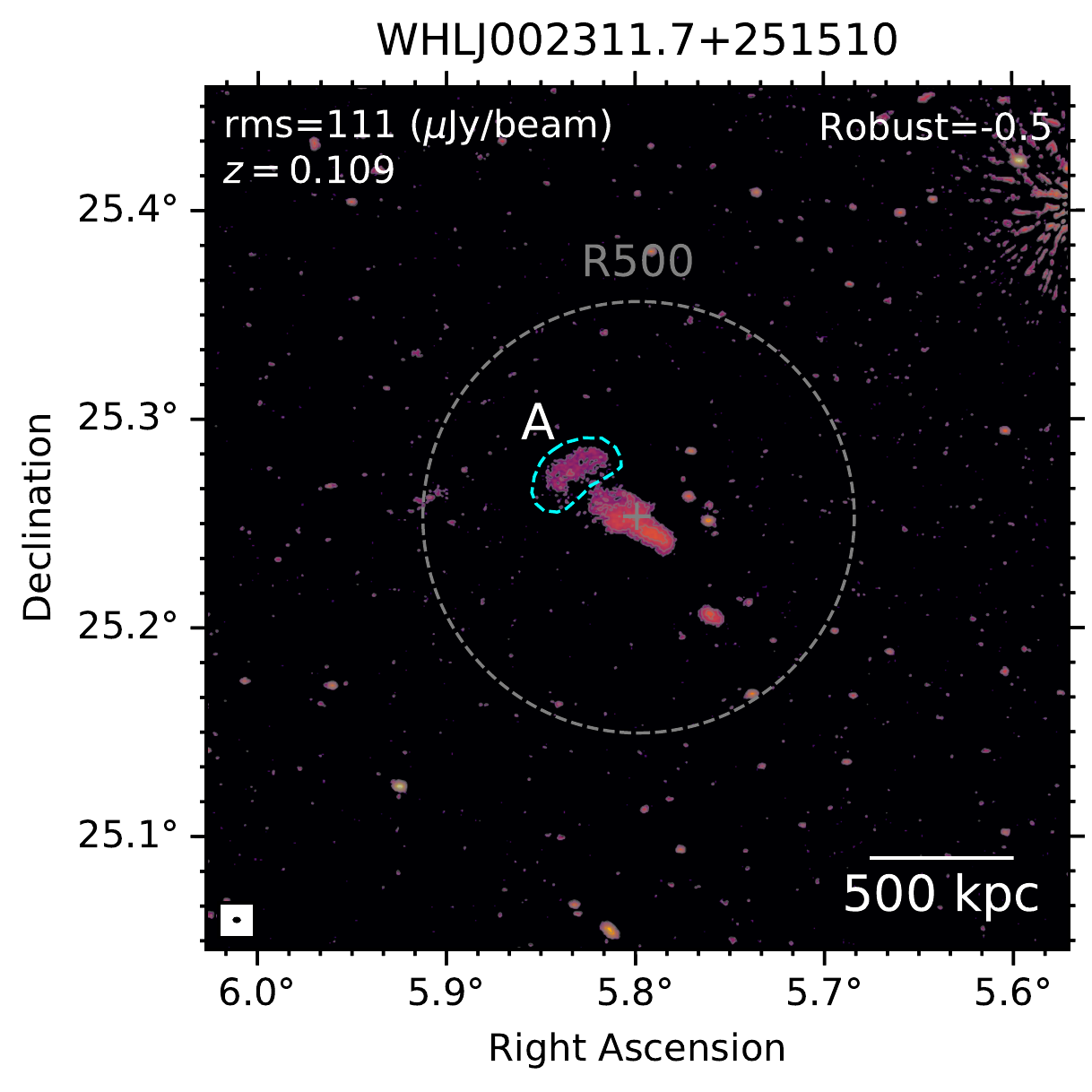}  \hfil
			\includegraphics[width=0.33\textwidth]{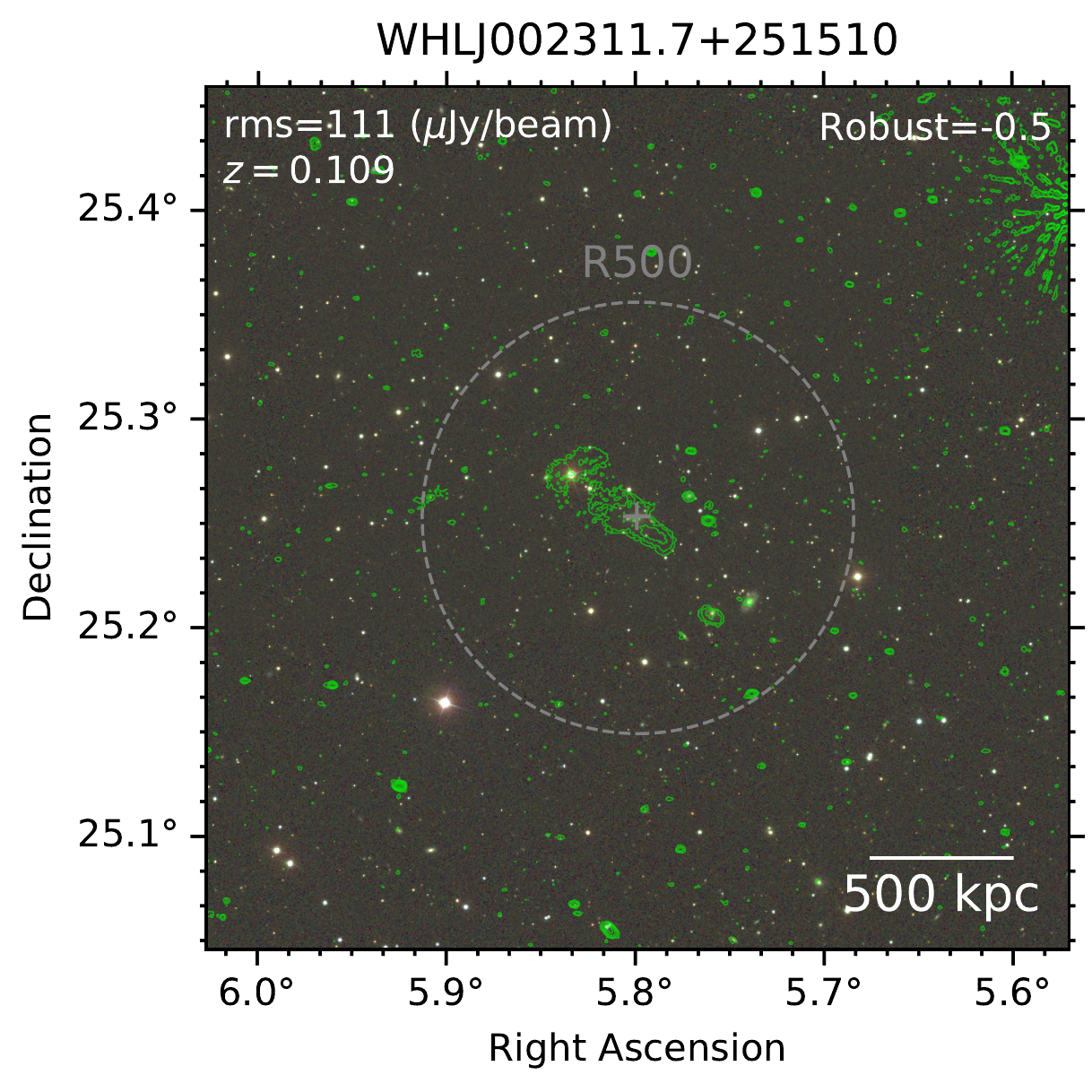}  \hfil
			\includegraphics[width=0.33\textwidth]{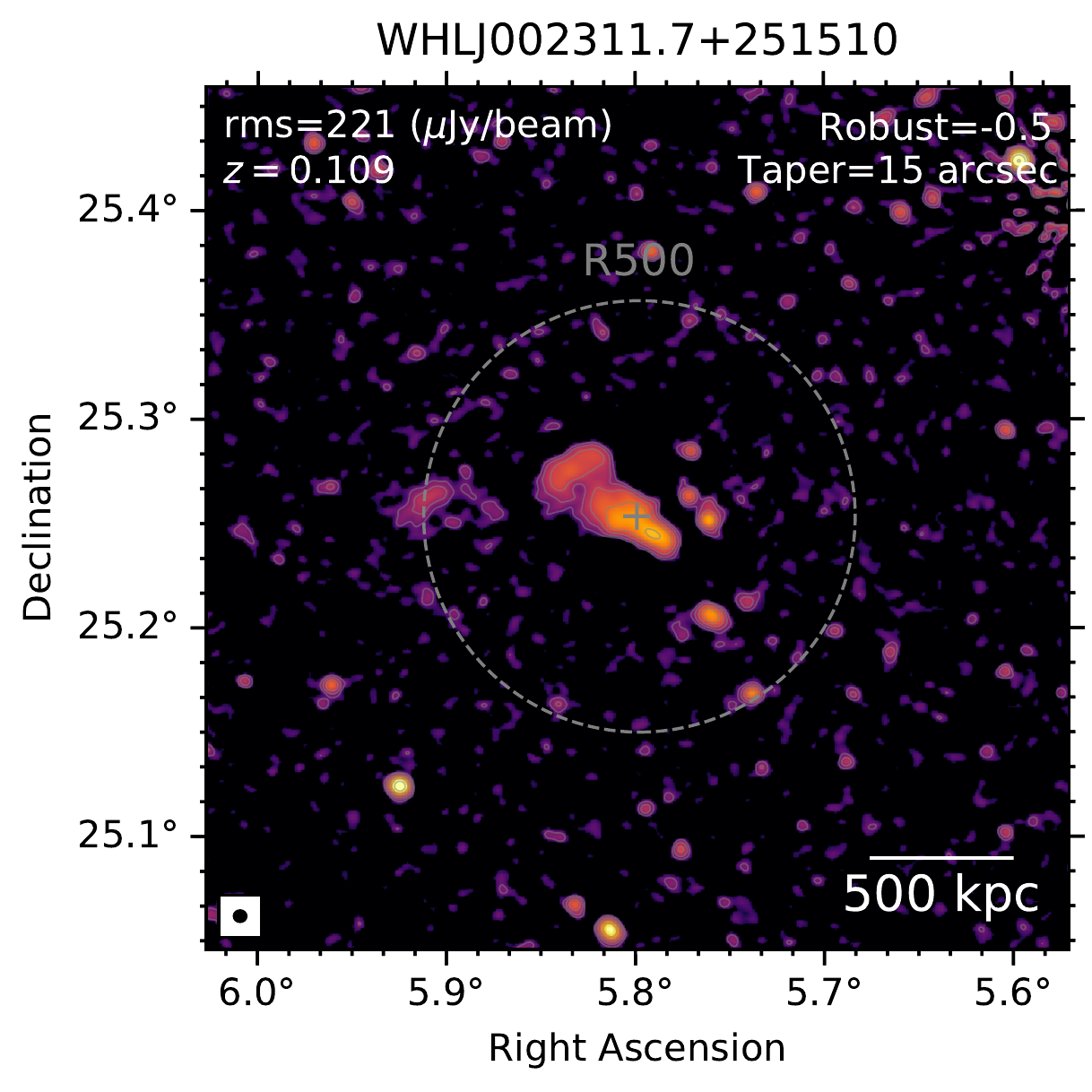}   };
		\draw (-7.7, -1.7) node {\color{white} (a)};
		\draw (-1.5, -1.7) node {\color{white} (b)};
		\draw (4.6, -1.55) node {\color{white} (c)};
	\end{tikzpicture}
	\caption{WHL~J002311.7+251510. Image description is the same as that in Fig.~\ref{fig:abell84}.
	}
	\label{fig:WHLJ002311.7+251510}
\end{figure*}

In Fig.~\ref{fig:WHLJ002311.7+251510}, a tailed radio galaxy is detected with LOFAR in the centre of the galaxy cluster WHL~J002311.7+251510 ($z=0.109$). At the end of the tail excess diffuse emission is detected, with a projected size of 230~kpc$\times$120~kpc. In the high-resolution image, the diffuse source is detached from the tailed galaxy and does not have an SDSS optical counterpart. However, we cannot exclude that it is related to the radio galaxy. The flux density of the diffuse source at 144~MHz is $47.9\pm4.9$~mJy but its nature remains unknown.

\subsection{WHL~J085608.5+541855}
\label{sec:WHLJ085608.5+541855}

\begin{figure*}[!ht]
	\centering
	\begin{tikzpicture}
		\draw (0, 0) node[inner sep=0] {\includegraphics[width=0.33\textwidth]{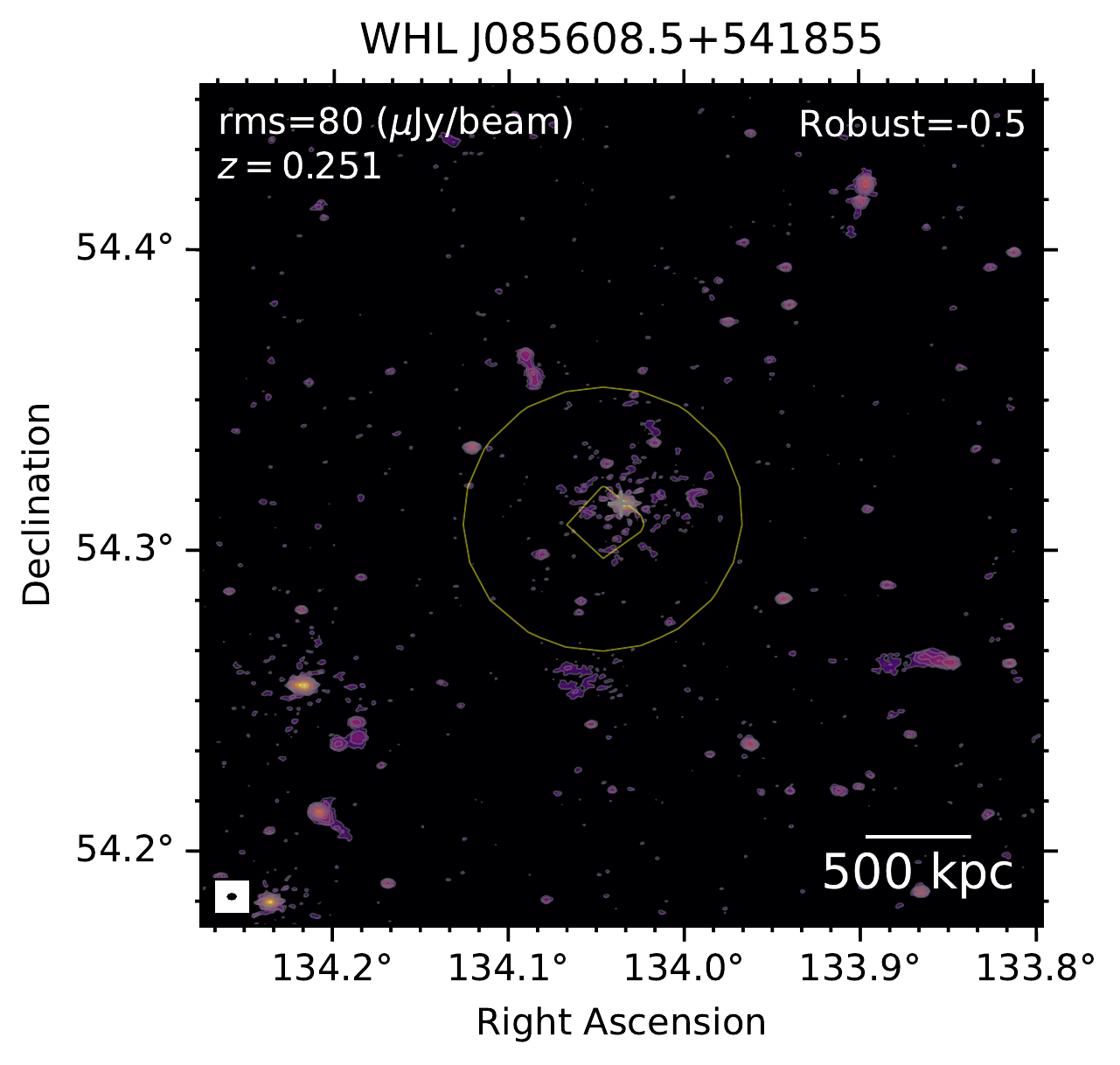}  \hfil
			\includegraphics[width=0.33\textwidth]{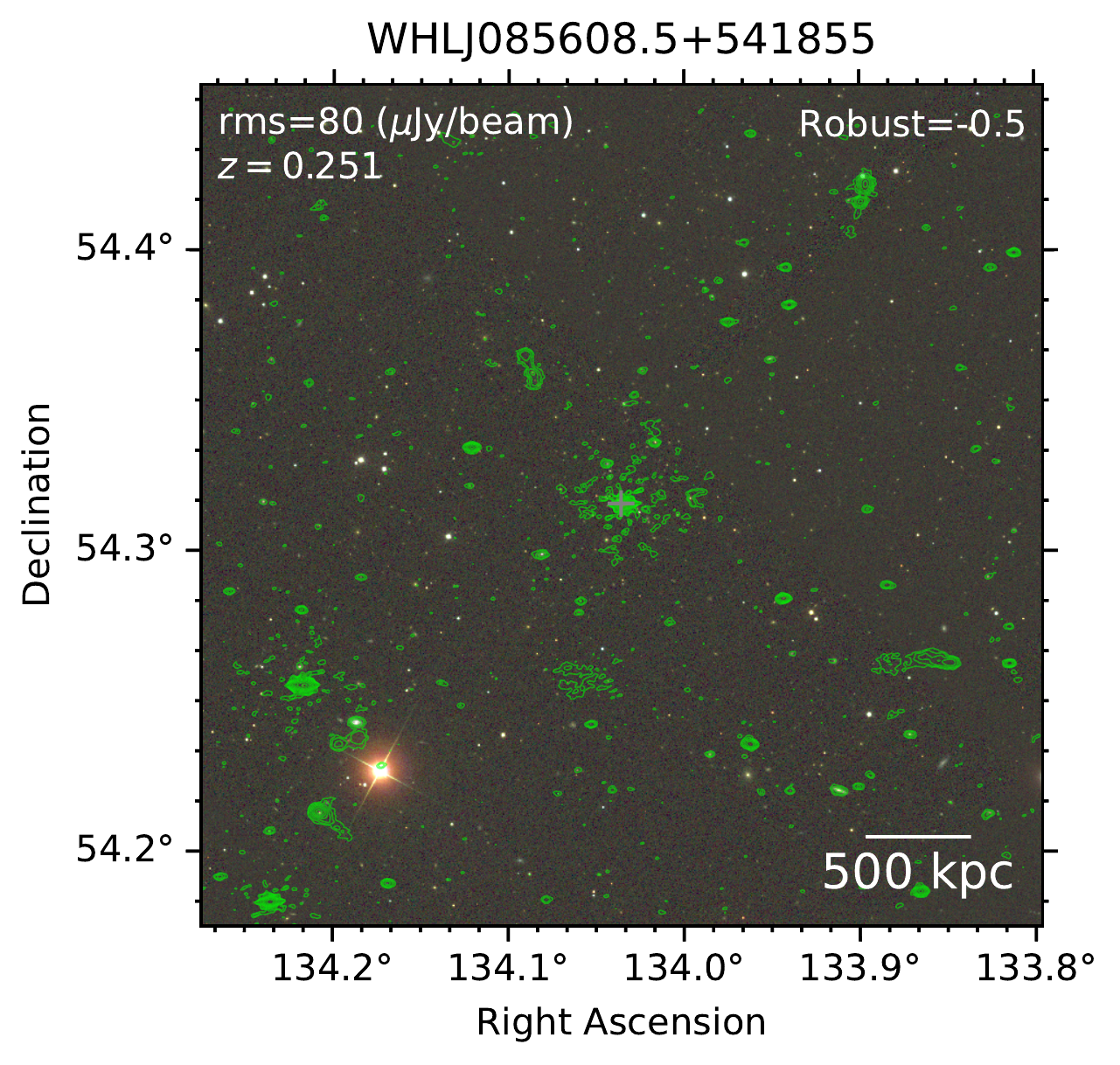}  \hfil
			\includegraphics[width=0.33\textwidth]{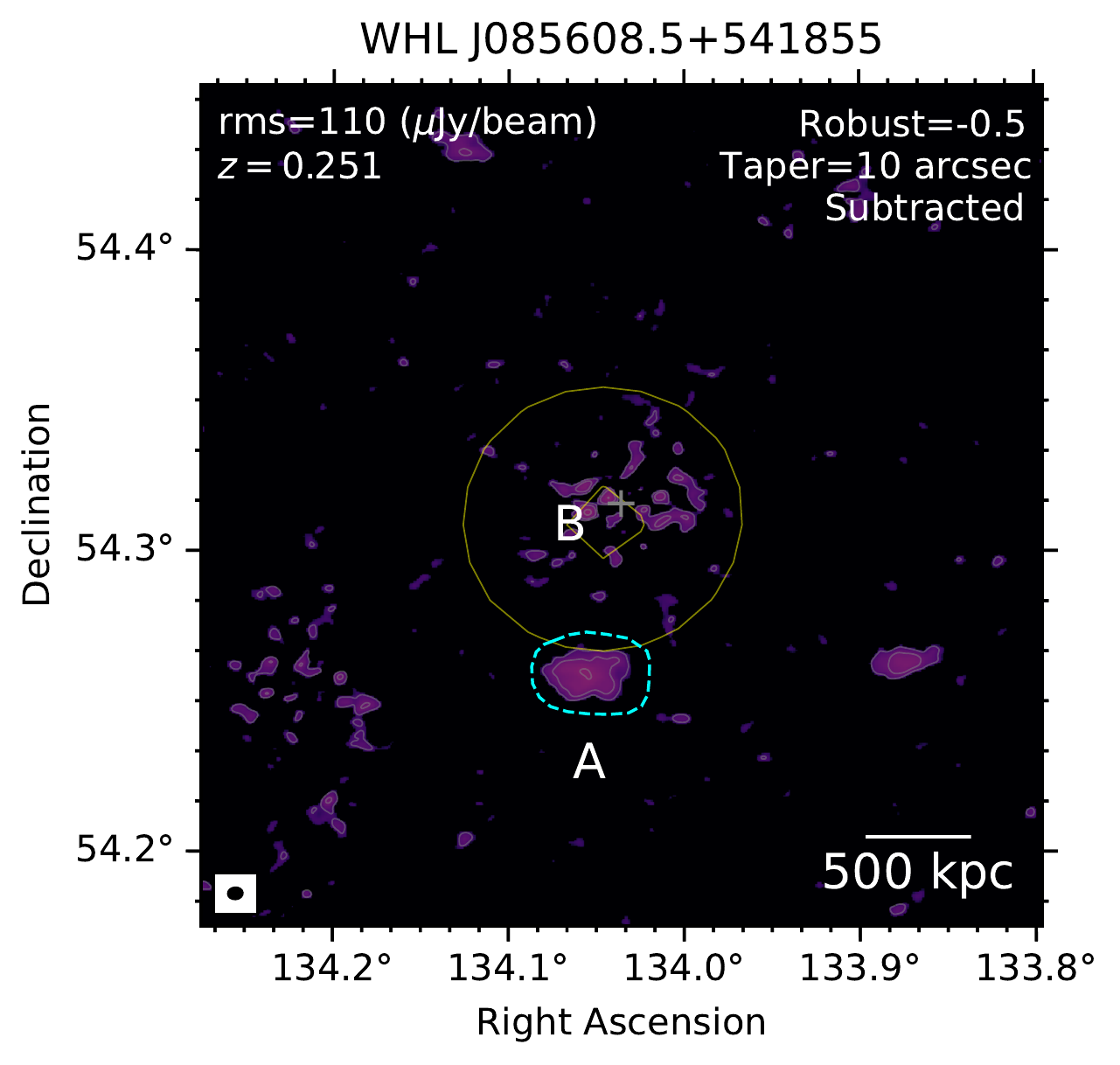}   };
		\draw (-7.7, -1.7) node {\color{white} (a)};
		\draw (-1.5, -1.7) node {\color{white} (b)};
		\draw (4.6, -1.55) node {\color{white} (c)};
	\end{tikzpicture}
	\caption{WHL~J085608.5+541855.  Image description is the same as that in Fig.~\ref{fig:abell84}.
	}
	\label{fig:WHLJ085608.5+541855}
\end{figure*}

LOFAR images of WHL~J085608.5+541855 ($z=0.251$) are shown in Fig. \ref{fig:WHLJ085608.5+541855}. The cluster hosts a newly detected 380~kpc$\times$210~kpc diffuse source in the southern region, 900~kpc from the cluster centre. The diffuse source is patchy at high resolution and is not clearly associated with any SDSS sources. The source could be the remnant from AGN activities. We measure the flux density and radio power of the diffuse source and present the measurements in Table \ref{tab:sources}. In the cluster central region, multiple small-scale radio sources are detected around the central radio galaxy. These sources are likely artefacts around the radio galaxy in the cluster centre.

\subsection{WHL~J091721.4+524607}
\label{sec:WHLJ091721.4+524607}

\begin{figure*}[!ht]
	\centering
	\begin{tikzpicture}
		\draw (0, 0) node[inner sep=0] {\includegraphics[width=0.33\textwidth]{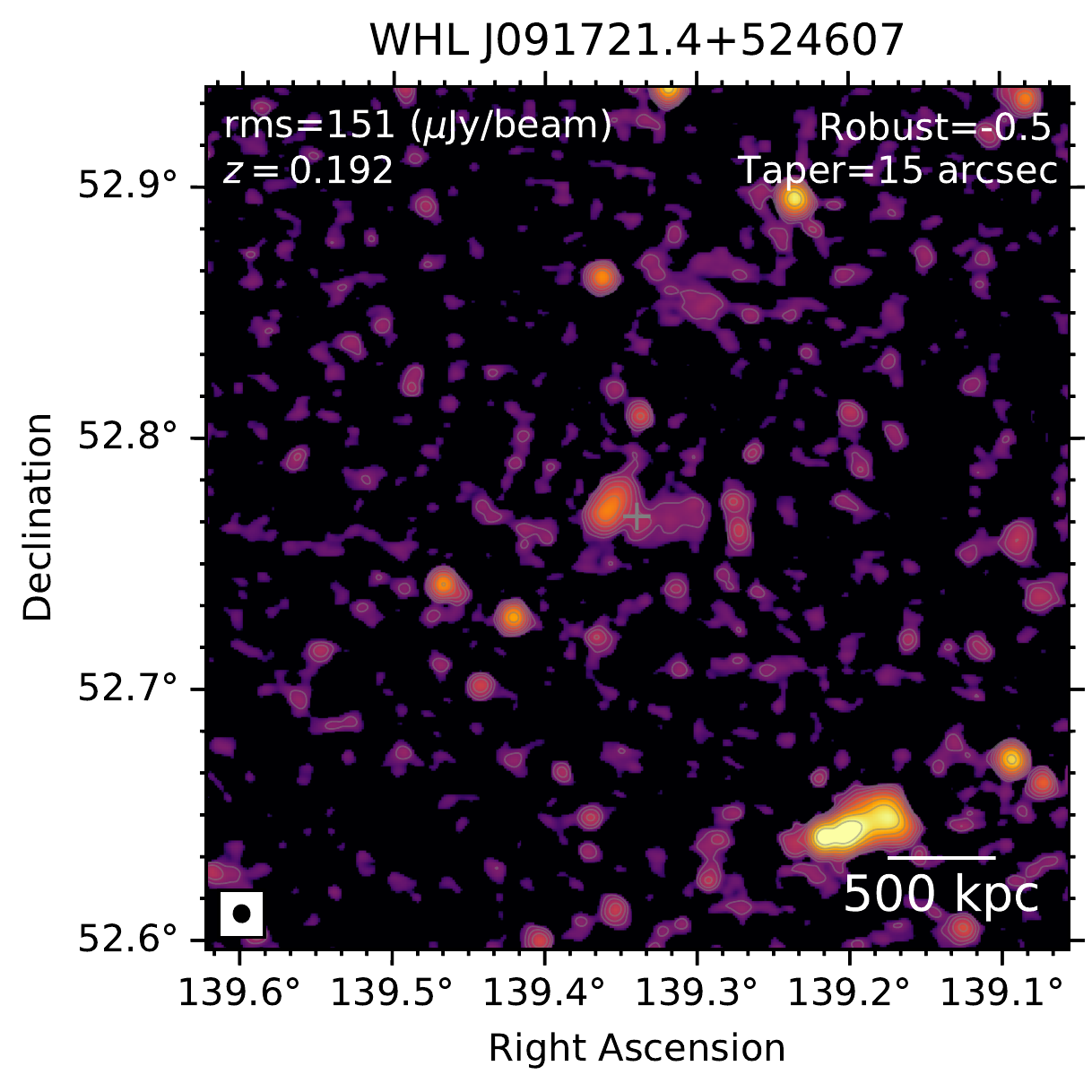}  \hfil
			\includegraphics[width=0.33\textwidth]{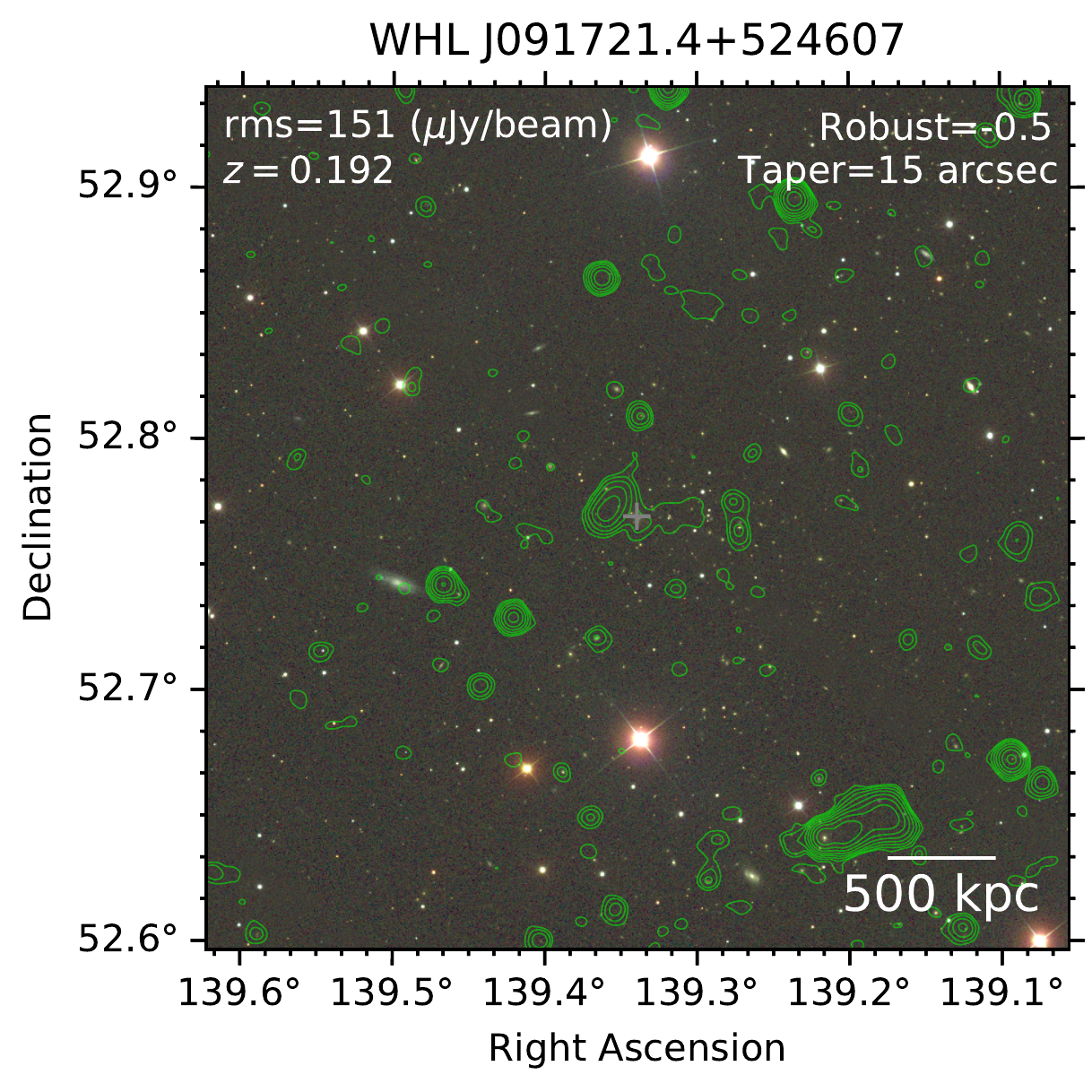}  \hfil
			\includegraphics[width=0.33\textwidth]{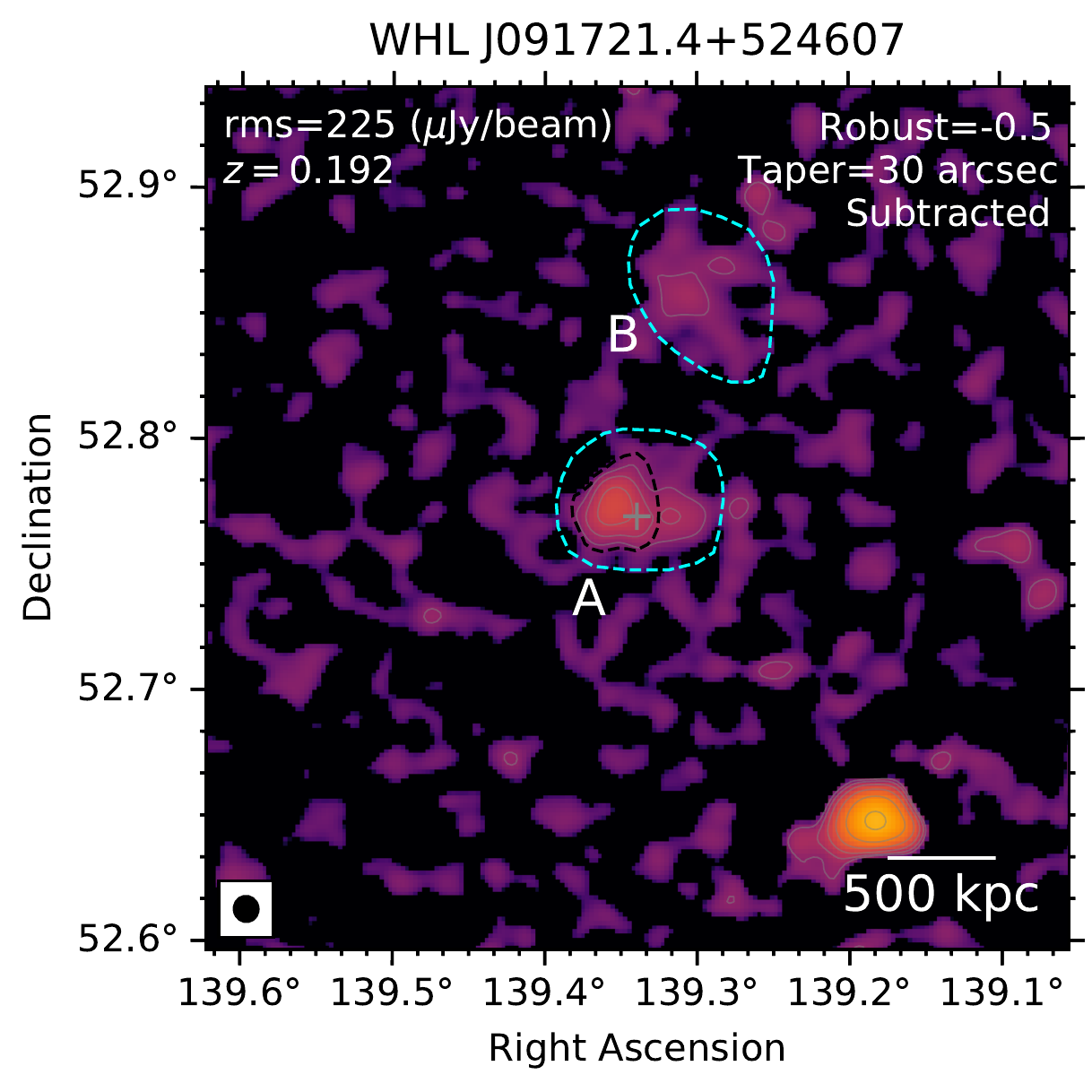}   };
		\draw (-7.7, -1.7) node {\color{white} (a)};
		\draw (-1.5, -1.7) node {\color{white} (b)};
		\draw (4.6, -1.55) node {\color{white} (c)};
	\end{tikzpicture}
	\caption{WHL~J091721.4+524607. Image description is the same as that in Fig.~\ref{fig:abell84}.
	}
	\label{fig:WHLJ091721.4+524607}
\end{figure*}

LOFAR images uncover the presence of multiple radio sources in the galaxy cluster WHL~J091721.4+524607 ($z=0.192$), seen in Fig.~\ref{fig:WHLJ091721.4+524607}. The diffuse emission in the cluster centre, labelled as A, has a projected size of 290~kpc$\times$170~kpc, orienting in the E-W direction. It is unclear whether A is connected to a radio galaxy 270~kpc to the east or they are seen in projection. Due to its small detected size and the possible connection with a radio galaxy, the classification of source A is difficult. To the north of the cluster centre (1~Mpc) a diffuse source, named B, is seen in the low-resolution image in the panel (c) of Fig.~\ref{fig:WHLJ091721.4+524607}. Source B with a projected size of 500~kpc$\times$300~kpc is detected at $2\sigma$. The morphology of source B at high resolution (panel a) implies that it may connect to an FR-I galaxy which does not have an SDSS optical counterpart.

\subsection{WHL~J101350.8+344251}
\label{sec:WHLJ101350.8+344251}

\begin{figure*}[!ht]
	\centering
	\begin{tikzpicture}
		\draw (0, 0) node[inner sep=0] {\includegraphics[width=0.33\textwidth]{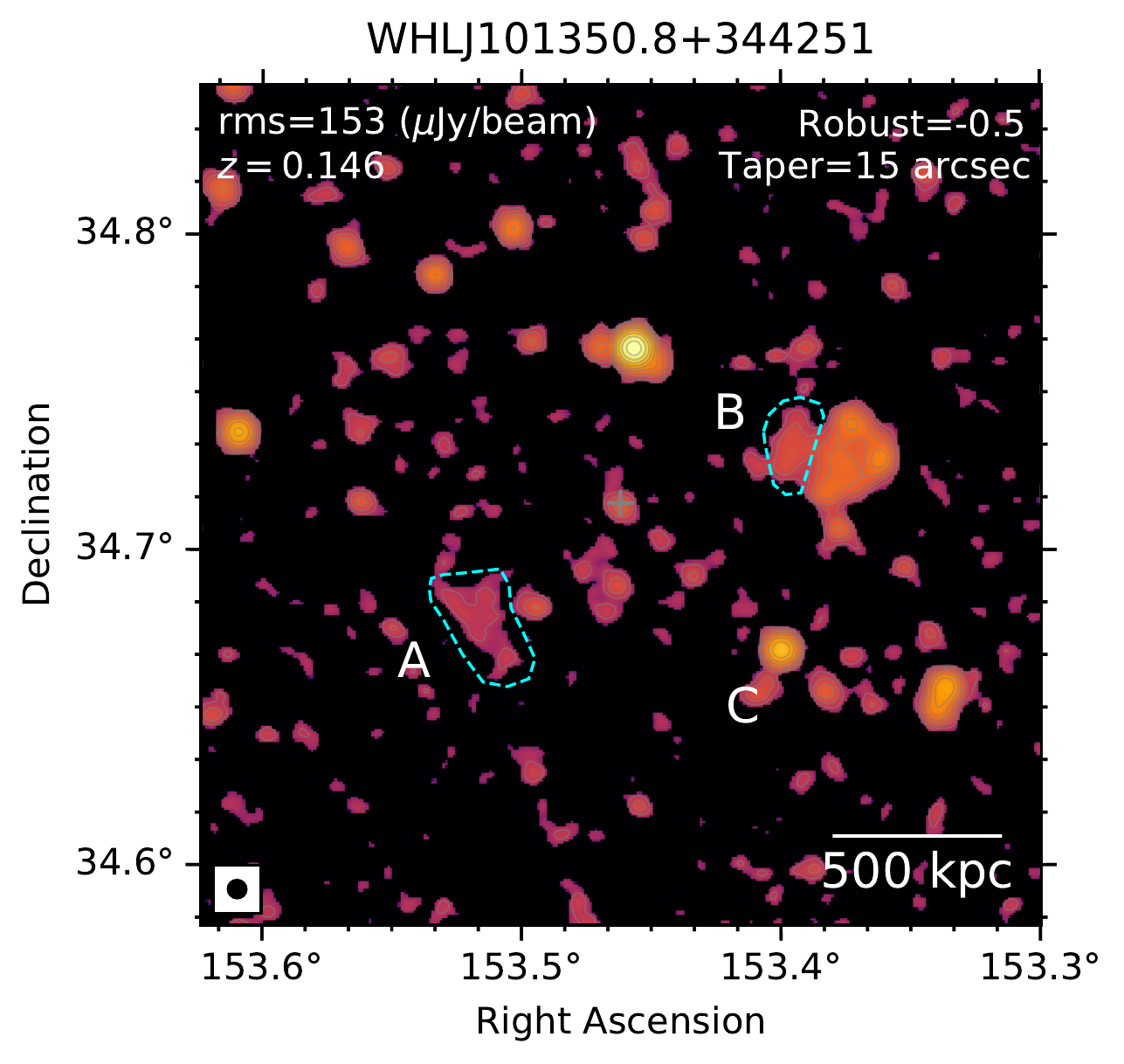}  \hfil
			\includegraphics[width=0.33\textwidth]{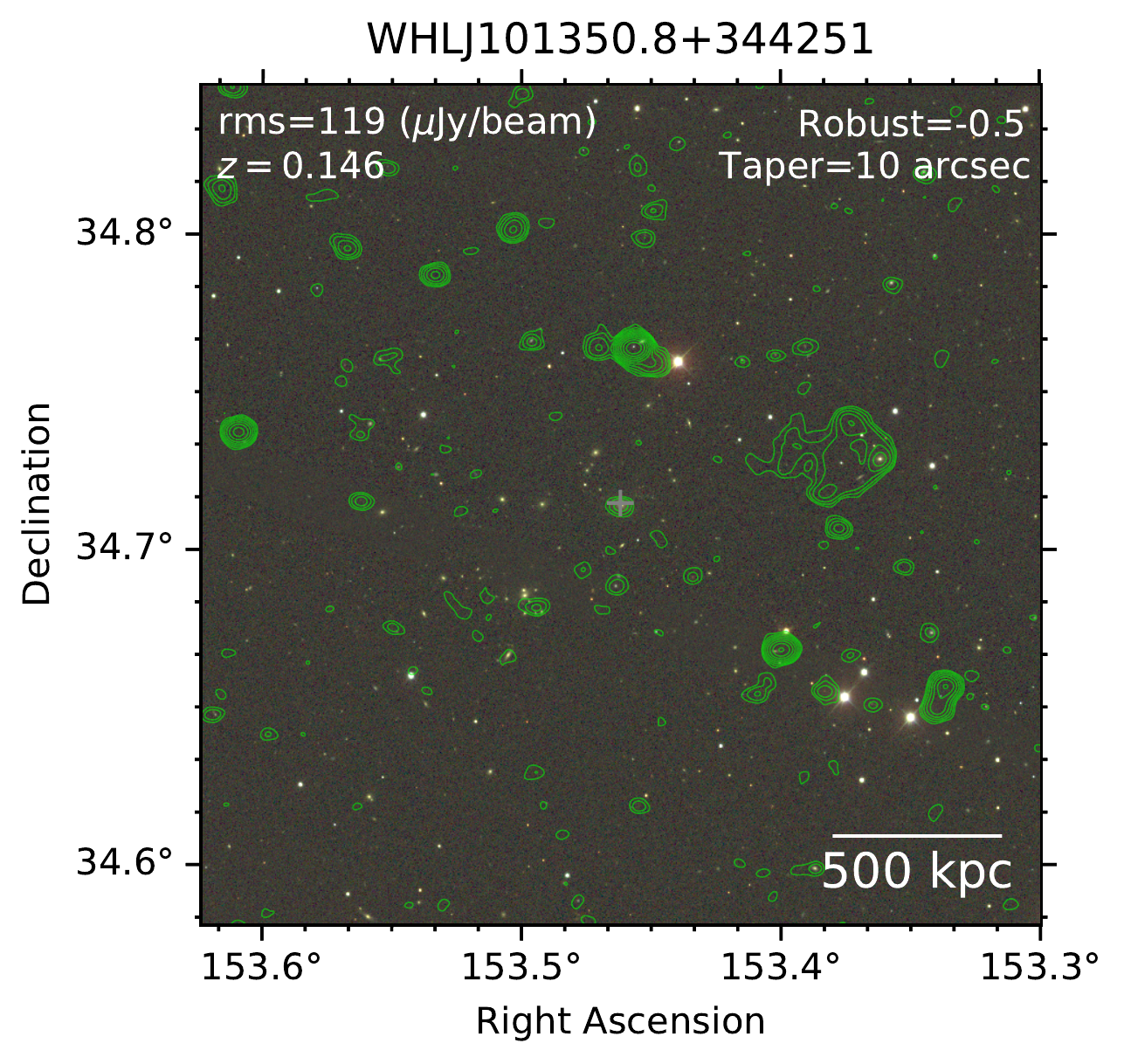}  \hfil
			\includegraphics[width=0.33\textwidth]{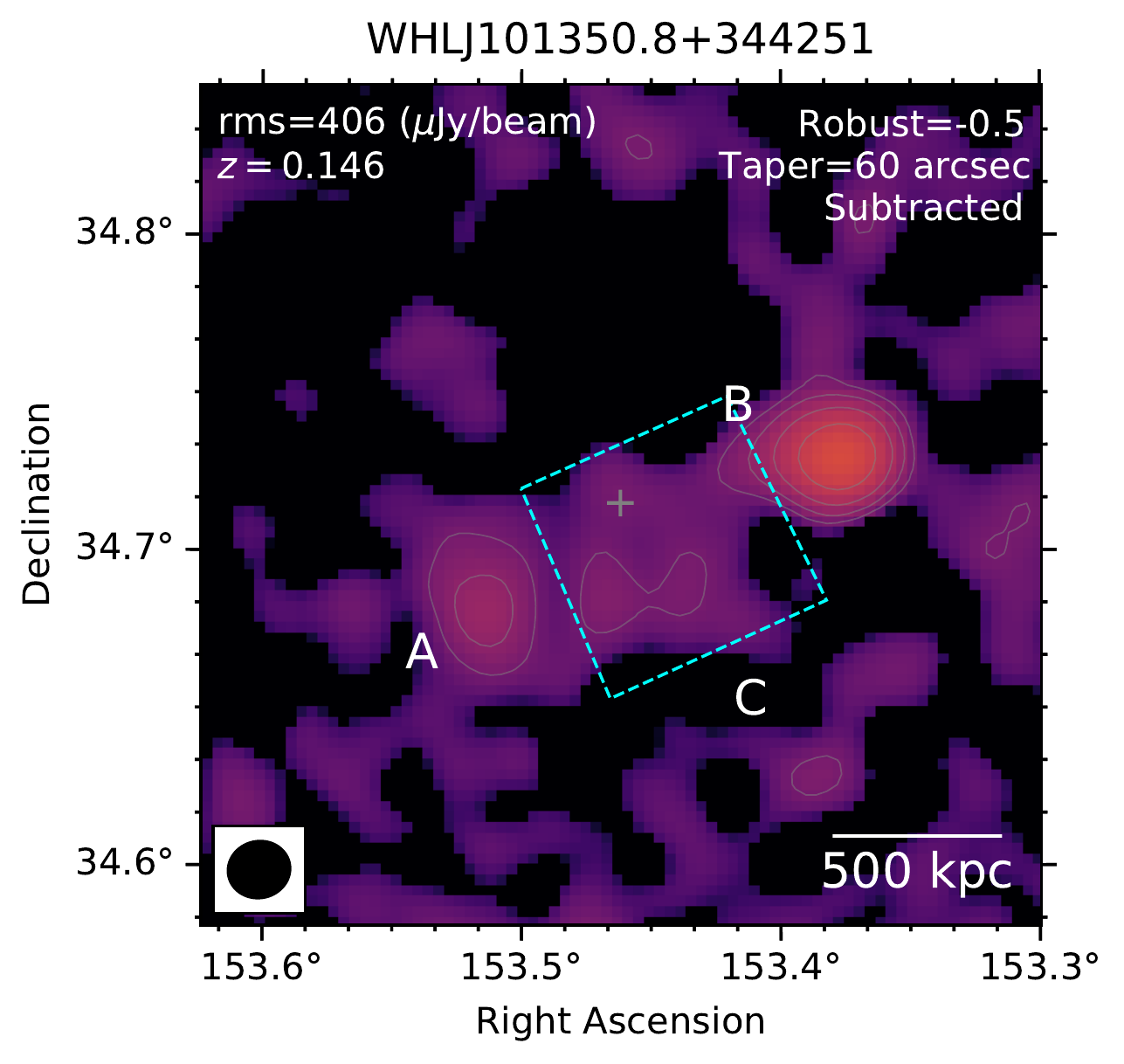}   };
		\draw (-7.7, -1.7) node {\color{white} (a)};
		\draw (-1.5, -1.7) node {\color{white} (b)};
		\draw (4.6, -1.55) node {\color{white} (c)};
	\end{tikzpicture}
	\caption{WHL~J101350.8+344251. Image description is the same as that in Fig.~\ref{fig:abell84}.
	}
	\label{fig:WHLJ101350.8+344251}
\end{figure*}

In Fig.~\ref{fig:WHLJ101350.8+344251}, LOFAR images of WHL~J101350.8+344251 ($z=0.146$) show the new detection of multiple diffuse sources, labelled as A--C. A diffuse source (A) is detected about 540~kpc from the cluster centre towards the SE direction. Source A has a projected size of 310~kpc$\times$150~kpc. The major axis of the source is perpendicular to the axis connecting the cluster centre and the source. There is no optical counterpart to A in the SDSS data and we classify it as a radio relic. To the west of the cluster, a WAT radio galaxy is detected and it has an SDSS optical counterpart at the central bright part. Interestingly, excess diffuse emission (B) is seen in the eastern region behind the tailed galaxy. The nature of B is still unknown. In the cluster central region, faint emission (C) is detected at $2\sigma$ connecting A and B. It spans a projected area of 900~kpc$\times$300~kpc. We classify it as a candidate radio halo. The flux density and radio power of the sources are given in Table \ref{tab:sources}.

\subsection{WHL~J130503.5+314255}
\label{sec:WHLJ130503.5+314255}

\begin{figure*}[!ht]
	\centering
	\begin{tikzpicture}
		\draw (0, 0) node[inner sep=0] {\includegraphics[width=0.33\textwidth]{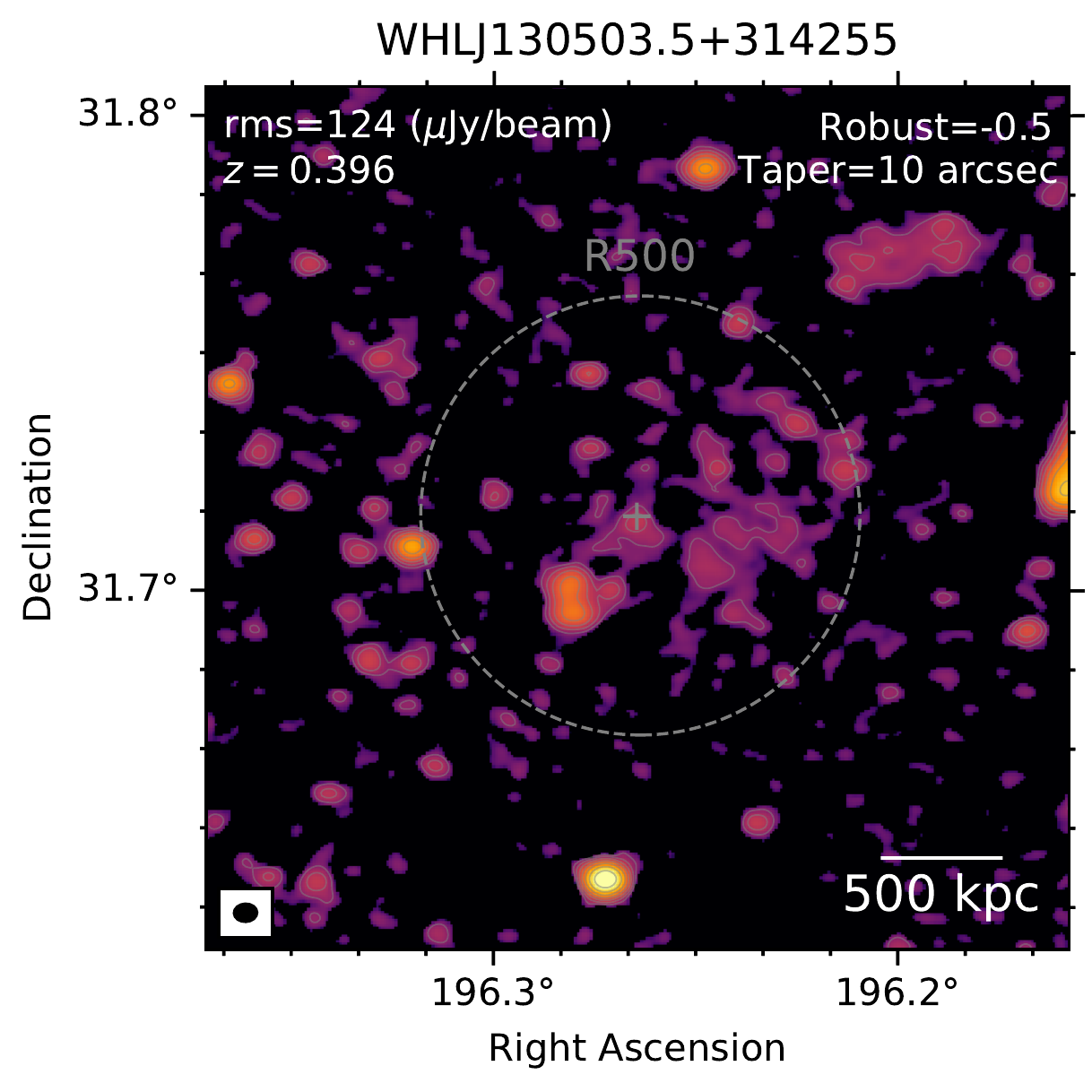}  \hfil
			\includegraphics[width=0.33\textwidth]{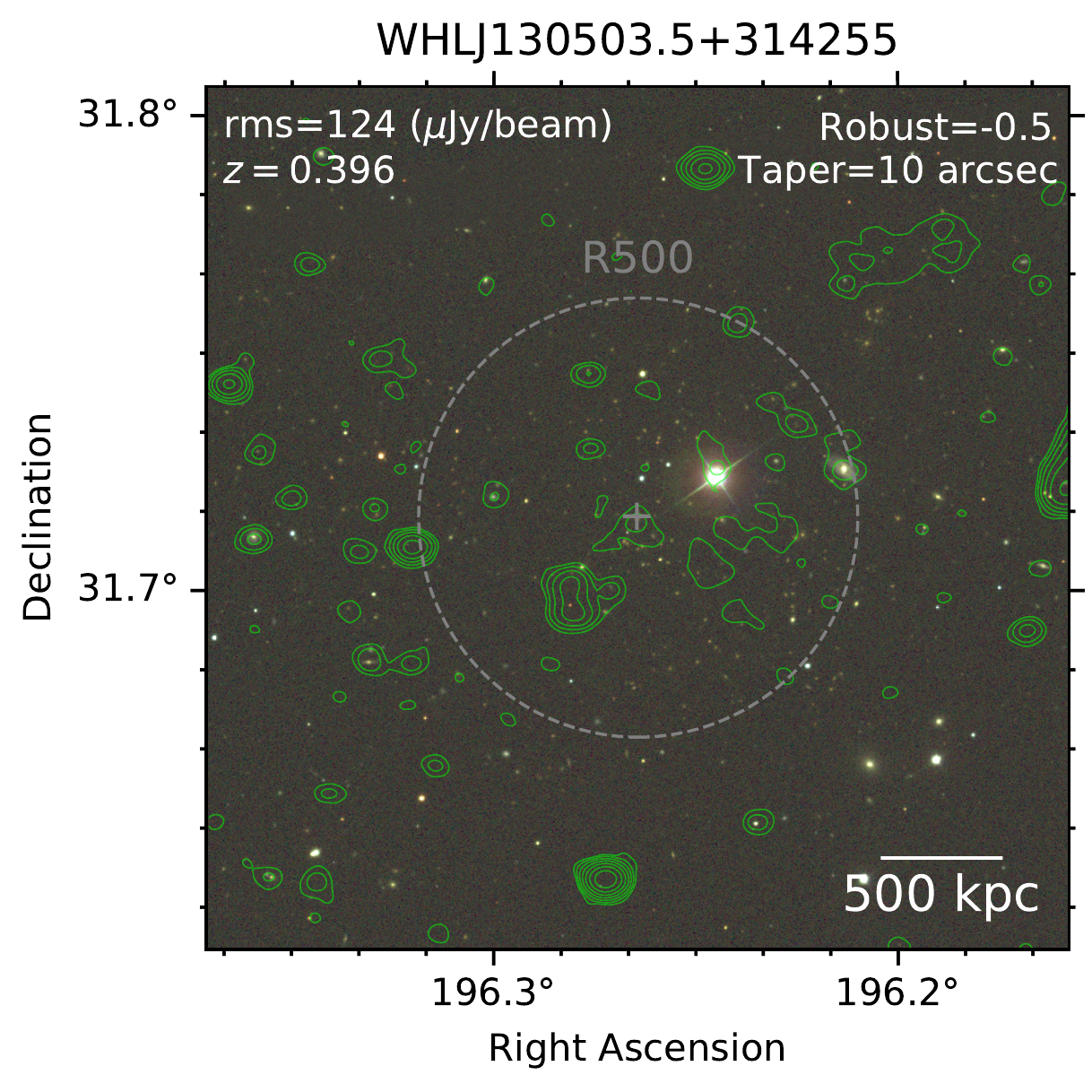}  \hfil
			\includegraphics[width=0.33\textwidth]{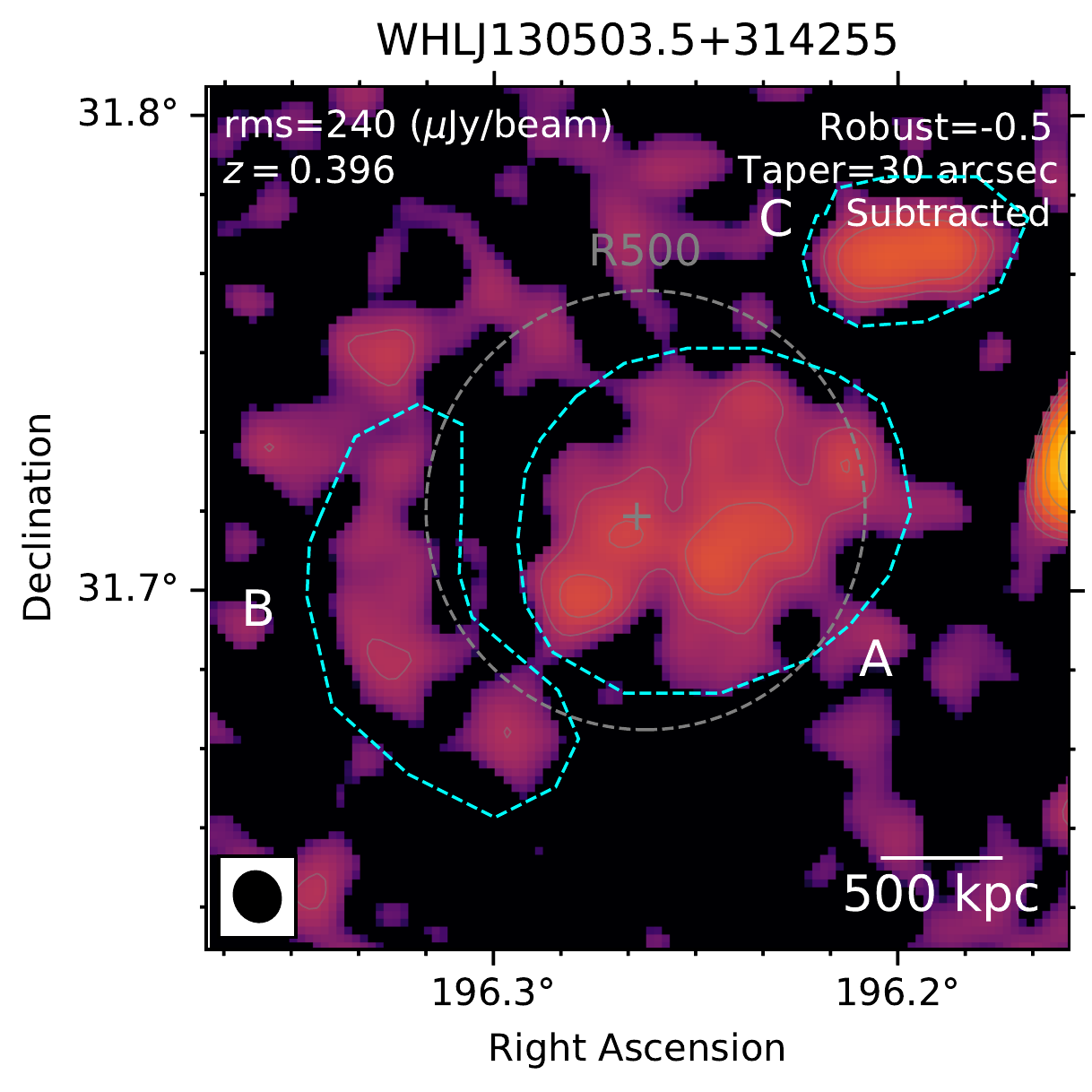}   };
		\draw (-7.7, -1.7) node {\color{white} (a)};
		\draw (-1.5, -1.7) node {\color{white} (b)};
		\draw (4.6, -1.55) node {\color{white} (c)};
	\end{tikzpicture}
	\caption{WHL~J130503.5+314255. Image description is the same as that in Fig.~\ref{fig:abell84}.
	}
	\label{fig:WHLJ130503.5+314255}
\end{figure*}

In WHL~J130503.5+314255 ($z=0.396$), three new diffuse sources are detected with LOFAR, labelled as A, B and C in Fig.~\ref{fig:WHLJ130503.5+314255}. In the cluster centre, a faint diffuse source (A) is detected at $>2\sigma$ and has a projected size of 1400~kpc$\times$800~kpc, oriented in the EW direction. Multiple radio sources from individual galaxies are embedded in the region of source A. As source A is unlikely to be a combination of these discrete sources, we classify it as a radio halo. In the eastern direction, an arc-like diffuse (1400~kpc$\times$320~kpc) source is also detected at $2\sigma$ which we classify as a candidate radio relic. In the NW region (1.3~Mpc) we detect a diffuse (720~kpc$\times$360~kpc) source (C) that has no clear counterpart in the SDSS image. The major axis of source B is in the E-W direction. It is unclear how source B is formed.

\subsection{WHL~J165540.4+334422}
\label{sec:WHLJ165540.4+334422}

\begin{figure*}[!ht]
	\centering
	\begin{tikzpicture}
		\draw (0, 0) node[inner sep=0] {\includegraphics[width=0.33\textwidth]{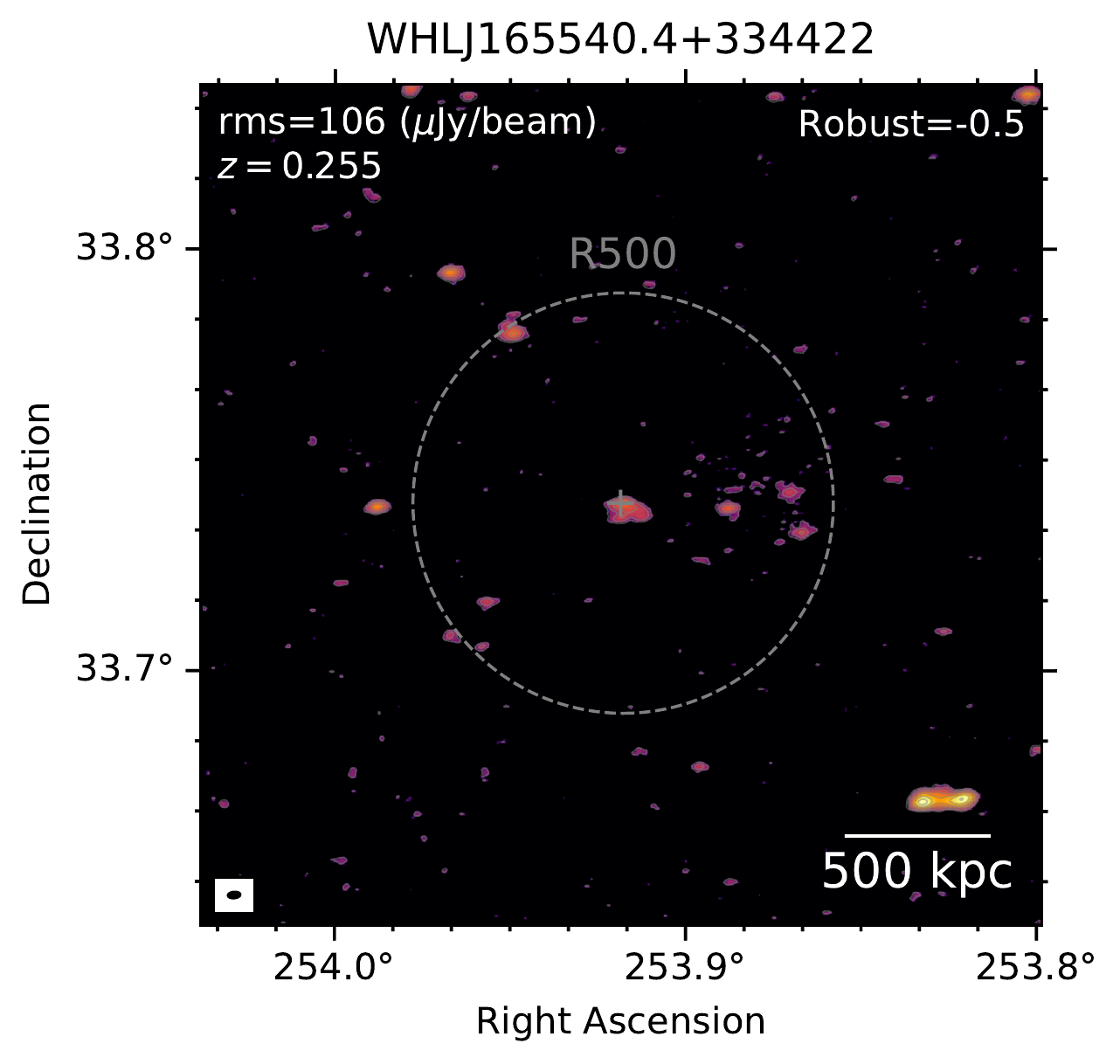}  \hfil
			\includegraphics[width=0.33\textwidth]{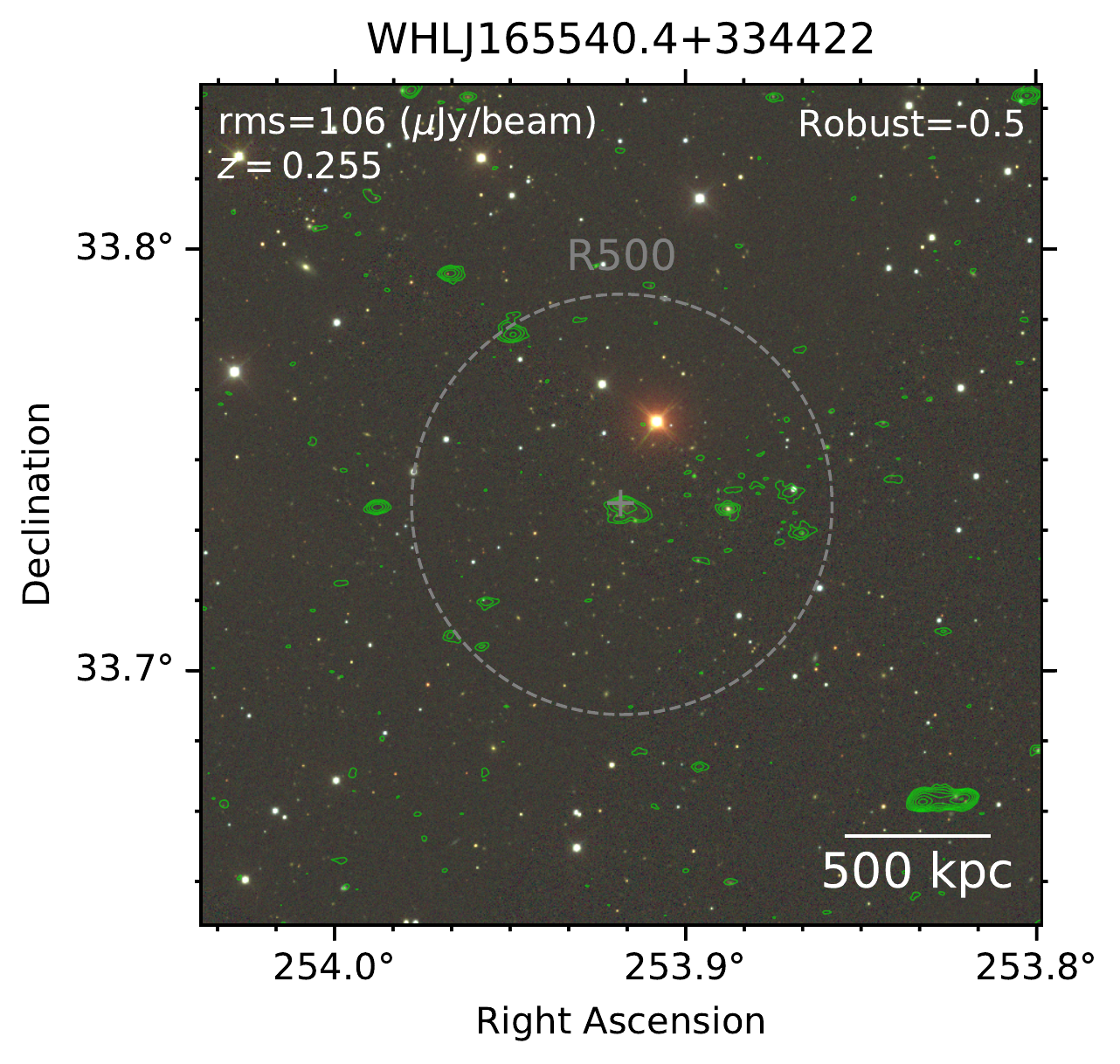}  \hfil
			\includegraphics[width=0.33\textwidth]{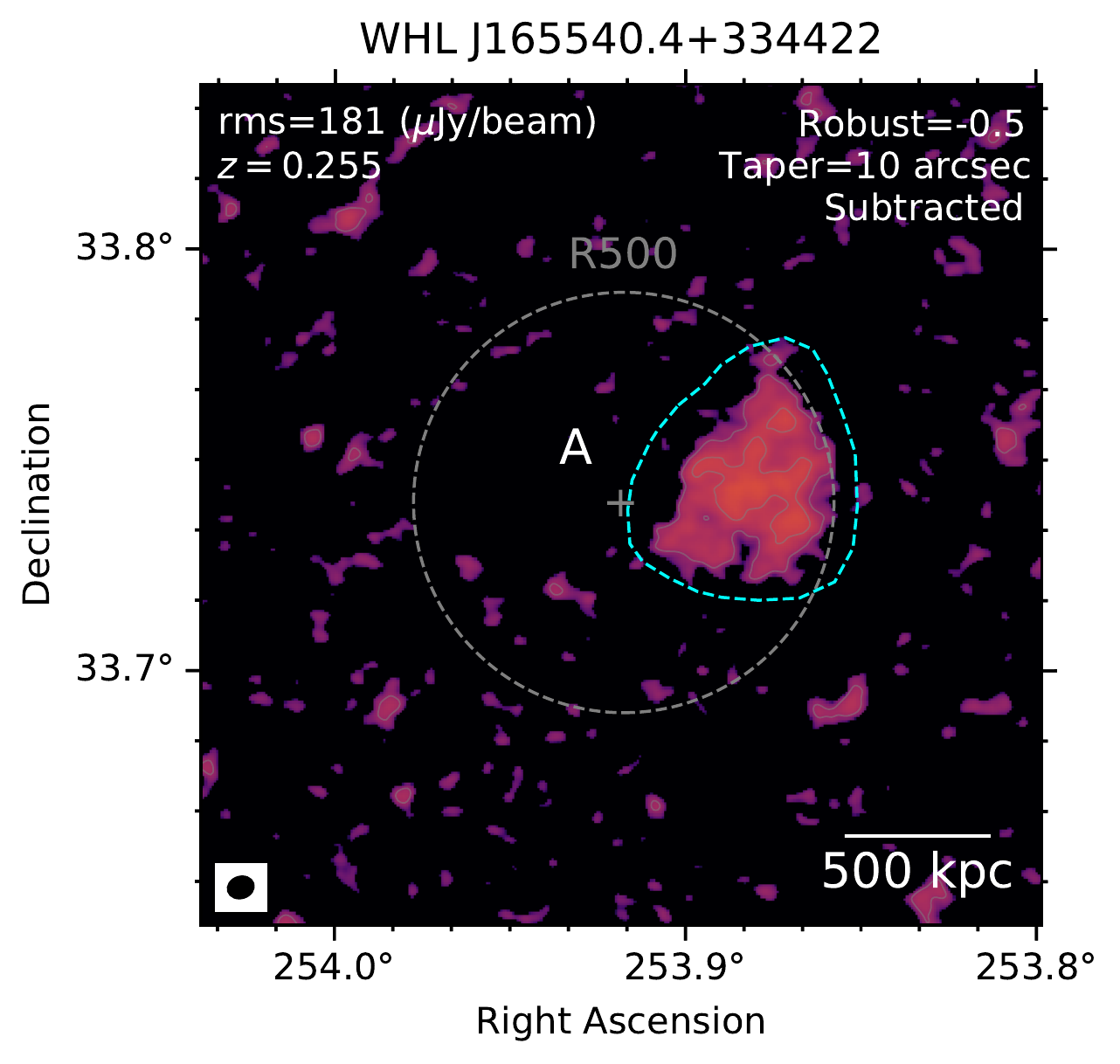}   };
		\draw (-7.7, -1.7) node {\color{white} (a)};
		\draw (-1.5, -1.7) node {\color{white} (b)};
		\draw (4.6, -1.55) node {\color{white} (c)};
	\end{tikzpicture}
	\caption{WHL~J165540.4+334422. Image description is the same as that in Fig.~\ref{fig:abell84}.
	}
	\label{fig:WHLJ165540.4+334422}
\end{figure*}

Fig.~\ref{fig:WHLJ165540.4+334422} shows the new detection of a diffuse source in the western region of WHL~J165540.4+334422 ($z=0.255$). Its projected size is 630~kpc$\times$400~kpc. The projected distance from the diffuse source to the reported cluster centre from the \cite{Wen2015} catalogue is $2\arcmin$ (480~kpc). We classify the diffuse source as fossil plasma originating from radio galaxies. 

\subsection{WHL~J172125.4+294144}
\label{sec:WHLJ172125.4+294144}

\begin{figure*}[!ht]
	\centering
	\begin{tikzpicture}
		\draw (0, 0) node[inner sep=0] {\includegraphics[width=0.33\textwidth]{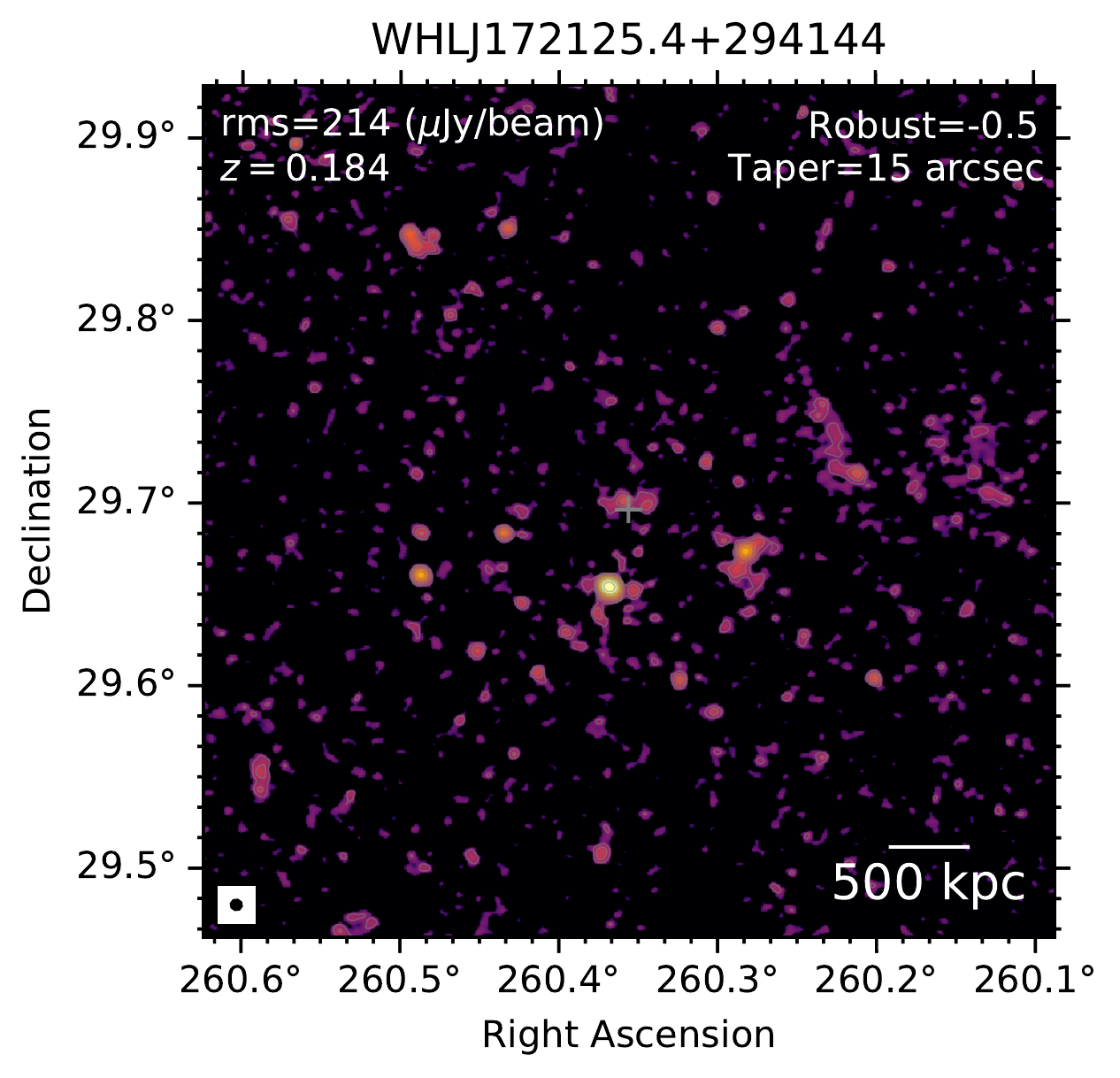}  \hfil
			\includegraphics[width=0.33\textwidth]{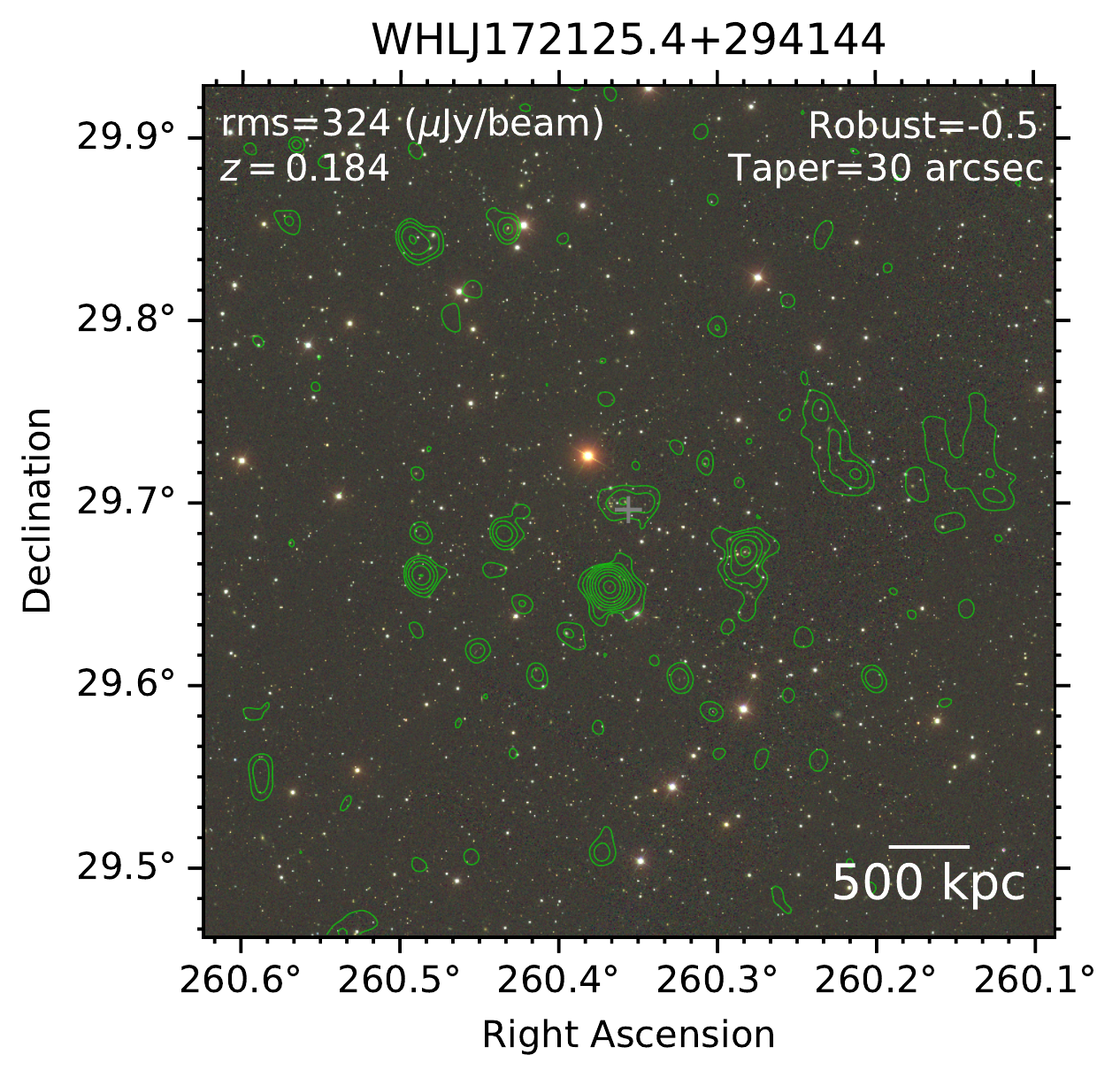}  \hfil
			\includegraphics[width=0.33\textwidth]{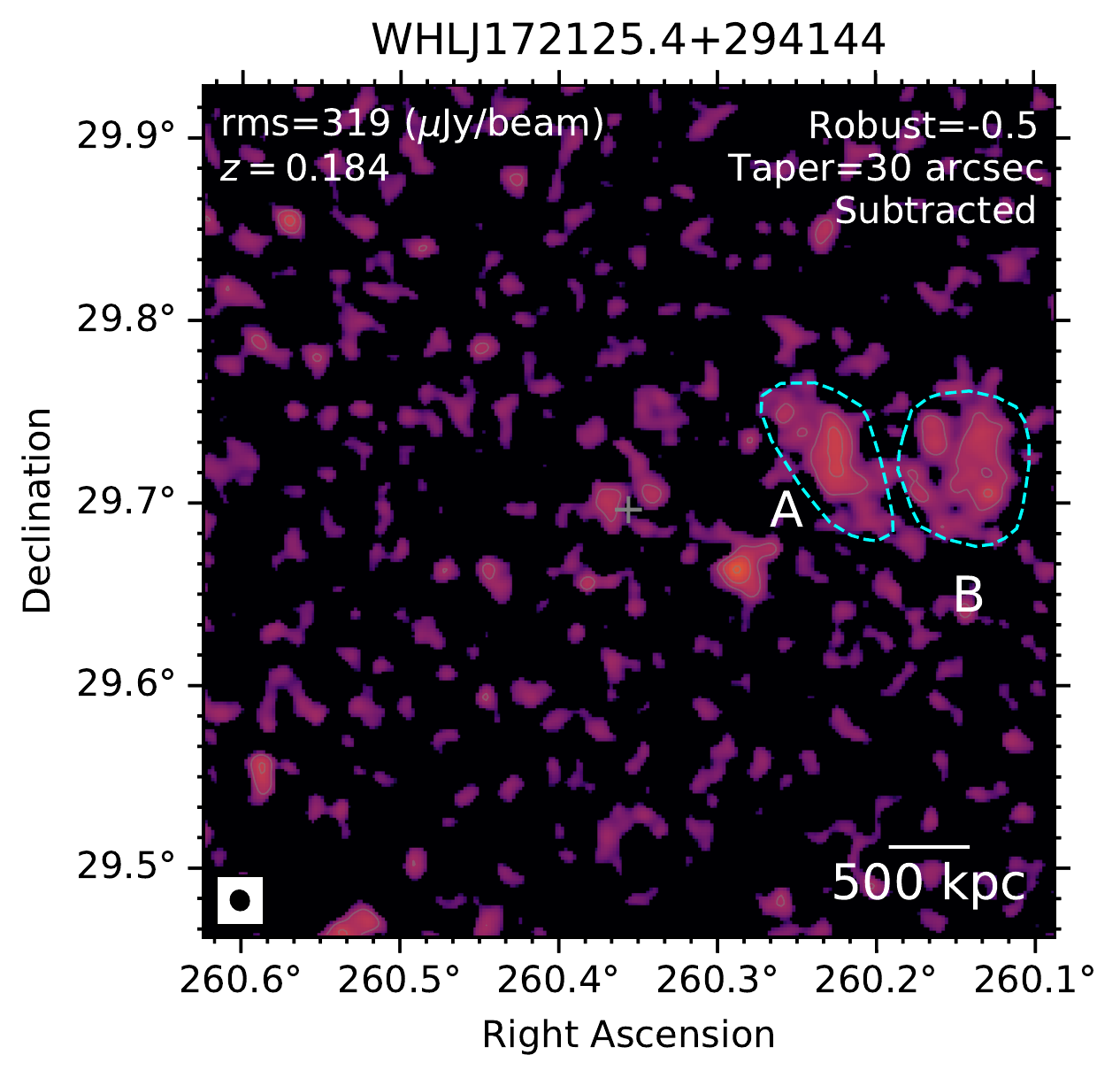}   };
		\draw (-7.7, -1.7) node {\color{white} (a)};
		\draw (-1.5, -1.7) node {\color{white} (b)};
		\draw (4.6, -1.55) node {\color{white} (c)};
	\end{tikzpicture}
	\caption{WHL~J172125.4+294144. Image description is the same as that in Fig.~\ref{fig:abell84}.
	}
	\label{fig:WHLJ172125.4+294144}
\end{figure*}

As seen in Fig.~\ref{fig:WHLJ172125.4+294144}, LOFAR observations of WHL~J172125.4+294144 ($z=0.184$) detect two diffuse sources, labelled as A and B, in the western region of the cluster. They are separated by a distance of 1.3~Mpc and 2.2~Mpc, respectively, from the cluster centre. The sizes of these sources are 560~kpc$\times$260~kpc and 600~kpc$\times$340~kpc. There is an optical source near the peak of the northern part of source A, but the optical source is not likely to generate the large structure of source A. It is unclear how source A is formed but they could be related to fossil plasma from AGN activities. The surface brightness of source B decreases sharply in the outside region which is consistent with it being a relic. No diffuse emission is seen in the central region of the cluster with the LOFAR observations.

\subsection{WHL~J173424.0+332526}
\label{sec:WHLJ173424.0+332526}

\begin{figure*}[!ht]
	\centering
	\begin{tikzpicture}
		\draw (0, 0) node[inner sep=0] {\includegraphics[width=0.33\textwidth]{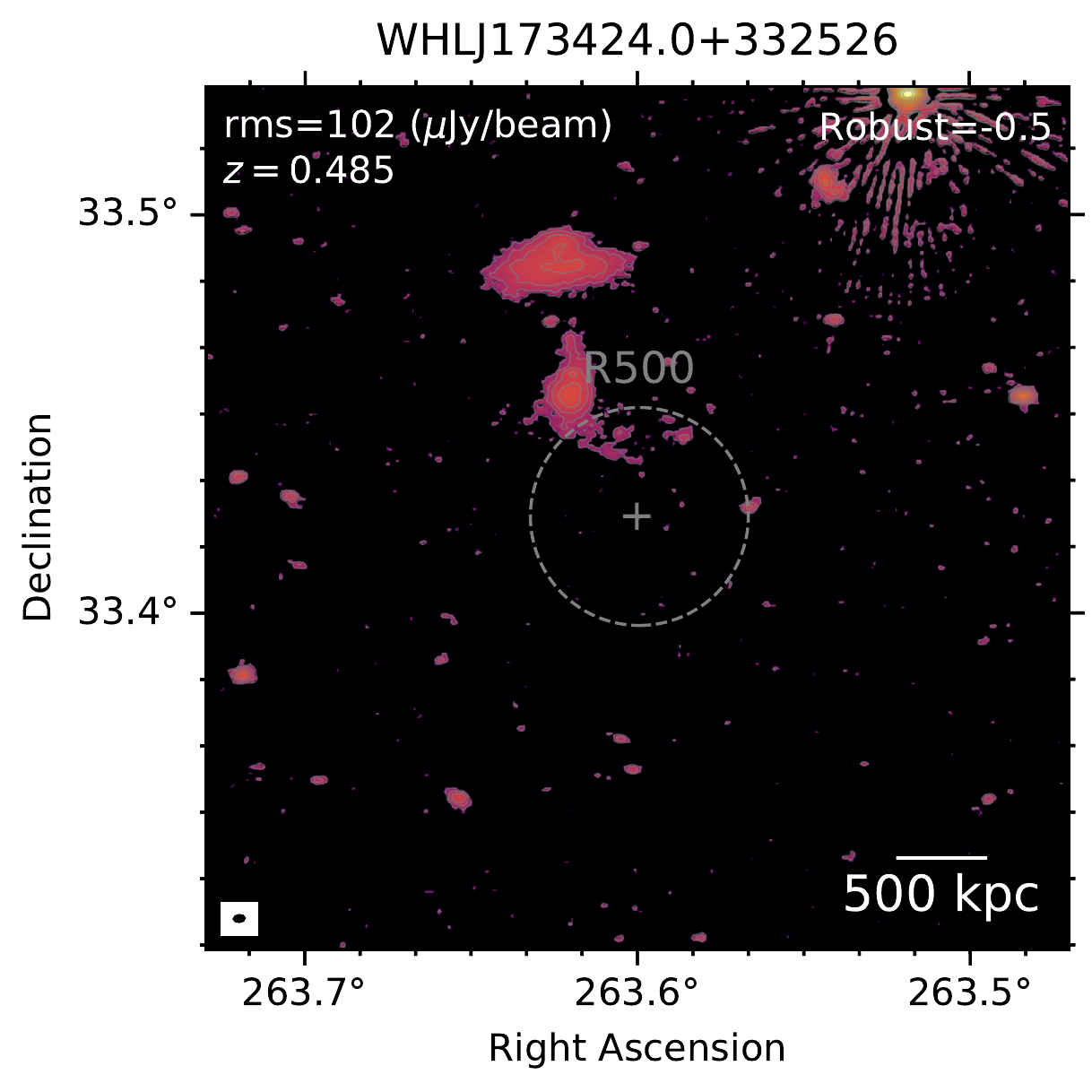}  \hfil
			\includegraphics[width=0.33\textwidth]{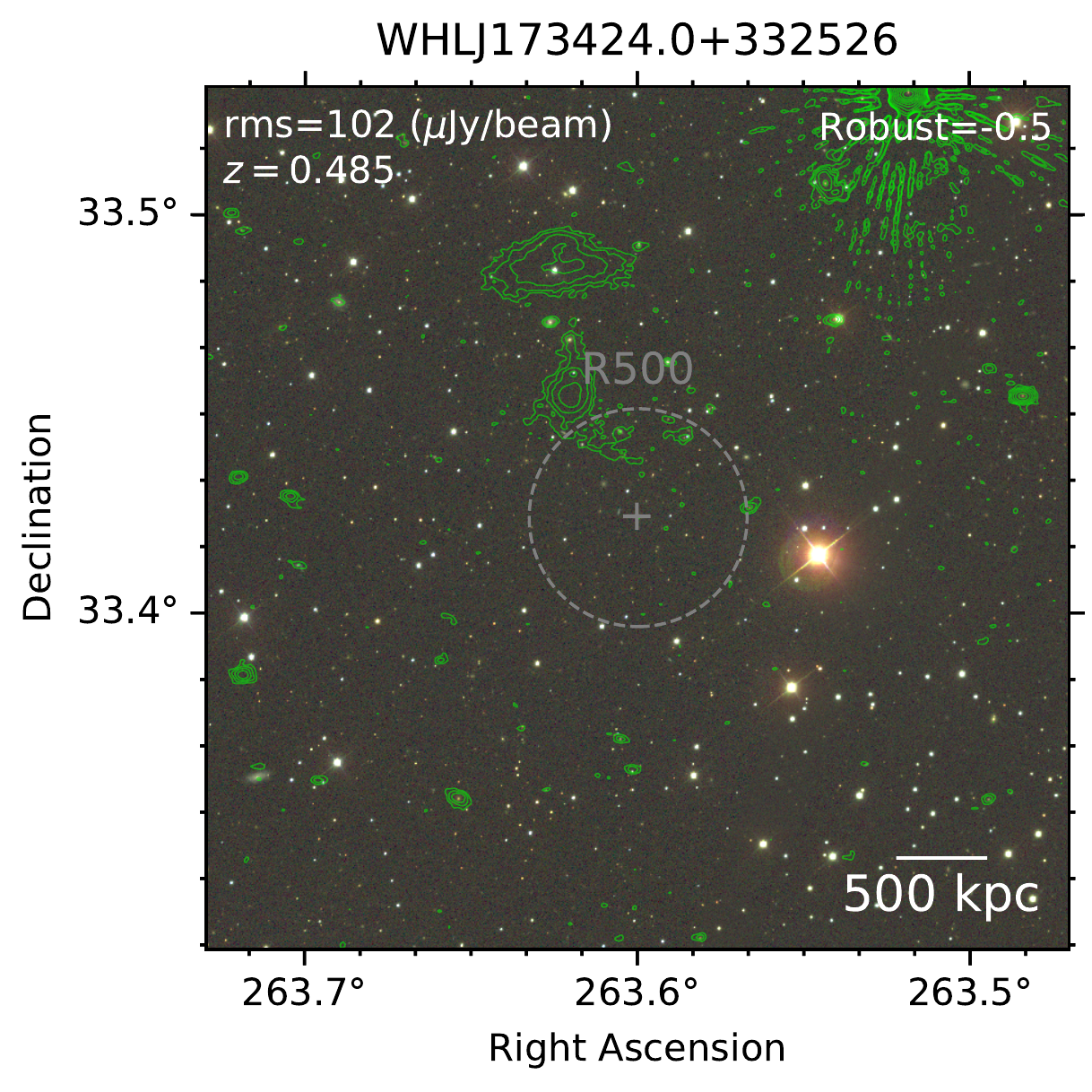}  \hfil
			\includegraphics[width=0.33\textwidth]{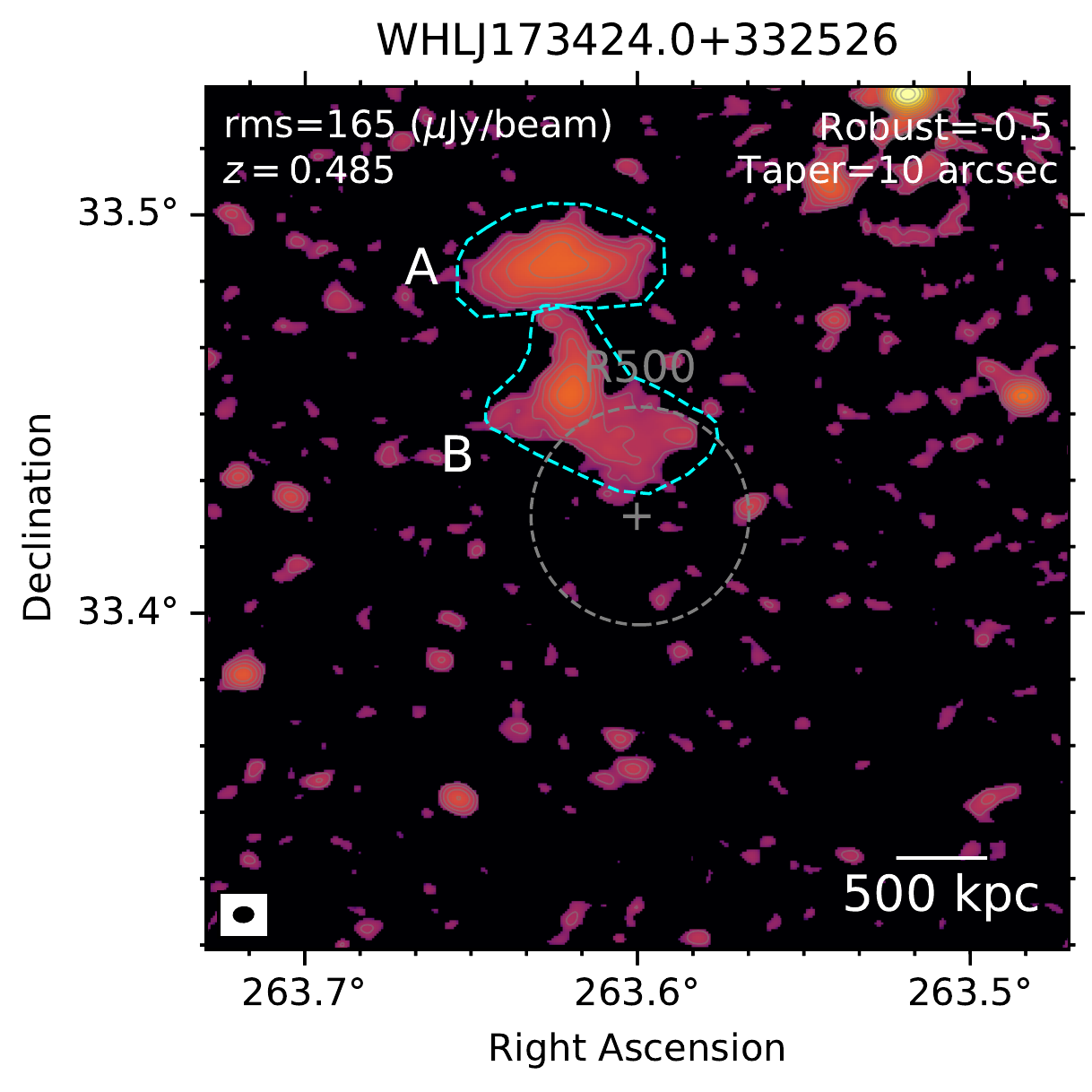}   };
		\draw (-7.7, -1.7) node {\color{white} (a)};
		\draw (-1.5, -1.7) node {\color{white} (b)};
		\draw (4.6, -1.55) node {\color{white} (c)};
	\end{tikzpicture}
	\caption{WHL~J173424.0+332526. Image description is the same as that in Fig.~\ref{fig:abell84}.
	}
	\label{fig:WHLJ173424.0+332526}
\end{figure*}

In Fig.~\ref{fig:WHLJ173424.0+332526}, LOFAR observations show diffuse sources in the northern region of the galaxy cluster WHL~J173424.0+332526 ($z=0.485$). The diffuse source, labelled as A, might be associated with an SDSS optical galaxy to the north. The southern side of A is more extended towards the centre of the cluster. Further to the north, there is a cap-shaped source (named B). It is unclear whether these diffuse sources are connected and their origin is unknown.

\begin{table*}[ht]
	\centering
	\scriptsize
	\caption{Classification of cluster diffuse sources}
		\begin{tabular}{llccccccccc}  
\hline\hline \\
Cluster & Source & VC & DTC & RA   & DEC & $S_{2\sigma}$ & $P_{2\sigma}$ &  $S_{fit}$ & $P_{fit}$ & Size\\
  &  & &  & [deg] & [deg] & [mJy]  & $[\times10^{23} W\,Hz^{-1}]$ & [mJy]  & $[\times10^{23} W\,Hz^{-1}]$ & [kpc$\times$kpc] \\ \hline
Abell~84 & A & cH & H & 10.4794 & 21.3771 & $91.9\pm14.3$ & $24.6\pm3.8$ & $131.9\pm15.4$ & $35.6\pm4.2$ & 1500$\times$740 \\
Abell~373 & A & R & R &  40.7516 & 28.0640 & $50.3\pm5.2$ & $30.8\pm3.2$ & -- & -- & 770$\times$200 \\
 & B & R & R &  40.9168 & 27.9990 & $27.6\pm2.9$ & $16.9\pm1.8$ & -- & -- & 770$\times$200 \\
 & C & H & H &  40.8332 & 28.0333 & $23.9\pm3.3$ & $14.6\pm2.0$  & -- & -- & 1000$\times$800 \\
Abell~1213 & A & U & NDE &  169.2839 & 29.3074 & $71.7\pm7.3$ & $3.9\pm0.4$ & -- & -- & 260$\times$80 \\
Abell~1330 & A & cH & cH &  174.5626 & 49.5312 & $13.7\pm1.8$ & -- & $13.0\pm1.9$ & -- & -- \\
Abell~1889 & A & R & R &  214.0442 & 30.8462 & $141.3\pm14.6$ & $142.1\pm14.7$ & -- & -- & 1650$\times$830 \\
 & B & R & R &  214.2344 & 30.6948 & $112.3\pm11.7$ & $112.9\pm11.7$ & -- & -- & 1650$\times$570 \\
Abell~1943 & A & cR/Rem & cR &  219.4436 & 30.1821 & $5.0\pm0.7$ & $7.8\pm1.1$ & -- & -- & 370$\times$170 \\
Abell~1963 & A & cR/Rem & NDE &  221.2542 & 31.5131 & $36.9\pm3.8$ & $55.2\pm5.6$ & -- & -- & 200$\times$200 \\
DESI~201 & A & U & U &  213.3675 & 43.9744 & $66.0\pm6.7$ & $278.0\pm28.1$ & -- & -- & 620$\times$270 \\
DESI~296 & A & U & U &  138.2187 & 41.9881 & $24.8\pm2.6$ & $16.7\pm1.8$ & -- & -- & 550$\times$200 \\
MCXC~J0928.6+3747 &  A & R & R & 142.1828 & 37.7782 & $35.2\pm3.6$ & $68.2\pm7.0$ & -- & -- & 350$\times$250 \\
 & B & R & R & 142.1169 &  37.7550 & $46.9\pm4.7$ & $90.8\pm9.2$ & -- & -- & 300$\times$200 \\
 & C & H & H & 142.1433 &  37.7846 & $68.1\pm6.9$ & $131.9\pm13.4$ & $88.7\pm9.0$ & $52.3\pm10.3$ & 850$\times$400 \\
MCXC~J0943.1+4659 & A & R &  R & 145.8396 & 47.0212 & $18.6\pm2.0$ & $116.7\pm12.2$ & -- & -- & 530$\times$460 \\
 & B & H & H &  145.7581 & 47.0013 & $56.0\pm6.2$ & $350.8\pm38.6$ & $70.8\pm7.4$ & $4.4\pm1.4$ & 1200$\times$1200 \\
MCXC~J1020.5+3922 & A & Rem & U &  155.1381 & 39.3548 & $122.1\pm12.4$ & $68.8\pm7.0$ & -- & -- & 360$\times$230 \\
PSZRX~G095.27+48.27 & A & R & cR &  230.9914 & 59.8315 & $50.1\pm5.2$ & $237.8\pm24.4$ & -- & -- & 1300$\times$450 \\
 & B & cH & cH &  230.8795 & 59.9022 & $20.4\pm2.5$ & $96.7\pm12.1$  & -- & -- & 1200$\times$610 \\
PSZRX~G100.21-30.38 & A & H & H &  350.5623 & 28.5211 & $31.4\pm3.6$ & $175.7\pm19.9$ & $41.4\pm4.9$ & $232.0\pm27.5$ & 600$\times$380 \\
PSZRX~G102.17+48.88 & A & cH/Rem & cH &  223.7520 & 62.9871 & $19.6\pm2.0$ & $56.0\pm5.8$ & $18.6\pm2.0$ & $53.0\pm5.8$ & 460$\times$340 \\
PSZRX~G116.06+80.14 & A & cH/Rem & U &  194.3397 & 36.9133 & $4.4\pm0.6$ & $50.2\pm7.0$  & -- & -- & 340$\times$250 \\
PSZRX~G181.53+21.43 & A & R & R &  110.4594 & 36.8719 & $91.8\pm9.3$ & $81.0\pm8.2$ & -- & -- & 1000$\times$250 \\
 & B & U & NDE &  110.2951 & 36.6618 & $300.2\pm30.1$ & $264.8\pm26.5$ & -- & -- & 600$\times$300 \\
 & C & AGN & NDE &  110.4509 & 36.5985 & $16.4\pm2.0$ & $14.5\pm1.8$ & -- & -- & 900$\times$90 \\
 & D & cH & H &  110.3850 & 36.7665 & -- & -- & -- & -- & -- \\
PSZRX~G195.91+62.83 & A & H & NDE &  162.0178 & 31.6440 & $16.5\pm1.8$ & $185.2\pm19.8$ & $15.5\pm1.8$ & $173.0\pm20.0$ & 600$\times$600 \\
WHL~J002056.4+221752 & A & U & NDE &  5.2696 & 22.3358 & $155.2\pm15.8$ & $183.1\pm18.6$ & -- & -- & 1000$\times$1000 \\
 & B & U & NDE &  5.2817 & 22.3614 & $84.3\pm8.7$ & $99.5\pm10.2$ & -- & -- & 1000$\times$400 \\
WHL~J002311.7+251510 & A & U & NDE &  5.8322 & 25.2764 & $47.9\pm4.9$ & $15.0\pm1.5$ & -- & -- & 230$\times$120 \\
WHL~J085608.5+541855 & A & Rem & U &  134.0545 & 54.2582 & $15.5\pm1.6$ & $31.0\pm3.3$ & -- & -- & 380$\times$210 \\
WHL~J091721.4+524607 & A & U & NDE &  139.3118 & 52.7691 & $11.8\pm2.0$ & $12.8\pm2.2$ & -- & -- & 290$\times$170 \\
 & B & AGN & NDE &  139.3072 & 52.8599 & $6.1\pm0.9$ & $6.7\pm1.0$ & -- & -- & 350$\times$170 \\
WHL~J101350.8+344251 & A & R & cR &  153.5143 & 34.6818 & $4.7\pm0.7$ & $2.8\pm0.4$ & -- & -- & 310$\times$150 \\
 & B & U & NDE &  153.4087 & 34.7283 & $7.1\pm0.8$ & $4.2\pm0.5$ & -- & -- & 220$\times$140 \\
 & C & cH & cH &  153.4531 & 34.7043 & $9.7\pm1.1$ & $5.7\pm0.6$  & -- & -- & 900$\times$300 \\
WHL~J130503.5+314255 & A & H & cH &  196.2470 & 31.7164 & $22.9\pm2.8$ & $134.3\pm16.7$ & $19.7\pm2.6$ & $116.0\pm15.4$ & 1400$\times$800 \\
 & B & cR & cR &  196.3271 & 31.7141 & $2.8\pm0.6$ & $16.6\pm3.5$ & -- & -- & 1400$\times$320 \\
 & C & U & cR &  196.1985 & 31.7700 & $8.6\pm1.0$ & $50.3\pm6.1$ & -- & -- & 720$\times$360 \\
WHL~J165540.4+334422 & A & Rem & U &  253.8790 & 33.7450 & $35.4\pm3.7$ & $73.3\pm7.7$ & $35.9\pm3.9$ & $74.6\pm8.1$ & 630$\times$400 \\
WHL~J172125.4+294144 & A & U & U &  260.2237 & 29.7245 & $19.3\pm2.3$ & $19.0\pm2.3$ & -- & -- & 560$\times$260 \\
 & B & R & cR &  260.1413 & 29.7298 & $23.1\pm2.8$ & $22.8\pm2.7$ & -- & -- & 600$\times$340 \\
WHL~J173424.0+332526 & A & U & NDE &  263.6228 & 33.4883 & $137.8\pm13.8$ & $1332.8\pm133.7$ & -- & -- & 800$\times$330 \\
 & B & U & NDE &  263.6175 & 33.4510 & $94.9\pm9.6$ & $917.6\pm92.6$ & -- & -- & 840$\times$290 \\
\hline \\
    \end{tabular} \\
\label{tab:sources}
Notes: Col. 1: cluster name. Col. 2: source name. Col. 3: visual classification (VC). Col. 4: decision-tree classification (DTC; R: relic, H: halo, cR: candidate relic, cH: candidate halo, U: unclassified/uncertain, Rem: AGN remnant, NDE: no diffuse emission). Col. 5: Right Ascension. Col. 6: Declination. Col. 7: 144~MHz flux density measured within $2\sigma$ contour. Col. 8: radio power estimated from $S_{2\sigma}$. Col. 9: 144~MHz flux density obtained by fitting the SB profile with elliptical model. Col. 10: radio power at 144~MHz estimated from $S_{fit}$. Col. 11: projected size of the source.
\end{table*}

\section {Discussion}
\label{sec:disc}

\subsection{Source classification}
\label{sec:class}

Diffuse radio sources, particularly halos and relics, are not well studied in low-mass ($M_{500}\lesssim4\times10^{14}\,{\rm M_{\odot}}$) clusters. One reason is their expected lower luminosity which runs into sensitivity limits of current instruments. The other reason is the steep-spectrum ($\alpha<-1$) nature of diffuse sources that makes them difficult to be detected at high frequencies (above $\sim$1~GHz). These limitations imply that diffuse sources may be best detected with deep observations at low radio frequencies such as those between 120~MHz and 168~MHz provided by LoTSS-DR2. Diffuse radio sources associated with disturbed clusters that are less massive than $4\times10^{14}\,{\rm M_{\odot}}$ have only been found in a handful of cases. For radio halos, these systems include Abell~3562 \citep{Venturi2003,Giacintucci2005}; Abell~2061 \citep{Rudnick2009}; PSZ1~G018.75+23.57 \citep{Bernardi2016}; Abell~2146 \citep{Hlavacek-Larrondo2017,Hoang2019b}; RXC~J1825.3+3026 \cite{Botteon2019c}; Abell~1775 \citep{Botteon2021a}), Ant cluster \citep{Botteon2021}, Abell~990 \citep{Hoang2021a}, MCXC~J1036.1+5713 \citep{Osinga2021}, PSZ2~G040.58+77.12, PSZ2~G031.93+78.71, PSZ2~G112.48+56.99, PSZ2~G189.31+59.24, PSZ2~G048.10+57.16, PSZ2~G192.18+56.12, PSZ2~G049.32+44.37, PSZ2~G179.09+60.12, and 
PSZ2~G172.63+35.15 \citep{Botteon2022a}. For relics, the low-mass systems hosting relic(s) that have been reported in literature are ZwCl~0008.8+5215 \citep{Feretti2012a,VanWeeren2011e}, Abell~1240 \citep{Kempner2001,Bonafede2009,Hoang2018}, Abell~3376 \citep{Bagchi2006}, PSZ1~G096.89+24.17 \citep{deGrasperin2014,Jones2021}, Abell~168 \citep{Dwarakanath2018}, Abell~1904 \citep{vanWeeren2021}, PSZ2~G080.16+57.65, PSZ2~G099.48+55.60, PSZ2~G048.10+57.16, PSZ2~G165.46+66.15, and PSZ2~G057.61+34.93 \citep{Botteon2022a}.

In this paper, we report the discovery of diffuse radio sources in 28 non-PSZ2 galaxy clusters in the LoTSS-DR2 fields. More specifically, radio halos are detected at high confidence in six clusters and there are a further seven tentative detections of halos. Eleven clusters host one or more radio relics (11 confirmed and three candidate relics). Among these clusters hosting relics, three systems have two relics either on opposite sides of the cluster centre or at 90-degree angles. Seven clusters host, both, a confirmed or candidate radio halo and radio relics. Five diffuse radio sources in the cluster sample are connected to tailed radio galaxies. Thirteen diffuse sources in 11 clusters are  unclassified.

The galaxy clusters in our sample are mostly low-mass ($M_{500}\lesssim4\times10^{14}\,{M_{\odot}}$) systems, except for PSZRX~G100.21-30.38 ($M_{500}=5.71\times10^{14}\,{M_{\odot}}$), PSZRX~G116.06+80.14 ($M_{500}=6.74\times10^{14}\,{M_{\odot}}$), and PSZRX~G195.91+62.83 ($M_{500}=5.55\times10^{14}\,{M_{\odot}}$). Furthermore, eleven of 13 halos or tentative halos are new detections; eleven of 13 relics or tentative relics are newly discovered. Whilst we have attempted to carefully classify the sources mainly based on their morphology and location obtained from the LOFAR maps, many sources (13) remain unclassified for a number of reasons. Firstly, the lack of X-ray data prevents us from comparing thermal to non-thermal emission. For instance, the radio emission from halos is known to correlate with the thermal X-ray emission of the ICM. However, X-ray archival data is only available for 13 clusters, nine of which have data from low-resolution, shallow observations with ROSAT (see Table~\ref{tab:sample}). The lack of X-ray emission from the other clusters (i.e. Abell~1330, PSZRX~G095.27+48.27, WHL~J101350.8+344251, and WHL~J165540.4+334422) that potentially host a radio halo prevents us from firmly classifying the diffuse sources. Similarly, the relative location of elongated relics with respect to the thermal ICM that can be used to aid the classification of the such sources. However, sensitive X-ray data is missing for several candidate relics (i.e. Abell~1943, Abell~1963,  WHL~J130503.5+314255, WHL~J173424.0+332526). Secondly, ancillary radio data at different frequencies that might provide crucial information on the spectral and polarimetric properties of the diffuse sources are largely unavailable. Thirdly, the observed sizes of the diffuse radio sources of some of our low-mass clusters are not as large as those in massive systems and make the distinction between diffuse sources and radio galaxies more difficult. Hence, to understand the nature of the diffuse sources, multi-wavelength X-ray and high-frequency polarimetric radio observations are needed. In addition, deeper low-frequency radio observations might also help with the classification of diffuse sources of small angular sizes. 

Finally, we note that our visual classification (VC) of diffuse radio sources has been a conscientious procedure and even then there is some ambiguity regarding classifications. Alternative approaches such as by \cite{Botteon2022a} make use of a decision tree classification (DTC) to render classifications more reproducible. We apply the DTC to classify the diffuse sources in our sample. The resulting classification is given in Table~\ref{tab:sources} (i.e. Col. 4). In general, our VC is roughly in agreement with the results from DTC, especially for halos and relics. The sources are classified as no diffuse emission (NDE) by DTC  if they are clearly associated with AGN or unclassified (U) in some cases in our VC. For instance, source C in PSZRX G181.53+21.43 is classified as AGN in our VC, but it is listed as NDE by DTC; source A in Abell~1213 is unclassified (U) by VC, but it is NDE in the DTC. Nevertheless, from many different morphologies, brightness and levels of contamination our study highlights that classification will remain a challenge with deep, low frequency surveys where our target sources are often embedded in a sea of abundant low-energy electrons that are particularly abundant at low frequencies. Techniques based on machine learning may be promising but even then it is apparent that robust classifications are a formidable challenge \citep[see, e.g.][]{Aniyan2017,Lukic2019,Vavilova2021}.

\subsection{Cluster mass calibration}
\label{sec:mass}

\begin{table}[t]
	\centering
	\scriptsize
	\caption{Best-fit parameters for cluster mass calibration.}
	\begin{tabular}{llcccccccc}  
		\hline\hline \\
		Catalogues & $a$ & $b$ & $R^2$\\ \hline
		PSZ2 - DESI     &  $0.95\pm0.01$   & $0.02$   &  0.92\\
		PSZ2 - MCXC    &  $0.85\pm0.02$   &  $0.17\pm0.02$ & 0.68   \\
		PSZ2 - comPRASS &  $0.96\pm0.02$  & $0.04\pm0.02$  &  0.56 \\
		PSZ2 - WHL &  $0.87\pm0.05$  & $0.04\pm0.04$  &  0.46 \\
		\hline \\
	\end{tabular}	
	\label{tab:mass}
\end{table}

\begin{figure*}[!ht]
	\centering
	\begin{tikzpicture}
		\draw (0, 0) node[inner sep=0] {\includegraphics[width=0.24\textwidth]{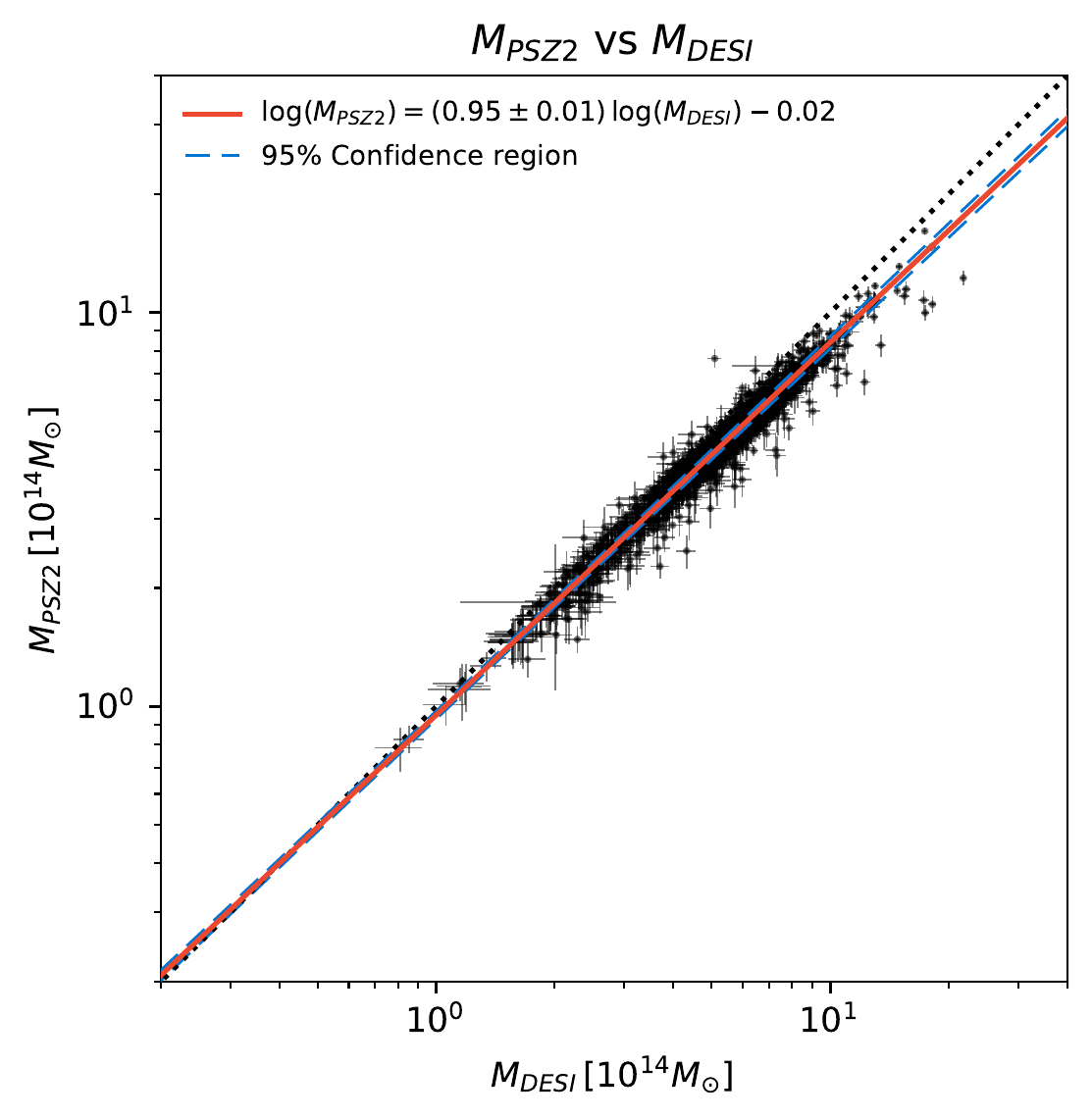}  \hfil
			\includegraphics[width=0.24\textwidth]{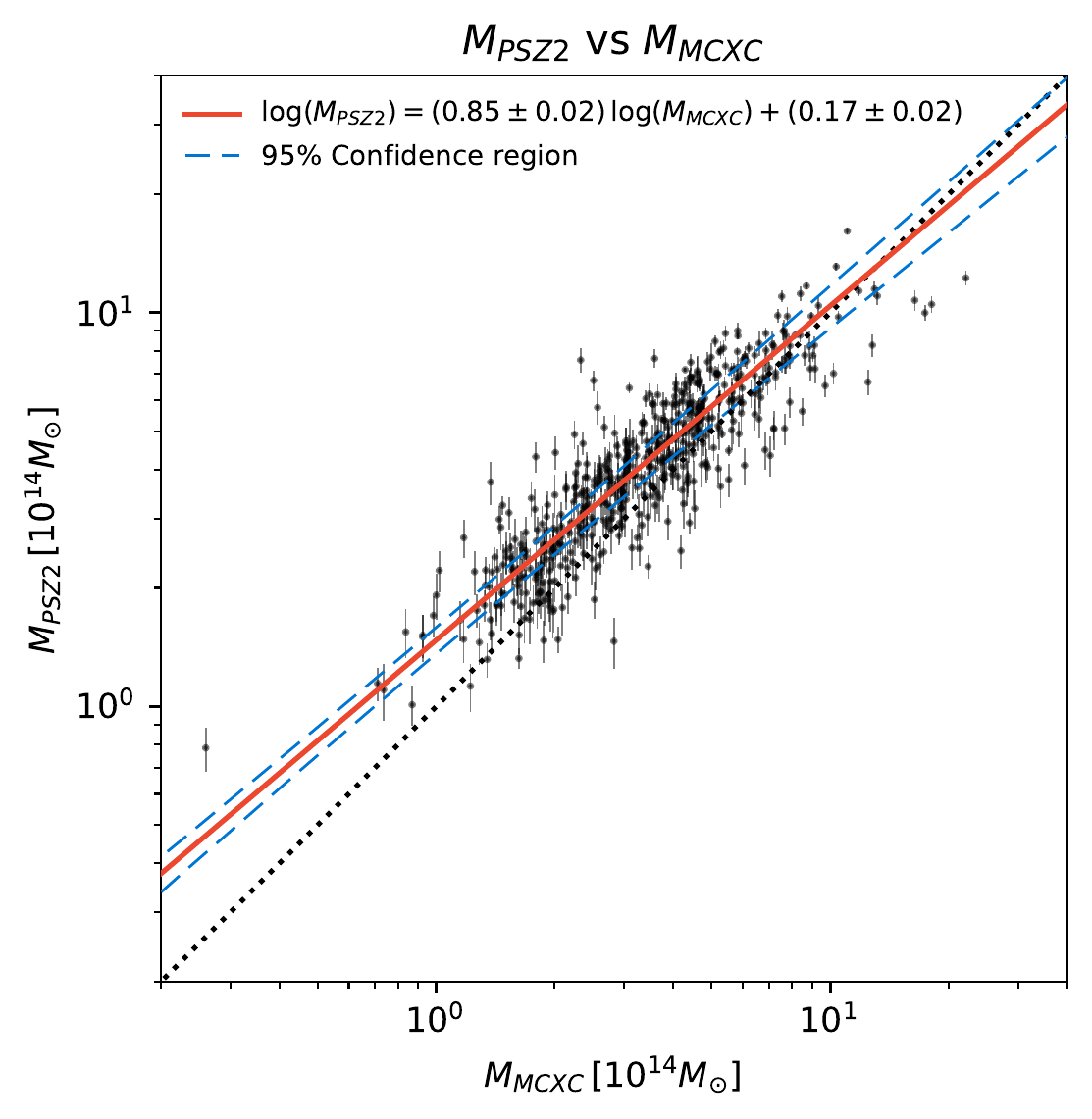}  \hfil
			\includegraphics[width=0.24\textwidth]{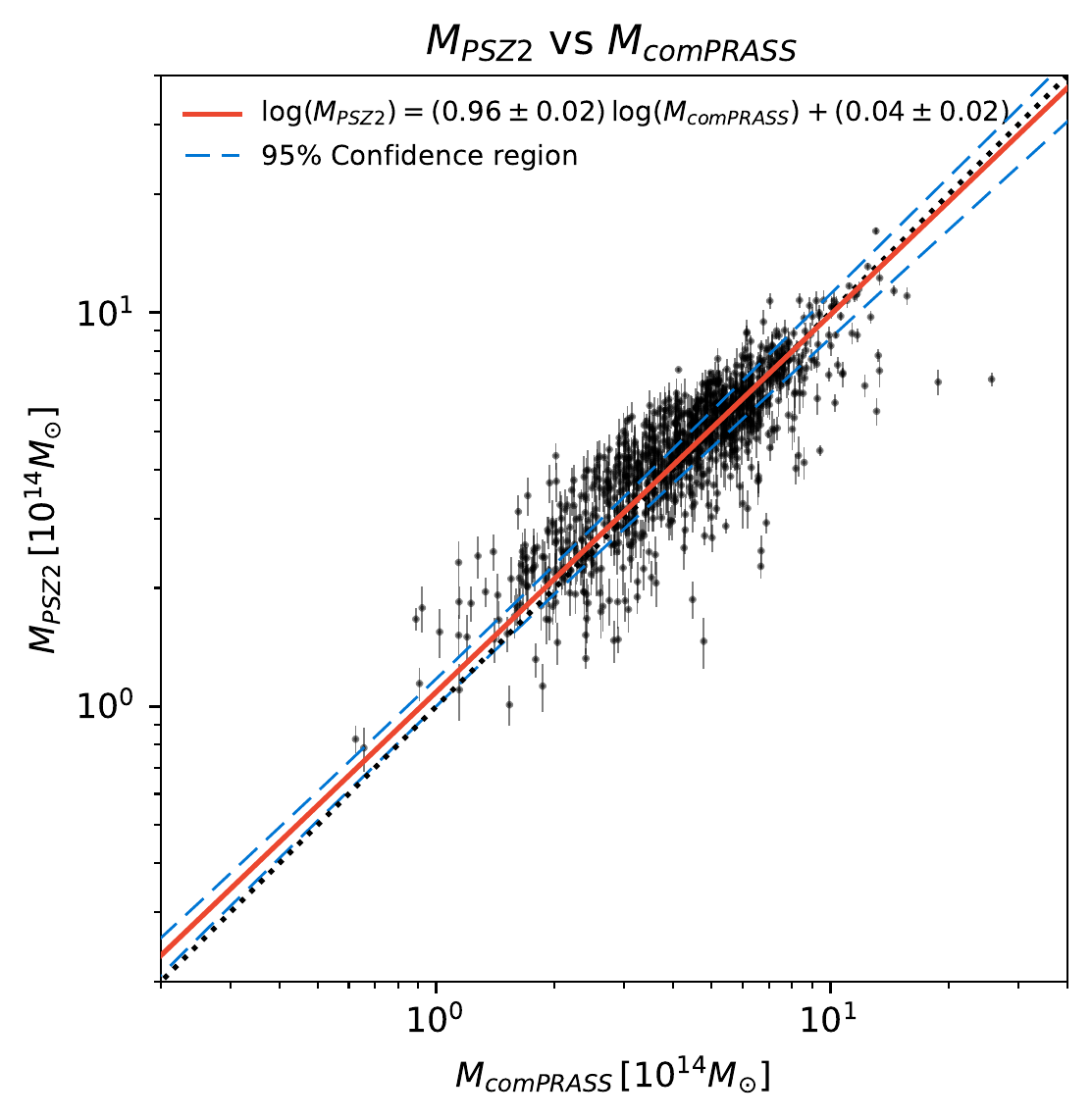}   \hfil
			\includegraphics[width=0.24\textwidth]{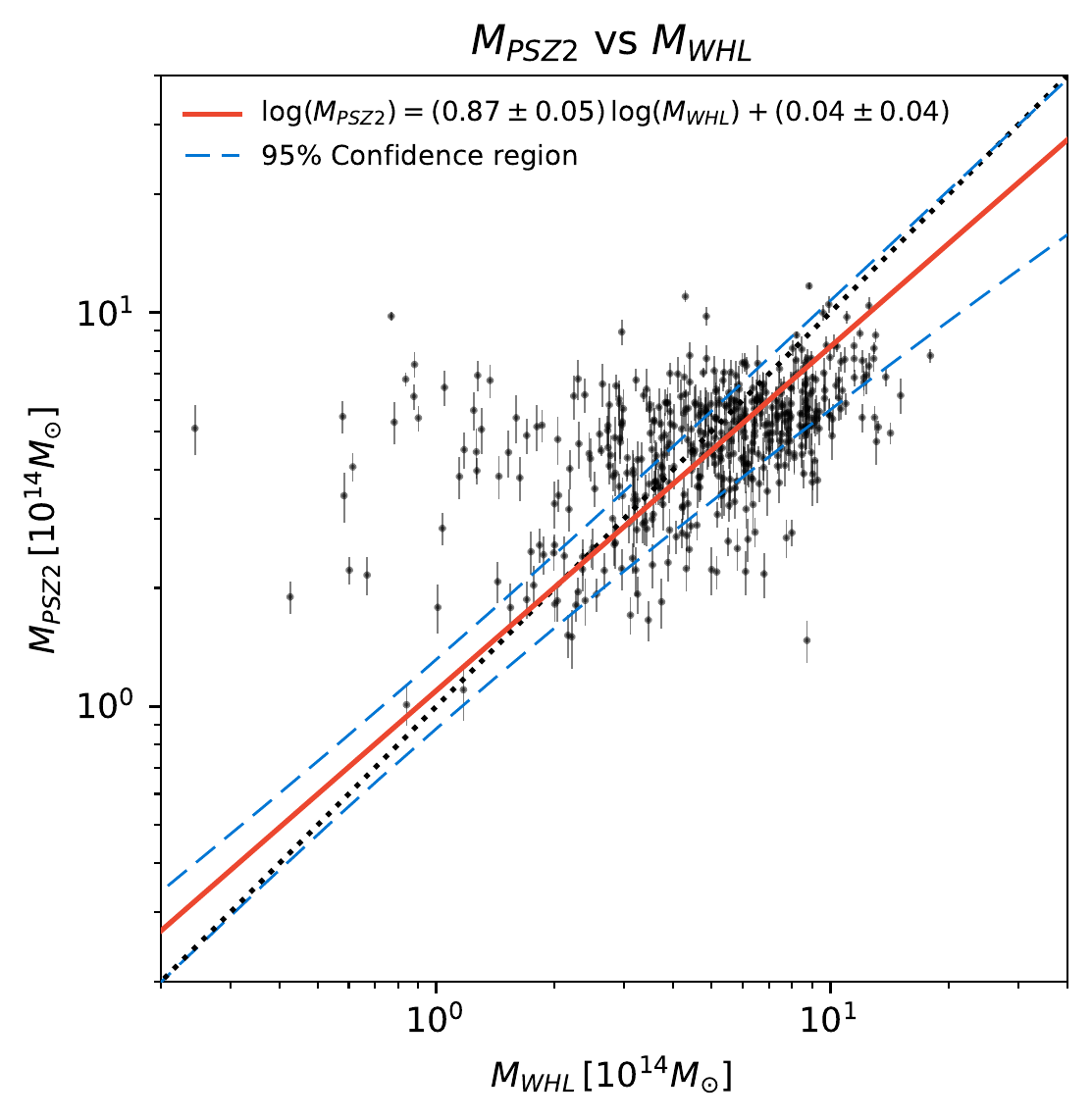}
		};
		%
	\end{tikzpicture}
	\caption{Scatter plots for the cluster mass reported in the PSZ2 catalogue and that reported in the DESI, MCXC, comPRASS, and WHL catalogues (from left to right). The red line indicate the best-fit correlation obtained by the orthogonal distance regression. The dashed blue lines show the 95\% confidence region. The dotted black line shows the diagonal line.
	}
	\label{fig:mass}
\end{figure*}

The masses of the galaxy clusters in the sample are shown in Table \ref{tab:sample} (Col. 7). For the DESI, MCXC, and comPRASS clusters we use the estimates reported in the corresponding catalogues. The masses of the WHL clusters presented in Table \ref{tab:sample} are not available, but they have richness estimates. This allows us to obtain $M_{500}$ of the WHL clusters using the $M_{500}$--richness scaling relation in Eq. 17 of \cite{Wen2015}. 

We note that there may also be unknown systematic uncertainties associating with the masses for the clusters in our sample because they are estimated with different methods using multi-wavelength data sets. We therefore search for such offsets and re-calibrate the mass estimates against the PSZ2 catalogues to make them consistent. To do this, we first cross-match within $5\arcmin$ (i.e. roughly equal to the resolution of the PSZ2 survey) the PSZ2 detected clusters and those in the DESI, MCXC, comPRASS, and WHL catalogues using \texttt{TOPCAT} \citep{Taylor2005}. In the case that a PSZ2 cluster is matched with multiple clusters in the other catalogues, the closest pair with smallest angular separation is chosen. In addition, we select only the pairs that have mass reported in both cross-matched catalogues. With these selection criteria, we find that the number of PSZ2 clusters in common with these catalogues are 1094, 527, 939, and 460, respectively. Using \texttt{Scipy} package\footnote{https://scipy.org}, we  perform the orthogonal distance regression \citep{Brown1990} to the cluster masses that are in, both, PSZ2 and other catalogues, i.e.:
\begin{equation}
    \log {M_{\rm PSZ2}} = a \log {M_{\rm x}} + b, 
    \label{eq:M}
\end{equation}
where $M_{\rm PSZ2}$ is the $M_{500}$ mass for the clusters in the PSZ2 catalogue; $M_{\rm x}$ is the $M_{500}$ mass for the DESI, MCXC, comPRASS, and WHL clusters; finally, $a$ and $b$ are free parameters. We make use of all common clusters in the cross-matched catalogues to estimate the best-fit parameters that are summarized in Table \ref{tab:mass}. The best-fit correlations are presented in Fig.~\ref{fig:mass}. The mass for the clusters in the DESI, MCXC, and comPRASS catalogues is well correlated with that reported in the PSZ2 catalogue. For the WHL clusters, their mass does not tightly follows that of the PSZ2 clusters, especially at low-mass regime ($M_{500}<2\times 10^{14}M_\odot$). 

We now use the best-fit relation in Eq. \ref{eq:M} to re-scale the mass of the galaxy clusters in our sample to that in the PSZ2 catalogue and calculate the scaled mass ($M_{500}^{\rm scaled}$) for these clusters. The resulting scaled mass is listed in Table~\ref{tab:sample} (Col. 8). In the following analysis, we use the scaled mass, instead of the original estimates reported in the DESI, MCXC, and comPRASS catalogues and the WHL mass calculated from the $M_{500}$--richness relation. 

The galaxy clusters from the optical Abell catalogues do not have mass estimates. We found that five Abell clusters in the sample are reported in the WHL catalogue. Two of them Abell 84 and Abell 1889 with Cluster sequence numbers (Seq) of 52 and 893, respectively, have mass estimates. For the other Abell clusters (i.e. Abell 1330, Abell 1943, and Abell 1963) mass estimate is not available. We follow the procedure for the WHL clusters described above to calculate the scaled mass ($M_{500}^{\rm scaled}$) of these Abell clusters (see Table~\ref{tab:sample}, Col. 8). 

\subsection{Scaling of luminosity of diffuse sources with cluster masses}
\label{sec:scaling}

\begin{figure}[!t]
	\centering
	\includegraphics[width=1\columnwidth]{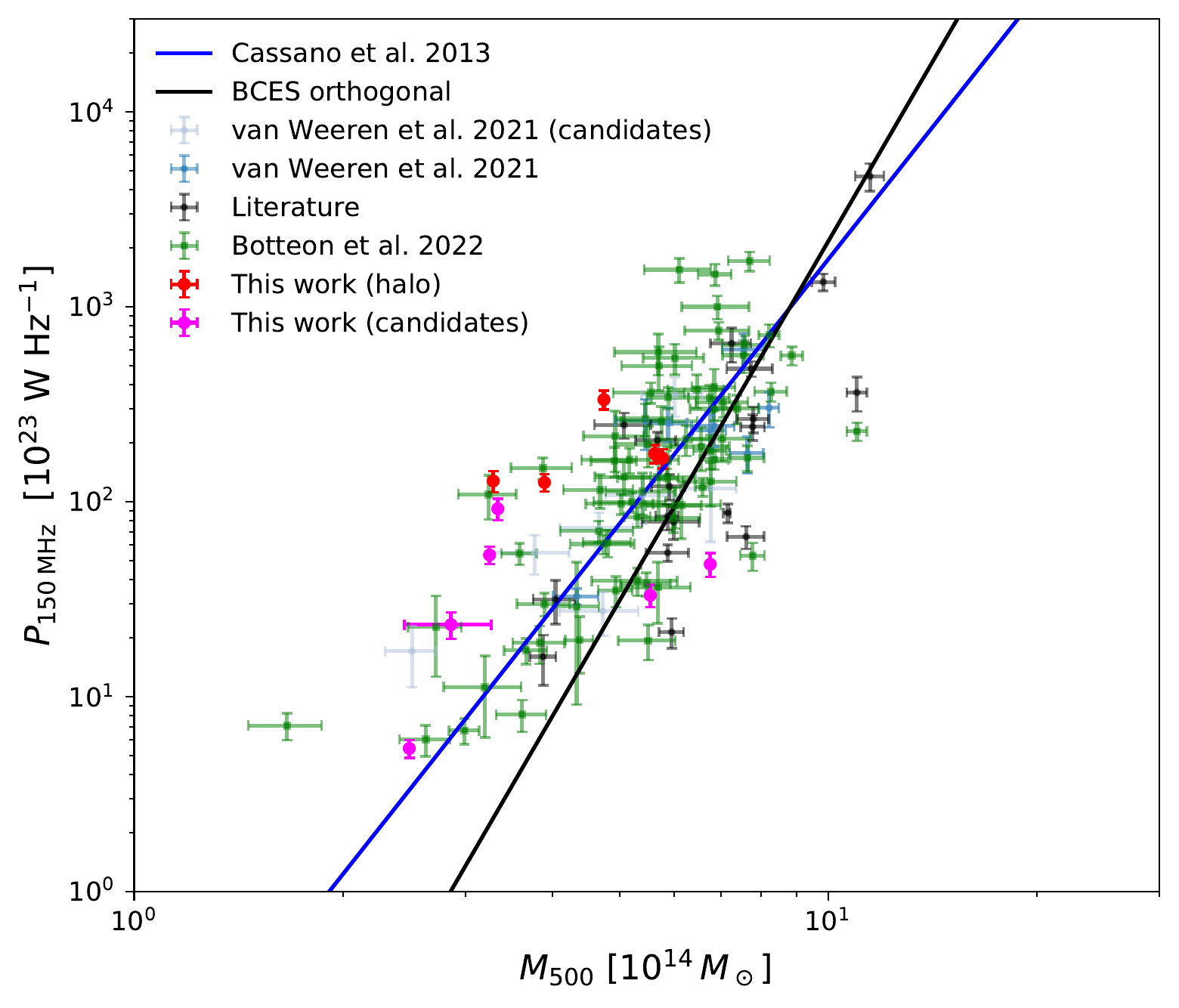}
	\caption{150 MHz radio power versus (scaled) mass plot for radio halos. To be consistent with the estimates reported in literature, the halo radio power in our sample is scaled to 150 MHz, assuming a spectral index of $-1.2$.  The plot is adapted from \cite{vanWeeren2021}.}
	\label{fig:PM}
\end{figure}

In Fig.~\ref{fig:PM}, we show the scaling of the radio power versus cluster masses for radio halos, adapting the plot from \cite{vanWeeren2021}. The radio power for the halos in our sample is scaled to that at 150~MHz, using the 144~MHz estimates in Table \ref{tab:sources} and assuming an index of -1.2 for the radio power spectrum. Abell 373 does not have mass estimate and the power for the tentative halo in PSZRX~G181.53+21.43 is not measured due to low SNR detection. These two clusters are not shown in the plot. We also include the measurements for the PSZ2 clusters from \cite{Botteon2022a} that has the 12 common halos with \cite{vanWeeren2021}. We display only the values for these common clusters reported in \cite{Botteon2022a}. In Fig.~\ref{fig:PM}, the black line shows the BCES (Bivariate Correlated Errors and intrinsic Scatter) orthogonal fit in \cite{vanWeeren2021} with a slope of 6.13 and the blue line is the result from \cite{Cassano2013a} with a slope of 4.51. The confirmed halos from this paper are shown in red and the tentative halos are in magenta. As expected the halos from our sample lie at lower masses than the halos that were taken from the literature. The radio powers of the halos lie above the correlation with two exceptions (i.e. the candidate halos in Abell~1330 and PSZRX~G116.06+80.14). However, since our sample does not obey any clearly defined selection criteria, it suffers from a bias as the more luminous halos are easier to find. It will take a carefully defined cluster sample at low masses in order to study the occurrence at lower masses and to determine whether the correlation is flatter at lower masses.

The total energy released in a merger between two clusters of mass $M$ is given by $E \propto M^2/R_{\rm vir}$, where $R_{\rm vir} \propto M^{1/3}$. Assuming that the radio luminosity scales as $P\propto  E/t_{\rm h}$, where $t_{\rm h}$ is the life time of the radio halo, we would obtain $P \propto M^{5/3}$ if $t_{\rm h}$ was independent of the cluster mass. If the luminosity-mass correlation is steeper than 5/3, this would imply that halos in higher-mass clusters had a shorter life time. Our new data may suggest that the correlation may not be as steep as the fit from \cite{vanWeeren2021}. In favour of this picture, the measurements for the halos in the PSZ2 clusters by \cite{Botteon2022a} (i.e. the green data points in Fig. \ref{fig:PM}) indicate a flatter slope. A detail study of the $P-M$ correlation is being carried out by Cuciti et al. (in prep.). 


\begin{figure*}[!ht]
	\centering
	\begin{tikzpicture}
		\draw (0, 0) node[inner sep=0] {\includegraphics[width=0.49\textwidth]{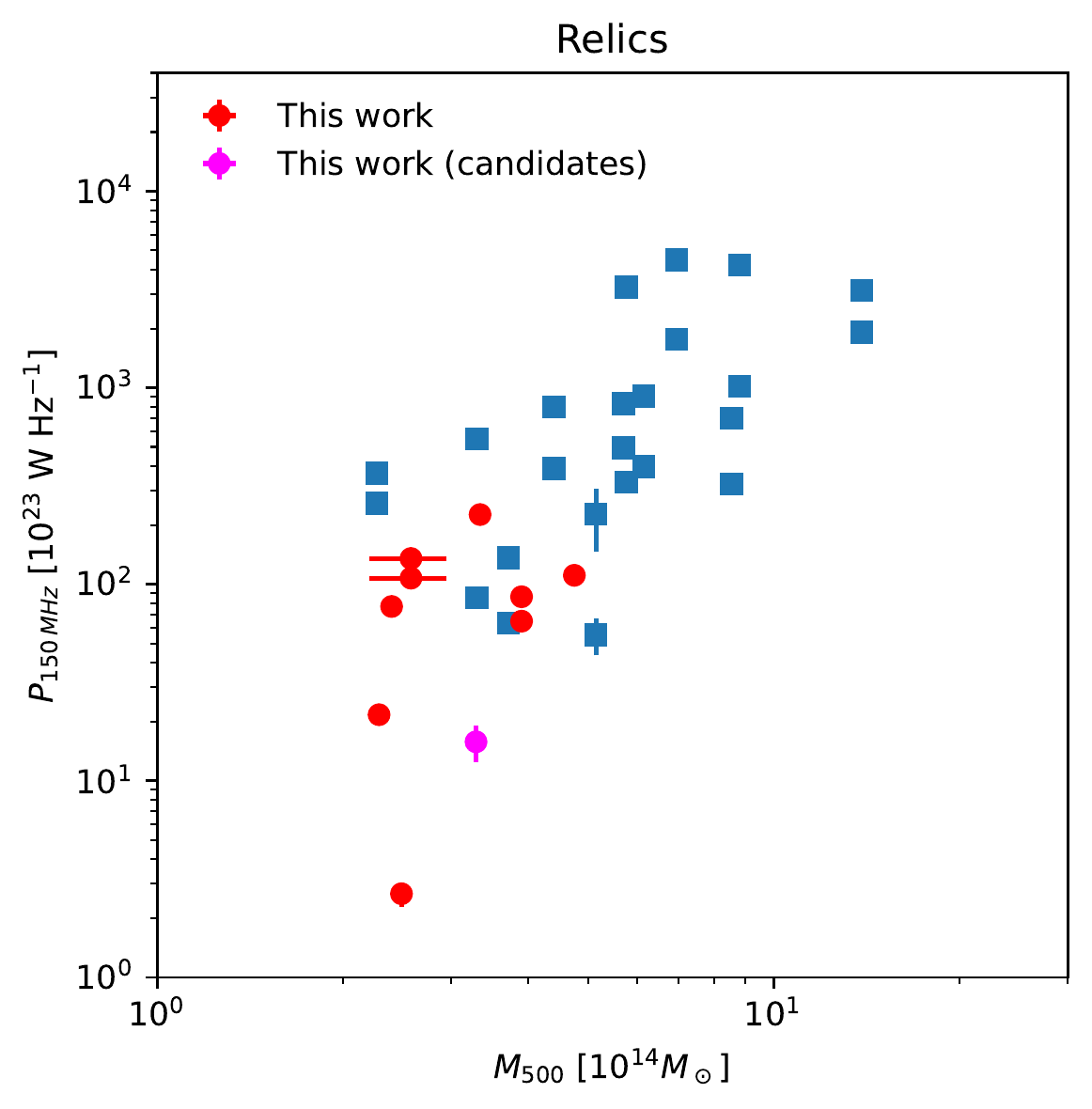}  \hfil
			\includegraphics[width=0.49\textwidth]{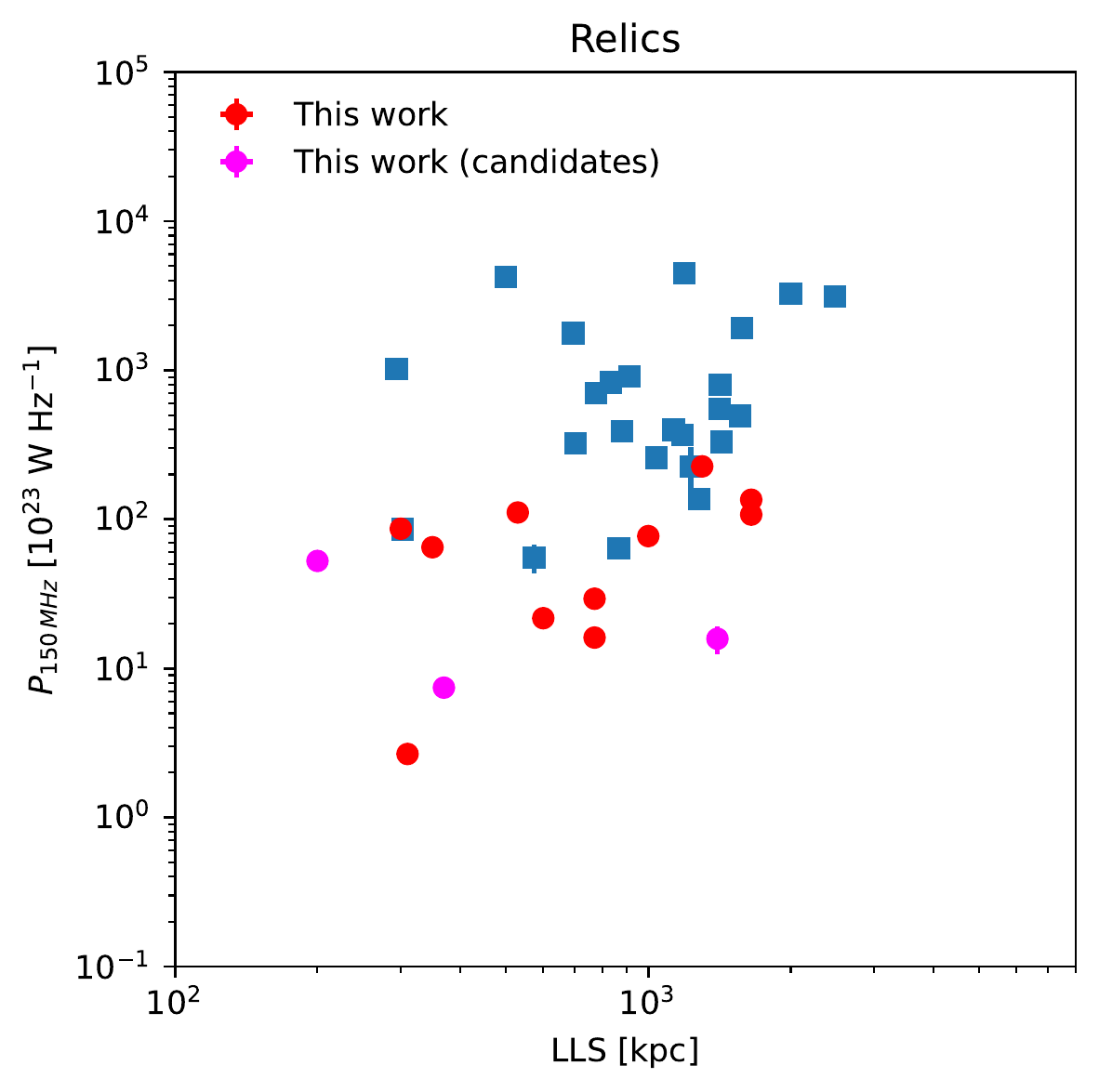}  
		};
		\draw (-7.25, -3.2) node {\color{black} (a)};
		\draw (2, -3.2) node {\color{black} (b)};
	\end{tikzpicture}
	\caption{Radio power, $P_{\rm 150\, MHz}$, vs scaled mass (a) and $P_{\rm 150\, MHz}$ versus largest linear size (LSS) (b) for radio relics. Data for other clusters are taken from \cite{deGrasperin2014}.
	}
	\label{fig:PM_relics}
\end{figure*}

In Fig.~\ref{fig:PM_relics} we plot the radio power, $P_{\rm 150\, MHz}$, vs mass and $P_{\rm 150\, MHz}$ versus largest linear size (LSS) for radio relics. Data for other clusters are taken from \cite{deGrasperin2014} and are scaled to 150~MHz using the spectral indices therein. For the relics without spectral index estimate, we assume a value of $-1.2$. The radio power of relics in lower-mass clusters appears to follow the trend of the higher-mass clusters, i.e. higher mass clusters host more powerful radio relics. However, the LLS of relics in the low-mass clusters do not appear to show any significant trends, just like the relics in higher-mass clusters. The scaling relations for a larger sample of relics in the LoTSS-DR2 fields are studied in more details in Jones et al. (in prep.).

\section{Conclusions}
\label{sec:conc}

In this paper we present LOFAR detections of diffuse emission from 28 non-PSZ2 galaxy clusters that are in the LoTSS-DR2 sky coverage. In this sample, there are 13 radio halos (seven tentative detection), 14 relics (three tentative detection), five diffuse sources clearly associating with AGN remnants (four tentative detection), and 13 unclassified diffuse sources. We measure the flux densities and radio powers directly from the LOFAR maps. For radio halos, we fitted the SB with elliptical models and derive the flux densities from the best-fit model.

In our paper we make several interesting points:

\begin{enumerate}

\item Diffuse sources are much harder to classify for lower-mass clusters because ancillary radio and other data is sparser and the source sizes are smaller, thus making them harder to distinguish from other radio sources, e.g. originating from AGN activities. For example, more than half of the halos in the sample are tentative detection.

\item A significant number of diffuse sources show an apparent connection to radio galaxies, as for example in  WHL~J002311.7+251510. There is also evidence for this at higher masses \citep[e.g.][]{vanWeeren2021,Botteon2022a} and it remains unclear what role radio galaxies play in supplying relativistic electrons to the acceleration mechanisms that power diffuse radio sources, in particular radio relics. Our sample adds to the growing number of examples suggesting that there may be a causal connection between radio galaxies and certain types of relics. Follow-up work, using surveys such as LoLSS  \citep[LOFAR LBA Sky Survey;][]{DeGasperin2021a}, LoDeSS (LOFAR Decametre Sky Survey), APERTIF (APERture Tile In Focus; Hess et al., in prep.) will be important to study the spectral properties of halos and relics and to investigate the effects of AGN.

\item Radio halos in lower-mass clusters are brighter than expected from the power--mass relation, even though this needs follow-up work before conclusions on the statistics of radio halos can be drawn. eROSITA (extended ROentgen Survey with an Imaging Telescope Array) surveys will provide a large X-ray flux limited sample of clusters and groups of galaxies that may help in testing scaling relations for diffuse radio sources \citep[e.g.][]{Pasini2021,Bulbul2021}. A recent example of joint LOFAR and eROSITA observations of galaxy clusters is given, e.g. by \cite{Ghirardini2021a}.

\end{enumerate}

\begin{acknowledgements}
      DNH, AB, and EB acknowledge support from the ERC through the grant ERC-Stg DRANOEL n. 714245. 
      MB acknowledges funding by the Deutsche Forschungsgemeinschaft (DFG, German Research Foundation) under Germany’s Excellence Strategy – EXC 2121 ‘Quantum Universe’ – 390833306.      
      A.S. is supported by the Women In Science Excel (WISE) programme of the Netherlands Organisation for Scientific Research (NWO), and acknowledges the World Premier Research Center Initiative (WPI) and the Kavli IPMU for the continued hospitality. SRON Netherlands Institute for Space Research is supported financially by NWO. 
      RJvW acknowledges support from the ERC Starting Grant ClusterWeb 804208.
      AD acknowledges support by the BMBF Verbundforschung under the grant 05A20STA.
      FG acknowledges support from INAF mainstream project 'Galaxy Clusters Science with LOFAR' 1.05.01.86.05. 
      AB acknowledges support from the VIDI research programme with project number 639.042.729, which is financed by the Netherlands Organisation for Scientific Research (NWO). 
      LOFAR (van Haarlem et al. 2013) is the Low Frequency Array designed and constructed by ASTRON. It has observing, data processing, and data storage facilities in several countries, which are owned by various parties (each with their own funding sources), and that are collectively operated by the ILT foundation under a joint scientific policy. The ILT resources have benefited from the following recent major funding sources: CNRS-INSU, Observatoire de Paris and Universit\'e d'Orl\'eans, France; BMBF, MIWF-NRW, MPG, Germany; Science Foundation Ireland (SFI), Department of Business, Enterprise and Innovation (DBEI), Ireland; NWO, The Netherlands; The Science and Technology Facilities Council, UK; Ministry of Science and Higher Education, Poland; The Istituto Nazionale di Astrofisica (INAF), Italy.
      This research made use of the Dutch national e-infrastructure with support of the SURF Cooperative (e-infra 180169) and the LOFAR e-infra group. The J\"ulich LOFAR Long Term Archive and the German LOFAR network are both coordinated and operated by the J\"ulich Supercomputing Centre (JSC), and computing resources on the supercomputer JUWELS at JSC were provided by the Gauss Centre for Supercomputing e.V. (grant CHTB00) through the John von Neumann Institute for Computing (NIC).
      This research made use of the University of Hertfordshire high-performance computing facility and the LOFAR-UK computing facility located at the University of Hertfordshire and supported by STFC [ST/P000096/1], and of the Italian LOFAR IT computing infrastructure supported and operated by INAF, and by the Physics Department of Turin university (under an agreement with Consorzio Interuniversitario per la Fisica Spaziale) at the C3S Supercomputing Centre, Italy.
      %
      %
\end{acknowledgements}

%
\bibliographystyle{aa} 
\bibliography{./library.bib} 

\begin{thebibliography}{99}
\expandafter\ifx\csname natexlab\endcsname\relax\def\natexlab#1{#1}\fi

\bibitem[{Abell {et~al.}(1989)Abell, {Corwin, Harold G.}, \&
  Olowin}]{Abell1989}
Abell, G.~O., {Corwin, Harold G.}, J., \& Olowin, R.~P. 1989, Astrophys. J.
  Suppl. Ser., 70, 1

\bibitem[{Ackermann {et~al.}(2016)Ackermann, Ajello, Albert, Atwood, Baldini,
  Ballet, Barbiellini, Bastieri, Bechtol, Bellazzini, Bissaldi, Blandford,
  Bloom, Bonino, Bottacini, Bregeon, Bruel, Buehler, Caliandro, Cameron,
  Caragiulo, Caraveo, Casandjian, Cavazzuti, Cecchi, Charles, Chekhtman,
  Chiaro, Ciprini, Cohen-Tanugi, Conrad, Cutini, D'Ammando, de~Angelis,
  de~Palma, Desiante, Digel, Venere, Drell, Favuzzi, Fegan, Fukazawa, Funk,
  Fusco, Gargano, Gasparrini, Giglietto, Giordano, Giroletti, Godfrey, Green,
  Grenier, Guiriec, Hays, Hewitt, Horan, J{\'{o}}hannesson, Kuss, Larsson,
  Latronico, Li, Li, Longo, Loparco, Lovellette, Lubrano, Madejski, Maldera,
  Manfreda, Mayer, Mazziotta, Michelson, Mitthumsiri, Mizuno, Monzani,
  Morselli, Moskalenko, Murgia, Nuss, Ohsugi, Orienti, Orlando, Ormes, Paneque,
  Pesce-Rollins, Petrosian, Piron, Pivato, Porter, Rain{\`{o}}, Rando, Razzano,
  Reimer, Reimer, S{\'{a}}nchez-Conde, Sgr{\`{o}}, Siskind, Spada, Spandre,
  Spinelli, Tajima, Takahashi, Thayer, Tibaldo, Torres, Tosti, Troja, Vianello,
  Wood, Zimmer, \& Rephaeli}]{Ackermann2016b}
Ackermann, M., Ajello, M., Albert, A., {et~al.} 2016, ApJ, 819, 149

\bibitem[{Ade {et~al.}(2016)Ade, Aghanim, Arnaud, Ashdown, Aumont, Baccigalupi,
  Banday, Barreiro, Barrena, Bartlett, Bartolo, Battaner, Battye, Benabed,
  Beno{\^{i}}t, Benoit-L{\'{e}}vy, Bernard, Bersanelli, Bielewicz, Bikmaev,
  B{\"{o}}hringer, Bonaldi, Bonavera, Bond, Borrill, Bouchet, Bucher, Burenin,
  Burigana, Butler, Calabrese, Cardoso, Carvalho, Catalano, Challinor,
  Chamballu, Chary, Chiang, Chon, Christensen, Clements, Colombi, Colombo,
  Combet, Comis, Couchot, Coulais, Crill, Curto, Cuttaia, Dahle, Danese,
  Davies, Davis, de~Bernardis, de~Rosa, de~Zotti, Delabrouille, D{\'{e}}sert,
  Dickinson, Diego, Dolag, Dole, Donzelli, Dor{\'{e}}, Douspis, Ducout, Dupac,
  Efstathiou, Eisenhardt, Elsner, En{\ss}lin, Eriksen, Falgarone, Fergusson,
  Feroz, Ferragamo, Finelli, Forni, Frailis, Fraisse, Franceschi, Frejsel,
  Galeotta, Galli, Ganga, G{\'{e}}nova-Santos, Giard, Giraud-H{\'{e}}raud,
  Gjerl{\o}w, Gonz{\'{a}}lez-Nuevo, G{\'{o}}rski, Grainge, Gratton, Gregorio,
  Gruppuso, Gudmundsson, Hansen, Hanson, Harrison, Hempel,
  Henrot-Versill{\'{e}}, Hern{\'{a}}ndez-Monteagudo, Herranz, Hildebrandt,
  Hivon, Hobson, Holmes, Hornstrup, Hovest, Huffenberger, Hurier, Jaffe, Jaffe,
  Jin, Jones, Juvela, Keih{\"{a}}nen, Keskitalo, Khamitov, Kisner, Kneissl,
  Knoche, Kunz, Kurki-Suonio, Lagache, Lamarre, Lasenby, Lattanzi, Lawrence,
  Leonardi, Lesgourgues, Levrier, Liguori, Lilje, Linden-V{\o}rnle,
  L{\'{o}}pez-Caniego, Lubin, Mac{\'{i}}as-P{\'{e}}rez, Maggio, Maino, Mak,
  Mandolesi, Mangilli, Martin, Mart{\'{i}}nez-Gonz{\'{a}}lez, Masi, Matarrese,
  Mazzotta, McGehee, Mei, Melchiorri, Melin, Mendes, Mennella, Migliaccio,
  Mitra, Miville-Desch{\^{e}}nes, Moneti, Montier, Morgante, Mortlock, Moss,
  Munshi, Murphy, Naselsky, Nastasi, Nati, Natoli, Netterfield,
  N{\o}rgaard-Nielsen, Noviello, Novikov, Novikov, Olamaie, Oxborrow, Paci,
  Pagano, Pajot, Paoletti, Pasian, Patanchon, Pearson, Perdereau, Perotto,
  Perrott, Perrotta, Pettorino, Piacentini, Piat, Pierpaoli, Pietrobon,
  Plaszczynski, Pointecouteau, Polenta, Pratt, Pr{\'{e}}zeau, Prunet, Puget,
  Rachen, Reach, Rebolo, Reinecke, Remazeilles, Renault, Renzi, Ristorcelli,
  Rocha, Rosset, Rossetti, Roudier, Rozo, Rubi{\~{n}}o-Mart{\'{i}}n, Rumsey,
  Rusholme, Rykoff, Sandri, Santos, Saunders, Savelainen, Savini, Schammel,
  Scott, Seiffert, Shellard, Shimwell, Spencer, Stanford, Stern, Stolyarov,
  Stompor, Streblyanska, Sudiwala, Sunyaev, Sutton, Suur-Uski, Sygnet, Tauber,
  Terenzi, Toffolatti, Tomasi, Tramonte, Tristram, Tucci, Tuovinen, Umana,
  Valenziano, Valiviita, {Van Tent}, Vielva, Villa, Wade, Wandelt, Wehus,
  White, Wright, Yvon, Zacchei, \& Zonca}]{Planck2016}
Ade, P. A.~R., Aghanim, N., Arnaud, M., {et~al.} 2016, Astron. Astrophys., 594,
  A27

\bibitem[{Alam {et~al.}(2015)Alam, Albareti, Prieto, Anders, Anderson,
  Anderton, Andrews, Armengaud, Aubourg, Bailey, Basu, Bautista, Beaton, Beers,
  Bender, Berlind, Beutler, Bhardwaj, Bird, Bizyaev, Blake, Blanton, Blomqvist,
  Bochanski, Bolton, Bovy, Bradley, Brandt, Brauer, Brinkmann, Brown,
  Brownstein, Burden, Burtin, Busca, Cai, Capozzi, Rosell, Carr, Carrera,
  Chambers, Chaplin, Chen, Chiappini, Chojnowski, Chuang, Clerc, Comparat,
  Covey, Croft, Cuesta, Cunha, Costa, Rio, Davenport, Dawson, Lee, Delubac,
  Deshpande, Dhital, Dutra-Ferreira, Dwelly, Ealet, Ebelke, Edmondson,
  Eisenstein, Ellsworth, Elsworth, Epstein, Eracleous, Escoffier, Esposito,
  Evans, Fan, Fern{\'{a}}ndez-Alvar, Feuillet, Ak, Finley, Finoguenov,
  Flaherty, Fleming, Font-Ribera, Foster, Frinchaboy, Galbraith-Frew,
  Garc{\'{i}}a, Garc{\'{i}}a-Hern{\'{a}}ndez, P{\'{e}}rez, Gaulme, Ge,
  G{\'{e}}nova-Santos, Georgakakis, Ghezzi, Gillespie, Girardi, Goddard,
  Gontcho, Hern{\'{a}}ndez, Grebel, Green, Grieb, Grieves, Gunn, Guo, Harding,
  Hasselquist, Hawley, Hayden, Hearty, Hekker, Ho, Hogg, Holley-Bockelmann,
  Holtzman, Honscheid, Huber, Huehnerhoff, Ivans, Jiang, Johnson, Kinemuchi,
  Kirkby, Kitaura, Klaene, Knapp, Kneib, Koenig, Lam, Lan, Lang, Laurent, Goff,
  Leauthaud, Lee, Lee, Licquia, Liu, Long, L{\'{o}}pez-Corredoira,
  Lorenzo-Oliveira, Lucatello, Lundgren, Lupton, Iii, Mahadevan, Maia,
  Majewski, Malanushenko, Malanushenko, Manchado, Manera, Mao, Maraston,
  Marchwinski, Margala, Martell, Martig, Masters, Mathur, McBride, McGehee,
  McGreer, McMahon, M{\'{e}}nard, Menzel, Merloni, M{\'{e}}sz{\'{a}}ros,
  Miller, Miralda-Escud{\'{e}}, Miyatake, Montero-Dorta, More, Morganson,
  Morice-Atkinson, Morrison, Mosser, Muna, Myers, Nandra, Newman, Neyrinck,
  Nguyen, Nichol, Nidever, Noterdaeme, Nuza, O'Connell, O'Connell, O'Connell,
  Ogando, Olmstead, Oravetz, Oravetz, Osumi, Owen, Padgett, Padmanabhan,
  Paegert, Palanque-Delabrouille, Pan, Parejko, P{\^{a}}ris, Park,
  Pattarakijwanich, Pellejero-Ibanez, Pepper, Percival, P{\'{e}}rez-Fournon,
  P{\'{e}}rez-Ra'Fols, Petitjean, Pieri, Pinsonneault, Mello, Prada, Prakash,
  Price-Whelan, Protopapas, Raddick, Rahman, Reid, Rich, Rix, Robin, Rockosi,
  Rodrigues, Rodr{\'{i}}guez-Torres, Roe, Ross, Ross, Rossi, Ruan,
  Rubi{\~{n}}o-Mart{\'{i}}n, Rykoff, Salazar-Albornoz, Salvato, Samushia,
  S{\'{a}}nchez, Santiago, Sayres, Schiavon, Schlegel, Schmidt, Schneider,
  Schultheis, Schwope, Sc{\'{o}}ccola, Scott, Sellgren, Seo, Serenelli, Shane,
  Shen, Shetrone, Shu, Aguirre, Sivarani, Skrutskie, Slosar, Smith, Sobreira,
  Souto, Stassun, Steinmetz, Stello, Strauss, Streblyanska, Suzuki, Swanson,
  Tan, Tayar, Terrien, Thakar, Thomas, Thomas, Thompson, Tinker, Tojeiro,
  Troup, Vargas-Maga{\~{n}}a, Vazquez, Verde, Viel, Vogt, Wake, Wang, Weaver,
  Weinberg, Weiner, White, Wilson, Wisniewski, Wood-Vasey, Ye'Che, York,
  Zakamska, Zamora, Zasowski, Zehavi, Zhao, Zheng, Zhou, Zhou, Zou, \&
  Zhu}]{Alam2015}
Alam, S., Albareti, F.~D., Prieto, C.~A., {et~al.} 2015, Astrophys. Journal,
  Suppl. Ser., 219

\bibitem[{Aniyan \& Thorat(2017)}]{Aniyan2017}
Aniyan, A.~K. \& Thorat, K. 2017, Astrophys. J. Suppl. Ser., 230, 20

\bibitem[{Bagchi {et~al.}(2006)Bagchi, Durret, Neto, \& Paul}]{Bagchi2006}
Bagchi, J., Durret, F., Neto, G. B. L. G. B.~L., \& Paul, S. 2006, Science
  (80-. )., 314, 791

\bibitem[{Basu(2012)}]{Basu2012a}
Basu, K. 2012, Mon. Not. R. Astron. Soc. Lett., 421, 1

\bibitem[{Bender {et~al.}(2016)Bender, Kennedy, Ade, Basu, Bertoldi, Burkutean,
  Clarke, Dahlin, Dobbs, Ferrusca, Flanigan, Halverson, Holzapfel, Horellou,
  Johnson, Kermish, Klein, Kneissl, Lanting, Lee, Mehl, Menten, Muders,
  Nagarajan, Pacaud, Reichardt, Richards, Schaaf, Schwan, Sommer, Spieler,
  Tucker, \& Westbrook}]{Bender2016}
Bender, A.~N., Kennedy, J., Ade, P. A.~R., {et~al.} 2016, Mon. Not. R. Astron.
  Soc., 460, 3432

\bibitem[{Bernardi {et~al.}(2016)Bernardi, Venturi, Cassano, Dallacasa,
  Brunetti, Cuciti, Johnston-Hollitt, Oozeer, Parekh, \&
  Smirnov}]{Bernardi2016}
Bernardi, G., Venturi, T., Cassano, R., {et~al.} 2016, Mon. Not. R. Astron.
  Soc., 456, 1259

\bibitem[{Bonafede {et~al.}(2009)Bonafede, Giovannini, Feretti, Govoni, \&
  Murgia}]{Bonafede2009}
Bonafede, A., Giovannini, G., Feretti, L., Govoni, F., \& Murgia, M. 2009,
  Astron. Astrophys., 494, 429

\bibitem[{Botteon {et~al.}(2019)Botteon, Cassano, Eckert, Brunetti, Dallacasa,
  Shimwell, van Weeren, Gastaldello, Bonafede, Br{\"{u}}ggen, B{\^{i}}rzan,
  Clavico, Cuciti, de~Gasperin, {De Grandi}, Ettori, Ghizzardi, Rossetti,
  R{\"{o}}ttgering, \& Sereno}]{Botteon2019c}
Botteon, A., Cassano, R., Eckert, D., {et~al.} 2019, Astron. Astrophys., 630,
  A77

\bibitem[{Botteon {et~al.}(2021{\natexlab{a}})Botteon, Cassano, van Weeren,
  Shimwell, Bonafede, Br{\"{u}}ggen, Brunetti, Cuciti, Dallacasa, de~Gasperin,
  {Di Gennaro}, Gastaldello, Hoang, Rossetti, \&
  R{\"{o}}ttgering}]{Botteon2021}
Botteon, A., Cassano, R., van Weeren, R.~J., {et~al.} 2021{\natexlab{a}},
  Astrophys. J. Lett., 914, L29

\bibitem[{Botteon {et~al.}(2021{\natexlab{b}})Botteon, Giacintucci,
  Gastaldello, Venturi, Brunetti, {Van Weeren}, Shimwell, Rossetti, Akamatsu,
  Br{\"{u}}ggen, Cassano, Cuciti, {De Gasperin}, Drabent, Hoeft, Mandal,
  R{\"{o}}ttgering, \& Tasse}]{Botteon2021a}
Botteon, A., Giacintucci, S., Gastaldello, F., {et~al.} 2021{\natexlab{b}},
  Astron. Astrophys., 649, 1

\bibitem[{Botteon {et~al.}(2022)Botteon, Shimwell, Cassano, Cuciti, Zhang,
  Bruno, Camillini, Natale, Jones, Gastaldello, Simionescu, Rossetti, Akamatsu,
  van Weeren, Brunetti, Br{\"{u}}ggen, Groeneveld, Hoang, Hardcastle, Ignesti,
  {Di Gennaro}, Bonafede, Drabent, R{\"{o}}ttgering, Hoeft, \&
  de~Gasperin}]{Botteon2022a}
Botteon, A., Shimwell, T.~W., Cassano, R., {et~al.} 2022, Astron. Astrophys.,
  660, A78

\bibitem[{Boxelaar {et~al.}(2021)Boxelaar, van Weeren, \&
  Botteon}]{Boxelaar2021}
Boxelaar, J.~M., van Weeren, R.~J., \& Botteon, A. 2021, Astron. Comput., 35,
  100464

\bibitem[{Brown \& Fuller(1990)}]{Brown1990}
Brown, P.~J. \& Fuller, W.~A., eds. 1990, Contemporary Mathematics, Vol. 112,
  {Statistical Analysis of Measurement Error Models and Applications}
  (Providence, Rhode Island: American Mathematical Society)

\bibitem[{Br{\"{u}}ggen \& Vazza(2020)}]{Bruggen2020}
Br{\"{u}}ggen, M. \& Vazza, F. 2020, Mon. Not. R. Astron. Soc., 493, 2306

\bibitem[{Brunetti {et~al.}(2004)Brunetti, Blasi, Cassano, \&
  Gabici}]{Brunetti2004}
Brunetti, G., Blasi, P., Cassano, R., \& Gabici, S. 2004, Mon. Not. R. Astron.
  Soc., 350, 1174

\bibitem[{Brunetti {et~al.}(2012)Brunetti, Blasi, Reimer, Rudnick, Bonafede, \&
  Brown}]{Brunetti2012}
Brunetti, G., Blasi, P., Reimer, O., {et~al.} 2012, Mon. Not. R. Astron. Soc.,
  426, 956

\bibitem[{Brunetti {et~al.}(2009)Brunetti, Cassano, Dolag, \&
  Setti}]{Brunetti2009}
Brunetti, G., Cassano, R., Dolag, K., \& Setti, G. 2009, Astron. Astrophys.,
  507, 661

\bibitem[{Brunetti {et~al.}(2008)Brunetti, Giacintucci, Cassano, Lane,
  Dallacasa, Venturi, Kassim, Setti, Cotton, \& Markevitch}]{Brunetti2008}
Brunetti, G., Giacintucci, S., Cassano, R., {et~al.} 2008, Nature, 455, 944

\bibitem[{Brunetti \& Jones(2014)}]{Brunetti2014}
Brunetti, G. \& Jones, T.~W. 2014, Int. J. Mod. Phys. D, 23, 1430007

\bibitem[{Brunetti \& Lazarian(2007)}]{Brunetti2007a}
Brunetti, G. \& Lazarian, A. 2007, Mon. Not. R. Astron. Soc., 378, 245

\bibitem[{Brunetti \& Lazarian(2011)}]{Brunetti2011b}
Brunetti, G. \& Lazarian, A. 2011, MNRAS, 412, 817

\bibitem[{Brunetti {et~al.}(2001)Brunetti, Setti, Feretti, \&
  Giovannini}]{Brunetti2001}
Brunetti, G., Setti, G., Feretti, L., \& Giovannini, G. 2001, MNRAS, 320, 365

\bibitem[{Brunetti {et~al.}(2017)Brunetti, Zimmer, \& Zandanel}]{Brunetti2017}
Brunetti, G., Zimmer, S., \& Zandanel, F. 2017, MNRAS, 472, 1506

\bibitem[{Bulbul {et~al.}(2021)Bulbul, Liu, Pasini, Comparat, Hoang, Klein,
  Ghirardini, Salvato, Merloni, Seppi, Wolf, Anderson, Bahar, Brusa, Brueggen,
  Buchner, Dwelly, Ibarra-Medel, Chitham, Liu, Nandra, Ramos-Ceja, Sanders, \&
  Shen}]{Bulbul2021}
Bulbul, E., Liu, A., Pasini, T., {et~al.} 2021 [\eprint[arXiv]{2110.09544}]

\bibitem[{Cassano {et~al.}(2013)Cassano, Ettori, Brunetti, Giacintucci, Pratt,
  Venturi, Kale, Dolag, \& Markevitch}]{Cassano2013a}
Cassano, R., Ettori, S., Brunetti, G., {et~al.} 2013, ApJ, 777, 141

\bibitem[{Cassano {et~al.}(2010)Cassano, Ettori, Giacintucci, Brunetti,
  Markevitch, Venturi, \& Gitti}]{Cassano2010}
Cassano, R., Ettori, S., Giacintucci, S., {et~al.} 2010, Astrophys. J., 721,
  L82

\bibitem[{Cuciti {et~al.}(2021)Cuciti, Cassano, Brunetti, Dallacasa,
  de~Gasperin, Ettori, Giacintucci, Kale, Pratt, van Weeren, \&
  Venturi}]{Cuciti2021}
Cuciti, V., Cassano, R., Brunetti, G., {et~al.} 2021, Astron. Astrophys., 647,
  A51

\bibitem[{Cuciti {et~al.}(2015)Cuciti, Cassano, Brunetti, Dallacasa, Kale,
  Ettori, \& Venturi}]{Cuciti2015}
Cuciti, V., Cassano, R., Brunetti, G., {et~al.} 2015, Astron. Astrophys., 580,
  A97

\bibitem[{de~Gasperin {et~al.}(2019)de~Gasperin, Dijkema, Drabent, Mevius,
  Rafferty, van Weeren, Br{\"{u}}ggen, Callingham, Emig, Heald, Intema,
  Morabito, Offringa, Oonk, Orr{\`{u}}, R{\"{o}}ttgering, Sabater, Shimwell,
  Shulevski, \& Williams}]{DeGasperin2019}
de~Gasperin, F., Dijkema, T.~J., Drabent, A., {et~al.} 2019, Astron.
  Astrophys., 622, A5

\bibitem[{de~Gasperin {et~al.}(2022)de~Gasperin, Rudnick, Finoguenov, Wittor,
  Akamatsu, Br{\"{u}}ggen, Chibueze, Clarke, Cotton, Cuciti,
  Dom{\'{i}}nguez-Fern{\'{a}}ndez, Knowles, O'Sullivan, \&
  Sebokolodi}]{deGasperin2022a}
de~Gasperin, F., Rudnick, L., Finoguenov, A., {et~al.} 2022, Astron.
  Astrophys., 659, A146

\bibitem[{{De Gasperin} {et~al.}(2014){De Gasperin}, {Van Weeren},
  Br{\"{u}}ggen, Vazza, Bonafede, Intema, Bruggen, Vazza, Bonafede, Intema,
  Br{\"{u}}ggen, Vazza, Bonafede, \& Intema}]{deGrasperin2014}
{De Gasperin}, F., {Van Weeren}, R.~J., Br{\"{u}}ggen, M., {et~al.} 2014,
  MNRAS, 444, 3130

\bibitem[{{De Gasperin} {et~al.}(2021){De Gasperin}, Williams, Best,
  Br{\"{u}}ggen, Brunetti, Cuciti, Dijkema, Hardcastle, Norden, Offringa,
  Shimwell, van Weeren, Bomans, Bonafede, Botteon, Callingham, Cassano,
  {Chy{\.{z}} Y}, Emig, Edler, Haverkorn, Heald, Heesen, Iacobelli, Intema,
  Kadler, Ma{\l}ek, Mevius, Miley, Mingo, Morabito, Sabater, Morganti,
  Orr{\'{u}}, Pizzo, Prandoni, Shulevski, Tasse, Vaccari, Zarka,
  R{\"{o}}ttgering, Chy{\.{z}}y, Emig, Edler, Haverkorn, Heald, Heesen,
  Iacobelli, Intema, Kadler, Ma{\l}ek, Mevius, Miley, Mingo, Morabito, Sabater,
  Morganti, Orr{\'{u}}, Pizzo, Prandoni, Shulevski, Tasse, Vaccari, Zarka,
  R{\"{o}}ttgering, {Chy{\.{z}} Y}, Emig, Edler, Haverkorn, Heald, Heesen,
  Iacobelli, Intema, Kadler, Ma{\l}ek, Mevius, Miley, Mingo, Morabito, Sabater,
  Morganti, Orr{\'{u}}, Pizzo, Prandoni, Shulevski, Tasse, Vaccari, Zarka, \&
  R{\"{o}}ttgering}]{DeGasperin2021a}
{De Gasperin}, F., Williams, W.~L., Best, P., {et~al.} 2021, Astron.
  Astrophys., 648, A104

\bibitem[{{Di Gennaro} {et~al.}(2021){Di Gennaro}, van Weeren, Rudnick, Hoeft,
  Br{\"{u}}ggen, Ryu, R{\"{o}}ttgering, Forman, Stroe, Shimwell, Kraft, Jones,
  \& Hoang}]{DiGennaro2021}
{Di Gennaro}, G., van Weeren, R.~J., Rudnick, L., {et~al.} 2021, Astrophys. J.,
  911, 3

\bibitem[{Dwarakanath {et~al.}(2018)Dwarakanath, Parekh, Kale, \&
  George}]{Dwarakanath2018}
Dwarakanath, K.~S., Parekh, V., Kale, R., \& George, L.~T. 2018, Mon. Not. R.
  Astron. Soc., 477, 957

\bibitem[{Ensslin {et~al.}(1997)Ensslin, Biermann, Klein, \&
  Kohle}]{EnBlin1998}
Ensslin, T.~A., Biermann, P.~L., Klein, U., \& Kohle, S. 1997, A{\&}A, 409, 395

\bibitem[{Fanti {et~al.}(1982)Fanti, Fanti, Feretti, Ficarra, Gioia,
  Giovannini, Gregorini, Mantovani, Marano, Padrielli, Parma, Tomasi, \&
  Vettolani}]{Fanti1982}
Fanti, C., Fanti, R., Feretti, L., {et~al.} 1982, $\backslash$aap, 105, 200

\bibitem[{Feretti {et~al.}(2012)Feretti, Giovannini, Govoni, \&
  Murgia}]{Feretti2012a}
Feretti, L., Giovannini, G., Govoni, F., \& Murgia, M. 2012, {Clusters of
  galaxies: Observational properties of the diffuse radio emission}

\bibitem[{Gal {et~al.}(2000)Gal, de~Carvalho, Brunner, Odewahn, \&
  Djorgovski}]{Gal2000}
Gal, R.~R., de~Carvalho, R.~R., Brunner, R., Odewahn, S.~C., \& Djorgovski,
  S.~G. 2000, Astron. J., 120, 540

\bibitem[{Ghirardini {et~al.}(2021)Ghirardini, Bulbul, Hoang, Klein, Okabe,
  Biffi, Br{\"{u}}ggen, Ramos-Ceja, Comparat, Oguri, Shimwell, Basu, Bonafede,
  Botteon, Brunetti, Cassano, de~Gasperin, Dennerl, Gatuzz, Gastaldello,
  Intema, Merloni, Nandra, Pacaud, Predehl, Reiprich, Robrade,
  R{\"{o}}ttgering, Sanders, van Weeren, \& Williams}]{Ghirardini2021a}
Ghirardini, V., Bulbul, E., Hoang, D.~N., {et~al.} 2021, Astron. Astrophys.,
  647, A4

\bibitem[{Giacintucci {et~al.}(2005)Giacintucci, Venturi, Brunetti, Bardelli,
  Dallacasa, Ettori, Finoguenov, Rao, \& Zucca}]{Giacintucci2005}
Giacintucci, S., Venturi, T., Brunetti, G., {et~al.} 2005, Astron. Astrophys.,
  440, 867

\bibitem[{Giovannini {et~al.}(2009)Giovannini, Bonafede, Feretti, Govoni,
  Murgia, Ferrari, \& Monti}]{Giovannini2009}
Giovannini, G., Bonafede, A., Feretti, L., {et~al.} 2009, Astron. Astrophys.,
  507, 1257

\bibitem[{Giovannini {et~al.}(2011)Giovannini, Feretti, Girardi, Govoni,
  Murgia, Vacca, \& Bagchi}]{Giovannini2011}
Giovannini, G., Feretti, L., Girardi, M., {et~al.} 2011, Astron. Astrophys.,
  530, 5

\bibitem[{Govoni {et~al.}(2012)Govoni, Ferrari, Feretti, Vacca, Murgia,
  Giovannini, Perley, \& Benoist}]{Govoni2012}
Govoni, F., Ferrari, C., Feretti, L., {et~al.} 2012, Astron. Astrophys., 545,
  A74

\bibitem[{Hardcastle {et~al.}(2021)Hardcastle, Shimwell, Tasse, Best, Drabent,
  Jarvis, Prandoni, R{\"{o}}ttgering, Sabater, \& Schwarz}]{Hardcastle2021}
Hardcastle, M.~J., Shimwell, T.~W., Tasse, C., {et~al.} 2021, Astron.
  Astrophys., 648, 1

\bibitem[{Hlavacek-Larrondo {et~al.}(2018)Hlavacek-Larrondo, Gendron-Marsolais,
  Fecteau-Beaucage, van Weeren, Russell, Edge, Olamaie, Rumsey, King, Fabian,
  McNamara, Hogan, Mezcua, \& Taylor}]{Hlavacek-Larrondo2017}
Hlavacek-Larrondo, J., Gendron-Marsolais, M.-L.~L., Fecteau-Beaucage, D.,
  {et~al.} 2018, Mon. Not. R. Astron. Soc., 475, 2743

\bibitem[{Hoang {et~al.}(2021)Hoang, Shimwell, Osinga, Bonafede, Br{\"{u}}ggen,
  Botteon, Brunetti, Cassano, Cuciti, Drabent, Jones, R{\"{o}}ttgering, \& {Van
  Weeren}}]{Hoang2021a}
Hoang, D.~N., Shimwell, T.~W., Osinga, E., {et~al.} 2021, Mon. Not. R. Astron.
  Soc., 501, 576

\bibitem[{Hoang {et~al.}(2018)Hoang, Shimwell, van Weeren, Intema,
  R{\"{o}}ttgering, Andrade-Santos, Akamatsu, Bonafede, Brunetti, Dawson,
  Golovich, Best, Botteon, Br{\"{u}}ggen, Cassano, de~Gasperin, Hoeft, Stroe,
  \& White}]{Hoang2018}
Hoang, D.~N., Shimwell, T.~W., van Weeren, R.~J., {et~al.} 2018, Mon. Not. R.
  Astron. Soc., 478, 2218

\bibitem[{Hoang {et~al.}(2019)Hoang, Shimwell, van Weeren, R{\"{o}}ttgering,
  Botteon, Brunetti, Br{\"{u}}ggen, Cassano, Hlavacek-Larrondo,
  Gendron-Marsolais, \& Stroe}]{Hoang2019b}
Hoang, D.~N., Shimwell, T.~W., van Weeren, R.~J., {et~al.} 2019, Astron.
  Astrophys., 622, A21

\bibitem[{Huber(1981)}]{Huber1981}
Huber, P.~J. 1981, {Robust Statistics}, Wiley Series in Probability and
  Statistics (Hoboken, NJ, USA: John Wiley {\&} Sons, Inc.)

\bibitem[{Jones {et~al.}(2021)Jones, de~Gasperin, Cuciti, Hoang, Botteon,
  Br{\"{u}}ggen, Brunetti, Finner, Forman, Jones, Kraft, Shimwell, \& van
  Weeren}]{Jones2021}
Jones, A., de~Gasperin, F., Cuciti, V., {et~al.} 2021, Mon. Not. R. Astron.
  Soc., 505, 4762

\bibitem[{Jones \& Forman(1999)}]{Jones1999}
Jones, C. \& Forman, W. 1999, Astrophys. J., 511, 65

\bibitem[{Kale {et~al.}(2013)Kale, Venturi, Giacintucci, Dallacasa, Cassano,
  Brunetti, Macario, \& Athreya}]{Kale2013}
Kale, R., Venturi, T., Giacintucci, S., {et~al.} 2013, Astron. Astrophys., 557,
  A99

\bibitem[{Kang \& Ryu(2011)}]{Kang2011a}
Kang, H. \& Ryu, D. 2011, ApJ, 734, 18

\bibitem[{Kang {et~al.}(2012)Kang, Ryu, \& Jones}]{Kang2012}
Kang, H., Ryu, D., \& Jones, T.~W. 2012, Astrophys. J., 756, 97

\bibitem[{Kempner \& Sarazin(2001)}]{Kempner2001}
Kempner, J.~C. \& Sarazin, C.~L. 2001, Astrophys. J., 548, 639

\bibitem[{Ledlow {et~al.}(2003)Ledlow, Voges, Owen, \& Burns}]{Ledlow2003}
Ledlow, M.~J., Voges, W., Owen, F.~N., \& Burns, J.~O. 2003, Astron. J., 126,
  2740

\bibitem[{Liang {et~al.}(2000)Liang, Hunstead, Birkinshaw, \&
  Andreani}]{Liang2000}
Liang, H., Hunstead, R.~W., Birkinshaw, M., \& Andreani, P. 2000, Astrophys.
  J., 544, 686

\bibitem[{Lukic {et~al.}(2019)Lukic, Br{\"{u}}ggen, Mingo, Croston, Kasieczka,
  \& Best}]{Lukic2019}
Lukic, V., Br{\"{u}}ggen, M., Mingo, B., {et~al.} 2019, Mon. Not. R. Astron.
  Soc., 487, 1729

\bibitem[{Markevitch {et~al.}(2005)Markevitch, Govoni, Brunetti, \&
  Jerius}]{Markevitch2005}
Markevitch, M., Govoni, F., Brunetti, G., \& Jerius, D. 2005, ApJ, 627, 733

\bibitem[{Murgia {et~al.}(2009)Murgia, Govoni, Markevitch, Feretti, Giovannini,
  Taylor, \& Carretti}]{Murgia2009}
Murgia, M., Govoni, F., Markevitch, M., {et~al.} 2009, A{\&}A, 499, 679

\bibitem[{Osinga {et~al.}(2021)Osinga, {Van Weeren}, Boxelaar, Brunetti,
  Botteon, Br{\"{u}}ggen, Shimwell, Bonafede, Best, Bonato, Cassano,
  Gastaldello, {Di Gennaro}, Hardcastle, Mandal, Rossetti, R{\"{o}}ttgering,
  Sabater, \& Tasse}]{Osinga2021}
Osinga, E., {Van Weeren}, R.~J., Boxelaar, J.~M., {et~al.} 2021, Astron.
  Astrophys., 648, 1

\bibitem[{Pasini {et~al.}(2021)Pasini, Br{\"{u}}ggen, Hoang, Ghirardini,
  Bulbul, Klein, Liu, Shimwell, Hardcastle, Williams, Botteon, Gastaldello, van
  Weeren, Merloni, de~Gasperin, Bahar, Pacaud, \& Ramos-Ceja}]{Pasini2021}
Pasini, T., Br{\"{u}}ggen, M., Hoang, D.~N., {et~al.} 2021, Astron. Astrophys.,
  1

\bibitem[{Pearce {et~al.}(2017)Pearce, van Weeren, Andrade-Santos, Jones,
  Forman, Br{\"{u}}ggen, Bulbul, Clarke, Kraft, Medezinski, Mroczkowski,
  Nonino, Nulsen, Randall, Umetsu, Weeren, Jones, Forman, Bulbul, Clarke,
  Kraft, Medezinski, Mroczkowski, Nonino, Nulsen, Randall, Umetsu, van Weeren,
  Andrade-Santos, Jones, Forman, Br{\"{u}}ggen, Bulbul, Clarke, Kraft,
  Medezinski, Mroczkowski, Nonino, Nulsen, Randall, Umetsu, Weeren, Jones,
  Forman, Bulbul, Clarke, Kraft, Medezinski, Mroczkowski, Nonino, Nulsen,
  Randall, \& Umetsu}]{Pearce2017}
Pearce, C. J.~J., van Weeren, R.~J., Andrade-Santos, F., {et~al.} 2017,
  Astrophys. J., 845, 81

\bibitem[{Petrosian(2001)}]{Petrosian2001a}
Petrosian, V. 2001, ApJ, 557, 560

\bibitem[{Piffaretti {et~al.}(2011)Piffaretti, Arnaud, Pratt, Pointecouteau, \&
  Melin}]{Piffaretti2011a}
Piffaretti, R., Arnaud, M., Pratt, G.~W., Pointecouteau, E., \& Melin, J.-B.~B.
  2011, Astron. Astrophys., 534, A109

\bibitem[{Pinzke {et~al.}(2017)Pinzke, Oh, \& Pfrommer}]{Pinzke2017a}
Pinzke, A., Oh, S.~P., \& Pfrommer, C. 2017, Mon. Not. R. Astron. Soc., 465,
  4800

\bibitem[{Rajpurohit {et~al.}(2021)Rajpurohit, Hoeft, Wittor, van Weeren,
  Vazza, Rudnick, Forman, Riseley, Brienza, Bonafede, Rajpurohit,
  Dom{\'{i}}nguez-Fern{\'{a}}ndez, Rajpurohit, Eilek, Bonnassieux,
  Br{\"{u}}ggen, Loi, R{\"{o}}ttgering, Drabent, Locatelli, Botteon, Brunetti,
  \& Clarke}]{Rajpurohit2021b}
Rajpurohit, K., Hoeft, M., Wittor, D., {et~al.} 2021, 1

\bibitem[{Roettiger {et~al.}(1999)Roettiger, Burns, \& Stone}]{Roettiger1999a}
Roettiger, K., Burns, J.~O., \& Stone, J.~M. 1999, Astrophys. J., 518, 603

\bibitem[{Roger {et~al.}(1973)Roger, Costain, \& Bridle}]{Roger1973}
Roger, R.~S., Costain, C.~H., \& Bridle, A.~H. 1973, Astron. J., 78, 1030

\bibitem[{Rudnick \& Lemmerman(2009)}]{Rudnick2009}
Rudnick, L. \& Lemmerman, J.~A. 2009, Astrophys. J., 697, 1341

\bibitem[{Scaife \& Heald(2012)}]{Scaife2012}
Scaife, A. M.~M. \& Heald, G.~H. 2012, Mon. Not. R. Astron. Soc. Lett., 423, 30

\bibitem[{Sen(1968)}]{Sen1968}
Sen, P.~K. 1968, J. Am. Stat. Assoc., 63, 1379

\bibitem[{Shimwell {et~al.}(2022)Shimwell, Hardcastle, Tasse, Best,
  R{\"{o}}ttgering, Williams, Botteon, Drabent, Mechev, Shulevski, van Weeren,
  Bester, Br{\"{u}}ggen, Brunetti, Callingham, Chy{\.{z}}y, Conway, Dijkema,
  Duncan, de~Gasperin, Hale, Haverkorn, Hugo, Jackson, Mevius, Miley, Morabito,
  Morganti, Offringa, Oonk, Rafferty, Sabater, Smith, Schwarz, Smirnov,
  O'Sullivan, Vedantham, White, Albert, Alegre, Asabere, Bacon, Bonafede,
  Bonnassieux, Brienza, Bilicki, Bonato, {Calistro Rivera}, Cassano, Cochrane,
  Croston, Cuciti, Dallacasa, Danezi, Dettmar, {Di Gennaro}, Edler, En{\ss}lin,
  Emig, Franzen, Garc{\'{i}}a-Vergara, Grange, G{\"{u}}rkan, Hajduk, Heald,
  Heesen, Hoang, Hoeft, Horellou, Iacobelli, Jamrozy, Jeli{\'{c}}, Kondapally,
  Kukreti, Kunert-Bajraszewska, Magliocchetti, Mahatma, Ma{\l}ek, Mandal,
  Massaro, Meyer-Zhao, Mingo, Mostert, Nair, Nakoneczny,
  Nikiel-Wroczy{\'{n}}ski, Orr{\'{u}}, Pajdosz-{\'{S}}mierciak, Pasini,
  Prandoni, van Piggelen, Rajpurohit, Retana-Montenegro, Riseley, Rowlinson,
  Saxena, Schrijvers, Sweijen, Siewert, Timmerman, Vaccari, Vink, West,
  Wo{\l}owska, Zhang, \& Zheng}]{Shimwell2022a}
Shimwell, T.~W., Hardcastle, M.~J., Tasse, C., {et~al.} 2022, Astron.
  Astrophys., 659, A1

\bibitem[{Shimwell {et~al.}(2017)Shimwell, R{\"{o}}ttgering, Best, Williams,
  Dijkema, de~Gasperin, Hardcastle, Heald, Hoang, Horneffer, Intema, Mahony,
  Mandal, Mechev, Morabito, Oonk, Rafferty, Retana-Montenegro, Sabater, Tasse,
  van Weeren, Br{\"{u}}ggen, Brunetti, Chy{\.{z}}y, Conway, Haverkorn, Jackson,
  Jarvis, McKean, Miley, Morganti, White, Wise, van Bemmel, Beck, Brienza,
  Bonafede, {Calistro Rivera}, Cassano, Clarke, Cseh, Deller, Drabent, van
  Driel, Engels, Falcke, Ferrari, Fr{\"{o}}hlich, Garrett, Harwood, Heesen,
  Hoeft, Horellou, Israel, Kapi{\'{n}}ska, Kunert-Bajraszewska, McKay, Mohan,
  Orr{\'{u}}, Pizzo, Prandoni, Schwarz, Shulevski, Sipior, Smith, Sridhar,
  Steinmetz, Stroe, Varenius, van~der Werf, Zensus, \& Zwart}]{Shimwell2017}
Shimwell, T.~W., R{\"{o}}ttgering, H. J.~A., Best, P.~N., {et~al.} 2017,
  A{\&}A, 598, A104

\bibitem[{Shimwell {et~al.}(2019)Shimwell, Tasse, Hardcastle, Mechev, Williams,
  Best, R{\"{o}}ttgering, Callingham, Dijkema, de~Gasperin, Hoang, Hugo,
  Mirmont, Oonk, Prandoni, Rafferty, Sabater, Smirnov, van Weeren, White,
  Atemkeng, Bester, Bonnassieux, Br{\"{u}}ggen, Brunetti, Chy{\.{z}}y,
  Cochrane, Conway, Croston, Danezi, Duncan, Haverkorn, Heald, Iacobelli,
  Intema, Jackson, Jamrozy, Jarvis, Lakhoo, Mevius, Miley, Morabito, Morganti,
  Nisbet, Orr{\'{u}}, Perkins, Pizzo, Schrijvers, Smith, Vermeulen, Wise,
  Alegre, Bacon, {Van Bemmel}, Beswick, Bonafede, Botteon, Bourke, Brienza,
  {Calistro Rivera}, Cassano, Clarke, Conselice, Dettmar, Drabent, Dumba, Emig,
  En{\ss}lin, Ferrari, Garrett, G{\'{e}}nova-Santos, Goyal, G{\"{u}}rkan, Hale,
  Harwood, Heesen, Hoeft, Horellou, Jackson, Kokotanekov, Kondapally,
  Kunert-Bajraszewska, Mahatma, Mahony, Mandal, McKean, Merloni, Mingo,
  Miskolczi, Mooney, Nikiel-Wroczy{\'{n}}ski, O'Sullivan, Quinn, Reich,
  Roskowi{\'{n}}ski, Rowlinson, Savini, Saxena, Schwarz, Shulevski, Sridhar,
  Stacey, Urquhart, {Van Der Wiel}, Varenius, Webster, Wilber, Chy, Cochrane,
  Conway, Croston, Danezi, Duncan, Haverkorn, Heald, Iacobelli, Intema,
  Jackson, Jamrozy, Jarvis, Lakhoo, Mevius, Miley, Morabito, Morganti, Nisbet,
  Orr{\'{u}}, Perkins, Pizzo, Schrijvers, Smith, Vermeulen, Wise, Alegre,
  Bacon, {Van Bemmel}, Beswick, Bonafede, Botteon, Bourke, Brienza, {Calistro
  Rivera}, Cassano, Clarke, Conselice, Dettmar, Drabent, Dumba, Emig,
  En{\ss}lin, Ferrari, Garrett, G{\'{e}}nova-Santos, Goyal, G{\"{u}}rkan, Hale,
  Harwood, Heesen, Hoeft, Horellou, Jackson, Kokotanekov, Kondapally,
  Kunert-Bajraszewska, Mahatma, Mahony, Mandal, McKean, Merloni, Mingo,
  Miskolczi, Mooney, Nikiel-Wroczy{\'{n}}ski, O'Sullivan, Quinn, Reich,
  Roskowi{\'{n}}ski, Rowlinson, Savini, Saxena, Schwarz, Shulevski, Sridhar,
  Stacey, Urquhart, {Van Der Wiel}, Varenius, Webster, Wilber, Chy{\.{z}}y,
  Cochrane, Conway, Croston, Danezi, Duncan, Haverkorn, Heald, Iacobelli,
  Intema, Jackson, Jamrozy, Jarvis, Lakhoo, Mevius, Miley, Morabito, Morganti,
  Nisbet, Orr{\'{u}}, Perkins, Pizzo, Schrijvers, Smith, Vermeulen, Wise,
  Alegre, Bacon, {Van Bemmel}, Beswick, Bonafede, Botteon, Bourke, Brienza,
  {Calistro Rivera}, Cassano, Clarke, Conselice, Dettmar, Drabent, Dumba, Emig,
  En{\ss}lin, Ferrari, Garrett, G{\'{e}}nova-Santos, Goyal, G{\"{u}}rkan, Hale,
  Harwood, Heesen, Hoeft, Horellou, Jackson, Kokotanekov, Kondapally,
  Kunert-Bajraszewska, Mahatma, Mahony, Mandal, McKean, Merloni, Mingo,
  Miskolczi, Mooney, Nikiel-Wroczy{\'{n}}ski, O'Sullivan, Quinn, Reich,
  Roskowi{\'{n}}ski, Rowlinson, Savini, Saxena, Schwarz, Shulevski, Sridhar,
  Stacey, Urquhart, {Van Der Wiel}, Varenius, Webster, Wilber, Chy, Cochrane,
  Conway, Croston, Danezi, Duncan, Haverkorn, Heald, Iacobelli, Intema,
  Jackson, Jamrozy, Jarvis, Lakhoo, Mevius, Miley, Morabito, Morganti, Nisbet,
  Orr{\'{u}}, Perkins, Pizzo, Schrijvers, Smith, Vermeulen, Wise, Alegre,
  Bacon, {Van Bemmel}, Beswick, Bonafede, Botteon, Bourke, Brienza, {Calistro
  Rivera}, Cassano, Clarke, Conselice, Dettmar, Drabent, Dumba, Emig,
  En{\ss}lin, Ferrari, Garrett, G{\'{e}}nova-Santos, Goyal, G{\"{u}}rkan, Hale,
  Harwood, Heesen, Hoeft, Horellou, Jackson, Kokotanekov, Kondapally,
  Kunert-Bajraszewska, Mahatma, Mahony, Mandal, McKean, Merloni, Mingo,
  Miskolczi, Mooney, Nikiel-Wroczy{\'{n}}ski, O'Sullivan, Quinn, Reich,
  Roskowi{\'{n}}ski, Rowlinson, Savini, Saxena, Schwarz, Shulevski, Sridhar,
  Stacey, Urquhart, {Van Der Wiel}, Varenius, Webster, Wilber, Others, Hoang,
  Hugo, Mirmont, Oonk, Prandoni, Rafferty, Sabater, Smirnov, van Weeren, White,
  Atemkeng, Bester, Bonnassieux, Br{\"{u}}ggen, Brunetti, Chy{\.{z}}y,
  Cochrane, Conway, Croston, Danezi, Duncan, Haverkorn, Heald, Iacobelli,
  Intema, Jackson, Jamrozy, Jarvis, Lakhoo, Mevius, Miley, Morabito, Morganti,
  Nisbet, Orr{\'{u}}, Perkins, Pizzo, Schrijvers, Smith, Vermeulen, Wise,
  Alegre, Bacon, {Van Bemmel}, Beswick, Bonafede, Botteon, Bourke, Brienza,
  {Calistro Rivera}, Cassano, Clarke, Conselice, Dettmar, Drabent, Dumba, Emig,
  En{\ss}lin, Ferrari, Garrett, G{\'{e}}nova-Santos, Goyal, G{\"{u}}rkan, Hale,
  Harwood, Heesen, Hoeft, Horellou, Jackson, Kokotanekov, Kondapally,
  Kunert-Bajraszewska, Mahatma, Mahony, Mandal, McKean, Merloni, Mingo,
  Miskolczi, Mooney, Nikiel-Wroczy{\'{n}}ski, O'Sullivan, Quinn, Reich,
  Roskowi{\'{n}}ski, Rowlinson, Savini, Saxena, Schwarz, Shulevski, Sridhar,
  Stacey, Urquhart, {Van Der Wiel}, Varenius, Webster, Wilber, Chy, Cochrane,
  Conway, Croston, Danezi, Duncan, Haverkorn, Heald, Iacobelli, Intema,
  Jackson, Jamrozy, Jarvis, Lakhoo, Mevius, Miley, Morabito, Morganti, Nisbet,
  Orr{\'{u}}, Perkins, Pizzo, Schrijvers, Smith, Vermeulen, Wise, Alegre,
  Bacon, {Van Bemmel}, Beswick, Bonafede, Botteon, Bourke, Brienza, {Calistro
  Rivera}, Cassano, Clarke, Conselice, Dettmar, Drabent, Dumba, Emig,
  En{\ss}lin, Ferrari, Garrett, G{\'{e}}nova-Santos, Goyal, G{\"{u}}rkan, Hale,
  Harwood, Heesen, Hoeft, Horellou, Jackson, Kokotanekov, Kondapally,
  Kunert-Bajraszewska, Mahatma, Mahony, Mandal, McKean, Merloni, Mingo,
  Miskolczi, Mooney, Nikiel-Wroczy{\'{n}}ski, O'Sullivan, Quinn, Reich,
  Roskowi{\'{n}}ski, Rowlinson, Savini, Saxena, Schwarz, Shulevski, Sridhar,
  Stacey, Urquhart, {Van Der Wiel}, Varenius, Webster, Wilber, \&
  Others}]{Shimwell2019}
Shimwell, T.~W., Tasse, C., Hardcastle, M.~J., {et~al.} 2019, Astron.
  Astrophys., 622, A1

\bibitem[{Sommer \& Basu(2014)}]{Sommer2014}
Sommer, M.~W. \& Basu, K. 2014, Mon. Not. R. Astron. Soc., 437, 2163

\bibitem[{Strazzullo {et~al.}(2005)Strazzullo, Paolillo, Longo, Puddu,
  Djorgovski, {De Carvalho}, \& Gal}]{Strazzullo2005}
Strazzullo, V., Paolillo, M., Longo, G., {et~al.} 2005, Mon. Not. R. Astron.
  Soc., 359, 191

\bibitem[{Struble \& Rood(1999)}]{Struble1999}
Struble, M.~F. \& Rood, H.~J. 1999, Astrophys. J. Suppl. Ser., 125, 35

\bibitem[{Tarr{\'{i}}o {et~al.}(2019)Tarr{\'{i}}o, Melin, \&
  Arnaud}]{Tarrio2019}
Tarr{\'{i}}o, P., Melin, J.-B., \& Arnaud, M. 2019, Astron. Astrophys., 626, A7

\bibitem[{Tasse {et~al.}(2021)Tasse, Shimwell, Hardcastle, O'Sullivan, van
  Weeren, Best, Bester, Hugo, Smirnov, Sabater, Calistro-Rivera, de~Gasperin,
  Morabito, R{\"{o}}ttgering, Williams, Bonato, Bondi, Botteon, Br{\"{u}}ggen,
  Brunetti, Chyay, Garrett, G{\"{u}}rkan, Jarvis, Kondapally, Mandal, Prandoni,
  Repetti, Retana-Montenegro, Schwarz, Shulevski, Wiaux, O'Sullivan, van
  Weeren, Best, Bester, Hugo, Smirnov, Sabater, Calistro-Rivera, de~Gasperin,
  Morabito, R{\"{o}}ttgering, Williams, Bonato, Bondi, Botteon, Br{\"{u}}ggen,
  Brunetti, Chy{\.{z}}y, Garrett, G{\"{u}}rkan, Jarvis, Kondapally, Mandal,
  Prandoni, Repetti, Retana-Montenegro, Schwarz, Shulevski, \&
  Wiaux}]{Tasse2021}
Tasse, C., Shimwell, T., Hardcastle, M.~J., {et~al.} 2021, Astron. Astrophys.,
  648, 1

\bibitem[{Taylor(2005)}]{Taylor2005}
Taylor, M.~B. 2005, Astron. Data Anal. Softw. Syst. XIV - ASP Conf. Ser., 347,
  29

\bibitem[{van Weeren {et~al.}(2011)van Weeren, Br{\"{u}}ggen, R{\"{o}}ttgering,
  Hoeft, Nuza, Intema, \& van Weeren}]{VanWeeren2011e}
van Weeren, R.~J., Br{\"{u}}ggen, M., R{\"{o}}ttgering, H. J.~A., {et~al.}
  2011, Astron. Astrophys., 533, A35

\bibitem[{van Weeren {et~al.}(2016{\natexlab{a}})van Weeren, Brunetti,
  Br{\"{u}}ggen, Andrade-Santos, Ogrean, Williams, R{\"{o}}ttgering, Dawson,
  Forman, de~Gasperin, Hardcastle, Jones, Miley, Rafferty, Rudnick, Sabater,
  Sarazin, Shimwell, Bonafede, Best, Birzan, Cassano, Chy{\.{z}}y, Croston,
  Dijkema, En{\ss}lin, Ferrari, Heald, Hoeft, Horellou, Jarvis, Kraft, Mevius,
  Intema, Murray, Orr{\'{u}}, Pizzo, Sridhar, Simionescu, Stroe, van~der Tol,
  \& White}]{VanWeeren2016b}
van Weeren, R.~J., Brunetti, G., Br{\"{u}}ggen, M., {et~al.}
  2016{\natexlab{a}}, Astrophys. J., 818, 204

\bibitem[{van Weeren {et~al.}(2019)van Weeren, de~Gasperin, Akamatsu,
  Br{\"{u}}ggen, Feretti, Kang, Stroe, \& Zandanel}]{VanWeeren2019b}
van Weeren, R.~J., de~Gasperin, F., Akamatsu, H., {et~al.} 2019, Space Sci.
  Rev., 215, 16

\bibitem[{van Weeren {et~al.}(2010)van Weeren, Rottgering, Bruggen, \&
  Hoeft}]{VanWeeren2010a}
van Weeren, R.~J., Rottgering, H. J.~A., Bruggen, M., \& Hoeft, M. 2010,
  Science (80-. )., 330, 347

\bibitem[{van Weeren {et~al.}(2021)van Weeren, Shimwell, Botteon, Brunetti,
  Br{\"{u}}ggen, Boxelaar, Cassano, {Di Gennaro}, Andrade-Santos, Bonnassieux,
  Bonafede, Cuciti, Dallacasa, de~Gasperin, Gastaldello, Hardcastle, Hoeft,
  Kraft, Mandal, Rossetti, R{\"{o}}ttgering, Tasse, \& Wilber}]{vanWeeren2021}
van Weeren, R.~J., Shimwell, T.~W., Botteon, A., {et~al.} 2021, Astron.
  Astrophys., 651, A115

\bibitem[{van Weeren {et~al.}(2016{\natexlab{b}})van Weeren, Williams,
  Hardcastle, Shimwell, Rafferty, Sabater, Heald, Sridhar, Dijkema, Brunetti,
  Br{\"{u}}ggen, Andrade-Santos, Ogrean, R{\"{o}}ttgering, Dawson, Forman,
  de~Gasperin, Jones, Miley, Rudnick, Sarazin, Bonafede, Best, Bîrzan,
  Cassano, Chy{\.{z}}y, Croston, Ensslin, Ferrari, Hoeft, Horellou, Jarvis,
  Kraft, Mevius, Intema, Murray, Orr{\'{u}}, Pizzo, Simionescu, Stroe, van~der
  Tol, \& White}]{VanWeeren2016a}
van Weeren, R.~J., Williams, W.~L., Hardcastle, M.~J., {et~al.}
  2016{\natexlab{b}}, Astrophys. J. Suppl. Ser., 223, 2

\bibitem[{Vavilova {et~al.}(2021)Vavilova, Dobrycheva, Vasylenko, Elyiv,
  Melnyk, \& Khramtsov}]{Vavilova2021}
Vavilova, I.~B., Dobrycheva, D.~V., Vasylenko, M.~Y., {et~al.} 2021, Astron.
  Astrophys., 648, A122

\bibitem[{Venturi {et~al.}(2003)Venturi, Bardelli, Dallacasa, Brunetti,
  Giacintucci, Hunstead, \& Morganti}]{Venturi2003}
Venturi, T., Bardelli, S., Dallacasa, D., {et~al.} 2003, Astron. Astrophys.,
  402, 913

\bibitem[{Venturi {et~al.}(2008)Venturi, Giacintucci, Dallacasa, Cassano,
  Brunetti, Bardelli, \& Setti}]{Venturi2008a}
Venturi, T., Giacintucci, S., Dallacasa, D., {et~al.} 2008, A{\&}A, 484, 327

\bibitem[{Wen \& Han(2015)}]{Wen2015}
Wen, Z.~L. \& Han, J.~L. 2015, Astrophys. J., 807, 178

\bibitem[{Wen {et~al.}(2012)Wen, Han, \& Liu}]{Wen2012}
Wen, Z.~L., Han, J.~L., \& Liu, F.~S. 2012, Astrophys. J. Suppl. Ser., 199, 34

\bibitem[{Williams {et~al.}(2016)Williams, van Weeren, R{\"{o}}ttgering, Best,
  Dijkema, de~Gasperin, Hardcastle, Heald, Prandoni, Sabater, Shimwell, Tasse,
  van Bemmel, Br{\"{u}}ggen, Brunetti, Conway, En{\ss}lin, Engels, Falcke,
  Ferrari, Haverkorn, Jackson, Jarvis, Kapi{\'{n}}ska, Mahony, Miley, Morabito,
  Morganti, Orr{\'{u}}, Retana-Montenegro, Sridhar, Toribio, White, Wise, \&
  Zwart}]{Williams2016}
Williams, W.~L., van Weeren, R.~J., R{\"{o}}ttgering, H. J.~A., {et~al.} 2016,
  Mon. Not. R. Astron. Soc., 460, 2385

\bibitem[{Wu {et~al.}(1998)Wu, Fang, \& Xu}]{Wu1998a}
Wu, X.-P., Fang, L.-Z., \& Xu, W. 1998, Astron. Astrophys., 338, 813

\bibitem[{Zandanel {et~al.}(2014)Zandanel, Pfrommer, \& Prada}]{Zandanel2014}
Zandanel, F., Pfrommer, C., \& Prada, F. 2014, MNRAS, 438, 124

\bibitem[{Zou {et~al.}(2021)Zou, Gao, Xu, Zhou, Ma, Zhou, Zhang, Nie, Wang, \&
  Xue}]{Zou2021a}
Zou, H., Gao, J., Xu, X., {et~al.} 2021, Astrophys. J. Suppl. Ser., 253, 56

\end{thebibliography}
%

%
%
%
%

\begin{appendix} 
\section{Halo model fitting}
\label{sec:bestfit}

In Table~\ref{tab:bestfit}, we present the best-fit parameters obtained by the elliptical model for the candidate/detected radio halos. Details of the procedure are described in Sec. \ref{sec:flux}. 

\begin{table*}[t]
	\centering
	\scriptsize
	\caption{Best-fit parameters for the elliptical model.}
		\begin{tabular}{llcccccccc}  
\hline\hline \\
Cluster & $I_0$   & $x_0$  & $y_0$  & $r1$  & $r2$  & $\phi$ & $\chi^2_{\rm red}$\\
  & $[\mu$ Jy arcsec$^{-2}]$  & [Deg] & [Deg] & [kpc] & [kpc] & [Deg] & \\ \hline
Abell~84 & $1.0^{+0.1}_{-0.2}$ & $10.4321^{+0.0050}_{-0.0047}$ & $21.3776^{+0.0036}_{-0.0037}$ & $428^{+40}_{-44}$ & $222^{+17}_{-18}$ & $6^{+4}_{-4}$ & $1.2$ \\
Abell~1330 & $2.1^{+0.4}_{-0.6}$ & $173.6790^{+0.0016}_{-0.0017}$ & $49.4580^{+0.0018}_{-0.0017}$ & $163^{+25}_{-26}$ & $81^{+17}_{-18}$ & $125^{+9}_{-10}$ & $1.2$ \\
MCXC~J0928.6+3747 & $9.9^{+0.2}_{-0.2}$ & $142.1478^{+0.0002}_{-0.0002}$ & $37.7854^{+0.0002}_{-0.0002}$ & $137^{+2}_{-2}$ & $85^{+2}_{-2}$ & $138^{+1}_{-1}$ & $1.7$ \\
MCXC~J0943.1+4659 & $4.1^{+0.1}_{-0.1}$ & $145.7555^{+0.0005}_{-0.0005}$ & $46.9954^{+0.0005}_{-0.0005}$ & $340^{+10}_{-11}$ & $298^{+9}_{-10}$ & $126^{+9}_{-9}$ & $1.2$ \\
PSZRX~G100.21-30.38 & $9.1^{+0.6}_{-0.7}$ & $350.5652^{+0.0005}_{-0.0005}$ & $28.5196^{+0.0005}_{-0.0005}$ & $29^{+13}_{-13}$ & $132^{+10}_{-10}$ & $39^{+8}_{-8}$ & $1.6$ \\
PSZRX~G102.17+48.88 & $10.5^{+0.6}_{-0.6}$ & $223.7549^{+0.0003}_{-0.0003}$ & $62.9857^{+0.0003}_{-0.0003}$ & $100^{+5}_{-6}$ & $67^{+4}_{-4}$ & $55^{+5}_{-6}$ & $1.2$ \\
PSZRX~G195.91+62.83 & $13.5^{+1.0}_{-1.0}$ & $162.0197^{+0.0003}_{-0.0003}$ & $31.6425^{+0.0003}_{-0.0003}$ & $110^{+7}_{-8}$ & $79^{+6}_{-6}$ & $37^{+8}_{-9}$ & $0.7$ \\
WHL~J130503.5+314255 & $2.1^{+0.2}_{-0.2}$ & $196.2459^{+0.0010}_{-0.0010}$ & $31.7092^{+0.0011}_{-0.0012}$ & $252^{+23}_{-26}$ & $212^{+22}_{-23}$ & $98^{+28}_{-35}$ & $0.8$ \\
WHL J165540.4+334422 & $6.1^{+0.3}_{-0.3}$ & $253.8812^{+0.0005}_{-0.0005}$ & $33.7441^{+0.0005}_{-0.0005}$ & $159^{+8}_{-8}$ & $115^{+5}_{-5}$ & $63^{+6}_{-5}$ & $1.3$ \\
\hline \\
		\end{tabular}	
	\label{tab:bestfit}
\end{table*}

\end{appendix}
  	
\end{document}